# Temporal Logic of Composable Distributed Components

JEREMIAH GRIFFIN, University of California, Riverside
MOHSEN LESANI, University of California, Riverside
NARGES SHADAB, University of California, Riverside
XIZHE YIN, University of California, Riverside

Distributed systems are critical to reliable and scalable computing; however, they are complicated in nature and prone to bugs. To modularly manage this complexity, network middleware has been traditionally built in layered stacks of components. We present a novel approach to compositional verification of distributed stacks to verify each component based on only the specification of lower components. We present TLC (Temporal Logic of Components), a novel temporal program logic that offers intuitive inference rules for verification of both safety and liveness properties of functional implementations of distributed components. To support compositional reasoning, we define a novel transformation on the assertion language that lowers the specification of a component to be used as a subcomponent. We prove the soundness of TLC and the lowering transformation with respect to the operational semantics for stacks of distributed components. We successfully apply TLC to compose and verify a stack of fundamental distributed components.

## 1 INTRODUCTION

Distributed systems are the backbone of the modern computing infrastructure. They support the reliable, scalable and responsive execution of Internet services and replicated aviation control systems, and are at the core of crypto-currencies. However, due to their combinatorially large state spaces, and node and network failures, distributed systems are complicated and prone to bugs. Therefore, they repeatedly suffer data and currency loss, and service outages [Guo et al. 2013; Web 2018a,b,c]. Several projects [Dragoi et al. 2016; Hawblitzel et al. 2015; Lesani et al. 2016; Padon et al. 2016; Rahli 2012; Sergey et al. 2017; Wilcox et al. 2015] have been recently successful in verification of various distributed systems. However, they either do not benefit from a *program logic* and carry out verification in the semantic domain [Hawblitzel et al. 2015; Lesani et al. 2016; Wilcox et al. 2015], do not consider *compositional reasoning* [Dragoi et al. 2016; Hawblitzel et al. 2015; Padon et al. 2016; Rahli 2012], or do not verify liveness properties [Sergey et al. 2017].

Both operating systems, and and network middleware have been traditionally built in *layers* [Gu et al. 2016; Peterson and Davie 2003]. Each node hosts a stack of protocol layers and communicates with other nodes by the communication primitives at the bottom layers. This modular approach brings separation of the implementation from the interface. A higher layer only uses the interface and is separate from the implementation of the lower layers. Similarly, modular verification of each layer using only the specification of the lower layers reduces the proof engineering effort and brings scalability to the development of reliable distributed systems. Layers can be verified separately and composed to build verified stacks of distributed systems. Further, a component remains correct if one of its subcomponents is replaced with a new component with the same specification.

This paper presents a novel *framework for compositional specification and verification of distributed system stacks*. A central goal of this project is intuitiveness to urge adoption. Protocol designers and practitioners do reason about their distributed systems. We observe that they often state the properties of a protocol as natural language statements on events, assume the properties of the sub-protocols, and argue about correctness using *intuitive arguments about the temporal precedence of the events* that the protocol and the sub-protocols exchange [Cachin et al. 2011]. They find it more natural to state properties about the past events rather than add ghost state. Similarly, they prove liveness properties by simple reasoning about future events. This observation led us to the following questions: Can we capture the properties in a temporal logic for components? Can we capture the use of the specifications of subcomponents as a sound transformation? Can we formalize the principles used in these intuitive proofs as logical inference rules? Program logics have been traditionally developed as extensions of the classical Floyd-Hoare logic. In the past decade, the community has witnessed increasingly complicated Hoare logics for concurrent programs that can be effectively used by experts. This project takes a distinct approach and strives to keep the formal techniques as close as possible to the practitioner language. It presents a new compositional and temporal approach to verification of stacks of distributed protocols.

We present a *layered programming model* to separately capture *functional implementations* of distributed components. Layers of components communicate through the interface of request and indication *events*: request events are input from the higher layer and output to the lower layers while indication events are input from the lower layers and output to the higher layer. We present a *temporal assertion language* on event traces to specify properties of components. The assertion language can naturally capture *both safety and liveness* properties of components in terms of their incoming requests and outgoing indications. We present a novel program logic called TLC (Temporal Logic of Components) that features *intuitive inference rules* to directly reason about implementations of distributed components. We want to *compositionally* verify each component based on only its own implementation and the specifications of its subcomponents. Thus, we present a novel syntactic transformation to *lower the temporal specifications* of a component to be used as a subcomponent. We present an *operational semantics* for stacks of distributed components where events propagate across layers in a node, nodes communicate via the bottom link layer and nodes may fail. We prove the *soundness* of TLC and the lowering transformation with respect to the operational semantics. We successfully applied the programming model, the lowering transformation and TLC to compose and verify stacks of *fundamental distributed components* including stubborn links, perfect links, best-effort broadcast, uniform reliable broadcast, epoch consensus, epoch change and Paxos consensus. Further, we present our progress in proof *mechanization* towards building certified middleware.

In summary, the main contribution of this paper is the novel program logic TLC and the lowering transformation for compositional verification of both safety and liveness properties of distributed components. More precisely, the paper makes the following contributions: (1) a compositional programming model for distributed stacks (§ 2) and its semantics (§ 6), (2) a temporal assertion language on event traces to specify safety and liveness properties of components (§ 3), and a sound composable verification technique based on lowering specifications (§ 4), (3) a program logic to reason about components (§ 5) and its soundness with respect to the semantics (§ 7) and (4) verification of stacks of fundamental distributed components (the appendix [Appendix 2020] § 5.2) and encoding of the programming model, logic and lowering in Coq to mechanize the proofs (§ 8). We start with an overview.

## 2 OVERVIEW

In this section, we illustrate composable verification with a simple component that uses a subcomponent. We present the specification of the properties of the component and the subcomponent, lower the specification of the subcomponent and then apply TLC to verify a property of the component.

**Component Composition.** We define the type of components Comp as a parametric record that is represented in Fig. 1. A component is parametric for the type of the events at the top and the bottom of the component as depicted in Fig. 2.(a). The events at the top are the interface of the component. They are the input requests of type *IR* and the output indications of type *OI*. A component may have multiple subcomponents. The events at the bottom are the output requests of types $\overline{OR}$ to and the input indications of types $\overline{II}$ from the subcomponents. We use the overline notation to denote multiple instances; for example, we use $\overline{OR}$ to denote multiple output request types, one per subcomponent. A component defines the State type and its initial value per node as the function init. It also defines three handler functions, request, indication and periodic, that are called in response to input request, input indication and periodic events. Periodic events are automatically issued regularly on correct nodes. Nodes may have crash-stop failures. A node is correct if it does not crash. The periodic handlers usually react to certain conditions; for example, when enough acknowledgements are received, an output indication is issued. Each of the three functions get the current node identifier (of type $\mathbb{N}$) and the pre-state of the component (of type State) as parameters. As the next parameter, the request function gets the input request from the higher component and the indication function gets the input indication from one of the subcomponents. The handler functions return the post-state, a list of output requests (to subcomponents) and a list of output indications (to the parent component).

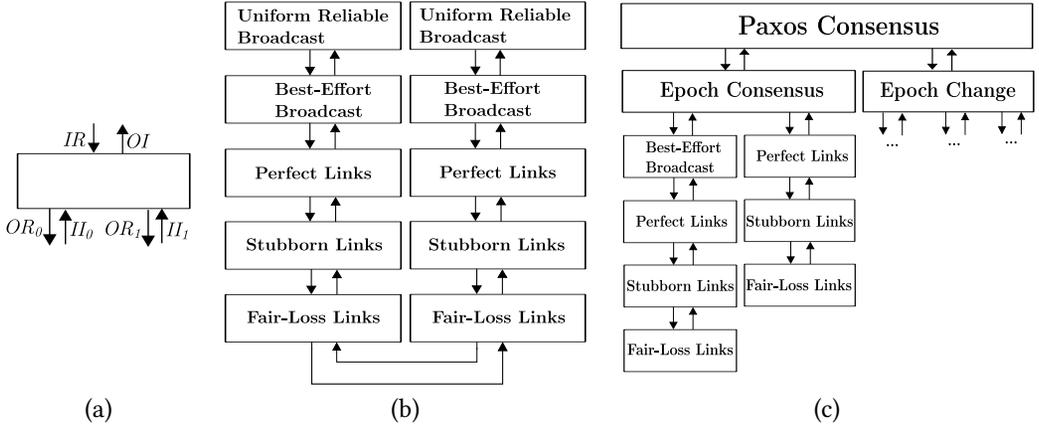

Fig. 2. (a) Events. IR: Input Request, OR: Output request, II: Input Indication, OI: Output Indication (b) Uniform Reliable Broadcast stack at two nodes and (c) Paxos Consensus stack

As Fig. 1 presents, we define the stack of components Stack as an inductive type that is parametrized on the interface of the stack. The interface of a stack is the top events *IR* and *OI* of its top component. A stack is constructed by either a component and its matching substacks (as the inductive case) or is a bottom link (as the base case). Basic links are the weakest components at the leaves of a stack. They accept requests $send_l(n, m)$ of type $Req_l$ to send message *m* to node *n* and issue indications $deliver_l(n, m)$ of type $Ind_l$ to deliver message *m* from node *n*. The semantics of a link can drop messages. However, it does not unfairly drop a particular message that is repeatedly sent. If a sender keeps resending a message and the receiver is not failed, the message is eventually delivered.

As Fig. 2.(b) and (c) show, increasingly stronger components can be built on top of basic links: stubborn links, perfect links, best-effort broadcast, uniform reliable broadcast, epoch consensus, epoch change, and Paxos (uniform) consensus. We have spent extensive effort to write proofs of correctness for these components. The implementation, properties and detailed proofs of all these components are available in the appendix [Appendix 2020] § 4 and 5.3. Fig. 2.(b) shows the stack of the uniform reliable broadcast. Two identical stacks are drawn to show replication at two different

PLC: Component $\text{Req}_{pl}$ $\text{Ind}_{pl}$ ($\text{Req}_{sl}$, $\text{Ind}_{sl}$) :=
$L_1$    let $slc := 0$ in
$L_2$    ⟨State := ⟨counter: Nat,
$L_3$            received: Set[⟨ℕ, Nat⟩]⟩,
$L_4$    init := $\lambda n$. ⟨0, ∅⟩,
$L_5$
$L_6$    request := $\lambda\ n, s, ir$.
$L_7$      let ⟨$c, r$⟩ := $s$ in
$L_8$      match $ir$ with
$L_9$      | $\text{send}_{pl}(n', m) \Rightarrow$
$L_{10}$        let $c' := c + 1$ in
$L_{11}$        let $or := (slc, \text{send}_{sl}(n', \langle c', m\rangle))$ in
$L_{12}$        ⟨⟨$c', r$⟩, [$or$], []⟩,
$L_{13}$      end
$L_{14}$
$L_{15}$    indication := $\lambda\ n', s, ii$.
$L_{16}$      let ⟨$c, r$⟩ =: $s$ in
$L_{17}$      match $ii$ with
$L_{18}$      | $(slc, \text{deliver}_{sl}(n, \langle c', m\rangle)) \Rightarrow$
$L_{19}$        if $(\langle n, c'\rangle \in r)$
$L_{20}$          ⟨$s$, [], []⟩
$L_{21}$        else
$L_{22}$          let $r' := r \cup \{\langle n, c'\rangle\}$ in
$L_{23}$          let $oi := \text{deliver}_{pl}(n, m)$ in
$L_{24}$          ⟨⟨$c, r'$⟩, [], [$oi$]⟩
$L_{25}$      end
$L_{26}$
$L_{27}$    periodic := $\lambda\ n, s$. ⟨$s$, [], []⟩ ⟩
(a)

$\text{SL}_1$ (Stubborn delivery):
$n \in \text{Correct} \land n' \in \text{Correct} \rightarrow$
$(n \bullet \top \downarrow \text{send}_{sl}(n', m)) \Rightarrow \Box\Diamond(n' \bullet \top \uparrow \text{deliver}_{sl}(n, m))$
If a correct node $n$ sends a message $m$ to a correct node $n'$, then $n'$ delivers $m$ infinitely often.

$\text{SL}_2$ (No-forge):
$(n \bullet \top \uparrow \text{deliver}_{sl}(n', m)) \leftsquigarrow (n' \bullet \top \downarrow \text{send}_{sl}(n, m))$
If a node $n$ delivers a message $m$ with sender $n'$, then $m$ was previously sent to $n$ by $n'$.
(b)

$\text{PL}_1$ (Reliable delivery):
$n \in \text{Correct} \land n' \in \text{Correct} \rightarrow$
$(n \bullet \top \downarrow \text{send}_{pl}(n', m)) \rightsquigarrow (n' \bullet \top \uparrow \text{deliver}_{pl}(n, m))$
If a correct node $n$ sends a message $m$ to a correct node $n'$, then $n'$ will eventually deliver $m$.

$\text{PL}_2$ (No-duplication):
$[n' \bullet \top \downarrow \text{send}_{pl}(n, m) \Rightarrow$
   $\hat{\boxminus}\neg(n' \bullet \top \downarrow \text{send}_{pl}(n, m))] \rightarrow$
$[n \bullet \top \uparrow \text{deliver}_{pl}(n', m) \Rightarrow$
   $\hat{\boxminus}\neg(n \bullet \top \uparrow \text{deliver}_{pl}(n', m))]$
If a message is sent at most once, it will be delivered at most once.

$\text{PL}_3$ (No-forge):
$(n \bullet \top \uparrow \text{deliver}_{pl}(n', m)) \leftsquigarrow (n' \bullet \top \downarrow \text{send}_{pl}(n, m))$
If a node $n$ delivers a message $m$ with sender $n'$, then $m$ was previously sent to $n$ by node $n'$.
(c)

Fig. 3. (a) Perfect Link Component PLC. (b) Stubborn Links Specification. (c) Perfect Links Specification. It is notable that $(p \Rightarrow \hat{\boxminus}\neg p) \rightarrow (p \Rightarrow \hat{\Box}\neg p)$; hence the $\hat{\Box}$ conjunct is omitted in the no-duplication property.

nodes. The bottom horizontal lines show the low-level message passing by the basic link. Fig. 2.(c) shows the Paxos consensus [Lamport 1998] stack. The Paxos consensus component is at the top and uses epoch change and epoch consensus as its two subcomponents. In the epoch consensus component, a leader tries to impose a value to the correct nodes. The epoch change component initiates the next epoch with a new leader if the current one fails. The two subcomponents are horizontally composed and epoch consensus is vertically composed on top of them.

The stubborn link repeatedly resends messages by the basic link so that they are eventually delivered. However, retransmission results in multiple deliveries that may not be desired by the higher-level components. Thus, the perfect link component is built on top of the stubborn link to eliminate duplicate messages. It keeps track of delivered messages and ignores duplicates. Fig. 3.(a) presents the perfect link component, PLC. It provides the perfect link interface and uses a substack with the stubborn link interface. The state of each node stores the number of messages sent by the current node, counter, initialized to zero and the set of received message identifiers, received, initialized to empty (at $L_2$-$L_4$). The counter is used to assign unique numbers to messages that the node sends. Each message can be uniquely identified by the pair of the sender node identifier and the number of the message in that node. Upon a request to send a message (at $L_6$-$L_9$), the counter is incremented (at $L_{10}$) and the message is sent together with the new counter value using the stubborn link subcomponent (at $L_{11}$-$L_{12}$). Upon a delivery indication of a message from the stubborn link subcomponent (at $L_{15}$-$L_{18}$), if the message is already received, it is ignored (at $L_{19}$-$L_{20}$). Otherwise, the message identifier is added to the received set and a delivery indication event is issued (at $L_{21}$-$L_{24}$). PLC does not need a periodic handler (at $L_{27}$).

**Semantics.** In § 6, we define the operational semantics of distributed components. It models the propagation of events across the stack, message passing across nodes in partially synchronous networks and node failures. Here, we illustrate the structure of a stack and a fragment of a round of a trace in Fig. 4. Components are represented as boxes and the orientation $o$ of request, periodic and indication events are shown as $\downarrow$, $\natural$ and $\uparrow$ respectively. Incoming events are executed on the component itself and outgoing events are issued to be executed on other components. The distinct location identifier $d$ of a component in the tree of a distributed stack is the *reverse* list of branch indices from the top component to that component. Going down and up the tree simply corresponds to adding and removing a subcomponent index at the head of this list. For example, the identifier $d$ for the top component $C_1$ is $[\,]$, for its left child $C_2$ is $[0]$ and for its right grandchild $C_5$ is $[1,0]$. The interface of each component is the events immediately above it and they share its identifier. For example, the identifier $d$ of the right child $C_3$ and its interface events are both $[1]$. Similarly, the location identifier of a substack is the location identifier of its top component. For example, the left substack rooted at $C_2$ is at location $[0]$. A simple trace is shown on the right of Fig. 4 where the lines show a sequence of events from left to right at different interface levels. The execution of an event at a component updates the state of the component and may issue other request and indication events. The issued events are subsequently executed. The trace starts with $e_1$, a request $\downarrow$ at the top $[\,]$ from the client. When $e_1$ is processed in the top component $C_1$ at $[\,]$, $e_2$, a request $\downarrow$ on the left child component $C_2$ at $[0]$ is issued. When $e_2$ is processed in $C_2$, in turn, $e_3$, a request $\downarrow$ to its right child $C_5$ at $[1,0]$, is issued. Processing of $e_3$ on $C_5$ issues $e_4$, an indication $\uparrow$ event at $[1,0]$. When $e_4$ is processed in the parent component $C_2$ at $[0]$, $e_5$ that is an indication $\uparrow$ at $[0]$, is issued. (We note that $C_2$ could instead issue another request like $e_3$ to one of its children.) When $e_5$ is executed at the parent component $C_1$ at $[\,]$, finally, $e_6$ that is an indication $\uparrow$ at the top level $[\,]$, is issued (that is executed in the client). An (infinite) sequence of event labels is an execution trace. Given a stack, the semantics defines its set of execution traces.

$$
\begin{aligned}
&\text{Comp } IR\ OI\ \overline{(OR, II)} := \\
&\quad \langle\, \text{State: Type,} \\
&\qquad \text{init: } \mathbb{N} \to \text{State,} \\
&\qquad \text{let Out} = \text{State} \times \text{List } (\Sigma\ \overline{OR}) \times \text{List } OI \text{ in} \\
&\qquad \text{request: } \mathbb{N} \to \text{State} \to IR \to \text{Out,} \\
&\qquad \text{indication: } \mathbb{N} \to \text{State} \to \Sigma\ \overline{II} \to \text{Out,} \\
&\qquad \text{periodic: } \mathbb{N} \to \text{State} \to \text{Out}\, \rangle
\end{aligned}
$$

$$
\begin{aligned}
&\text{Stack: Type} \to \text{Type} \to \text{Type} := \\
&\mid \text{stack: Comp } IR\ OI\ \overline{(OR, II)} \to \\
&\qquad \prod\,(\overline{\text{Stack } OR\ II}) \to \\
&\qquad \text{Stack } IR\ OI \\
&\mid \text{link: Stack Req}_l\ \text{Ind}_l
\end{aligned}
$$

Fig. 1. Component and Stack. Let $\Sigma$ and $\Pi$ be parametric sum and product types. In component descriptions, we write a sum term constructed from a term $t$ of the $i$-th type parameter as $(i, t)$ for brevity.

**Assertion Language.** To represent the specifications of distributed component stacks, we define a temporal assertion language that can describe traces of events across the stack. It features specific variables for the properties of the handler calls and a location variable to distinguish the unique places of the composed components in the stack. It can concisely capture safety and liveness properties of distributed components. In § 3, we will describe the assertion language and here, briefly describe the parts that we use in the overview. The always assertion $\Box \mathcal{A}$ states that the assertion $\mathcal{A}$ holds at every event in the future including the current event. The always in the past assertion $\boxminus \mathcal{A}$ states that $\mathcal{A}$ holds at every event in the past including the current event. The eventually assertion $\Diamond \mathcal{A}$ states that $\mathcal{A}$ holds at some future event. The eventually in the past assertion $\Diamondminus \mathcal{A}$ states that $\mathcal{A}$ holds at some past event. The strict versions $\hat{\Box}$, $\hat{\Boxminus}$, $\hat{\Diamond}$ and $\hat{\Diamondminus}$ exclude the current event. The strong implication $\mathcal{A} \Rightarrow \mathcal{A}'$ is syntactic sugar for $\Box(\mathcal{A} \to \mathcal{A}')$ where $\to$ is the logical implication. The leads-to assertion $\mathcal{A} \leadsto \mathcal{A}'$ is syntactic sugar for $\Box(\mathcal{A} \to \Diamond \mathcal{A}')$. Similarly, the preceded-by assertion $\mathcal{A} \leftsquigarrow \mathcal{A}'$ is syntactic sugar for $\Box(\mathcal{A} \to \Diamondminus \mathcal{A}')$. An assertion is non-temporal if it does not include any temporal operators.

The user events are the event objects that the protocol handlers take as argument and issue, for example $\mathsf{send}_{\mathsf{pl}}(n,m)$. A trace event represents the execution of a user event by a handler. The assertion language can describe event traces across the stack. Variables are partitioned into rigid and flexible variables. A rigid variable has the same value in all events of an execution, while a flexible variable may assume dif-ferent values in different events. We represent the flexible variables with the bold face. The flexible variables for an event are the identifier **n** of the node that executes the event, the round number **r** that executes the event, location identifier **d** that the event is executed at, the orientation **o** of the event, the user event **e** that is processed, the output requests **ors** and output indications **ois** that the event issues, the pre-state **s** of the event, and the post-state **s'** of the event. The pre and the post-state represent functions from node identifiers to the state of the component at the nodes. We call the top component self as it is the top component that is verified assuming the correctness of the subcomponents. We use the syntactic sugar assertion self to describe events that are applied to the top component. A self event is either a request or periodic event at the top or an indication event from a subcomponent at the second level. The constants Correct represents the set of correct node identifiers.

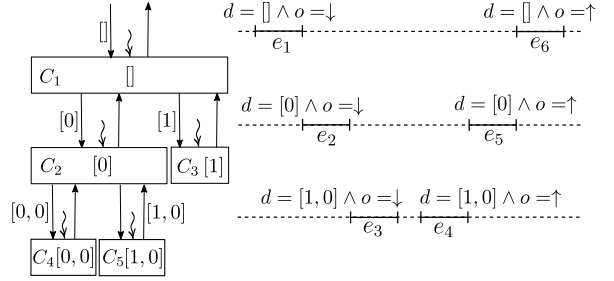

Fig. 4. Semantics of Component Stacks.

The syntactic sugar assertion $n \bullet \mathcal{A}$ (where $\bullet$ is used as a separator) is sugar for $\mathbf{n} = n \wedge \mathcal{A}$; it describes an event that is executed at node $n$ and satisfies $\mathcal{A}$. The syntactic sugar assertion $\top o\, e$ stands for $\mathbf{d} = [\,] \wedge \mathbf{o} = o \wedge \mathbf{e} = e$; it describes an event that is at the top ($\top$) level interface $[\,]$, its orientation is $o$ (either the constant $\downarrow$ for requests, $\wr$ for periodics or $\uparrow$ for indications) and its user event is $e$. For example, the assertion $(n \bullet \top \downarrow \mathsf{send}_{\mathsf{sl}}(n', m))$ describes an event at node $n$ at the top level interface $[\,]$ where the request ($\downarrow$) event $\mathsf{send}_{\mathsf{sl}}(n', m)$ is executed. As the periodic handler is not called with a user event, we use the constant per to represent periodic user events. Similarly, the syntactic sugar assertion $i\, o\, e$ stands for $\mathbf{d} = [i] \wedge \mathbf{o} = o \wedge \mathbf{e} = e$; it describes an event that is at the interface of the $i$-th subcomponent at location $[i]$, its orientation is $o$ and its user event is $e$. For example, the assertion $(n \bullet 1 \uparrow \mathsf{deliver}_{\mathsf{sl}}(n', m))$ describes an event at node $n$ at the interface location $[1]$ where the indication ($\uparrow$) event $\mathsf{deliver}_{\mathsf{sl}}(n', m)$ is executed. It is notable that as a pleasant result of compositional reasoning, we only need to refer to the events at the top and events at the second level. Therefore, we defined syntactic sugar for only the first two levels.

**Specifications.** Fig. 3 shows the specification of stubborn links and perfect links that are written almost verbatim from their natural language descriptions. A stubborn link stubbornly retransmits messages. The stubborn delivery property $\mathsf{SL}_1$ states that once a message is sent, it is delivered infinitely often. The no-forge property $\mathsf{SL}_2$ states that a stubborn link never forges a message. The properties $\mathsf{SL}_1$ and $\mathsf{SL}_2$ are liveness and safety properties respectively. Intuitively, a safety property states that a bad state never happens and a liveness property states that a good state eventually happens. The reliable delivery property $PL_1$ states that perfect links can reliably transmit messages between correct nodes. The no-duplication property $PL_2$ states that perfect links do not redundantly deliver messages. The no-forge property $PL_3$ states that perfect links do not forge messages. The property $PL_1$ is a liveness and the properties $PL_2$ and $PL_3$ are safety properties.

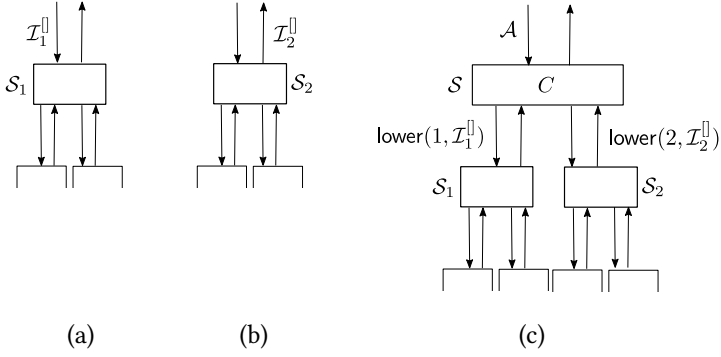

Fig. 5. Illustration of Lowering. (a) Stack $\mathcal{S}_1$ and its specification $\mathcal{I}_1^{[]}$. (b) Stack $\mathcal{S}_2$ and its specification $\mathcal{I}_2^{[]}$. (c) Stack $\mathcal{S}$ that composes $\mathcal{S}_1$ and $\mathcal{S}_2$ as subcomponents. To prove the specification $\mathcal{A}$ of $\mathcal{S}$, the lowering of $\mathcal{I}_1^{[]}$ and $\mathcal{I}_2^{[]}$ can be assumed.

**Lowering Specifications.** We now showcase lowering specifications and the program logic inference rules with the short proof of the no-forge property of the perfect link component PLC.

PLC uses the stubborn link interface and relies on its properties. The specification of the stubborn link in Fig. 3.(b) is stated on its interface as the top-level component but PLC uses it as a subcomponent. When a component is used as a subcomponent, its events are at lower locations and are also interleaved with the events of the parent and sibling components. Given the top-level specification of the stubborn link, how can we transform it to be used as the specification of a subcomponent for the perfect link? Not every assertion can be lowered. In § 4, we present a subset of the assertion language that can be lowered by a syntactic transformation lower.

Fig. 5 illustrates lowering. The specification of each stack $\mathcal{S}_i$ is given as an invariant $\mathcal{I}_i^{[]}$ (Fig. 5.(a) and (b)). Consider that we have a stack $\mathcal{S}$ with the component $c$ at the top and the substacks $\overline{\mathcal{S}_i}$ i.e. $\mathcal{S} = \text{stack}(c, \overline{\mathcal{S}_i})$. We want to verify that $\mathcal{S}$ satisfies its specification $\mathcal{A}$. What can we assume for each subcomponents $\mathcal{S}_i$? (Fig. 5.(c)) We define the translation function lower on invariants and show that to prove the validity of $\mathcal{A}$ for $\mathcal{S}$, it is sufficient to assume $\overline{\text{lower}(i, \mathcal{I}_i^{[]})}$ and derive $\mathcal{A}$ in TLC.

The $SL_2$ assertion is in the invariant sub-language. Applying the lower transformation to $SL_2$ to use it as the 0-th subcomponent results in the following:

$$SL_2' = \text{lower}(0, SL_2) = \text{lower}(0, (n \bullet \top \uparrow \text{deliver}_{sl}(n', m)) \leftsquigarrow (n' \bullet \top \downarrow \text{send}_{sl}(n, m))) = \\ (n \bullet 0 \uparrow \text{deliver}_{sl}(n', m)) \leftsquigarrow (n' \bullet 0 \downarrow \text{send}_{sl}(n, m)) \quad (1)$$

We will see the transformation details later in § 4, but notice here that ⊤ is changed to 0. The lowering transformation can be similarly applied to $SL_1$ to result in $SL_1'$.

The judgements of the logic are of the form $\Gamma \vdash_c \mathcal{A}$ that states that under assumptions $\Gamma$, the assertion $\mathcal{A}$ holds for the component $c$. The two lowered assertions are assumed; thus, we have $\Gamma = SL_1', SL_2'$.

**Program Logic.** We now showcase the program logic using a simple example. The no-forge property of perfect links states that a perfect link delivery event is preceded by a corresponding

OR′
    ⊢$_c$ $n \bullet i \downarrow e \Rightarrow \hat{\diamond}(n \bullet (i,e) \in \mathbf{ors} \wedge \mathsf{self})$

OI′
    ⊢$_c$ $n \bullet \top \uparrow e \Rightarrow \hat{\diamond}(n \bullet e \in \mathbf{ois} \wedge \mathsf{self})$

INVL
    $\forall e.\ \top \downarrow e \wedge \mathsf{request}_c(\mathbf{n},\mathbf{s}(\mathbf{n}),e) = (\mathbf{s}'(\mathbf{n}),\mathbf{ois},\mathbf{ors}) \to \mathcal{A}$
    $\forall e,i.\ i \uparrow e \wedge \mathsf{indication}_c(\mathbf{n},\mathbf{s}(\mathbf{n}),(i,e)) = (\mathbf{s}'(\mathbf{n}),\mathbf{ois},\mathbf{ors}) \to \mathcal{A}$
    $\top \nmid \mathsf{per} \wedge \mathsf{periodic}_c(\mathbf{n},\mathbf{s}(\mathbf{n})) = (\mathbf{s}'(\mathbf{n}),\mathbf{ois},\mathbf{ors}) \to \mathcal{A}$
    $\mathcal{A}$ non-temporal
    ―――――――――――――――
    ⊢$_c$ $\mathsf{self} \Rightarrow \mathcal{A}$

TRANS$\diamond$
$(\mathcal{A} \Rightarrow \diamond\mathcal{A}' \wedge \mathcal{A}' \Rightarrow \mathcal{A}'') \to$
$\mathcal{A} \Rightarrow \diamond\mathcal{A}''$

TRANS$\diamond\diamond$
$(\mathcal{A} \Rightarrow \diamond\mathcal{A}' \wedge \mathcal{A}' \Rightarrow \diamond\mathcal{A}'') \to$
$\mathcal{A} \Rightarrow \diamond\mathcal{A}''$

Fig. 6. Three Selected TLC Inference Rules and Two Basic Temporal Logic Lemmas

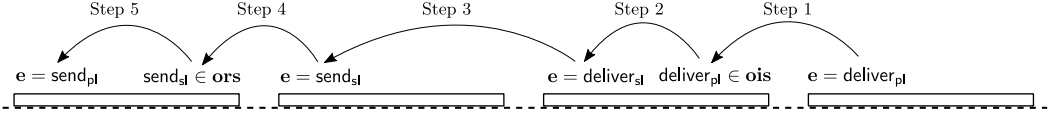

Fig. 7. Illustration of the Proof Steps. The trace is a sequence of events from left to right. Each rectangle represents an event. In a sequence of steps, the proof shows that a perfect link deliver event deliver$_{\mathsf{pl}}$ is preceded by a perfect link send event send$_{\mathsf{pl}}$.

perfect link send event. We want to apply TLC to prove the following judgement that states that assuming Γ, the no-forge property is valid for PLC.

$$\Gamma \vdash_{\mathsf{PLC}} (n \bullet \top \uparrow \mathsf{deliver}_{\mathsf{pl}}(n',m)) \leftsquigarrow (n' \bullet \top \downarrow \mathsf{send}_{\mathsf{pl}}(n,m))$$

At a high-level level, the proof shows a precedence sequence that transitively imply the desired precedence. The proof steps are illustrated in Fig. 7. Step 1: A perfect link delivery event is executed; hence, the event should have been previously issued. Step 2: By the component implementation (in Fig. 3.(a)), a perfect link delivery event is issued by only the indication handler function. Thus, the issuing event is a stubborn link delivery event. Step 3: By the no-forge property of stubborn links, a stubborn link delivery is preceded by a stubborn link send. Step 4: A stubborn link send event is executed before; thus, it should have been previously issued. Step 5: By the component implementation (in Fig. 3.(a)), a stubborn link send event is issued by only the request handler function. Thus, the issuing event is a perfect link send event. By the transitivity of precedence, it is concluded form the above steps that A perfect link delivery is preceded by a perfect link send.

The rules and lemmas that we use for this proof are presented in Fig. 6. We will look at the rules closely in § 5. Here, we use two basic rules: rule OR′ and rule OI′, one derived rule: rule INVL and two basic temporal logic lemmas TRANS$\diamond$ and TRANS$\diamond\diamond$. Intuitively, the two rules OR′ and OI′ state that if an event is executed, it should have been previously issued. The rule OR′ states that if at a node $n$ and the subcomponent $i$, a request ↓ event $e$ is processed, then in the past, at the same node $n$, the request $(i,e)$ is issued by a self event. Similarly, the rule OI′ states that if at a node $n$ and at the top level ⊤, an output indication ↑ event $e$ is processed, then in the past, at the same node $n$, the indication $e$ is issued by a self event. The rule INVL states that if a non-temporal assertion holds for all the three handler functions of the component, request, periodic and indication, then the assertion holds in every self event. It is notable that INVL reduces a temporal global assertion to non-temporal local proof obligations: each premise of this rule is a non-temporal assertion about a single handler function. Thus, the functional implementation of the component can be directly used to infer its properties. The two temporal logic lemmas TRANS$\diamond$ and TRANS$\diamond\diamond$ state basic

temporal transitivity properties. By rule OI′, we have

$$\text{Step 1:} \quad \Gamma \vdash_{\text{PLC}} (n \bullet \top \uparrow \text{deliver}_{\text{pl}}(n',m)) \Rightarrow \Diamond(n \bullet \text{deliver}_{\text{pl}}(n',m) \in \textbf{ois} \land \text{self}) \tag{2}$$

that states that if a perfect link indication is executed, it is previously issued by a self event. We now prove that it is issued only when a stubborn link delivery is executed. We use rule InvL with

$$\mathcal{A} = n \bullet \text{deliver}_{\text{pl}}(n',m) \in \textbf{ois} \to \exists c.\ (n \bullet 0 \uparrow \text{deliver}_{\text{sl}}(n',\langle c,m \rangle)) \tag{3}$$

Considering the implementation in Fig. 3.(a), the two cases for request and periodic are straightforward as **ois** = [] in both and the premise is refuted. The other case is for indication where the stubborn link indication is executed. Thus, by rule InvL (and then reducing two implications to one), we have:

$$\text{Step 2:} \quad \Gamma \vdash_{\text{PLC}} (\text{self} \land n \bullet \text{deliver}_{\text{pl}}(n',m) \in \textbf{ois}) \Rightarrow \exists c.\ (n \bullet 0 \uparrow \text{deliver}_{\text{sl}}(n',\langle c,m \rangle)) \tag{4}$$

By Lemma Trans$\Diamond$ on Eq. 2 and Eq. 4, and existential elimination for $c$, we have

$$\Gamma \vdash_{\text{PLC}} (n \bullet \top \uparrow \text{deliver}_{\text{pl}}(n',m)) \Rightarrow \Diamond(n \bullet 0 \uparrow \text{deliver}_{\text{sl}}(n',\langle c,m \rangle)) \tag{5}$$

that states that the perfect link delivery event is preceded by a stubborn link delivery event.

From $\Gamma$, and Eq. 1 (lowered SL$'_2$), instantiating $m$ with $\langle c,m \rangle$ and unfolding $\leadsto$, we have

$$\text{Step 3:} \quad \Gamma \vdash_{\text{PLC}} (n \bullet 0 \uparrow \text{deliver}_{\text{sl}}(n',\langle c,m \rangle)) \Rightarrow \Diamond(n' \bullet 0 \downarrow \text{send}_{\text{sl}}(n,\langle c,m \rangle)) \tag{6}$$

that is the assumption that every stubborn link delivery event is preceded by a stubborn link send event. By rule OR′, we have

$$\text{Step 4:} \quad \Gamma \vdash_{\text{PLC}} (n' \bullet 0 \downarrow \text{send}_{\text{sl}}(n,\langle c,m \rangle)) \Rightarrow \Diamond(n' \bullet (0,\text{send}_{\text{sl}}(n,\langle c,m \rangle)) \in \textbf{ors} \land \text{self}) \tag{7}$$

that states that every executed stubborn link send event is previously issued by a self event. We use rule InvL again with

$$\mathcal{A} = n' \bullet (0,\text{send}_{\text{sl}}(n,\langle c,m \rangle)) \in \textbf{ors} \to (n' \bullet \top \downarrow \text{send}_{\text{pl}}(n,m))$$

Considering the implementation in Fig. 3.(a), the two cases indication and periodic are straightforward as **ors** = [] in both. The other case is request where the perfect link request is executed. Thus, we have

$$\text{Step 5:} \quad \Gamma \vdash_{\text{PLC}} (\text{self} \land n' \bullet (0,\text{send}_{\text{sl}}(n,\langle c,m \rangle)) \in \textbf{ors}) \Rightarrow (n' \bullet \top \downarrow \text{send}_{\text{pl}}(n,m)) \tag{8}$$

By Lemma Trans$\Diamond$ on Eq. 7 and Eq. 8, we have

$$\Gamma \vdash_{\text{PLC}} (n' \bullet 0 \downarrow \text{send}_{\text{sl}}(n,\langle c,m \rangle)) \Rightarrow \Diamond(n' \bullet \top \downarrow \text{send}_{\text{pl}}(n,m)) \tag{9}$$

From Lemma Trans$\Diamond\Diamond$ on Eq. 5, Eq. 6 and Eq. 9, we have

$$\Gamma \vdash_{\text{PLC}} (n \bullet \top \uparrow \text{deliver}_{\text{pl}}(n',m)) \Rightarrow \Diamond(n' \bullet \top \downarrow \text{send}_{\text{pl}}(n,m))$$

that is

$$\Gamma \vdash_{\text{PLC}} (n \bullet \top \uparrow \text{deliver}_{\text{pl}}(n',m)) \looparrowleft (n' \bullet \top \downarrow \text{send}_{\text{pl}}(n,m))$$

The implementations, specifications and proofs of the other components are available in the appendix [Appendix 2020] § 4 and 5.2. After this overview, we first define the assertion language (§ 3). Next, we define the lowering transformation and prove its soundness for compositional reasoning (§ 4). Then, we present TLC inference rules (§ 5). We finally present the mechanization framework (§ 8).

$$
\begin{aligned}
x &:= & &\text{Variable} \\
&\mid n \mid d \mid o \mid e & &\text{Rigid} \\
&\mid \textit{ors} \mid \textit{ois} \mid s \mid i & & \\
&\mid \mathbf{n} \mid \mathbf{r} \mid \mathbf{d} \mid \mathbf{o} \mid \mathbf{e} & &\text{Flexible} \\
&\mid \mathbf{ors} \mid \mathbf{ois} \mid \mathbf{s} \mid \mathbf{s}' & & \\
c &:= & &\text{Constant} \\
&\mid [] \mid \downarrow \mid \uparrow \mid \text{\textcent} \mid \text{per} & & \\
&\mid \mathbb{N} \mid \text{Correct} & & \\
f &:= + \mid :: \mid .. & &\text{Function} \\
t &:= & &\text{Term} \\
&\mid x \mid c & & \\
&\mid f(t_1,..,t_n) \mid x(t_1,..,t_n) & &\text{Function App} \\
p &:= & &\text{Predicate} \\
&\mid < \mid = \mid \in \mid \subseteq \mid .. & & \\
a &:= p(t_1,..,t_n) & &\text{Atom} \\
\mathcal{A} &:= a & &\text{Assertion} \\
&\mid \mathcal{A} \wedge \mathcal{A} \mid \neg \mathcal{A} & &\text{Proposition} \\
&\mid \forall x.\ \mathcal{A} & &\text{Quantification} \\
&\mid \hat{\Box}\mathcal{A} \mid \hat{\boxminus}\mathcal{A} & &\text{Temporal} \\
&\mid \hat{\Diamond}\mathcal{A} \mid \hat{\Diamondminus}\mathcal{A} \mid \bigcirc \mathcal{A} & &\text{Temporal} \\
&\mid \text{\textcircled{S}}\ \mathcal{A} & &\text{Self Assertion}
\end{aligned}
$$

Assertion for the stack at location $d$:
$\mathcal{A}^d := \mathcal{A}$ such that
$a := (\mathbf{n} = t \wedge \mathbf{d} = d' \wedge \mathbf{o} = t \wedge \mathbf{e} = t) \quad d' \supseteq d$
$\quad \mid t \in \text{Correct}$
and $\bigcirc$ and $\text{\textcircled{S}}$ are not used.

Invariant for the stack at location $d$:
$\mathcal{I}^d := \Box \mathcal{A}^d$

Invariant:
$\mathcal{I} := \Box \mathcal{A}$ such that
$\bigcirc$ and $\text{\textcircled{S}}$ are not used.

Syntactic Sugar:
$$
\begin{aligned}
n \bullet \mathcal{A} &\triangleq \mathbf{n} = n \wedge \mathcal{A} \\
\top o\ e &\triangleq \mathbf{d} = [] \wedge \mathbf{o} = o \wedge \mathbf{e} = e \\
i\ o\ e &\triangleq \mathbf{d} = [i] \wedge \mathbf{o} = o \wedge \mathbf{e} = e \\
\text{self} &\triangleq (\mathbf{d} = [] \wedge \mathbf{o} = \downarrow) \vee \\
& \quad (\mathbf{d} = [] \wedge \mathbf{o} = \text{\textcent}) \vee \\
& \quad (\exists i.\ \mathbf{d} = [i] \wedge \mathbf{o} = \uparrow) \\
\Box \mathcal{A} &\triangleq \mathcal{A} \wedge \hat{\Box}\mathcal{A} \\
\boxminus \mathcal{A} &\triangleq \mathcal{A} \wedge \hat{\boxminus}\mathcal{A} \\
\Diamond \mathcal{A} &\triangleq \mathcal{A} \vee \hat{\Diamond}\mathcal{A} \\
\Diamondminus \mathcal{A} &\triangleq \mathcal{A} \vee \hat{\Diamondminus}\mathcal{A} \\
\mathcal{A} \Rightarrow \mathcal{A}' &\triangleq \Box(\mathcal{A} \rightarrow \mathcal{A}') \\
\mathcal{A} \rightsquigarrow \mathcal{A}' &\triangleq \Box(\mathcal{A} \rightarrow \Diamond \mathcal{A}') \\
\mathcal{A} \leftsquigarrow \mathcal{A}' &\triangleq \Box(\mathcal{A} \rightarrow \Diamondminus \mathcal{A}')
\end{aligned}
$$

Fig. 8. Assertion Language

## 3 ASSERTION LANGUAGE

We now present the assertion language of TLC in Fig. 8. It is a temporal language on event traces of stacks composed of distributed components. It can concisely capture both safety and liveness properties. We have already seen parts of the language in § 2; we consider the rest here.

The constants $\mathbb{N}$ and Correct are the set of participating and correct node identifiers respectively. Similar to classical first-order logic, a term $t$ can be a variable, a constant or an application of a function $f$ to other terms. As the pre-state $\mathbf{s}$ and post-state $\mathbf{s}'$ variables have function values, a term can be constructed by applying a variable to terms as well.

An atomic assertion is an application of a predicate $p$ to terms. All propositional and quantified formula can be constructed as syntactic sugar to conjunction, negation and universal quantification. Similarly, the grammar only shows the strict versions of the temporal operators as the non-strict versions can be defined as syntactic sugar. The temporal operators that we did not introduce in § 2 are $\bigcirc$ and $\text{\textcircled{S}}$. The next operator $\bigcirc$ states that its operand assertion holds in the immediate next event. The self subtrace is the sequence of events executed on the top component. The self operator $\text{\textcircled{S}}$ allows stating assertions about the self subtrace. The assertion $\text{\textcircled{S}}\ \mathcal{A}$ asserts $\mathcal{A}$ on the self subtrace. The self operator is usually used as the outermost operator.

We use $\mathcal{I}^d$ to represent invariant assertions for the substack at location $d$. The specification of a component is written when it is at the top $[]$ as a top-level invariant $\mathcal{I}^{[]}$. An invariant $\mathcal{I}^d$ is of the form $\Box \mathcal{A}^d$. To support lowering, the atomic assertions used in an assertion $\mathcal{A}^d$ are constrained to have the form $\mathbf{n} = t \wedge \mathbf{d} = d' \wedge \mathbf{o} = t' \wedge \mathbf{e} = t''$ such that $d' \supseteq d$. The location variable $\mathbf{d}$ is explicitly equal to an extension $d'$ of $d$ i.e. the event is executed under the substack at location $d$. For example, the event at location $[2,1,0]$ is executed under the 0-th child of the top component, i.e. $[2,1,0] \supseteq [0]$.

DEFINITION 1 (LOWERING ASSERTIONS). $lower(i, \mathcal{I}^{[]})$
$$lower(i, \mathcal{I}^{[]}) \triangleq restrict(\mathbf{d} \supseteq [i], push(i, \mathcal{I}^{[]}))$$

DEFINITION 2 (PUSHING AN ASSERTION).
$push(i, \mathcal{A}^{[]})$:

$push(i, \mathbf{n} = t_1 \land \mathbf{d} = d \land \mathbf{o} = t_2 \land \mathbf{e} = t_3)$
$\triangleq$
$\mathbf{n} = t_1 \land \mathbf{d} = i ::: d \land \mathbf{o} = t_2 \land \mathbf{e} = t_3$

| | | |
|---|---|---|
| $push(i, t \in Correct)$ | $\triangleq$ | $t \in Correct$ |
| $push(i, \mathcal{A}_1^{[]} \land \mathcal{A}_2^{[]})$ | $\triangleq$ | $push(i, \mathcal{A}_1^{[]}) \land push(i, \mathcal{A}_2^{[]})$ |
| $push(i, \neg \mathcal{A}^{[]})$ | $\triangleq$ | $\neg push(i, \mathcal{A}^{[]})$ |
| $push(i, \forall x. \mathcal{A}^{[]})$ | $\triangleq$ | $\forall x. push(i, \mathcal{A}^{[]})$ |
| $push(i, \hat{\Box} \mathcal{A}^{[]})$ | $\triangleq$ | $\hat{\Box} push(i, \mathcal{A}^{[]})$ |
| $push(i, \hat{\boxminus} \mathcal{A}^{[]})$ | $\triangleq$ | $\hat{\boxminus} push(i, \mathcal{A}^{[]})$ |
| $push(i, \hat{\Diamond} \mathcal{A}^{[]})$ | $\triangleq$ | $\hat{\Diamond} push(i, \mathcal{A}^{[]})$ |
| $push(i, \hat{\diamondsuit} \mathcal{A}^{[]})$ | $\triangleq$ | $\hat{\diamondsuit} push(i, \mathcal{A}^{[]})$ |

DEFINITION 3 (RESTRICTING AN ASSERTION).
$restrict(\mathcal{A}', \mathcal{A})$:

| | | |
|---|---|---|
| $restrict(\mathcal{A}', a)$ | $\triangleq$ | $a$ |
| $restrict(\mathcal{A}', \mathcal{A}_1 \land \mathcal{A}_2)$ | $\triangleq$ | $restrict(\mathcal{A}', \mathcal{A}_1) \land restrict(\mathcal{A}', \mathcal{A}_2)$ |
| $restrict(\mathcal{A}', \neg \mathcal{A})$ | $\triangleq$ | $\neg restrict(\mathcal{A}', \mathcal{A})$ |
| $restrict(\mathcal{A}', \forall x. \mathcal{A})$ | $\triangleq$ | $\forall x. restrict(\mathcal{A}', \mathcal{A})$ |
| $restrict(\mathcal{A}', \hat{\Box} \mathcal{A})$ | $\triangleq$ | $\hat{\Box}(\mathcal{A}' \rightarrow restrict(\mathcal{A}', \mathcal{A}))$ |
| $restrict(\mathcal{A}', \hat{\boxminus} \mathcal{A})$ | $\triangleq$ | $\hat{\boxminus}(\mathcal{A}' \rightarrow restrict(\mathcal{A}', \mathcal{A}))$ |
| $restrict(\mathcal{A}', \hat{\Diamond} \mathcal{A})$ | $\triangleq$ | $\hat{\Diamond} restrict(\mathcal{A}', \mathcal{A})$ |
| $restrict(\mathcal{A}', \hat{\diamondsuit} \mathcal{A})$ | $\triangleq$ | $\hat{\diamondsuit} restrict(\mathcal{A}', \mathcal{A})$ |

Fig. 9. Lowering (Pushing and Restricting) Assertions. An atomic assertion is denoted by $a$.

The assertion also includes explicit equalities for the executing node **n**, the orientation **o** and the user-level event **e**. Further, invariant assertions do not use the next ○ or self ⓢ operators.

## 4 SPECIFICATION LOWERING

The presented programming model allows a component to be programmed using subcomponents. The goal of compositional verification is to verify the component using only the specifications (and not the implementations) of the subcomponents. The specification of each component is written when it is the main component at the top of the stack; however, it should be later used as the specification of a subcomponent. A fundamental question is how the specification of a component should be lowered to be used as a subcomponent. The lowered specifications of the subcomponents are used as assumptions to verify the specification of the new parent component (that is programmed on top of the subcomponents). In this section, we define the *lowering* transformation on specification assertions and prove its soundness. Lowering is not possible for every assertion. We observed that lowering specifications requires certain information, such as the location of events, to be present and certain operators, such as next, to be absent from the specification. We identify a subset of the assertion language that is both restrictive enough to allow the definition of the lowering transformation and expressive enough to represent specifications.

As we saw in Fig. 5, the specification of each stack $\mathcal{S}_i$ is given as an invariant $\mathcal{I}_i^{[]}$ (Fig. 5.(a) and (b)). We have a stack $\mathcal{S}$ with the component $c$ at the top and the substacks $\overline{\mathcal{S}_i}$ i.e. $\mathcal{S} = \text{stack}(c, \overline{\mathcal{S}_i})$. We want to verify that $\mathcal{S}$ satisfies its specification $\mathcal{A}$ (Fig. 5.(c)). We define the translation function lower on invariants and show that to prove the validity of $\mathcal{A}$ for $\mathcal{S}$, it is sufficient to assume $\overline{lower(i, \mathcal{I}_i^{[]})}$ and derive $\mathcal{A}$ in TLC. Fig. 9 represents the the function lower on the invariant sub-language $\mathcal{I}^{[]}$. It first *pushes* and then *restricts* the assertion. We visit each in turn.

**Pushing.** For the component at the top, the semantics of stacks models the most general client that may issue any request. However, when the component is used as a subcomponent, the parent component may only issue a subset of the possible requests. Therefore, if a stack is pushed from the top to a lower layer, its set of subtraces can only become smaller (or stay the same). Thus, the

$$\text{SL}_2' = \text{lower}(0, \text{SL}_2) = \tag{10}$$
$$\text{lower}(0, (n \bullet \top \uparrow \text{deliver}_{\text{sl}}(n', m)) \leftsquigarrow (n' \bullet \top \downarrow \text{send}_{\text{sl}}(n, m))) = \tag{11}$$
$$\text{restrict}(\mathbf{d} \supseteq [0], \text{push}(0, \Box[(n \bullet \top \uparrow \text{deliver}_{\text{sl}}(n', m)) \rightarrow \Diamond(n' \bullet \top \downarrow \text{send}_{\text{sl}}(n, m))])) = \tag{12}$$
$$\text{restrict}(\mathbf{d} \supseteq [0], \Box[(n \bullet 0 \uparrow \text{deliver}_{\text{sl}}(n', m)) \rightarrow \Diamond(n' \bullet 0 \downarrow \text{send}_{\text{sl}}(n, m))]) = \tag{13}$$
$$\Box[\mathbf{d} \supseteq [0] \rightarrow (n \bullet 0 \uparrow \text{deliver}_{\text{sl}}(n', m)) \rightarrow \Diamond(n' \bullet 0 \downarrow \text{send}_{\text{sl}}(n, m))] = \tag{14}$$
$$\Box[(n \bullet 0 \uparrow \text{deliver}_{\text{sl}}(n', m)) \rightarrow \Diamond(n' \bullet 0 \downarrow \text{send}_{\text{sl}}(n, m))]] = \tag{15}$$
$$(n \bullet 0 \uparrow \text{deliver}_{\text{sl}}(n', m)) \leftsquigarrow (n' \bullet 0 \downarrow \text{send}_{\text{sl}}(n, m)) \tag{16}$$

Fig. 10. Lowering Example

specification of a stack at the top level can serve as a starting point for its specification as the substack $i$. However, the stack is now used at a deeper level. The location of every event of the stack is now under branch $i$. For example, the location of its highest events is $[\,]$ when it is at the top and is $[i]$ when it is at the $i$-th substack. Similarly, an event at location $[1]$ is pushed to location $[1, i]$. Therefore, the first transformation is to *push* the locations under branch $i$. The function push is defined in Fig. 9. As we saw in the definition of the invariant sub-language $\mathcal{I}^d := \Box \mathcal{A}^d$, the location values $d' \supseteq d$ are explicit in assertions $\mathcal{A}^d$. Given a top-level assertion $\mathcal{A}^{[\,]}$ and a branch index $i$, the function push translates the location value from $d'$ to $i ::: d'$. Appending $i$ to $d'$ effectively pushes the events to branch $i$.

**Restricting.** When a stack is at the top, all events belong to that stack. However, when it is pushed to a substack, its events are interleaved with events from the top component and the sibling substacks. Therefore, the second transformation is to *restrict* the specification to remain valid on traces that are extended with interleaving events. Consider a specification $\Box \mathcal{A}$ for a stack. After pushing the assertion to the $i$-th substack, the resulting assertion $\Box \text{push}(i, \mathcal{A})$ does not necessarily remain valid because although the assertion $\text{push}(i, \mathcal{A})$ is valid on events under branch $i$, it may simply not be valid on events of the top component and the sibling substacks. Thus, the restricting condition of being under branch $i$ should be added and the assertion $\Box \text{push}(i, \mathcal{A})$ is translated to $\Box(\mathbf{d} \supseteq [i] \rightarrow \text{push}(i, \mathcal{A}))$. (The assertion $\text{push}(i, \mathcal{A})$ should be recursively translated as well.) As the definition of the function restrict in Fig. 9 shows, the other variants of the always operator are translated similarly. On the other hand, an eventually assertion $\Diamond \mathcal{A}$ remains the same. If an event will happen in the future, it will still happen if other events are interleaved before it. Similarly, the other variants of the eventually operator remain the same.

As an example, Fig. 10 elaborates lowering of $\text{SL}_2$ that we saw in Eq. 1. The property $\text{SL}_2$ is in the invariant language $\mathcal{I}^{[\,]}$ and can be easily lowered by the syntactic transformation. The steps follow the definition in Fig. 9. We only explain a couple of subtleties. In Eq. 12 - Eq. 13, the push function translates $\mathbf{d} = [\,]$ to $\mathbf{d} = [0]$. Thus, in the syntactic sugar, $\top$ is translated to $0$. In Eq. 14, the syntactic sugar $(n \bullet 0 \uparrow \text{deliver}_{\text{sl}}(n', m))$ includes the conjunct $\mathbf{d} = [0]$. From basic propositional logic, for any $\mathcal{A}$ and $\mathcal{A}'$, we have that $\mathbf{d} \supseteq [0] \rightarrow ((\mathbf{d} = [0] \land \mathcal{A}) \rightarrow \mathcal{A}')$ simplifies to $(\mathbf{d} \supseteq [0] \land \mathbf{d} = [0] \land \mathcal{A}) \rightarrow \mathcal{A}'$. Since $\mathbf{d} = [0]$ is stronger than $\mathbf{d} \supseteq [0]$, it further simplifies to $(\mathbf{d} = [0] \land \mathcal{A}) \rightarrow \mathcal{A}'$.

We note that if the location was not explicit in the assertion, the assertion could not be pushed and would remain too general. For example, consider the assertion $\mathbf{e} = \text{send}(n, m) \Rightarrow m > 0$ for a stack $\mathcal{S}_1$ that states that all the messages that it sends are positive. This assertion is too general for the stack $S$ that composes $S_1$ as a subcomponent because the top component or the sibling subcomponents may send negative messages. Further, if the assertions included the next operator $\bigcirc$, they could not be restricted to remain valid after the trace is interleaved with events of the top component and the sibling subcomponents. For example, consider the assertion $(\mathbf{d} = [\,] \wedge \mathbf{o} = \downarrow) \Rightarrow \bigcirc (\mathbf{d} = [\,] \wedge \mathbf{o} = \uparrow)$ for a stack $\mathcal{S}_1$ that states that every top-level request is immediately followed by a indication. The pushed assertion $(\mathbf{d} = [1] \wedge \mathbf{o} = \downarrow) \Rightarrow \bigcirc (\mathbf{d} = [1] \wedge \mathbf{o} = \uparrow)$ is not valid for the stack $S$ that composes $\mathcal{S}_1$ as a subcomponent because the events of the other components may be interleaved between the request and indication. Similarly, assertions that use the self operator $\textcircled{s}$ cannot be lowered. For example, consider the assertion $\textcircled{s}(\forall n.\ \mathbf{s}(n) > 0)$ for a stack $\mathcal{S}_1$ that states that the state of the top component of $\mathcal{S}_1$ remains positive. Obviously, this assertion does not necessarily hold for the new top component.

**Soundness.** The following theorem states the soundness of the lowering transformation for compositional reasoning. If a top-level invariant $\mathcal{I}_i^{[\,]}$ is valid for the stack $\mathcal{S}_i$ and $\mathcal{S}_i$ is a substack of the stack $\mathcal{S}$, then the lowered invariant $lower(i, \mathcal{I}_i^{[\,]})$ is valid for $\mathcal{S}$. We use the validity judgement $\vDash_\mathcal{S} \mathcal{A}$ that states that $\mathcal{A}$ is valid in every trace of $\mathcal{S}$. (Validity is defined more precisely in § 7.) The detailed proofs are available in the appendix [Appendix 2020] § 5.2.

THEOREM 1. *For all $\mathcal{S}$, $c$, and $\overline{\mathcal{S}_i}$, such that $\mathcal{S} = stack(c, \overline{\mathcal{S}_i})$, if $\vDash_{\mathcal{S}_i} \mathcal{I}_i^{[\,]}$ then $\vDash_\mathcal{S} lower(i, \mathcal{I}_i^{[\,]})$.*

We now state the *compositional proof technique* and its soundness. The specifications of substacks can be lowered and used to derive the specification of the stack. Judgements of TLC are of the form $\Gamma \vdash_c \mathcal{A}$ where $\Gamma$ is the assumed assertions and $\mathcal{A}$ is the deduced assertion. Consider valid top-level invariants $\overline{\mathcal{I}_i^{[\,]}}$ for stacks $\overline{\mathcal{S}_i}$, and a stack $\mathcal{S}$ built by the component $c$ on top of $\overline{\mathcal{S}_i}$. The following theorem states that assuming the lowered invariants $\overline{lower(i, \mathcal{I}_i^{[\,]})}$, any assertion that TLC deduces for $c$ is valid for $\mathcal{S}$.

COROLLARY 1 (COMPOSITION SOUNDNESS). *For all $\mathcal{S}$, $c$, and $\overline{\mathcal{S}_i}$ such that $\mathcal{S} = stack(c, \overline{\mathcal{S}_i})$, if $\overline{\vDash_{\mathcal{S}_i} \mathcal{I}_i^{[\,]}}$ and $\overline{lower(i, \mathcal{I}_i^{[\,]})} \vdash_c \mathcal{A}$ then $\vDash_\mathcal{S} \mathcal{A}$.*

## 5 TLC INFERENCE RULES

In this section, we present the basic and derived inference rules of TLC. The basic inference rules of TLC are intuitive and fit in half a page. Yet, they provide the basis for verification of full stacks such as Fig. 2.(b) and (c). We incrementally captured fundamental reasoning steps required for verification of the use-case protocols as the basic rules. Further, we captured the other common reasoning steps as derived rules. The sequent judgements are of the form $\Gamma \vdash_c \mathcal{A}$ where $c$ is the component, $\Gamma$ is a set of assumed assertions and $\mathcal{A}$ is the deduced assertion. The inference rules axiomatize the properties of the semantics and the low-level communication primitive. More importantly, they allow deducing assertions about execution traces from the functional definition of the component. Fig. 11 presents the basic inference rules of TLC. (We elide the standard rules of sequent logic to the appendix [Appendix 2020] § 1.1). Derived rules present higher-level reasoning steps than basic rules. We present a few derived rules in Fig. 12. A full list of derived rules are available in the appendix [Appendix 2020] § 1.3.

The first three rules IR, II and PE state that when an event is executed on the top component, the correspond handler function of the component is applied. These rules take the reasoning to the functional definitions of the component. (Rule IR): The rule IR (for input request) states that if at the top level $\top$, a request $\downarrow$ event $e$ is executed then the $request_c$ handler function is called. The $request_c$ function of the component $c$ represents a relation between its inputs: the stepping node $\mathbf{n}$, the pre-state $\mathbf{s}(\mathbf{n})$ of $\mathbf{n}$, and the user event $e$, and its outputs: the post-state $\mathbf{s}'(\mathbf{n})$, the issued output requests $\mathbf{ors}$ and the issued output indications $\mathbf{ois}$. The rules II (for input indication) and PE (for periodic) are similar. (Rule II): The rule II states that if an indication $\uparrow$ event $e$ from the $i$-th

IR
$\quad \vdash_c \top \downarrow e \Rightarrow (\mathbf{s}'(\mathbf{n}), \mathbf{ors}, \mathbf{ois}) = \text{request}_c(\mathbf{n}, \mathbf{s}(\mathbf{n}), e)$
II
$\quad \vdash_c i \uparrow e \Rightarrow (\mathbf{s}'(\mathbf{n}), \mathbf{ors}, \mathbf{ois}) = \text{indication}_c(\mathbf{n}, \mathbf{s}(\mathbf{n}), (i, e))$
PE
$\quad \vdash_c \top \text{\textperthousand} \text{ per} \Rightarrow (\mathbf{s}'(\mathbf{n}), \mathbf{ors}, \mathbf{ois}) = \text{periodic}_c(\mathbf{n}, \mathbf{s}(\mathbf{n}))$
OR
$\quad \vdash_c n \bullet (i, e) \in \mathbf{ors} \wedge \text{self} \Rightarrow \hat{\diamond}(n \bullet i \downarrow e)$
OI
$\quad \vdash_c n \bullet e \in \mathbf{ois} \wedge \text{self} \Rightarrow \hat{\diamond}(n \bullet \top \uparrow e)$
OR$'$
$\quad \vdash_c n \bullet i \downarrow e \Rightarrow \hat{\diamond}(n \bullet (i, e) \in \mathbf{ors} \wedge \text{self})$
OI$'$
$\quad \vdash_c n \bullet \top \uparrow e \Rightarrow \hat{\diamond}(n \bullet e \in \mathbf{ois} \wedge \text{self})$
APER
$\quad \vdash_c n \in \text{Correct} \leftrightarrow \square \diamond (n \bullet \top \text{\textperthousand} \text{ per})$
ASELF
$\quad \vdash_c \text{\textcircled{S}} \square \text{ self}$
SINV
$\quad \vdash_c (\text{\textcircled{S}} \mathcal{I}) \leftrightarrow \text{restrict}(\text{self}, \mathcal{I})$
INIT
$\quad \vdash_c \text{\textcircled{S}} (\mathbf{s} = \lambda n. \text{ init}_c(n))$
POSTPRE
$\quad \vdash_c \text{\textcircled{S}} (\mathbf{s}' = s \Leftrightarrow \bigcirc \mathbf{s} = s)$

SEQ
$\quad \vdash_c \mathbf{n} \neq n \Rightarrow \mathbf{s}'(n) = \mathbf{s}(n)$
RSEQ
$\quad \vdash_c \mathbf{r} = r \Rightarrow \hat{\boxminus}(\mathbf{r} \leq r)$
GST
$\quad \vdash_c n' \in \text{Correct} \wedge r \geq r_{GST} \wedge$
$\quad\quad (n \bullet d \downarrow \text{send}_l(n', m) \wedge \mathbf{r} = r) \Rightarrow$
$\quad\quad \diamond(n' \bullet d \uparrow \text{deliver}_l(n, m) \wedge \mathbf{r} = r)$
FDUP
$\quad \vdash_c \square\diamond(n' \bullet d \uparrow \text{deliver}_l(n, m)) \rightarrow$
$\quad\quad \square\diamond(n \bullet d \downarrow \text{send}_l(n', m))$
NFORGE
$\quad \vdash_c (n' \bullet d \uparrow \text{deliver}_l(n, m)) \Rightarrow$
$\quad\quad \diamondsuit(n \bullet d \downarrow \text{send}_l(n', m))$
NODE
$\quad \vdash_c \square \, \mathbf{n} \in \mathbb{N}$
UNIOR
$\quad \vdash_c (\text{occ}(\mathbf{ors}, e) \leq 1 \wedge$
$\quad\quad \hat{\boxminus}(\mathbf{n} = n \wedge \text{self} \rightarrow (i, e) \notin \mathbf{ors}) \wedge$
$\quad\quad \hat{\square}(\mathbf{n} = n \wedge \text{self} \rightarrow (i, e) \notin \mathbf{ors})) \Rightarrow$
$\quad\quad (n \bullet i \downarrow e) \Rightarrow$
$\quad\quad \hat{\boxminus}\neg(n \bullet i \downarrow e) \wedge \hat{\square}\neg(n \bullet i \downarrow e)$

Fig. 11. TLC Basic Inference Rules. We use $\text{request}_c$, $\text{indication}_c$ and $\text{periodic}_c$ to refer to the handler functions of the component $c$. The function call $\text{occ}(l, e)$ counts the number of the element $e$ in the list $l$.

subcomponent is executed then the $\text{indication}_c$ handler function is called. An indication event $e$ from a subcomponent $i$ is passed to the $\text{indication}_c$ function as the sum term $(i, e)$. (Rule PE): The rule PE states that if at the top $\top$, a periodic $\text{\textperthousand}$ event per is executed then the $\text{periodic}_c$ handler function is called.

The next four rules axiomatize the relation of issued and executed events. The rules OR and OI state that an event that is issued for a component is eventually executed on the component, and the rules OR$'$ and OI$'$ state that executed events are previously issued. These rules let the reasoning follow a chain of actions. (Rule OR): An output request from the top component is a downward ($\downarrow$) event $e$ to a subcomponent $i$ that is issued as the sum term $(i, e)$. The rule OR (for output request) states that if at a node $n$, an output request $(i, e)$ is issued by a self event, then eventually at $n$ and the subcomponent $i$, the request $\downarrow$ event $e$ is executed. (Rule OI): Similarly, the rule OI (for output indication) states that if at a node $n$, an output indication $e$ is issued by a self event, then eventually at $n$ and the top level $\top$, the indication $\uparrow$ event $e$ is executed. The rules OR$'$ and OI$'$ state the relation of issued and executed events in the opposite direction of the rules OR and OI. (Rule OR$'$): The rule OR$'$ states that if at a node $n$ and a subcomponent $i$, a request $\downarrow$ event $e$ is executed, then in the past, at that node $n$, the request event for that subcomponent $(i, e)$ is issued by a self event. (Rule OI$'$): Similarly, the rule OI$'$ states that if at a node $n$ and the top level $\top$, an output indication $\uparrow$ event $e$ is executed, then in the past, at that node $n$, that indication event $e$ is issued by a self event. (Rule APER): The rule APER (for always periodic) states that if a node is correct, it infinitely often executes the periodic event.

(Rule ASelf): The self subtrace is the sequence of events executed on the top component. The rule ASelf (for always self) states that every event in the self subtrace is a self event. (Rule SInv): The rule SInv (for self invariant) states that an invariant for the self subtrace can be transformed to an invariant for the whole trace using the restrict function that we defined earlier in Fig. 9. The restriction condition is the self assertion as the invariant continues to hold in the self events. The rules ASelf and SInv lead to the derived rule InvLSe (in Fig. 12). (Rule InvLSe): The rule InvLSe states that if a non-temporal assertion holds for all the three handler functions, request, indication and periodic, then the assertion holds in every self event. We note that this rule reduces a global temporal assertion to local non-temporal proof obligations about the handler functions. To derive the rule InvLSe, the self assertion in the rule ASelf is expanded to the disjunction of three events. Then, for the three cases, the rule SInv is applied to the rules IR, II and PE to derive the same assertions for the self subtrace. Similarly, the rule InvL that we saw in the overview section (in Fig. 6) is in fact derived by applying the rule SInv to the rule InvLSe (in Fig. 12).

(Rule Init): The rule Init states that at the beginning of the execution, the state **s** of every node $n$ is the state defined by the $init_c$ function of the implementation. (Rule PostPre): The rule PostPre states that in the self subtrace, the post-state **s'** of every event is the pre-state **s** of the next event. We note that this assertion does not hold on the whole trace as the events of different components are interleaved. (Rule SEq): The rule SEq (for state equality) states that if the stepping node **n** is not a node $n$, then the state of $n$ stays unchanged ie. its pre-state $\mathbf{s}(n)$ and post-state $\mathbf{s'}(n)$ are equal.

The above four rules derive inductive inference rules for the state of the top component. Let us consider the derived rule InvSSe' in Fig. 12. (Rule InvSSe'): The rule InvSSe' states that if a state predicate $S$ holds in the initial state and all the three handler functions, request, indication and periodic preserve $S$, then $S$ always holds in self events. This rule is used to prove state invariants. Similar to the rule InvLSe, this rule reduces a global temporal assertion to local non-temporal assertions about the handler functions. To derive this rule, the rule Init is used in the base case. For the inductive case, the rule PostPre brings the invariant $S$ on the state from the post-state of an event to the pre-state of the next event. Then, if the node is not stepping, the rule SEq is used. If it is stepping, expanding self in the rule ASelf leads to a case-analysis for the handler functions.

(Rule RSeq): The rule RSeq (for round sequence) states that the round numbers are non-decreasing. Messages are transmitted using basic links at the leaves of the stack. A basic link is a weak communication primitive; it can drop, reorder and duplicate messages. The next three rules GST, FDup and NForge. state the properties of basic links. Stronger communication primitives can be programmed based on these properties. (Rule GST): The rule GST states that after the round $r_{GST}$ if a message is sent to a correct node, it is delivered in the same round. It axiomatizes the partial synchrony of the network and is used to prove liveness properties. In particular, it is used to show that the eventual failure detector component eventually suspects no correct node. The rule GST also derives the rule FLoss (in Fig. 12). (Rule FLoss): The rule FLoss (for fair-loss) states that links are fair in dropping messages in the sense that they do not systematically drop any particular message. If a node sends a message infinitely often and the receiver is correct, then the message is delivered to the receiver infinitely often. The rule FLoss is used to verify stubborn links that are implemented on top of basic links. (Rule FDup): The rule FDup (for finite duplication) states that links duplicate a message only a finite number of times. If the same message is delivered to a node infinitely often, then it is sent infinitely often. (Rule NForge): The rule NForge (for no-forge) states that links do not forge messages. If a message is delivered, it is previously sent.

(Rule UniOR): The rule UniOR (for unique output request) states that if a request is issued at most once, then it is executed at most once. If an output request $e$ is issued at most once at the current event and it is not issued at any other event, then if $e$ is executed at an event, it is never

INVLSE
$\forall e. \ \top \downarrow e \land \text{request}_c(\mathbf{n}, \mathbf{s}(\mathbf{n}), e) = (\mathbf{s}'(\mathbf{n}), \mathbf{ois}, \mathbf{ors}) \to \mathcal{A}$
$\forall e, i. \ i \uparrow e \land \text{indication}_c(\mathbf{n}, \mathbf{s}(\mathbf{n}), (i, e)) = (\mathbf{s}'(\mathbf{n}), \mathbf{ois}, \mathbf{ors}) \to \mathcal{A}$
$\top \updownarrow \text{per} \land \text{periodic}_c(\mathbf{n}, \mathbf{s}(\mathbf{n})) = (\mathbf{s}'(\mathbf{n}), \mathbf{ois}, \mathbf{ors}) \to \mathcal{A}$
$\mathcal{A}$ non-temporal
$\vdash_c \mathcal{S} \ \Box \mathcal{A}$

INVSSE'
$S(\text{init}_c(n))$
$\forall s, e, s'. \ S(s) \land \text{request}_c(n, s, e) = (s', \_, \_) \to S(s')$
$\forall s, i, e, s'. \ S(s) \land \text{indication}_c(n, s, (i, e)) = (s', \_, \_) \to S(s')$
$\forall s, s'. \ S(s) \land \text{periodic}_c(n, s) = (s', \_, \_) \to S(s')$
$\vdash_c \mathcal{S} \ \Box S(\mathbf{s}(n))$

FLOSS
$\vdash_c n' \in \text{Correct} \to \Box\Diamond(n \bullet d \downarrow \text{send}_l(n', m)) \to \Box\Diamond(n' \bullet d \uparrow \text{deliver}_l(n, m))$

QUORUM
$|\text{Correct}| > t_1 \ \vdash_c \ N \subseteq \mathbb{N} \land |N| > t_2 \land t_1 + t_2 \geq |\mathbb{N}| \Rightarrow \exists n. \ n \in N \land n \in \text{Correct}$

Fig. 12. A subset of the TLC Derived Inference Rules. $S$ is a predicate on $\text{State}_c$. $N$ is a set.

executed before or after that event. The rule UNIOI (unique output indication) states a similar fact for indications and is elided.

(Rule NODE): The rule NODE states that the current node $\mathbf{n}$ is always in the set of executing nodes $\mathbb{N}$. This rule is used to reason about the subsets of nodes such as quorums. (Rule QUORUM): The derived rule QUORUM states that assuming that the number of correct nodes is more than $t_1$, if the size of a subset $N$ of nodes (called a quorum) is more than $t_2$ and the sum of $t_1$ and $t_2$ is more than the total number of nodes, then there is at least one correct node in $N$. Usually $t_1$ and $t_2$ are both half the number of nodes $\mathbb{N}$. Intuitively, this rule holds because the two sets are large enough to have at least one common element. Quorums are the basis of many distributed protocols and the QUORUM rule presents an intuitive reasoning principle for them. We illustrate the use of the quorum rules in verification of the uniform reliable broadcast in the appendix [Appendix 2020] § 2.

## 6 DISTRIBUTED STACK SEMANTICS

In this section, we present the semantics of distributed components. It is a novel *operational semantics* that models the interaction of *composed components* in a *partially synchronous* network. The transitions are labeled with traces of events. Thus, the operational semantics leads to a trace semantics for composed components. TLC is proved sound with respect to this semantics in § 7.

Given a stack $\mathcal{S}$ of components, the semantics defines the transitions that nodes deploying $\mathcal{S}$ take. It models propagation and processing of downward request and periodic events, and upward indication events *across layers* of components. It also models the crash-stop failure of nodes, and loss and duplication of messages. After a node crashes, it does not take any steps. A node is called correct if it does not crash.

The semantics models *partially synchronous* networks. In synchronous networks, a fixed upper bound on message delivery time is known. In contrast, no such bound exists for asynchronous networks and many distributed computing abstractions including consensus are impossible in this model. However, most practical networks including the Internet fall in the partially synchronous model. In these networks, either a bound holds but is not known a priori, or a bound eventually holds. Partial synchrony [Dwork et al. 1988] presented the basic round model for partially synchronous networks. Our semantics follows the basic round model. In this model, each round consists of sending, delivering and processing messages. In each round, only a subset of sent messages may be delivered; the rest are lost. However, after a round $r_{GST}$ called the Global Stabilization Time (GST), every message that a correct node sends to another correct node is delivered in the same round. After this round, the network stabilizes and protocols can rely on its synchrony. In practice, the network will eventually remain stable long enough for the protocol to achieve its goal.

The variables used in the operational semantics are defined in Fig. 13. We denote the set of node identifiers by $\mathbb{N}$. As mentioned before, we uniquely identify each component in a stack by the (reverse) list of branch numbers in the path from the top to that component. With the reverse list, moving up and down the tree corresponds to adding and removing an index at the head of the list. We call this sequence the distinct location $d \in \mathcal{D}$ of the component. The substack at location $d$ of a stack $\mathcal{S}$ is denoted by $\mathcal{S}(d)$. The definition of each component declares the component state type $\text{State}_c$. The state of a distributed component $s \in S$ is a mapping from $\mathbb{N}$ to $\text{State}_c$. The state of a distributed stack $\sigma \in \Sigma$ is a heterogeneous map from $\mathcal{D}$ to $S$ types. A message $m$ is a tuple of the sender node, the receiver node, the location of the receiver component and the payload. We use $ms \in \mathcal{M}$ to denote a multi-set of messages. The state of the transition system $w \in \mathcal{W}$ (for world) is a tuple $(\sigma, ms, f, r)$ where $\sigma$ is the state of the distributed stack, $ms$ is the multi-set (or bag) of in-transit messages, $f$ is the set of failed nodes and $r$ is the round number. Given a stack $\mathcal{S}$, the initial state $w_0(\mathcal{S})$ maps the state of every node and location to the state that the init function of the component at that location returns, assigns the empty sets to the initial set of messages $ms$ and failed nodes $f$, and the initial round $r_0$ to $r$. We use $e$ to denote user-level events. The orientation $o$ of an event is either $\downarrow$ for request, $\uparrow$ for indication or $\circlearrowleft$ for periodic events.

| | | |
|---|---|---|
| $n :$ | $\mathbb{N}$ | Node ID |
| $d :$ | $\mathcal{D} = \text{List Nat}$ | Distinct Location |
| $cs :$ | $\text{State}_c$ | Component State |
| $s :$ | $S = \text{Map } \mathbb{N} \text{ State}_c$ | Dist. Comp. State |
| $\sigma :$ | $\Sigma = \text{Map } \mathcal{D} \, S$ | Stack state |
| $m :$ | $M$ | Message Payload |
| $ms :$ | $\mathcal{M} = \text{MultiSet } (\mathbb{N} \times \mathbb{N} \times \mathcal{D} \times M)$ | Messages |
| $f :$ | $\mathbb{N}$ | Failed nodes |
| $r :$ | $\mathcal{R}$ | Round |
| $w :$ | $\mathcal{W} = \Sigma \times \mathcal{M} \times \mathbb{N} \times \mathcal{R}$ | World |
| $w_0(\mathcal{S}) =$ | $\langle (\lambda d, n. \text{ let } (c, \_) = \mathcal{S}(d) \text{ in } \text{init}_c(n)), \emptyset, \emptyset, r_0 \rangle$ | Initial World |
| $e, oi :$ | $E$ | User Event |
| $or :$ | $IE = \text{Nat} \times E$ | Output request |
| $fe :$ | $FE$ | Event or Fail |
| $::=$ | $e \mid \text{fail}$ | |
| $o :$ | $O$ | Orientation |
| $::=$ | $\downarrow \mid \uparrow \mid \circlearrowleft$ | |
| $\ell :$ | $\mathbb{N} \times \mathcal{R} \times \mathcal{D} \times O \times FE \times \Sigma \times \Sigma \times IE \times E$ | Event Label |
| $\tau ::=$ | $\ell^*$ | Trace |

Fig. 13. Operational Semantics Variables

An event label $\ell$ is a record $(n, r, d, o, fe, \sigma, \sigma', or, oi)$ where $n$ is the node, $r$ is the round and $d$ is the location where the event is executed, $o$ is the orientation of the event, $fe$ is either fail or an executed user event $e$, $\sigma$ and $\sigma'$ are the stack pre-state and post-state, $or$ is the issued request event and $oi$ is the issued indication event. To access the fields of a label, we use functions with the same names as fields. An output request event $or$ is a tuple $(i, e)$ of the target subcomponent number $i$ and the user event $e$. (We note that to present a core semantics, the request, indication and periodic handler functions of the components return one rather than a list of request and indication events. A list of events can be similarly processed in sequence. We also elide the complication that $o$, $or$ and $oi$ are option values.) A trace $\tau$ is a sequence of label events. The $i$-th event of a trace $\tau$ is denoted by $\tau(i)$. The trace $\tau_{i..j}$ denotes the sub-trace of $\tau$ from and including location $i$ to and excluding location $j$ and the trace $\tau_{i..}$ denotes the sub-trace of $\tau$ from and including location $i$ onward. Given a predicate $p$ on event labels, the sub-trace $\tau|_p$ is the projection of $\tau$ for events that satisfy $p$. We use the overline notation to denote multiple instances; for example, we use $\overline{\tau_i}$ to denote multiple traces $\tau_i$ one for each index $i$ in the context. The trace $\tau \cdot \tau'$ denotes the concatenation of the traces $\tau$ and $\tau'$ and $\overline{\tau_i}$ denotes the concatenation of the traces $\tau_i$.

Given a stack of components $\mathcal{S}$, Fig. 14 presents the operational semantics for $\mathcal{S}$. The semantics is parametric in the round $r_{GST}$ (Global Stabilization Time). We start with an overview. A round $\rightarrow$ comprises two parts. The first part $\xrightarrow{\tau}{}_t^*$ is a finite sequence of (1) top-level request transitions (and their following request and indication transitions) and (2) node failure transitions. Send request transitions at the leaf layers result in messages. The second part $\xrightarrow{\tau'}{}_p$ is a transition that delivers and processes messages and executes the periodic handlers. The two parts result is a round transition $\xrightarrow{\tau \cdot \tau'}$. The trace semantics $T(\mathcal{S})$ of $\mathcal{S}$ is the set of traces $\tau$ of infinite transitions $\xrightarrow{\tau}{}^*$ starting from the initial state $w_0(\mathcal{S})$. We consider infinite traces to reason about liveness properties.

In the rules, we use _ as a place holder for variables that are not used in the context. The two rules Fail and Request make top-level transitions $\rightarrow_t$. The rule Request uses the helper transition $\rightarrow_{req}$. The transition $\rightarrow_{req}$ that processes request events is taken by the two rules Req and Req'. The rule Req uses the helper transition $\rightarrow_{ind}$. The transition $\rightarrow_{ind}$ that processes indication events is taken by the two rules Ind and Ind'. The rule Ind, in turn, uses the helper transition $\rightarrow_{req}$. The transitions $\rightarrow_{req}$ and $\rightarrow_{ind}$ are interdependent; the rule Req that makes the transition $\rightarrow_{req}$ uses the transition $\rightarrow_{ind}$ and the rule Ind that makes the transition $\rightarrow_{ind}$ uses the transition $\rightarrow_{req}$. The rule Periodic makes periodic transitions $\rightarrow_p$. It uses the helper transitions $\rightarrow_{msg}$ and $\rightarrow_{per}$. The transition $\rightarrow_{msg}$ that processes messages is taken by the rule Msg. The transition $\rightarrow_{per}$ that executes the periodic functions is taken by the rules Per and Per'. Next, we take a closer look at each rule.

The rule Fail makes a top-level transition with a fail event for a node that has not already failed and adds it to the set of failed nodes $f$. Similarly, the other rules require that the executing node has not already failed.

The rule Request executes a top-level request that, in turn, may result in a sequence of requests and indications. It makes a transition $\rightarrow_t$, if a request transition $\rightarrow_{req}$ can be taken with a trace starting with a top-level (request) event. The rule Req makes transitions $\rightarrow_{req}$ for request events on the internal (non-leaf) components. We take a close look at the rule Req and the other rules are similar to it. The first event of the trace represents that at a node $n$ and a component at location $d$, a request $\downarrow$ event $e$ is executed that takes the pre-state of the stack $\sigma$ to the post-state $\sigma'$, issues the request event $e_1$ to the $i$-th subcomponent and issues the indication event $e_2$. Let $c$ be the top component of the stack at location $d$. The state $s$ for the component $c$ is obtained from the stack state $\sigma$. The request function $\text{request}_c$ is called with the node identifier $n$, the pre-state for the node $s(n)$ and the request event $e$ and results in the post-state $s'_n$ for $n$, the request event $(i, e_1)$ i.e. the request $e_1$ for the $i$-th subcomponent and the indication event $e_2$. The state $s$ of the component is updated to $s'$ with the new state $s'_n$ for the node $n$ and the state of the stack $\sigma$ is updated to $\sigma'$ with the new state $s'$ at location $d$. The issued request event $(i, e_1)$ inductively results in a transition $\rightarrow_{req}$ at the $i$-th substack whose location is $i :: d$. Similarly the issued indication event $e_2$ inductively results in a transition $\rightarrow_{ind}$ at location $d$ for the parent component. The trace for the whole transition is the original request event concatenated with the two transition traces for the issued request and indication events.

Basic links are used as the leaves of a stack. If the component at the location $d$ is a link, the rule Req' makes a transition for a send request. The rule adds tuples to the in-transit messages that contain the sender and the receiver node identifiers, the component location and the message payload. A message can be duplicated in the network. Therefore, a finite number of duplicate tuples are added (with the overline notation). Further, we note that messages are added to an unordered multi-set of messages. Therefore, messages can be arbitrarily reordered.

The rule INd makes a transition $\rightarrow_{ind}$ for indication events from components except the topmost. (We note that the location of the first event in its trace is $i :: d$ to exclude the top.) The rule is similar to the rule REq in structure but executes an indication instead of a request event. The indication event is executed at the parent component that is at location $d$. The rule INd' makes a transition $\rightarrow_{ind}$ for indication events from the topmost component. As there is no explicit parent component, the rule simply records the issued indication in its label.

The rule Periodic delivers messages and executes periodic functions. It drops messages sent to failed nodes $f$ (using the drop function). In addition, before the round $r_{GST}$, it may drop some other messages and retain a subset of messages $ms'$. The set of remained messages are delivered by the message transition $\rightarrow_{msg}$ and calls to the periodic handlers are started at the top level by the period transition $\rightarrow_{per}$. The rule Msg that makes the transition $\rightarrow_{msg}$ delivers all the messages in its pre-state. For every message, it issues a delivery indication event at the recipient node and component location. The trace of the transition is the concatenation of the traces of all the indication transitions. The rule Per executes the periodic function of a component at a (non-leaf) location $d$ and recursively for every subcomponent $i$ at location $i :: d$. The rule Per is similar to the rule REq in structure but in addition to calling the periodic function on the component, propagates the periodic calls to lower-level components as well. At the leaf layers, the rule Per' makes trivial periodic transitions.

## 7 SOUNDNESS OF TLC

In this section, we define the semantics of assertions on execution traces and prove the soundness of TLC with respect to distributed stack semantics (defined in § 6). We note that TLC is independent of the distributed stack semantics and its soundness can be studied for alternative semantics.

We define a model $m = (\tau, i, I)$ as a tuple of a trace $\tau$, a position $i$ in the trace and an interpretation $I$. A trace $\tau$ is the sequence of event labels of an execution. To evaluate temporal operators, the model includes a position $i$ in the trace. The model also includes an interpretation $I$ that maps free rigid variables, and interpreted functions and predicates to concrete values, functions and predicates. We define the set of models $M(\mathcal{S})$ of a stack $\mathcal{S}$ as the set of tuples $(\tau, 0, I_0)$ where $\tau$ is a trace of $\mathcal{S}$, $0$ is the first position and $I_0$ is an initial interpretation that includes mappings for commonly used integer, list and set functions and predicates. The traces of a stack $\mathcal{S}$ is the set of traces of the executions of $\mathcal{S}$ (for any $r_{GST}$).

In Fig. 16, we define the models relation, $m \vDash \mathcal{A}$, that is read as the model $m$ models or satisfies the assertion $\mathcal{A}$. We also use the models relation for terms, $m \vDash t : v$, that is read as the model $m$ evaluates the term $t$ to the value $v$. We remember from the classical temporal logic [Manna and Pnueli 1992] that a rigid variable has the same value in all elements of a trace, while a flexible variable may assume distinct values in different elements. The rule VarM evaluates a rigid variable $x$ using the interpretation $I$. On the other hand, separate rules evaluate the flexible variables **n**, **r**, **d**, **o**, **e**, **s**, **s'**, **ois** and **ors** based on $\tau(i)$, the event at the $i$-th position in the trace $\tau$. For instance, the rule NM evaluates the flexible variable **n** for the current node to the first element $n$ of $\tau(i)$. The rule SM evaluates the flexible variable **s** (pre-state). We remember from the distributed stack semantics that if an event at location $d$ is a request or periodic event, then it is applied to the component at location $d$, but if it is an indication event, it is applied to the parent component at location tail($d$). Therefore, if the event at location $d$ has a downward orientation i.e. $\downarrow$ or $\updownarrow$ then the location $d$ is applied to the stack state $\sigma$ to obtain the component state. However, if it has an upward orientation i.e. $\uparrow$ then the location tail($d$) is applied to $\sigma$. The rule SM' for the post-state **s'** is similar. The rule CM evaluates constants except Correct by the interpretation $I$. The rule CSM evaluates the constant Correct to the set of nodes in $\mathbb{N}$ that do not have a fail transition in the trace. As the pre-state **s** and

PERIODIC
$$\frac{\begin{array}{c} \begin{cases} ms' = \mathsf{drop}(ms,f) & \text{if } r \geq r_{GST} \\ ms' \subseteq \mathsf{drop}(ms,f) & \text{else} \end{cases} \\ (\sigma, ms', f, r) \xrightarrow{\tau}_{msg} (\sigma_0, ms_0) \\ (\sigma_n, ms_n, f, r+1) \xrightarrow{\tau_n}_{per} (\sigma_{n+1}, ms_{n+1})_{n \in \mathbb{N} \smallsetminus \{f\}} \\ \overline{\tau_n = (n, r+1, [\,], \natural, \mathsf{per}, \_,\_,\_,\_) \cdot \tau'_n}_{n \in \mathbb{N} \smallsetminus \{f\}} \\ \overline{\sigma_n = \sigma_{n+1} \quad ms_n = ms_{n+1}}_{n \in f} \end{array}}{(\sigma, ms, f, r) \xrightarrow{\tau \cdot \overline{\tau_n}_{n \in \mathbb{N} \smallsetminus \{f\}}}_p (\sigma_{|\mathbb{N}|}, ms_{|\mathbb{N}|}, f, r+1)}$$

REQUEST
$$\frac{n \in \mathbb{N} \smallsetminus \{f\} \quad (\sigma, ms, f, r) \xrightarrow{\tau}_{req} (\sigma', ms') \quad \tau = (n, r, [\,], \_,\_,\_,\_,\_,\_) \cdot \tau'}{(\sigma, ms, f, r) \xrightarrow{\tau}_t (\sigma', ms', f, r)}$$

FAIL
$$\frac{n \in \mathbb{N} \smallsetminus \{f\}}{(\sigma, ms, f, r) \xrightarrow{(n, r, [\,], \bot, \mathsf{fail}, \sigma, \sigma, \bot, \bot)}_t (\sigma, ms', f \cup \{n\}, r)}$$

REQ
$$\frac{\begin{array}{c} n \in \mathbb{N} \smallsetminus \{f\} \quad \mathcal{S}(d) = \mathsf{stack}(c,\_) \quad \sigma(d) = s \\ \mathsf{request}_c(n, s(n), e) = (s'_n, (i, e_1), e_2) \\ s' = s[n \mapsto s'_n] \quad \sigma' = \sigma[d \mapsto s'] \\ (\sigma', ms, f, r) \xrightarrow{\tau_1}_{req} (\sigma_1, ms_1) \\ (\sigma_1, ms_1, f, r) \xrightarrow{\tau_2}_{ind} (\sigma_2, ms_2) \quad \tau_2 = (n, r, d, \uparrow, e_2, \_,\_,\_,\_) \cdot \tau'_2 \\ \tau = (n, r, d, \downarrow, e, \sigma, \sigma', (i, e_1), e_2) \cdot \tau_1 \cdot \tau_2 \end{array}}{(\sigma, ms, f, r) \xrightarrow{\tau}_{req} (\sigma_2, ms_2)}$$

REQ'
$$\frac{n \in \mathbb{N} \smallsetminus \{f\} \quad \mathcal{S}(d) = \mathsf{link}}{(\sigma, ms, f, r) \xrightarrow{(n, r, d, \downarrow, \mathsf{send}_l(n', m), \sigma, \sigma, \bot, \bot)}_{req} (\sigma', ms \uplus \{\overline{(n, n', d, m)}\})}$$

IND
$$\frac{\begin{array}{c} n \in \mathbb{N} \smallsetminus \{f\} \quad \mathcal{S}(d) = \mathsf{stack}(c,\_) \quad \sigma(d) = s \\ \mathsf{indication}_c(n, s(n), (i, e)) = (s'_n, (i', e_1), e_2) \\ s' = s[n \mapsto s'_n] \quad \sigma' = \sigma[d \mapsto s'] \\ (\sigma', ms, f, r) \xrightarrow{\tau_1}_{req} (\sigma_1, ms_1) \quad \tau_1 = (n, r, i' :: d, \downarrow, e_1, \_,\_,\_,\_) \cdot \tau'_1 \\ (\sigma_1, ms_1, f, r) \xrightarrow{\tau_2}_{ind} (\sigma_2, ms_2) \quad \tau_2 = (n, r, d, \uparrow, e_2, \_,\_,\_,\_) \cdot \tau'_2 \\ \tau = (n, r, i :: d, \uparrow, e, \sigma, \sigma', (i', e_1), e_2) \cdot \tau_1 \cdot \tau_2 \end{array}}{(\sigma, ms, f, r) \xrightarrow{\tau}_{ind} (\sigma_2, ms_2)}$$

IND'
$$\frac{n \in \mathbb{N} \smallsetminus \{f\}}{(\sigma, ms, f, r) \xrightarrow{(n, r, [\,], \uparrow, e, \sigma, \sigma, \bot, \bot)}_{ind} (\sigma, ms)}$$

MSG
$$\frac{\begin{array}{c} ms = \{\overline{(n_i, n'_i, d_i, m_i)}_{i \in I}\} \quad \sigma_0 = \sigma \quad ms_0 = \varnothing \\ (\sigma_i, ms_i, f, r) \xrightarrow{\tau_i}_{ind} (\sigma_{i+1}, ms_{i+1})_{i \in I} \\ \tau_i = (n'_i, r, d_i, \uparrow, \mathsf{deliver}_l(n_i, m_i), \_,\_,\_,\_) \cdot \tau'_i{}_{i \in I} \end{array}}{(\sigma, ms, f, r) \xrightarrow{\overline{\tau_i}_{i \in I}}_{msg} (\sigma_{|I|}, ms_{|I|})}$$

PER'
$$\frac{n \in \mathbb{N} \smallsetminus \{f\} \quad \mathcal{S}(d) = \mathsf{link}}{(\sigma, ms, f, r) \xrightarrow{(n, r, d, \natural, \bot, \sigma, \sigma, \bot, \bot)}_{per} (\sigma, ms)}$$

PER
$$\frac{\begin{array}{c} n \in \mathbb{N} \smallsetminus \{f\} \quad \mathcal{S}(d) = \mathsf{stack}(c, \overline{\mathcal{S}'}) \quad k = |\overline{\mathcal{S}'}| \quad \sigma(d) = s \\ \mathsf{periodic}_c(n, s(n)) = (s'_n, (i, e_1), e_2) \quad s' = s[n \mapsto s'_n] \quad \sigma' = \sigma[d \mapsto s'] \\ (\sigma', ms, f, r) \xrightarrow{\tau_1}_{req} (\sigma'', ms'') \quad \tau_1 = (n, r, i :: d, \downarrow, e_1, \_,\_,\_,\_) \cdot \tau'_1 \\ (\sigma'', ms'', f, r) \xrightarrow{\tau_2}_{ind} (\sigma_0, ms_0) \quad \tau_2 = (n, r, d, \uparrow, e_2, \_,\_,\_,\_) \cdot \tau'_2 \\ (\sigma_i, ms_i, f, r) \xrightarrow{\tau_i}_{per} (\sigma_{i+1}, ms_{i+1})_{i \in \{0..k-1\}} \quad \tau_i = (n, r, i :: d, \natural, \_,\_,\_,\_,\_) \cdot \tau'_i{}_{i \in \{0..k-1\}} \\ \tau = (n, r, d, \natural, \mathsf{per}, \sigma, \sigma', (i, e_1), e_2) \cdot \tau_1 \cdot \tau_2 \cdot \overline{\tau_i}_{i \in \{0..k-1\}} \end{array}}{(\sigma, ms, f, r) \xrightarrow{\tau}_{per} (\sigma_k, ms_k)}$$

Fig. 14. Semantics of Distributed Stacks.

post-state **s'** variables take function values, not only a function but also a variable can be applied to terms. The two applications are evaluated in FUNM and FUNM' respectively.

The rule PREDM evaluates predicates using the interpretation $I$. The definition of models for conjunction, negation and quantification is standard. The strict always operator $\hat{\Box}$ requires the assertion to hold in every future position starting from the position after the current. The strict always in the past operator $\hat{\boxminus}$ requires a similar condition in the past. The strict eventual operator $\hat{\diamondsuit}$ requires the assertion to hold in at least one future position starting from the position after the current. The strict eventual in the past operator $\hat{\diamondsuit}$ requires a similar condition in the past. As we defined the non-strict temporal operators as syntactic sugar for strict ones, their semantics are derived from the above semantics. The next operator $\bigcirc$ requires the assertion to hold in the position after the current. As defined at the bottom of Fig. 16, we say that an event label $\ell$ is on the self component mself($\ell$), if it is a request ($\downarrow$) or periodic ($\xi$) at the top location ([ ]) or is an indication ($\uparrow$) at the second level (at location [$i$] for some $i$). The self operator $\circledS$ requires the assertion to hold on the self subtrace i.e. the projection of the trace over mself. More precisely, the self operator requires the assertion to hold on the first self position after the current position.

We are now ready to state the soundness of TLC. An assertion $\mathcal{A}$ is valid for a stack $\mathcal{S}$, written as $\vDash_{\mathcal{S}} \mathcal{A}$, if and only if every model of $\mathcal{S}$ satisfies $\mathcal{A}$.

Definition 4 (Valid Assertion). *For all $\mathcal{S}$ and $\mathcal{A}$, $\vDash_{\mathcal{S}} \mathcal{A}$ iff for all $m \in M(\mathcal{S}), m \vDash \mathcal{A}$.*

We say that a set of assertions $\Gamma$ entail an assertion $\mathcal{A}$ if and only if every model $m$ of $\mathcal{S}$ that models $\Gamma$ also models $\mathcal{A}$.

Definition 5 (Models Relation). *For all $\Gamma$, $\mathcal{S}$ and $\mathcal{A}$, $\Gamma \vDash_{\mathcal{S}} \mathcal{A}$ iff $\forall m \in M(\mathcal{S}), m \vDash \Gamma \rightarrow m \vDash \mathcal{A}$.*

The following theorem states the soundness of TLC. If assuming assertions $\Gamma$, TLC derives an assertion $\mathcal{A}$ then $\Gamma$ entails $\mathcal{A}$.

Theorem 2 (Soundness). *For all $\Gamma, \mathcal{S}, c, \overline{\mathcal{S}'}, \mathcal{A}$, such that $\mathcal{S} = stack(c, \overline{\mathcal{S}'})$, if $\Gamma \vdash_c \mathcal{A}$, then $\Gamma \vDash_{\mathcal{S}} \mathcal{A}$.*

The detailed proofs are available in the appendix [Appendix 2020] § 5.1. The following corollary is immediately derived. It states that if assuming valid assertions, TLC derives an assertion then that assertion is valid as well. In other words, TLC derives only valid assertions from valid assertions.

Corollary 2. *For all $\Gamma, c, \mathcal{S}, \overline{\mathcal{S}'}, \mathcal{A}$ such that $\mathcal{S} = stack(c, \overline{\mathcal{S}'})$, if $\vDash_{\mathcal{S}} \Gamma$ and $\Gamma \vdash_c \mathcal{A}$, then $\vDash_{\mathcal{S}} \mathcal{A}$.*

## 8 MECHANIZATION

The ultimate goal of this project is mechanized distributed middleware. This goal is a huge undertaking and fully achieving it may take multiple years. The main topic of this paper is TLC, its compositionality and applicability. We have finished all the proofs of the components in the appendix to ensure that TLC is comprehensive. Nonetheless, we have been mechanizing the proofs in Coq. The TLC Coq framework provides a deep embedding of an enriched lambda calculus for defining functional components, an evaluation engine for embedded terms, an inductive definition of TLC, and a set of tactics for constructing TLC proof terms. We have used this library to successfully mechanize the verification of the stubborn link and the perfect link components and we are extending mechanization to the other components.

**Embedding Approaches.** We tried different approaches for encoding TLC in Coq. The earliest attempts were shallow embeddings of TLC. The intent was to utilize Coq's Gallina functional programming language to capture component definitions and its Ltac proof language to construct proof terms. These approaches proved unsuccessful due to the syntactic nature of proofs in TLC. The syntactic rules of TLC require recursive analysis of the syntax of terms, which cannot be done directly on Gallina terms. We define a deep embedding of a minimal functional programming language to program component terms. This embedding is an untyped lambda calculus enriched with pattern matching terms, externally defined functions, value literals, value constructors, and locally nameless parameters. Similarly, we define a deep embedding of the syntax of TLC as well.

**Embedding.** The syntax of terms is defined as the inductive type presented in Fig. 15.(a). The TParameter, TAbstraction, and TApplication terms come directly from untyped lambda calculus. We adopted the locally nameless representation [Charguéraud 2012] to separately define parameters and variables. We chose this encoding instead of implementing capture-avoiding substitution of arbitrarily named variables. Coq requires all recursive functions to be structurally decreasing. Algorithms for capture-avoiding substitution are not strictly decreasing and are rejected by Coq. Bound variables, represented by the TParameter constructor, are referenced using deBruijn indices. Free variables, represented by the TVariable constructor, are named strings.

The TConstructor term represents a constructor of an inductive type. The TLiteral term represents a literal value of a Coq type, such as the natural numbers. The TFunction term represents a function that is not defined explicitly in the term language, such as recursively defined functions. These terms are lifted into Coq, evaluated, and the result is lowered into the embedded term type. The TMatch constructor represents a pattern matching expression. The TFailure term is the empty term, produced when an ill-formed term is evaluated.

To simplify the definition of terms, we have defined a library of Coq notations for the term language. The library allows for the relatively direct translation of the implementation of the components into embedded terms. Similarly, Coq notations are provided for the assertion language and the sequent judgements. These notations allow writing judgements in the commonly used form. The context of a sequent is the set of variable names that are universally quantified along with the list of assumed assertions. As an example, Fig. 15.(b) shows the statement of the stubborn delivery property: if a message is sent, it is infinitely often delivered. The context declares the list of free variable n, n' and m and no assumed assertions [::]. The conclusion is read as follows: if two nodes n and n' are correct and n at the top level [] sends a request event -> to send the message m to n', then infinitely often at n', at the top level [], the indication event <- that delivers the message m from n is executed.

```
Inductive term :=
| TParameter (p : parameter) (* Nameless bound params *)
| TVariable (v : variable) (* Named free variables *)
| TAbstraction (tb : term) (* Function abstraction *)
| TApplication (tf ta : term) (* Function application *)
| TConstructor (c : constr) (* Value constructors *)
| TLiteral (l : literal) (* Value literals *)
| TFunction (f : function) (* External functions *)
| TFailure (* Computation error *)
| TMatch (ta : term) (cs : cases) (* Pattern matching *)
(* Cases of pattern matching *)
with acase :=
| TCase (p : pattern) (t : term)
with cases :=
| TCNil
| TCCons (c : acase) (cs : cases).
```
(a)

```
Theorem SL_1 :
  Context [:: V "m"; V "n'"; V "n"] [::]
  |- stubborn_link, {A:
  correct "n" /\ correct "n'" ->
  on "n", event []-> CSLSend $ "n'" $ "m" =>>
  always eventually
    on "n'", event []<- CSLDeliver $ "n" $ "m" }.
```
(b)

```
Inductive derives : context -> assertion -> Prop :=
| DAEvaluateP Delta Gamma Ap Ap' Ac :
  (* Replaces the head premise with its evaluation *)
  [[A Ap]] = Success Ap' ->
  Context Delta (Ap' :: Gamma) |- Ac ->
  Context Delta (Ap :: Gamma) |- Ac
  (* ... *)
| DSCut Delta Gamma Ap Ac :
  (* Add a proven assertion to the proof context *)
  Context Delta Gamma |- Ap ->
  Context Delta (Ap :: Gamma) |- Ac ->
  Context Delta Gamma |- Ac
  (* ... *)
| DPIR ctx :
  ctx |- {A: forall: "?e": event []-> "?e" =>>
    ("Fs'" $ "Fn", "Fors", "Fois") =
    request C $ "Fn" $ ("Fs" $ "Fn") $ "?e"}
```
(c)

Fig. 15. Mechanizing TLC

Definition 6 (Model relation).

| | | | | |
|---|---|---|---|---|
| VarM | $(\tau,i,I)$ | $\vDash$ | $x:I(x)$ | if $x$ rigid |
| NM | $(\tau,i,I)$ | $\vDash$ | $\boldsymbol{n}:n(\tau(i))$ | |
| RM | $(\tau,i,I)$ | $\vDash$ | $\boldsymbol{r}:r(\tau(i))$ | |
| DM | $(\tau,i,I)$ | $\vDash$ | $\boldsymbol{d}:d(\tau(i))$ | |
| OM | $(\tau,i,I)$ | $\vDash$ | $\boldsymbol{o}:o(\tau(i))$ | |
| EM | $(\tau,i,I)$ | $\vDash$ | $\boldsymbol{e}:e(\tau(i))$ | |
| SM | $(\tau,i,I)$ | $\vDash$ | $\boldsymbol{s}:\sigma(\tau(i))(d')$ where $d' = \begin{cases} d(\tau(i)) & \text{if } o(\tau(i)) = \downarrow \vee o(\tau(i)) = \natural \\ tail(d(\tau(i))) & \text{else} \end{cases}$ | |
| SM' | $(\tau,i,I)$ | $\vDash$ | $\boldsymbol{s'}:\sigma'(\tau(i))(d')$ where $d' = \begin{cases} d(\tau(i)) & \text{if } o(\tau(i)) = \downarrow \vee o(\tau(i)) = \natural \\ tail(d(\tau(i))) & \text{else} \end{cases}$ | |
| ORSM | $(\tau,i,I)$ | $\vDash$ | $\boldsymbol{ors}:ors(\tau(i))$ | |
| OISM | $(\tau,i,I)$ | $\vDash$ | $\boldsymbol{ois}:ois(\tau(i))$ | |
| CM | $(\tau,i,I)$ | $\vDash$ | $c:I(c)$ | if $c \ne Correct$ |
| CSM | $(\tau,i,I)$ | $\vDash$ | $Correct: \{n \mid n \in \mathbb{N} \wedge \nexists j \ge 0.\ \tau(j) = (n,[],\bot,\mathit{fail},\_,\_,\_,\_)\}$ | |
| FunM | $m$ | $\vDash$ | $f(t_1,..,t_n):f'(v_1,..,v_n)$ | if $I(f) = f'$, $m \vDash t_1:v_1$, .., $m \vDash t_n:v_n$ |
| FunM' | $m$ | $\vDash$ | $x(t_1,..,t_n):f'(v_1,..,v_n)$ | if $m \vDash x:f'$, $m \vDash t_1:v_1$, .., $m \vDash t_n:v_n$ |
| PredM | $m$ | $\vDash$ | $p(t_1,..,t_n)$ | iff $I(p) = p'$, |
| | | | | $m \vDash t_1:v_1$, .. $m \vDash t_n:v_n$, |
| | | | | $p'(v_1,..,v_n) = true$ |
| | | | | $\tau(j) \ne (n,[],\bot,\mathit{fail},\_,\_,\_,\_),\ \text{for all } j \ge 0$ |
| AndM | $m$ | $\vDash$ | $\mathcal{A} \wedge \mathcal{A}'$ | iff $m \vDash \mathcal{A}$ and $m \vDash \mathcal{A}'$ |
| NotM | $m$ | $\vDash$ | $\neg \mathcal{A}$ | iff $m \nvDash \mathcal{A}$ |
| ForallM | $(\tau,i,I)$ | $\vDash$ | $\forall x.\ \mathcal{A}$ | iff $(\tau,i,I[x \mapsto v]) \vDash \mathcal{A}$ for all $v \in dom(I)$ |
| AlwaysM | $(\tau,i,I)$ | $\vDash$ | $\hat{\Box} \mathcal{A}$ | iff $(\tau,j,I) \vDash \mathcal{A}$ for all $j,\ j > i$ |
| PAlwaysSM | $(\tau,i,I)$ | $\vDash$ | $\hat{\boxminus} \mathcal{A}$ | iff $(\tau,j,I) \vDash \mathcal{A}$ for all $j,\ 0 \le j < i$ |
| EventualSM | $(\tau,i,I)$ | $\vDash$ | $\hat{\Diamond} \mathcal{A}$ | iff $(\tau,j,I) \vDash \mathcal{A}$ there exists $j,\ j > i$ |
| PEventualSM | $(\tau,i,I)$ | $\vDash$ | $\hat{\diamondsuit} \mathcal{A}$ | iff $(\tau,j,I) \vDash \mathcal{A}$ there exists $j,\ 0 \le j < i$ |
| NextM | $(\tau,i,I)$ | $\vDash$ | $\bigcirc \mathcal{A}$ | iff $(\tau,i+1,I) \vDash \mathcal{A}$ |
| SelfM | $(\tau,i,I)$ | $\vDash$ | $\circledS \mathcal{A}$ | iff $\tau'_1 = \tau_{0..i-1}\vert_{\mathit{mself}}$, $\tau'_2 = \tau_{i..}\vert_{\mathit{mself}}$, |
| | | | | $\tau' = \tau'_1 \cdot \tau'_2$, $i' = \vert\tau'_1\vert$, $(\tau',i',I) \vDash \mathcal{A}$ |

$$mself(\ell) \triangleq (d(\ell) = [] \wedge o(\ell) = \downarrow)\ \vee\ (d(\ell) = [] \wedge o(\ell) = \natural)\ \vee\ (\exists i.\ d(\ell) = [i] \wedge o(\ell) = \uparrow)$$

Fig. 16. Models Relation. $m \vDash \mathcal{A}$ and $m \vDash t:v$.

**Logic.** The basic rules and axioms of TLC are encoded as an inductive type. Fig. 15.(c) shows three constructors that are representative of the encoding of the rules and axioms of TLC. The rules are extended with rules specific to the implementation of the extended term language. The first constructor, DAEvaluateP, states that if the terms within the first premise can be simplified then proving the conclusion assuming the premise can be reduced to proving the conclusion assuming the simplified premise. The [[A Ap]] notation represents assertion evaluation, which replaces all computational terms within an assertion with the terms produced by their evaluation. Terms are evaluated recursively inside of a monad, which produces a failure case when a failure term is evaluated. The second constructor, DSCut, is a sequent logic rule that can be used to introduce an assertion as a premise. The third constructor, DPIR, is an axiom of the TLC program logic, the rule IR that we saw in § 5.

The framework provides tactics to facilitate applying TLC. For example, it provides a set of tactics that mirror a subset of Coq's Ltac tactics for producing Gallina proof terms, as well as a library of lemmas. These tactics allow proofs to be written in a more natural, Coq-like style. These are tactics such as d_left (d refers to the derives relation), d_right, d_splitp (p for premise), and d_splitc (c for conclusion), which mirror the primitive Coq tactics left, right, destruct (on conjunctive hypotheses), and split. In addition to these basic tactics, there are tactics that automate some multi-step common tasks. For example, the d_evalc imitates the Coq simpl tactic, evaluating all terms in the conclusion assertion, and the d_have tactic automates the application of the DSCut rule that we saw above.

## 9 RELATED WORK

High-level DSLs, language extensions and tools [Bakst et al. 2017; Biely et al. 2013; Burckhardt et al. 2012; Cejtin et al. 1995; Kato et al. 1993; Ketsman et al. 2019; Killian et al. 2007; Liu et al. 2012; Miller et al. 2016; Salvaneschi et al. 2019; Samanta et al. 2013; Weisenburger et al. 2018] have been used to raise the level of abstraction, and improve the reliability of distributed systems. Model checking has been extensively applied for bounded verification [Dutertre et al. 2018; Jackson 2006; John et al. 2013; Killian et al. 2007; Konnov et al. 2017; Marić et al. 2017; Musuvathi and Engler 2004; Yabandeh et al. 2009; Yang et al. 2009; Zave 2012] of distributed algorithms. Recently, domain specific logics and verification frameworks have gained momentum to establish the absence of bugs.

Temporal logic [Manna and Pnueli 1992] is a modal logic that can abstract and reason about time. It can be used to state and check properties of programs specially reactive programs [Alur et al. 2004; Cave et al. 2014; Cook et al. 2011; Das et al. 2018; Jeffrey 2012]. TLA (Temporal Logic of Actions) [Lamport 1994, 2000] is a logic for description, specification and verification of distributed protocols. The transition system of a protocol can be described as action assertions and its specification can be written as temporal logic assertions [Manna and Pnueli 1992]. It has been used [Chaudhuri et al. 2010; Lamport 2002] for model checking [Newcombe et al. 2015] and interactive verification [Chand et al. 2016] of distributed systems. A TLA protocol is described as a monolithic transition system. In contrast, TLC defines event interfaces between components and supports their composition. More importantly, it supports compositional verification of components. In addition, in contrast to TLA that requires the protocol to be described as a transition system, TLC supports functional implementation of protocols that can be directly executed. Further, in contrast to TLA, TLC defines an operational semantics to support the soundness of the logic.

I/O Automata [Lynch and Tuttle 1989] models specifications and protocols as transition systems and provides simulation proof techniques [Lynch and W. Vaandrager 1995] between automata. In contrast, TLC captures component implementations as functional programs and specifications as descriptive temporal assertions, and provides a program logic and a compositional proof technique to derive the specification for the implementation.

Both I/O Automata [Lynch and Tuttle 1989] and Reactive Modules [Alur and Henzinger 1999] model protocols as transition systems. They capture specifications in the semantic domain as either transitions systems or properties of execution traces. In contrast, TLC captures component implementations as functional programs and specifications as temporal assertions. I/O Automata provides simulation proof techniques [Lynch and W. Vaandrager 1995] between automata. The simulation proofs are written in the semantic domain. In contrast, TLC provides a program logic to derive the specification for the implementation. Both I/O Automata and and Reactive Modules support composition of interacting modules with matching input and outputs. They support assume-guarantee reasoning where each module can be verified based on the specification of the other module. TLC models distributed systems as structured stacks of components where each component composes with its subcomponents below and its parent component above. Similar to the assume-guarantee reasoning, it supports compositional verification of each component based on the specification of its subcomponents. However, no assumptions for the parent component is needed. The specification of each component is for the most general parent. A verified component can serve as the subcomponent of any parent component.

EventML [Rahli 2012] is a functional domain-specific language for distributed protocols. Protocols written in EventML can be translated to Nuprl [Constable et al. 1986] and then interactively verified. It has been used to verify monolithic replicated services [Rahli et al. 2018; Schiper et al. 2014]; however, it does not address compositional verification.

IronFleet [Hawblitzel et al. 2015] models a distributed system as a hierarchy of state transition systems at multiple levels of abstraction from the high-level specification to the low-level implementation. It proves a refinement [He et al. 1986; Lynch and W. Vaandrager 1995] between a layer and the layer immediately above it. However, it only considers monolithic protocols without subcomponents and the verification is carried out using refinement in contrast to a program logic.

Similarly, network refinement [Koh et al. 2019] presents specifications for a swap server that can be both tested and verified using observational refinement. The server is well-integrated with several other verified systems. To contrast, TLC is a temporal logic and can verify liveness in addition to safety properties. Further, TLC focuses on composition of distributed protocols and builds component stacks on basic links that are much weaker than TCP.

Verdi [Wilcox et al. 2015; Woos et al. 2016] models several network semantics and provides transformations from correct protocols in one semantics to another. It has been applied to verification of state machine replication. Similar to TLC, Verdi provides a form of vertical composition. However, its proofs are based on simulation [He et al. 1986; Lynch and W. Vaandrager 1995] in the semantic domain rather than a program logic.

Chapar [Lesani et al. 2016] presents an operational semantics and a proof technique for verification of causally consistent distributed stores. Similar to TLC, Chapar considers the interface between clients and store implementations; however, only for causal consistency. Further, verification is based on simulation rather than program logic.

PSync [Dragoi et al. 2016] is a DSL for distributed protocols based on the heard-of round-based model [Charron-Bost and Schiper 2009]. This lockstep model enables proof automation that has been successfully applied to verification of consensus variants. However, PSync left composition as future work.

Ivy [Padon et al. 2016] is an interactive tool that assists in finding inductive invariants. It has been applied to verification of a few distributed protocols. A follow-up work [Taube et al. 2018] lets the user split a protocol into logical modules with explicit invariants. Modules facilitate an assume-guarantee reasoning such that verification of each falls in a separate decidable theory. While the main focus of Ivy is automatic verification of separate parts of monolithic protocols, TLC's focus is compositional verification of stacks of protocols. Further, in contrast to Ivy, TLC supports verification of liveness properties.

Disel [Sergey et al. 2017; Wilcox et al. 2017] is a program logic for distributed protocols that provides Floyd-Hoare-style specification [Floyd 1967; Hoare 1969; Reynolds 2002] and proof rules for horizontal composition. In Disel, the specification of a component is written in terms of its state and the message pool. On the other hand, in TLC, the temporal specification is written in terms of the interface events almost verbatim from the natural language description. Disel and Hoare-style reasoning require definition of stable invariants and sometimes ghost variables. However, TLC does not require additional annotations on the components. Disel can state and prove safety properties while TLC can state and prove both safety and liveness properties. A follow-up work [García-Pérez et al. 2018] similarly applies the rely-guarantee reasoning to verification of a decomposition of Paxos [Boichat et al. 2003]. However, it does not consider the leader election subcomponent. This paper considers leader election and epoch change as well.

## 10 CONCLUSION

TLC is a temporal program logic for compositional verification of stacks of distributed components. Its assertion language can capture both safety and liveness properties. Using a transformation function that lowers the specification of components to be used as subcomponents, TLC supports compositional verification of components based on only the specification of their subcomponents. It features intuitive inference rules and induction principles that can deduce assertions about a component based on its functional implementation. TLC and the transformation are proved sound with respect to the operational semantics of distributed stacks. They have been successfully applied to verify a stack of fundamental distributed components as the first steps towards certified distributed system stacks.

# REFERENCES

Rajeev Alur, Kousha Etessami, and Parthasarathy Madhusudan. 2004. A temporal logic of nested calls and returns. In *International Conference on Tools and Algorithms for the Construction and Analysis of Systems*. Springer, 467–481.

Rajeev Alur and Thomas A Henzinger. 1999. Reactive modules. *Formal methods in system design* 15, 1 (1999), 7–48.

Appendix. 2020. *Submitted Supplement Document*.

Alexander Bakst, Klaus v. Gleissenthall, Rami Gokhan Kici, and Ranjit Jhala. 2017. Verifying Distributed Programs via Canonical Sequentialization. *Proc. ACM Program. Lang.* 1, OOPSLA, Article 110 (Oct. 2017), 27 pages. https://doi.org/10.1145/3133934

M. Biely, P. Delgado, Z. Milosevic, and A. Schiper. 2013. Distal: A framework for implementing fault-tolerant distributed algorithms. In *2013 43rd Annual IEEE/IFIP International Conference on Dependable Systems and Networks (DSN)*. 1–8. https://doi.org/10.1109/DSN.2013.6575306

Romain Boichat, Partha Dutta, Svend Frølund, and Rachid Guerraoui. 2003. Deconstructing paxos. *ACM Sigact News* 34, 1 (2003), 47–67.

Sebastian Burckhardt, Manuel Fähndrich, Daan Leijen, and Benjamin P Wood. 2012. Cloud types for eventual consistency. In *European Conference on Object-Oriented Programming*. Springer, 283–307.

Christian Cachin, Rachid Guerraoui, and Lus Rodrigues. 2011. *Introduction to Reliable and Secure Distributed Programming* (2nd ed.). Springer Publishing Company, Incorporated.

Andrew Cave, Francisco Ferreira, Prakash Panangaden, and Brigitte Pientka. 2014. Fair Reactive Programming. In *Proceedings of the 41st ACM SIGPLAN-SIGACT Symposium on Principles of Programming Languages (POPL âĂŹ14)*. Association for Computing Machinery, New York, NY, USA, 361âĂŞ372. https://doi.org/10.1145/2535838.2535881

Henry Cejtin, Suresh Jagannathan, and Richard Kelsey. 1995. Higher-order Distributed Objects. *ACM Trans. Program. Lang. Syst.* 17, 5 (Sept. 1995), 704–739. https://doi.org/10.1145/213978.213986

Saksham Chand, Yanhong A. Liu, and Scott D. Stoller. 2016. Formal Verification of Multi-Paxos for Distributed Consensus. In *FM 2016: Formal Methods*, John Fitzgerald, Constance Heitmeyer, Stefania Gnesi, and Anna Philippou (Eds.). Springer International Publishing, Cham, 119–136.

Arthur Charguéraud. 2012. The Locally Nameless Representation. *Journal of Automated Reasoning* 49, 3 (01 Oct 2012), 363–408. https://doi.org/10.1007/s10817-011-9225-2

Bernadette Charron-Bost and André Schiper. 2009. The Heard-Of model: computing in distributed systems with benign faults. *Distributed Computing* 22, 1 (01 Apr 2009), 49–71. https://doi.org/10.1007/s00446-009-0084-6

Kaustuv Chaudhuri, Damien Doligez, Leslie Lamport, and Stephan Merz. 2010. The TLA+ Proof System: Building a Heterogeneous Verification Platform. In *Theoretical Aspects of Computing – ICTAC 2010*, Ana Cavalcanti, David Deharbe, Marie-Claude Gaudel, and Jim Woodcock (Eds.). Springer Berlin Heidelberg, Berlin, Heidelberg, 44–44.

R. L. Constable, S. F. Allen, H. M. Bromley, W. R. Cleaveland, J. F. Cremer, R. W. Harper, D. J. Howe, T. B. Knoblock, N. P. Mendler, P. Panangaden, J. T. Sasaki, and S. F. Smith. 1986. *Implementing Mathematics with the Nuprl Proof Development System*. Prentice-Hall, Inc., Upper Saddle River, NJ, USA.

Byron Cook, Eric Koskinen, and Moshe Vardi. 2011. Temporal property verification as a program analysis task. In *International Conference on Computer Aided Verification*. Springer, 333–348.

Ankush Das, Jan Hoffmann, and Frank Pfenning. 2018. Parallel Complexity Analysis with Temporal Session Types. *arXiv preprint arXiv:1804.06013* (2018).

Cezara Dragoi, Thomas A Henzinger, and Damien Zufferey. 2016. PSYNC : A partially synchronous language for fault-tolerant distributed algorithms. *Popl* (2016), 1–16. https://doi.org/10.1145/nnnnnnn.nnnnnnn

Bruno Dutertre, Dejan Jovanovic, and Jorge A. Navas. 2018. Verification of Fault-Tolerant Protocols with Sally. In *NFM (Lecture Notes in Computer Science)*, Vol. 10811. Springer, 113–120.

Cynthia Dwork, Nancy Lynch, and Larry Stockmeyer. 1988. Consensus in the presence of partial synchrony. *Journal of the ACM (JACM)* 35, 2 (1988), 288–323.

Robert W. Floyd. 1967. Assigning Meanings to Programs. *Proceedings of Symposium on Applied Mathematics* 19 (1967), 19–32. http://laser.cs.umass.edu/courses/cs521-621.Spr06/papers/Floyd.pdf

Álvaro García-Pérez, Alexey Gotsman, Yuri Meshman, and Ilya Sergey. 2018. Paxos consensus, deconstructed and abstracted. In *European Symposium on Programming*. Springer, Cham, 912–939.

Ronghui Gu, Zhong Shao, Hao Chen, Xiongnan Wu, Jieung Kim, Vilhelm Sjöberg, and David Costanzo. 2016. CertiKOS: An Extensible Architecture for Building Certified Concurrent OS Kernels. In *Proceedings of the 12th USENIX Conference on Operating Systems Design and Implementation (OSDI'16)*. USENIX Association, Berkeley, CA, USA, 653–669. http://dl.acm.org/citation.cfm?id=3026877.3026928

Zhenyu Guo, Sean McDirmid, Mao Yang, Li Zhuang, Pu Zhang, Yingwei Luo, Tom Bergan, Peter Bodik, Madan Musuvathi, Zheng Zhang, and Lidong Zhou. 2013. Failure Recovery: When the Cure is Worse Than the Disease. In *Proceedings of the 14th USENIX Conference on Hot Topics in Operating Systems (HotOS'13)*. USENIX Association, Berkeley, CA, USA, 8–8. http://dl.acm.org/citation.cfm?id=2490483.2490491

Chris Hawblitzel, Jon Howell, Manos Kapritsos, Jacob R. Lorch, Bryan Parno, Michael L. Roberts, Srinath Setty, and Brian Zill. 2015. IronFleet: Proving Practical Distributed Systems Correct. In *Proceedings of the 25th Symposium on Operating Systems Principles (SOSP '15)*. ACM, New York, NY, USA, 1–17. https://doi.org/10.1145/2815400.2815428

Jifeng He, C. A. R. Hoare, and Jeff W. Sanders. 1986. Data Refinement Refined. In *Proc. ESOP*.

C. A. R. Hoare. 1969. An Axiomatic Basis for Computer Programming. *Commun. ACM* 12, 10 (Oct. 1969), 576–580. https://doi.org/10.1145/363235.363259

Daniel Jackson. 2006. *Software Abstractions: Logic, Language, and Analysis*. The MIT Press.

Alan Jeffrey. 2012. LTL types FRP: linear-time temporal logic propositions as types, proofs as functional reactive programs. In *Proceedings of the sixth workshop on Programming languages meets program verification*. 49–60.

Annu John, Igor Konnov, Ulrich Schmid, Helmut Veith, and Josef Widder. 2013. Parameterized model checking of fault-tolerant distributed algorithms by abstraction. In *Proc. FMCAD*.

Kazuhiko Kato, Atsushi Ohori, Takeo Murakami, and Takashi Masuda. 1993. Distributed C language based on a higher-order RPC technique. (1993).

Bas Ketsman, Aws Albarghouthi, and Paraschos Koutris. 2019. Distribution policies for datalog. *Theory of Computing Systems* (2019), 1–34.

Charles Edwin Killian, James W. Anderson, Ryan Braud, Ranjit Jhala, and Amin M. Vahdat. 2007. Mace: Language Support for Building Distributed Systems. In *Proc. PLDI*.

Nicolas Koh, Yao Li, Yishuai Li, Li-yao Xia, Lennart Beringer, Wolf Honoré, William Mansky, Benjamin C Pierce, and Steve Zdancewic. 2019. From C to interaction trees: specifying, verifying, and testing a networked server. In *Proceedings of the 8th ACM SIGPLAN International Conference on Certified Programs and Proofs*. ACM, 234–248.

Igor Konnov, Marijana Lazić, Helmut Veith, and Josef Widder. 2017. A Short Counterexample Property for Safety and Liveness Verification of Fault-tolerant Distributed Algorithms. In *Proceedings of the 44th ACM SIGPLAN Symposium on Principles of Programming Languages (POPL 2017)*. ACM, New York, NY, USA, 719–734. https://doi.org/10.1145/3009837.3009860

Leslie Lamport. 1994. The Temporal Logic of Actions. *ACM Trans. Program. Lang. Syst.* 16, 3 (May 1994), 872–923. https://doi.org/10.1145/177492.177726

Leslie Lamport. 1998. The Part-time Parliament. *ACM Trans. Comput. Syst.* 16, 2 (1998).

Leslie Lamport. 2000. Distributed Algorithms in TLA (Abstract). In *Proceedings of the Nineteenth Annual ACM Symposium on Principles of Distributed Computing (PODC '00)*. ACM, New York, NY, USA, 3–. https://doi.org/10.1145/343477.343497

Leslie Lamport. 2002. *Specifying Systems: The TLA+ Language and Tools for Hardware and Software Engineers*. Addison-Wesley Longman Publishing Co., Inc., Boston, MA, USA.

Mohsen Lesani, Christian J. Bell, and Adam Chlipala. 2016. Chapar: Certified Causally Consistent Distributed Key-value Stores. In *Proceedings of the 43rd Annual ACM SIGPLAN-SIGACT Symposium on Principles of Programming Languages (POPL '16)*. ACM, New York, NY, USA, 357–370. https://doi.org/10.1145/2837614.2837622

Yanhong A. Liu, Scott D. Stoller, Bo Lin, and Michael Gorbovitski. 2012. From Clarity to Efficiency for Distributed Algorithms. In *Proceedings of the ACM International Conference on Object Oriented Programming Systems Languages and Applications (OOPSLA '12)*. ACM, New York, NY, USA, 395–410. https://doi.org/10.1145/2384616.2384645

Nancy Lynch and Frits W. Vaandrager. 1995. Forward and Backward Simulations Part I: Untimed Systems. 121 (09 1995), 214–233.

Nancy A. Lynch and Mark R. Tuttle. 1989. An introduction to input/output automata. *CWI Quarterly* 2 (1989).

Zohar Manna and Amir Pnueli. 1992. *The Temporal Logic of Reactive and Concurrent Systems*. Springer-Verlag New York, Inc., New York, NY, USA.

Ognjen Marić, Christoph Sprenger, and David Basin. 2017. Cutoff Bounds for Consensus Algorithms. In *Computer Aided Verification*, Rupak Majumdar and Viktor Kunčak (Eds.). Springer International Publishing, Cham, 217–237.

Heather Miller, Philipp Haller, Normen Müller, and Jocelyn Boullier. 2016. Function Passing: A Model for Typed, Distributed Functional Programming. In *Proceedings of the 2016 ACM International Symposium on New Ideas, New Paradigms, and Reflections on Programming and Software (Onward! 2016)*. ACM, New York, NY, USA, 82–97. https://doi.org/10.1145/2986012.2986014

Madanlal Musuvathi and Dawson R. Engler. 2004. Model Checking Large Network Protocol Implementations. In *Proc. NSDI*.

Chris Newcombe, Tim Rath, Fan Zhang, Bogdan Munteanu, Marc Brooker, and Michael Deardeuff. 2015. How Amazon Web Services Uses Formal Methods. *Commun. ACM* 58, 4 (March 2015), 66–73. https://doi.org/10.1145/2699417

Oded Padon, Kenneth L. McMillan, Aurojit Panda, Mooly Sagiv, and Sharon Shoham. 2016. Ivy: Safety Verification by Interactive Generalization. In *Proceedings of the 37th ACM SIGPLAN Conference on Programming Language Design and Implementation (PLDI '16)*. ACM, New York, NY, USA, 614–630. https://doi.org/10.1145/2908080.2908118

Larry L. Peterson and Bruce S. Davie. 2003. *Computer Networks: A Systems Approach, 3rd Edition*. Morgan Kaufmann Publishers Inc., San Francisco, CA, USA.

Vincent Rahli. 2012. Interfacing with Proof Assistants for Domain Specific Programming Using EventML. (2012). 10th International Workshop on User Interfaces for Theorem Provers.

Vincent Rahli, Ivana Vukotic, Marcus Völp, and Paulo Esteves-Verissimo. 2018. Velisarios: Byzantine Fault-Tolerant Protocols Powered by Coq. In *Programming Languages and Systems*, Amal Ahmed (Ed.). Springer International Publishing, Cham, 619–650.

John C. Reynolds. 2002. Separation Logic: A Logic for Shared Mutable Data Structures. In *Proceedings of the 17th Annual IEEE Symposium on Logic in Computer Science (LICS '02)*. IEEE Computer Society, Washington, DC, USA, 55–74. http://dl.acm.org/citation.cfm?id=645683.664578

Guido Salvaneschi, Mirko Köhler, Daniel Sokolowski, Philipp Haller, Sebastian Erdweg, and Mira Mezini. 2019. Language-integrated privacy-aware distributed queries. *Proceedings of the ACM on Programming Languages* 3, OOPSLA (2019), 1–30.

Roopsha Samanta, Jyotirmoy V. Deshmukh, and Swarat Chaudhuri. 2013. Robustness Analysis of Networked Systems. In *Verification, Model Checking, and Abstract Interpretation*, Roberto Giacobazzi, Josh Berdine, and Isabella Mastroeni (Eds.). Springer Berlin Heidelberg, Berlin, Heidelberg, 229–247.

N. Schiper, V. Rahli, R. van Renesse, M. Bickford, and R.L. Constable. 2014. Developing Correctly Replicated Databases Using Formal Tools. In *Proc. DSN*.

Ilya Sergey, James R. Wilcox, and Zachary Tatlock. 2017. Programming and Proving with Distributed Protocols. *Proc. ACM Program. Lang.* 2, POPL, Article 28 (Dec. 2017), 30 pages. https://doi.org/10.1145/3158116

Marcelo Taube, Giuliano Losa, Kenneth McMillan, Oded Padon, Mooly Sagiv, Sharon Shoham, James R. Wilcox, , and Doug Woos. 2018. Modularity for Decidability of Deductive Verification with Applications to Distributed Systems. In *Proc. PLDI*.

Web. 2018a. Bitcoin Spinoff Hacked in Rare 51% Attack. http://fortune.com/2018/05/29/bitcoin-gold-hack/. (2018). Accessed: 2018-06-23.

Web. 2018b. Hackers have stolen about 14% of big digital currencies. http://www.latimes.com/business/la-fi-bitcoin-stolen-hackers-20180118-story.html. (2018). Accessed: 2018-06-23.

Web. 2018c. High Scalability. http://highscalability.com/blog/2011/5/2/the-updated-big-list-of-articles-on-the-amazon-outage.html. (2018). Accessed: 2018-06-23.

Pascal Weisenburger, Mirko Köhler, and Guido Salvaneschi. 2018. Distributed system development with ScalaLoci. *Proceedings of the ACM on Programming Languages* 2, OOPSLA (2018), 129.

James R. Wilcox, Ilya Sergey, and Zachary Tatlock. 2017. Programming Language Abstractions for Modularly Verified Distributed Systems. In *2nd Summit on Advances in Programming Languages (SNAPL 2017) (Leibniz International Proceedings in Informatics (LIPIcs))*, Benjamin S. Lerner, Rastislav Bodík, and Shriram Krishnamurthi (Eds.), Vol. 71. Schloss Dagstuhl–Leibniz-Zentrum fuer Informatik, Dagstuhl, Germany, 19:1–19:12. https://doi.org/10.4230/LIPIcs.SNAPL.2017.19

James R. Wilcox, Doug Woos, Pavel Panchekha, Zachary Tatlock, Xi Wang, Michael D. Ernst, and Thomas Anderson. 2015. Verdi: A framework for implementing and formally verifying distributed system. In *Proc. PLDI*.

Doug Woos, James R. Wilcox, Steve Anton, Zachary Tatlock, Michael D. Ernst, and Thomas Anderson. 2016. Planning for Change in a Formal Verification of the Raft Consensus Protocol. In *Proceedings of the 5th ACM SIGPLAN Conference on Certified Programs and Proofs (CPP 2016)*. ACM, New York, NY, USA, 154–165. https://doi.org/10.1145/2854065.2854081

Maysam Yabandeh, Nikola Knezevic, Dejan Kostic, and Viktor Kuncak. 2009. CrystalBall: Predicting and Preventing Inconsistencies in Deployed Distributed Systems. In *Proc. NSDI*.

Junfeng Yang, Tisheng Chen, Ming Wu, Zhilei Xu, Xuezheng Liu, Haoxiang Lin, Mao Yang, Fan Long, Lintao Zhang, and Lidong Zhou. 2009. MODIST: Transparent Model Checking of Unmodified Distributed Systems. In *Proc. NSDI*.

Pamela Zave. 2012. Using Lightweight Modeling to Understand Chord. *SIGCOMM Comput. Commun. Rev.* 42, 2 (March 2012), 49–57. https://doi.org/10.1145/2185376.2185383

# Temporal Logic of Composable Distributed Components Appendix

## Contents

# 1 Logic

## 1.1 Sequent Logic

$$\text{I} \over \mathcal{A} \vdash \mathcal{A}$$

$$\text{THIN} \quad {\Gamma \vdash \mathcal{A}' \over \Gamma, \mathcal{A} \vdash \mathcal{A}'}$$

$$\text{CONTRACTION} \quad {\Gamma, \mathcal{A}, \mathcal{A} \vdash \mathcal{A}' \over \Gamma, \mathcal{A} \vdash \mathcal{A}'}$$

$$\text{EXCHANGE} \quad {\Gamma, \mathcal{A}, \mathcal{A}' \vdash \mathcal{A}'' \over \Gamma, \mathcal{A}', \mathcal{A} \vdash \mathcal{A}''}$$

$$\text{CUT} \quad {\Gamma \vdash \mathcal{A} \quad \Gamma, \mathcal{A} \vdash \mathcal{A}' \over \Gamma \vdash \mathcal{A}'}$$

$$\neg l \quad {\Gamma \vdash \mathcal{A} \over \Gamma, \neg \mathcal{A} \vdash \mathcal{A}'}$$

$$\neg r \quad {\Gamma, \mathcal{A} \vdash \bot \over \Gamma \vdash \neg \mathcal{A}}$$

$$\wedge l \quad {\Gamma, \mathcal{A}, \mathcal{A}' \vdash \mathcal{A}'' \over \Gamma, \mathcal{A} \wedge \mathcal{A}' \vdash \mathcal{A}''}$$

$$\wedge r \quad {\Gamma \vdash \mathcal{A} \quad \Gamma \vdash \mathcal{A}' \over \Gamma \vdash \mathcal{A} \wedge \mathcal{A}'}$$

$$\vee l \quad {\Gamma, \mathcal{A} \vdash \mathcal{A}'' \quad \Gamma, \mathcal{A}' \vdash \mathcal{A}'' \over \Gamma, \mathcal{A} \vee \mathcal{A}' \vdash \mathcal{A}''}$$

$$\vee rl \quad {\Gamma \vdash \mathcal{A} \over \Gamma \vdash \mathcal{A} \vee \mathcal{A}'}$$

$$\vee rr \quad {\Gamma \vdash \mathcal{A}' \over \Gamma \vdash \mathcal{A} \vee \mathcal{A}'}$$

$$\to l \quad {\Gamma \vdash \mathcal{A} \quad \Gamma, \mathcal{A}' \vdash \mathcal{A}'' \over \Gamma, \mathcal{A} \to \mathcal{A}' \vdash \mathcal{A}''}$$

$$\to r \quad {\Gamma, \mathcal{A} \vdash \mathcal{A}' \over \Gamma \vdash \mathcal{A} \to \mathcal{A}'}$$

$$\forall l \quad {\Gamma, \mathcal{A}(x') \vdash \mathcal{A}' \over \Gamma, \forall x.\ \mathcal{A}(x) \vdash \mathcal{A}'}$$

$$\forall r \quad {\Gamma \vdash \mathcal{A}(x') \quad x' \text{ fresh} \over \Gamma \vdash \forall x.\ \mathcal{A}(x)}$$

$$\exists l \quad {\Gamma, \mathcal{A}(x) \vdash \mathcal{A}' \quad x' \text{ fresh} \over \Gamma, \exists x.\ \mathcal{A}(x) \vdash \mathcal{A}'}$$

$$\exists r \quad {\Gamma \vdash \mathcal{A}(x') \over \Gamma \vdash \exists x.\ \mathcal{A}(x)}$$

Figure 1: Basic rules

## 1.2 Basic Rules

Node
$\vdash_c \Box\ \mathsf{n} \in \mathbb{N}$

IR
$\vdash_c \top \downarrow e\ \Rightarrow\ (\mathsf{s'(n)}, \mathsf{ors}, \mathsf{ois}) = \mathsf{request}_c(\mathsf{n}, \mathsf{s(n)}, e)$

II
$\vdash_c i \uparrow e\ \Rightarrow\ (\mathsf{s'(n)}, \mathsf{ors}, \mathsf{ois}) = \mathsf{indication}_c(\mathsf{n}, \mathsf{s(n)}, (i, e))$

PE
$\vdash_c \top \wr \mathsf{per}\ \Rightarrow\ (\mathsf{s'(n)}, \mathsf{ors}, \mathsf{ois}) = \mathsf{periodic}_c(\mathsf{n}, \mathsf{s(n)})$

OR
$\vdash_c n \bullet (i, e) \in \mathsf{ors} \wedge \mathsf{self}\ \Rightarrow\ \hat{\Diamond}(n \bullet i \downarrow e)$

OI
$\vdash_c n \bullet e \in \mathsf{ois} \wedge \mathsf{self}\ \Rightarrow\ \hat{\Diamond}(n \bullet \top \uparrow e)$

OR'
$\vdash_c n \bullet i \downarrow e\ \Rightarrow\ \hat{\ominus}(n \bullet (i, e) \in \mathsf{ors} \wedge \mathsf{self})$

OI'
$\vdash_c n \bullet \top \uparrow e\ \Rightarrow\ \hat{\ominus}(n \bullet e \in \mathsf{ois} \wedge \mathsf{self})$

Init
$\vdash_c \mathbb{S}\ (\mathsf{s} = \lambda n.\ \mathsf{init}_c(n))$

PostPre
$\vdash_c \mathbb{S}\ (\mathsf{s'} = s\ \Leftrightarrow\ \bigcirc \mathsf{s} = s)$

SEq
$\vdash_c \mathsf{n} \neq n\ \Rightarrow\ \mathsf{s'}(n) = \mathsf{s}(n)$

ASelf
$\vdash_c \mathbb{S}\ \Box\ \mathsf{self}$

SInv
$\vdash_c (\mathbb{S}\ \mathcal{I})\ \leftrightarrow\ \mathsf{restrict}(\mathsf{self}, \mathcal{I})$

APer
$\vdash_c n \in \mathsf{Correct}\ \leftrightarrow\ \Box \Diamond(n \bullet \top \wr \mathsf{per})$

Figure 2: Program Logic

RSEQ
$$\vdash_c \mathsf{r} = r \;\Rightarrow\; \hat{\boxminus}(\mathsf{r} \leq r)$$

GST
$$\vdash_c n' \in \mathsf{Correct} \land r > r_{GST} \land$$
$$(n \bullet d \downarrow \mathsf{send}_\mathsf{l}(n', m) \land \mathsf{r} = r) \Rightarrow$$
$$\Diamond(n' \bullet d \uparrow \mathsf{deliver}_\mathsf{l}(n, m) \land \mathsf{r} = r)$$

FDUP
$$\vdash_c \Box\Diamond(n' \bullet d \uparrow \mathsf{deliver}_\mathsf{l}(n, m)) \to \Box\Diamond(n \bullet d \downarrow \mathsf{send}_\mathsf{l}(n', m))$$

NFORGE
$$\vdash_c (n' \bullet d \uparrow \mathsf{deliver}_\mathsf{l}(n, m)) \Rightarrow \hat{\Diamond}(n \bullet d \downarrow \mathsf{send}_\mathsf{l}(n', m))$$

UNIOR
$$\vdash_c (\mathsf{occ}(\mathsf{ors}, e) \leq 1 \land$$
$$\hat{\boxminus}(\mathsf{n} = n \land \mathsf{self} \to (i, e) \notin \mathsf{ors}) \land$$
$$\hat{\Box}(\mathsf{n} = n \land \mathsf{self} \to (i, e) \notin \mathsf{ors})) \Rightarrow$$
$$(n \bullet i \downarrow e) \Rightarrow$$
$$\hat{\boxminus}\neg(n \bullet i \downarrow e) \land \hat{\Box}\neg(n \bullet i \downarrow e)$$

UNIOI
$$\vdash_c (\mathsf{occ}(\mathsf{ois}, e) \leq 1 \land$$
$$\hat{\boxminus}(\mathsf{n} = n \land \mathsf{self} \to e \notin \mathsf{ois}) \land$$
$$\hat{\Box}(\mathsf{n} = n \land \mathsf{self} \to e \notin \mathsf{ois})) \Rightarrow$$
$$(n \bullet \top \uparrow e) \Rightarrow$$
$$\hat{\boxminus}\neg(n \bullet \top \uparrow e) \land \hat{\Box}\neg(n \bullet \top \uparrow e)$$

EXEORDEROR
$$\vdash_c [n \bullet (i, e) \in \mathsf{ors} \land \mathsf{self} \land$$
$$\hat{\Diamond}(n \bullet (i, e') \in \mathsf{ors} \land \mathsf{self})] \Rightarrow$$
$$\hat{\Diamond}[n \bullet i \downarrow e \Rightarrow$$
$$\hat{\Diamond}(n \bullet i \downarrow e)]$$

EXEORDEROI
$$\vdash_c [n \bullet e \in \mathsf{ois} \land \mathsf{self} \land$$
$$\hat{\Diamond}(n \bullet e' \in \mathsf{ois} \land \mathsf{self})] \Rightarrow$$
$$\hat{\Diamond}[n \bullet \top \uparrow e \Rightarrow$$
$$\hat{\Diamond}(n \bullet \top \uparrow e)]$$

EXEFEOI
$$\vdash_c \Box\neg(n \bullet e \in \mathsf{ois} \land \mathsf{self}) \Rightarrow$$
$$\Diamond\Box\neg(n \bullet i \uparrow e)$$

EXEFEOR
$$\vdash_c \Box\neg(n \bullet (i, e) \in \mathsf{ors} \land \mathsf{self}) \Rightarrow$$
$$\Diamond\Box\neg(n \bullet i \downarrow e)$$

Figure 3: Program Logic

## 1.3 Derived Rules

FLoss
$\vdash_c n' \in \mathsf{Correct} \rightarrow$
$\quad \Box\Diamond(n \bullet d \downarrow \mathsf{send}_l(n', m)) \rightarrow \Box\Diamond(n' \bullet d \uparrow \mathsf{deliver}_l(n, m))$

IRSe
$\vdash_c \circledS\, [\top \downarrow e \;\Rightarrow\; (\mathsf{s}'(\mathsf{n}), \mathsf{ois}, \mathsf{ors}) = \mathsf{request}_c(\mathsf{n}, \mathsf{s}(\mathsf{n}), e)]$

PeSe
$\vdash_c \circledS\, [\top \mathbin{\updownarrow} \mathsf{per} \;\Rightarrow\; (\mathsf{s}'(\mathsf{n}), \mathsf{ois}, \mathsf{ors}) = \mathsf{periodic}_c(\mathsf{n}, \mathsf{s}(\mathsf{n}))]$

IISe
$\vdash_c \circledS\, [i \uparrow e \;\Rightarrow\; (\mathsf{s}'(\mathsf{n}), \mathsf{ois}, \mathsf{ors}) = \mathsf{indication}_c(\mathsf{n}, \mathsf{s}(\mathsf{n}), (i, e))]$

ORSe
$\vdash_c \circledS\, [n \bullet (i, e) \in \mathsf{ors} \;\Rightarrow\; \hat{\Diamond}(n \bullet i \downarrow e)]$

OISe
$\vdash_c \circledS\, [n \bullet e \in \mathsf{ois} \;\Rightarrow\; \hat{\Diamond}(n \bullet \top \uparrow e)]$

ORSe'
$\vdash_c \circledS\, [n \bullet i \downarrow e \;\Rightarrow\; \hat{\ominus}(n \bullet (i, e) \in \mathsf{ors})]$

OISe'
$\vdash_c \circledS\, [n \bullet \top \uparrow e \;\Rightarrow\; \hat{\ominus}(n \bullet e \in \mathsf{ois})]$

Figure 4: Program Logic, Derived rules. $S$ is a predicate on state.

IROI
$$\frac{\forall s.\ S(s) \wedge \mathsf{request}_c(n,s,e) = (\_, \mathsf{ois}, \_) \to e' \in \mathsf{ois}}{\vdash_c n \bullet \top \downarrow e \wedge S(\mathsf{s}(n)) \;\Rightarrow\; \Diamond(n \bullet \top \uparrow e')}$$

IIOI
$$\frac{\forall s.\ S(s) \wedge \mathsf{indication}_c(n,s,(i,e)) = (\_, \mathsf{ois}, \_) \to e' \in \mathsf{ois}}{\vdash_c n \bullet i \uparrow e \wedge S(\mathsf{s}(n)) \;\Rightarrow\; \Diamond(n \bullet \top \uparrow e')}$$

PeOI
$$\frac{\forall s.\ S(s) \wedge \mathsf{periodic}_c(n,s) = (\_, \mathsf{ois}, \_) \to e' \in \mathsf{ois}}{\vdash_c n \bullet i \wr \mathsf{per} \wedge S(\mathsf{s}(n)) \;\Rightarrow\; \Diamond(n \bullet \top \uparrow e')}$$

IROR
$$\frac{\forall s.\ S(s) \wedge \mathsf{request}_c(n,s,e) = (\_, \_, \mathsf{ors}) \to (i,e') \in \mathsf{ors}}{\vdash_c n \bullet \top \downarrow e \wedge S(\mathsf{s}(n)) \;\Rightarrow\; \Diamond(n \bullet i \downarrow e')}$$

IIOR
$$\frac{\forall s.\ S(s) \wedge \mathsf{indication}_c(n,s,(i,e)) = (\_, \_, \mathsf{ors}) \to (i,e') \in \mathsf{ors}}{\vdash_c n \bullet i \uparrow e \wedge S(\mathsf{s}(n)) \;\Rightarrow\; \Diamond(n \bullet i \downarrow e')}$$

PeOR
$$\frac{\forall s.\ S(s) \wedge \mathsf{periodic}_c(n,s) = (\_, \_, \mathsf{ors}) \to (i,e') \in \mathsf{ors}}{\vdash_c n \bullet i \wr \mathsf{per} \wedge S(\mathsf{s}(n)) \;\Rightarrow\; \Diamond(n \bullet i \downarrow e')}$$

Figure 5: Program Logic, Derived rules. $S$ is a predicate on state.

UniOISe
$\vdash_c \ \textcircled{S} \ (\text{occ}(\text{ois}, e) \leq 1 \ \land$
$\quad \hat{\boxminus}(\mathsf{n} = n \to e \notin \text{ois}) \ \land$
$\quad \hat{\square}(\mathsf{n} = n \to e \notin \text{ois})) \Rightarrow$
$\quad (n \bullet \top \uparrow e) \Rightarrow \hat{\boxminus}\neg(n \bullet \top \uparrow e) \land \hat{\square}\neg(n \bullet \top \uparrow e)$

UniORSe
$\vdash_c \ \textcircled{S} \ (\text{occ}(\text{ors}, e) \leq 1 \ \land$
$\quad \hat{\boxminus}(\mathsf{n} = n \to (i, e) \notin \text{ors}) \ \land$
$\quad \hat{\square}(\mathsf{n} = n \to (i, e) \notin \text{ors})) \Rightarrow$
$\quad (n \bullet i \downarrow e) \Rightarrow \hat{\boxminus}\neg(n \bullet i \downarrow e) \land \hat{\square}\neg(n \bullet i \downarrow e)$

IROISe
$$\frac{\forall s.\ S(s) \land \text{request}_c(n, s, e) = (\_, \text{ois}, \_) \to e' \in \text{ois}}{\vdash_c \textcircled{S}\ n \bullet \top \downarrow e \land S(\mathsf{s}(n)) \ \Rightarrow\ \Diamond(n \bullet \top \uparrow e')}$$

IIOISe
$$\frac{\forall s.\ S(s) \land \text{indication}_c(n, s, (i,e)) = (\_, \text{ois}, \_) \to e' \in \text{ois}}{\vdash_c \textcircled{S}\ n \bullet i \uparrow e \land S(\mathsf{s}(n)) \ \Rightarrow\ \Diamond(n \bullet \top \uparrow e')}$$

PeOISe
$$\frac{\forall s.\ S(s) \land \text{periodic}_c(n, s) = (\_, \text{ois}, \_) \to e' \in \text{ois}}{\vdash_c \textcircled{S}\ n \bullet i \wr \text{per} \land S(\mathsf{s}(n)) \ \Rightarrow\ \Diamond(n \bullet \top \uparrow e')}$$

IRORSe
$$\frac{\forall s.\ S(s) \land \text{request}_c(n, s, e) = (\_, \_, \text{ors}) \to (i, e') \in \text{ors}}{\vdash_c \textcircled{S}\ n \bullet \top \downarrow e \land S(\mathsf{s}(n)) \ \Rightarrow\ \Diamond(n \bullet i \downarrow e')}$$

IIORSe
$$\frac{\forall s.\ S(s) \land \text{indication}_c(n, s, (i,e)) = (\_, \_, \text{ors}) \to (i, e') \in \text{ors}}{\vdash_c \textcircled{S}\ n \bullet i \uparrow e \land S(\mathsf{s}(n)) \ \Rightarrow\ \Diamond(n \bullet i \downarrow e')}$$

PeORSe
$$\frac{\forall s.\ S(s) \land \text{periodic}_c(n, s) = (\_, \_, \text{ors}) \to (i, e') \in \text{ors}}{\vdash_c \textcircled{S}\ n \bullet i \wr \text{per} \land S(\mathsf{s}(n)) \ \Rightarrow\ \Diamond(n \bullet i \downarrow e')}$$

APerSe
$\vdash_c \textcircled{S}\ n \in \text{Correct} \to \square\Diamond(n \bullet \top \wr \text{per})$

Quorum
$|\text{Correct}| > t_1 \ \vdash_c$
$N \subseteq \mathbb{N} \ \land\ |N| > t_2 \ \land\ t_1 + t_2 \geq |\mathbb{N}| \Rightarrow \exists n.\ n \in N \land n \in \text{Correct}$

Figure 6: Program Logic, Derived rules

SEQSE
$\vdash_c \circledS [\mathsf{n} \neq n \Rightarrow \mathsf{s}'(n) = \mathsf{s}(n)]$

CSELF
$n \bullet \mathsf{self} \Leftrightarrow$
$(\exists e.\ n \bullet \top \downarrow e) \vee (\exists i, e.\ n \bullet \top \uparrow (i, e)) \vee (n \bullet \top \wr \mathsf{per})$

INVSE
$\Gamma \vdash_c \circledS \forall e.\ \top \downarrow e \Rightarrow \mathcal{A}$
$\Gamma \vdash_c \circledS \forall i, e.\ i \uparrow e \Rightarrow \mathcal{A}$
$\Gamma \vdash_c \circledS \top \wr \mathsf{per} \Rightarrow \mathcal{A}$
$\overline{\Gamma \vdash_c \circledS \Box \mathcal{A}}$

INVSE'
$\Gamma \vdash_c \circledS \forall e.\ \top \downarrow e \wedge \hat{\boxminus} \mathcal{A} \Rightarrow \mathcal{A}$
$\Gamma \vdash_c \circledS \forall i, e.\ i \uparrow e \wedge \hat{\boxminus} \mathcal{A} \Rightarrow \mathcal{A}$
$\Gamma \vdash_c \circledS \top \wr \mathsf{per} \wedge \hat{\boxminus} \mathcal{A} \Rightarrow \mathcal{A}$
$\overline{\Gamma \vdash_c \circledS \Box \mathcal{A}}$

INVUSE
$\Gamma \vdash_c \circledS \forall e.\ \top \downarrow e \wedge (\mathsf{s}'(\mathsf{n}), \mathsf{ois}, \mathsf{ors}) = \mathsf{request}_c(\mathsf{n}, \mathsf{s}(\mathsf{n}), e) \wedge$
$\hat{\boxminus} \mathcal{A} \Rightarrow \mathcal{A}$
$\Gamma \vdash_c \circledS \forall e, i.\ i \uparrow e \wedge (\mathsf{s}'(\mathsf{n}), \mathsf{ois}, \mathsf{ors}) = \mathsf{indication}_c(\mathsf{n}, \mathsf{s}(\mathsf{n}), (i, e)) \wedge$
$\hat{\boxminus} \mathcal{A} \Rightarrow \mathcal{A}$
$\Gamma \vdash_c \circledS \top \wr \mathsf{per} \wedge (\mathsf{s}'(\mathsf{n}), \mathsf{ois}, \mathsf{ors}) = \mathsf{periodic}_c(\mathsf{n}, \mathsf{s}(\mathsf{n})) \wedge$
$\hat{\boxminus} \mathcal{A} \Rightarrow \mathcal{A}$
$\overline{\Gamma \vdash_c \circledS \Box \mathcal{A}}$

INVMSE
$\Gamma \vdash_c \circledS \forall e.\ \top \downarrow e \wedge \hat{\boxminus} \mathcal{A} \wedge \hat{\boxminus} \mathcal{A}' \Rightarrow \mathcal{A} \wedge \mathcal{A}'$
$\Gamma \vdash_c \circledS \forall e, i.\ i \uparrow e \wedge \hat{\boxminus} \mathcal{A} \wedge \hat{\boxminus} \mathcal{A}' \Rightarrow \mathcal{A} \wedge \mathcal{A}'$
$\Gamma \vdash_c \circledS \top \wr \mathsf{per} \wedge \hat{\boxminus} \mathcal{A} \wedge \hat{\boxminus} \mathcal{A}' \Rightarrow \mathcal{A} \wedge \mathcal{A}'$
$\overline{\Gamma \vdash_c \circledS \Box (\mathcal{A} \wedge \mathcal{A}')}$

INVMSE'
$\Gamma \vdash_c \circledS \forall e.\ \top \downarrow e \wedge (\mathsf{s}'(\mathsf{n}), \mathsf{ois}, \mathsf{ors}) = \mathsf{request}_c(\mathsf{n}, \mathsf{s}(\mathsf{n}), e) \wedge$
$\hat{\boxminus} \mathcal{A} \wedge \hat{\boxminus} \mathcal{A}' \Rightarrow \mathcal{A} \wedge \mathcal{A}'$
$\Gamma \vdash_c \circledS \forall e, i.\ i \uparrow e \wedge (\mathsf{s}'(\mathsf{n}), \mathsf{ois}, \mathsf{ors}) = \mathsf{indication}_c(\mathsf{n}, \mathsf{s}(\mathsf{n}), (i, e)) \wedge$
$\hat{\boxminus} \mathcal{A} \wedge \hat{\boxminus} \mathcal{A}' \Rightarrow \mathcal{A} \wedge \mathcal{A}'$
$\Gamma \vdash_c \circledS \top \wr \mathsf{per} \wedge (\mathsf{s}'(\mathsf{n}), \mathsf{ois}, \mathsf{ors}) = \mathsf{periodic}_c(\mathsf{n}, \mathsf{s}(\mathsf{n})) \wedge$
$\hat{\boxminus} \mathcal{A} \wedge \hat{\boxminus} \mathcal{A}' \Rightarrow \mathcal{A} \wedge \mathcal{A}'$
$\overline{\Gamma \vdash_c \circledS \Box (\mathcal{A} \wedge \mathcal{A}')}$

Figure 7: Program Logic, Derived rules. $S$ is a predicate on state.

INVLSE
$$\forall e.\ \top \downarrow e \wedge \mathsf{request}_c(\mathsf{n},\mathsf{s}(\mathsf{n}),e) = (\mathsf{s}'(\mathsf{n}),\mathsf{ois},\mathsf{ors}) \to \mathcal{A}$$
$$\forall e,i.\ i \uparrow e \wedge \mathsf{indication}_c(\mathsf{n},\mathsf{s}(\mathsf{n}),(i,e)) = (\mathsf{s}'(\mathsf{n}),\mathsf{ois},\mathsf{ors}) \to \mathcal{A}$$
$$\top \updownarrow \mathsf{per} \wedge \mathsf{periodic}_c(\mathsf{n},\mathsf{s}(\mathsf{n})) = (\mathsf{s}'(\mathsf{n}),\mathsf{ois},\mathsf{ors}) \to \mathcal{A}$$
$$\mathcal{A}\ \text{non-temporal}$$
$$\vdash_c \circledS\ \square \mathcal{A}$$

INVL
$$\forall e.\ \top \downarrow e \wedge \mathsf{request}_c(\mathsf{n},\mathsf{s}(\mathsf{n}),e) = (\mathsf{s}'(\mathsf{n}),\mathsf{ois},\mathsf{ors}) \to \mathcal{A}$$
$$\forall e,i.\ i \uparrow e \wedge \mathsf{indication}_c(\mathsf{n},\mathsf{s}(\mathsf{n}),(i,e)) = (\mathsf{s}'(\mathsf{n}),\mathsf{ois},\mathsf{ors}) \to \mathcal{A}$$
$$\top \updownarrow \mathsf{per} \wedge \mathsf{periodic}_c(\mathsf{n},\mathsf{s}(\mathsf{n})) = (\mathsf{s}'(\mathsf{n}),\mathsf{ois},\mathsf{ors}) \to \mathcal{A}$$
$$\mathcal{A}\ \text{non-temporal}$$
$$\vdash_c \mathsf{self} \Rightarrow \mathcal{A}$$

INVUSSE
$$\Gamma \vdash_c \circledS\ \forall e.\ n \bullet \top \downarrow e\ \wedge$$
$$(\mathsf{s}'(n),\mathsf{ois},\mathsf{ors}) = \mathsf{request}_c(n,\mathsf{s}(n),e)\ \wedge$$
$$S(\mathsf{s}(n)) \Rightarrow S(\mathsf{s}'(n))$$

$$\Gamma \vdash_c \circledS\ \forall e,i.\ n \bullet i \uparrow e\ \wedge$$
$$(\mathsf{s}'(n),\mathsf{ois},\mathsf{ors}) = \mathsf{indication}_c(n,\mathsf{s}(n),(i,e))\ \wedge$$
$$S(\mathsf{s}(n)) \Rightarrow S(\mathsf{s}'(n))$$

$$\Gamma \vdash_c \circledS\ n \bullet \top \updownarrow \mathsf{per}\ \wedge$$
$$(\mathsf{s}'(n),\mathsf{ois},\mathsf{ors}) = \mathsf{periodic}_c(n,\mathsf{s}(n))\ \wedge$$
$$S(\mathsf{s}(n)) \Rightarrow S(\mathsf{s}'(n))$$
$$\Gamma \vdash_c \circledS\ [S(\mathsf{s}(n)) \Rightarrow \square S(\mathsf{s}(n))]$$

INVUSSE$'$
$$S(\mathsf{init}_c(n))$$

$$\Gamma \vdash_c \circledS\ \forall e.\ n \bullet \top \downarrow e\ \wedge$$
$$(\mathsf{s}'(n),\mathsf{ois},\mathsf{ors}) = \mathsf{request}_c(n,\mathsf{s}(n),e)\ \wedge$$
$$S(\mathsf{s}(n)) \Rightarrow S(\mathsf{s}'(n))$$

$$\Gamma \vdash_c \circledS\ \forall e,i.\ n \bullet i \uparrow e\ \wedge$$
$$(\mathsf{s}'(n),\mathsf{ois},\mathsf{ors}) = \mathsf{indication}_c(n,\mathsf{s}(n),(i,e))\ \wedge$$
$$S(\mathsf{s}(n)) \Rightarrow S(\mathsf{s}'(n))$$

$$\Gamma \vdash_c \circledS\ n \bullet \top \updownarrow \mathsf{per}\ \wedge$$
$$(\mathsf{s}'(n),\mathsf{ois},\mathsf{ors}) = \mathsf{periodic}_c(n,\mathsf{s}(n))\ \wedge$$
$$S(\mathsf{s}(n)) \Rightarrow S(\mathsf{s}'(n))$$
$$\Gamma \vdash_c \circledS\ \square S(\mathsf{s}(n))$$

Figure 8: Program Logic, Derived rules

INVSSE
$$\frac{\begin{array}{c}\forall s,e,s'.\ \ S(s) \wedge \mathsf{request}_c(n,s,e) = (s',\_,\_) \rightarrow S(s')\\ \forall s,i,e,s'.\ \ S(s) \wedge \mathsf{indication}_c(n,s,(i,e)) = (s',\_,\_) \rightarrow S(s')\\ \forall s,s'.\ \ S(s) \wedge \mathsf{periodic}_c(n,s) = (s',\_,\_) \rightarrow S(s')\end{array}}{\vdash_c \text{\textcircled{S}}\,[S(\mathsf{s}(n)) \Rightarrow \Box S(\mathsf{s}(n))]}$$

INVS
$$\frac{\begin{array}{c}\forall s,e,s'.\ \ S(s) \wedge \mathsf{request}_c(n,s,e) = (s',\_,\_) \rightarrow S(s')\\ \forall s,i,e,s'.\ \ S(s) \wedge \mathsf{indication}_c(n,s,(i,e)) = (s',\_,\_) \rightarrow S(s')\\ \forall s,s'.\ \ S(s) \wedge \mathsf{periodic}_c(n,s) = (s',\_,\_) \rightarrow S(s')\end{array}}{\vdash_c \mathsf{self} \wedge S(\mathsf{s}(n)) \Rightarrow (\mathsf{self} \Rightarrow S(\mathsf{s}(n)))}$$

INVSSE'
$$\frac{\begin{array}{c}S(\mathsf{init}_c(n))\\ \forall s,e,s'.\ \ S(s) \wedge \mathsf{request}_c(n,s,e) = (s',\_,\_) \rightarrow S(s')\\ \forall s,i,e,s'.\ \ S(s) \wedge \mathsf{indication}_c(n,s,(i,e)) = (s',\_,\_) \rightarrow S(s')\\ \forall s,s'.\ \ S(s) \wedge \mathsf{periodic}_c(n,s) = (s',\_,\_) \rightarrow S(s')\end{array}}{\vdash_c \text{\textcircled{S}}\,\Box S(\mathsf{s}(n))}$$

INVS'
$$\frac{\begin{array}{c}S(\mathsf{init}_c(n))\\ \forall s,e,s'.\ \ S(s) \wedge \mathsf{request}_c(n,s,e) = (s',\_,\_) \rightarrow S(s')\\ \forall s,i,e,s'.\ \ S(s) \wedge \mathsf{indication}_c(n,s,(i,e)) = (s',\_,\_) \rightarrow S(s')\\ \forall s,s'.\ \ S(s) \wedge \mathsf{periodic}_c(n,s) = (s',\_,\_) \rightarrow S(s')\end{array}}{\vdash_c (\mathsf{self} \Rightarrow S(\mathsf{s}(n)))}$$

Figure 9: Program Logic, Derived rules

INVSSE″
$$\forall s, e, s'. \ S(s) \land \mathsf{request}_c(n, s, e) = (s', \_, \_) \to S(s')$$
$$\forall s, i, e, s'. \ S(s) \land \mathsf{indication}_c(n, s, (i, e)) = (s', \_, \_) \to S(s')$$
$$\forall s, s'. \ S(s) \land \mathsf{periodic}_c(n, s) = (s', \_, \_) \to S(s')$$
$$\vdash_c \circledS \ S(\mathsf{s}'(n)) \Rightarrow \hat{\Box} S(\mathsf{s}(n))$$

INVS″
$$\forall s, e, s'. \ S(s) \land \mathsf{request}_c(n, s, e) = (s', \_, \_) \to S(s')$$
$$\forall s, i, e, s'. \ S(s) \land \mathsf{indication}_c(n, s, (i, e)) = (s', \_, \_) \to S(s')$$
$$\forall s, s'. \ S(s) \land \mathsf{periodic}_c(n, s) = (s', \_, \_) \to S(s')$$
$$\vdash_c (\mathsf{self} \land S(\mathsf{s}'(n))) \Rightarrow \hat{\Box}(\mathsf{self} \to S(\mathsf{s}(n)))$$

INVSASE
$$\neg S(\mathsf{init}_c(n))$$

$$\forall e. \ n \bullet \top \downarrow e \land$$
$$\mathsf{request}_c(n, \mathsf{s}(n), e) = (\mathsf{s}'(n), \mathsf{ois}, \mathsf{ors}) \land$$
$$\neg S(\mathsf{s}(n)) \land S(\mathsf{s}'(n)) \to \mathcal{A}$$

$$\forall e, i. \ n \bullet i \uparrow e \land$$
$$\mathsf{indication}_c(n, \mathsf{s}(n), (i, e)) = (\mathsf{s}'(n), \mathsf{ois}, \mathsf{ors}) \land$$
$$\neg S(\mathsf{s}(n)) \land S(\mathsf{s}'(n)) \to \mathcal{A}$$

$$n \bullet \top \wr \mathsf{per} \land$$
$$\mathsf{periodic}_c(\mathsf{n}, \mathsf{s}(\mathsf{n})) = (\mathsf{s}'(\mathsf{n}), \mathsf{ois}, \mathsf{ors}) \land$$
$$\neg S(\mathsf{s}(n)) \land S(\mathsf{s}'(n)) \to \mathcal{A}$$

$$\mathcal{A} \ \text{non-temporal}$$
$$\vdash_c \circledS \ [S(\mathsf{s}(n)) \Rightarrow \hat{\diamondsuit}(n \bullet \mathcal{A})]$$

INVSA
$$\neg S(\mathsf{init}_c(n))$$

$$\forall e. \ n \bullet \top \downarrow e \land$$
$$\mathsf{request}_c(n, \mathsf{s}(n), e) = (\mathsf{s}'(n), \mathsf{ois}, \mathsf{ors}) \land$$
$$\neg S(\mathsf{s}(n)) \land S(\mathsf{s}'(n)) \to \mathcal{A}$$

$$\forall e, i. \ n \bullet i \uparrow e \land$$
$$\mathsf{indication}_c(n, \mathsf{s}(n), (i, e)) = (\mathsf{s}'(n), \mathsf{ois}, \mathsf{ors}) \land$$
$$\neg S(\mathsf{s}(n)) \land S(\mathsf{s}'(n)) \to \mathcal{A}$$

$$n \bullet \top \wr \mathsf{per} \land$$
$$\mathsf{periodic}_c(\mathsf{n}, \mathsf{s}(\mathsf{n})) = (\mathsf{s}'(\mathsf{n}), \mathsf{ois}, \mathsf{ors}) \land$$
$$\neg S(\mathsf{s}(n)) \land S(\mathsf{s}'(n)) \to \mathcal{A}$$

$$\mathcal{A} \ \text{non-temporal}$$
$$\vdash_c [\mathsf{self} \land S(\mathsf{s}(n))] \Rightarrow \hat{\diamondsuit}(n \bullet \mathcal{A})$$

Figure 10: Program Logic, Derived rules

INVSAG

$$\neg S(\mathsf{init}_c(n))$$

$$\forall e.\ n \bullet \top \downarrow e \wedge$$
$$\mathsf{request}_c(n, \mathsf{s}(n), e) = (\mathsf{s}'(n), \mathsf{ois}, \mathsf{ors}) \wedge$$
$$\neg S(\mathsf{s}(n)) \wedge S(\mathsf{s}'(n)) \to \mathcal{A}$$

$$\forall e, i.\ n \bullet i \uparrow e \wedge$$
$$\mathsf{indication}_c(n, \mathsf{s}(n), (i, e)) = (\mathsf{s}'(n), \mathsf{ois}, \mathsf{ors}) \wedge$$
$$\neg S(\mathsf{s}(n)) \wedge S(\mathsf{s}'(n)) \to \mathcal{A}$$

$$n \bullet \top \updownarrow \mathsf{per} \wedge$$
$$\mathsf{periodic}_c(n, \mathsf{s}(n)) = (\mathsf{s}'(n), \mathsf{ois}, \mathsf{ors}) \wedge$$
$$\neg S(\mathsf{s}(n)) \wedge S(\mathsf{s}'(n)) \to \mathcal{A}$$

$$\mathcal{A} \text{ non-temporal}$$

$$\overline{\vdash_c [\mathsf{self} \wedge \neg S(\mathsf{s}(n)) \wedge \Diamond S(\mathsf{s}(n))] \Rightarrow \hat{\Diamond}(n \bullet \mathcal{A})}$$

Figure 11: Program Logic, Derived rules

INVMSIASE
$$\forall x.\ \neg S(\mathsf{init}_c(n))$$

$$\Gamma \vdash_c \textcircled{S}\ \forall e.\ [\top \downarrow e \land \hat{\boxminus}\mathcal{A} \land (\forall x.\ S(\mathsf{s}(\mathsf{n})) \to \hat{\diamondsuit}\mathcal{A}')] \Rightarrow \\ \mathcal{A} \land (\forall x.\ S(\mathsf{s}'(\mathsf{n})) \to \diamondsuit\mathcal{A}')$$

$$\Gamma \vdash_c \textcircled{S}\ \forall i, e.\ [i \uparrow e \land \hat{\boxminus}\mathcal{A} \land (\forall x.\ S(\mathsf{s}(\mathsf{n})) \to \hat{\diamondsuit}\mathcal{A}')] \Rightarrow \\ \mathcal{A} \land (\forall x.\ S(\mathsf{s}'(\mathsf{n})) \to \diamondsuit\mathcal{A}')$$

$$\dfrac{\Gamma \vdash_c \textcircled{S}\ [\top \wr \mathsf{per} \land \hat{\boxminus}\mathcal{A} \land (\forall x.\ S(\mathsf{s}(\mathsf{n})) \to \hat{\diamondsuit}\mathcal{A}')] \Rightarrow \\ \mathcal{A} \land (\forall x.\ S(\mathsf{s}'(\mathsf{n})) \to \diamondsuit\mathcal{A}')}{\Gamma \vdash_c \textcircled{S}\ \Box\mathcal{A} \land (\forall x.\ S(\mathsf{s}(\mathsf{n})) \Rightarrow \hat{\diamondsuit}\mathcal{A}')}$$

ASASE
$$\dfrac{\Gamma \vdash_c \textcircled{S}\ \mathcal{A} \Rightarrow S(\mathsf{s}'(n)) \\ \Gamma \vdash_c \textcircled{S}\ S(\mathsf{s}(n)) \Rightarrow \Box\mathcal{A}'}{\Gamma \vdash_c \textcircled{S}\ \mathcal{A} \Rightarrow \hat{\Box}\mathcal{A}'}$$

ASA
$$\dfrac{\Gamma \vdash_c \mathsf{self} \land \mathcal{A} \Rightarrow S(\mathsf{s}'(n)) \\ \Gamma \vdash_c \mathsf{self} \land S(\mathsf{s}(n)) \Rightarrow (\mathsf{self} \Rightarrow \mathcal{A}') \\ \mathcal{A}\ \text{non-temporal}}{\Gamma \vdash_c \mathsf{self} \land \mathcal{A} \Rightarrow \hat{\Box}(\mathsf{self} \to \mathcal{A}')}$$

APERSA
$$\dfrac{S(\mathsf{s}(n)) \land \mathsf{periodic}_c(n, \mathsf{s}(n)) = (\mathsf{s}'(n), \mathsf{ois}, \mathsf{ors}) \to \mathcal{A} \\ \mathcal{A}\ \text{non-temporal}}{\Gamma \vdash_c n \in \mathsf{Correct} \to (\mathsf{self} \Rightarrow S(\mathsf{s}(n)))\ \Rightarrow\ \Box\diamondsuit\mathcal{A}}$$

Figure 12: Program Logic, Derived rules

## 2 Reasoning about Quorums

We have successfully applied the programming model, the lowering transformation and TLC to program and verify the stacks of distributed components shown in Figure 2.(b) and (c): stubborn links, perfect links, best-effort broadcast, uniform reliable broadcast and epoch consensus. The specification and implementation of these components are available in section 4 and the detailed proofs are available in subsection 5.3. In this section, we present the uniform reliable broadcast component and present how its proof of uniformity uses the quorum inference rule. Uniform reliable broadcast guarantees that if a message is delivered to a node (even a faulty one), then it is delivered to every correct node as well. Uniform reliable broadcast guarantees that the set of messages delivered to correct nodes is the same and is a superset of the messages delivered to faulty nodes; hence the name uniform.

Uniform reliable broadcast accepts requests $\mathsf{broadcast}_{\mathsf{urb}}(m)$ to broadcast the message $m$ and issues indications $\mathsf{deliver}_{\mathsf{urb}}(n, m)$ to deliver a message $m$ broadcast by a node $n$. Figure 13.(a) presents the specification of the uniform reliable broadcast. In particular, the uniform agreement property states that if a message is delivered to a node, a correct node does not miss it. More precisely, if a message $m$ is delivered to some node (whether correct or faulty), then $m$ is eventually (in the past or future) delivered to every correct node. In this section, we focus on this property. Uniform reliable broadcast also provides the following properties. Validity: a correct node eventually receives his own broadcast messages. No-duplication: Messages are not redundantly delivered. No-forge: delivered messages are previously broadcast.

Figure 13.(b) presents the uniform reliable broadcast component URBC. It assumes that a majority of nodes is correct i.e. more than $|\mathbb{N}|/2$ nodes are correct (where $\mathbb{N}$ is the set of nodes). The high-level idea of the protocol is that each node, before delivering a message, makes sure that a quorum (majority) of nodes have acknowledged the receipt of the message. As a majority of nodes are correct, there is at least one correct node in the acknowledging nodes. Even if the sender fails, that correct node rebroadcasts the message that leads to its delivery to every correct node.

The component uses a best-effort broadcast subcomponent bebc. It stores the number of messages sent by the current node *count*, the set of delivered messages *delivered*, the set of messages that are received but are pending for acknowledgement *pending*, and a mapping from each message to the set of nodes that have acknowledged the receipt of the message *ack*. Each message is uniquely identified by the identifier of the sender node and the number of the message in that node. Messages are transmitted and stored with their identifiers. The state *count* is initialized to zero, and the other sets are all initialized to ∅.

Upon a broadcast request, the counter is incremented, the message is added to the pending set and is broadcast using bebc together with the current node identifier and the new value of the counter. Upon a delivery indication by bebc, it is recorded in the acknowledgement map that the sender has acknowledged the receipt of the message and if the message is not already in the pending set, it is added to the pending set and broadcast by bebc. Thus, every node broadcasts a message by bebc only once when it receives it for the first time. In the periodic function, messages in the pending set are iterated, and if an acknowledgement for a message is received from a quorum of nodes, and it is not delivered before, then it is delivered.

We present a proof sketch for the uniform agreement property. Let $\Gamma$ be the assumption that a quorum of nodes is correct and the lowered specification of the best-effort broadcast.

$$\Gamma = |\mathsf{Correct}| > |\mathbb{N}|/2; \ \mathcal{A}_{\mathsf{bebc}} \tag{1}$$

The premise is that a message is delivered to a node. Thus the message is previously issued. The component issues delivery of a message only in the **periodic** function when the acknowledgement set

Assumption:
|Correct| > |ℕ|/2

URB$_1$ (Validity)
$n \in$ Correct →
$(n \bullet \top \downarrow \mathsf{broadcast}_{\mathsf{urb}}(m)) \rightsquigarrow$
$(n \bullet \top \uparrow \mathsf{deliver}_{\mathsf{urb}}(n, m))$
If a correct node $n$ broadcasts a message $m$,
then $n$ itself eventually delivers $m$.

URB$_2$ (No-duplication)
$[n \bullet \top \downarrow \mathsf{broadcast}_{\mathsf{urb}}(m) \Rightarrow$
  $\hat{\boxminus} \neg (n \bullet \top \downarrow \mathsf{broadcast}_{\mathsf{urb}}(m))] \rightarrow$
$[(n' \bullet \top \uparrow \mathsf{deliver}_{\mathsf{urb}}(n, m)) \Rightarrow$
  $\hat{\boxminus} \neg (n' \bullet \top \uparrow \mathsf{deliver}_{\mathsf{urb}}(n, m))]$
If a message is broadcast at most once,
it will be delivered at most once.

URB$_3$ (No-forge)
$(n \bullet \top \uparrow \mathsf{deliver}_{\mathsf{urb}}(n', m)) \leftarrow\!\sim$
$(n' \bullet \top \downarrow \mathsf{broadcast}_{\mathsf{urb}}(m))$
If a node delivers a message $m$ with sender $n'$,
then $m$ was previously broadcast by node $n'$.

URB$_4$ (Uniform Agreement)
$n \in$ Correct →
$(n' \bullet \top \uparrow \mathsf{deliver}_{\mathsf{urb}}(n'', m)) \Rightarrow$
$\Diamond (n \bullet \top \uparrow \mathsf{deliver}_{\mathsf{urb}}(n'', m)) \vee$
$\diamondsuit\!\!\!\!\!\leftarrow (n \bullet \top \uparrow \mathsf{deliver}_{\mathsf{urb}}(n'', m))$
If a message $m$ is delivered by some node
(whether correct or faulty), then $m$ is
eventually delivered by every correct node.

(a) Specification

URBC: Component Req$_{\mathsf{urb}}$ Ind$_{\mathsf{urb}}$ (Req$_{\mathsf{beb}}$, Ind$_{\mathsf{beb}}$) ≔
  let bebc ≔ 0 in
  ⟨State ≔
    ⟨$count$: Nat,
     $delivered$: Set[⟨$M$, ℕ, Nat⟩],
     $pending$: Set[⟨$M$, ℕ, Nat⟩],
     $ack$: Map[⟨$M$, ℕ, Nat⟩, Set[ℕ]]⟩
  init ≔ λ$n$. ⟨0, ∅, ∅, ∅⟩
  request ≔ λ $n, s, ir$.
    let ⟨$c, d, p, a$⟩ ≔ $s$ in
    match $ir$ with
    | broadcast$_{\mathsf{urb}}(m)$ ⇒
        let $c' ≔ c + 1$ in
        let $p' ≔ p \cup \{⟨m, n, c'⟩\}$ in
        let $or$ ≔ (bebc, broadcast$_{\mathsf{beb}}$⟨$m, n, c'$⟩) in
        ⟨⟨$c', d, p', a$⟩, [$or$], [ ]⟩
    end
  indication ≔ λ $n', s, ii$.
    let ⟨$c, d, p, a$⟩ ≔ $s$ in
    match $ii$ with
    | (bebc, deliver$_{\mathsf{beb}}(n'', ⟨m, n, c'⟩)$) ⇒
        let $a' ≔ a[⟨m, n, c'⟩ \mapsto a(⟨m, n, c'⟩) \cup \{n''\}]$ in
        if $⟨m, n, c'⟩ \notin p$
          let $p' ≔ p \cup \{⟨m, n, c'⟩\}$ in
          let $or$ ≔ (bebc, broadcast$_{\mathsf{beb}}$⟨$m, n, c'$⟩) in
          ⟨⟨$c, d, p', a'$⟩, [$or$], [ ]⟩
        else
          ⟨⟨$c, d, p, a'$⟩, [ ], [ ]⟩
    end
  periodic ≔ λ $n, s$.
    let ⟨$c, d, p, a$⟩ ≔ $s$ in
    let ⟨$d', ois$⟩ ≔ foldl (
      (λ ⟨$d', ois$⟩, ⟨$m, n', c'$⟩.
        if $|a(⟨m, n', c'⟩)| > |ℕ|/2 \wedge ⟨m, n', c'⟩ \notin d$
          ⟨$d' \cup \{⟨m, n', c'⟩\}, ois :: \mathsf{deliver}_{\mathsf{urb}}(n', m)$⟩
        else
          ⟨$d', ois$⟩),
      ⟨∅, [ ]⟩,
      $p$ ) in
    ⟨⟨$c, d \cup d', p, a$⟩, [ ], $ois$ ⟩ ⟩

(b) Implementation

Figure 13: Uniform Reliable Broadcast Component

of the message is a quorum. The other handler functions do not issue delivery events. Thus, by rule INVLSE, we have

$$\Gamma \vdash_{\mathsf{URBC}} \circledS\ (n' \bullet \mathsf{deliver}_{\mathsf{urb}}(n'', m) \in \mathsf{ois}) \Rightarrow \\ \exists c.\ |ack(\mathsf{s}(n'))(\langle m, n'', c\rangle)| > |\mathbb{N}|/2 \tag{2}$$

that states that when a delivery is issued for a message, the size of its acknowledgement set is more than half of the number of nodes. As a quorum of nodes are correct, there should be at least one correct node in the acknowledging set. We sketch the proof of this fact using the QUORUM. We can separately show that every element of the acknowledgement set is a node.

$$\Gamma \vdash_{\mathsf{URBC}} \circledS\ \Box\ ack(\mathsf{s}(n'))(\langle m, n_s, c\rangle) \subseteq \mathbb{N} \tag{3}$$

and obviously

$$\Gamma \vdash_{\mathsf{URBC}} \circledS\ \Box\ (\mathbb{N}/2 + \mathbb{N}/2 \geq \mathbb{N}) \tag{4}$$

By rule QUORUM on Equation 1, Equation 3 and Equation 4, we have

$$\Gamma \vdash_{\mathsf{URBC}} \circledS\ |ack(\mathsf{s}(n'))(\langle m, n'', c\rangle)| > |\mathbb{N}|/2 \Rightarrow \\ \exists n.\ n \in \mathsf{Correct} \land n \in ack(\mathsf{s}(n'))(\langle m, n'', c\rangle) \tag{5}$$

that states that if the size of the acknowledgement set for a message is more than half of the number of nodes, then a correct node is in the set. We note that reasoning about quorums is done in a simple single step. Thus, from Equation 2 and Equation 5, we have

$$\Gamma \vdash_{\mathsf{URBC}} \circledS\ (n' \bullet \mathsf{deliver}_{\mathsf{urb}}(n'', m) \in \mathsf{ois}) \Rightarrow \\ \exists c, n.\ n \in \mathsf{Correct} \land n \in ack(\mathsf{s}(n'))(\langle m, n'', c\rangle) \tag{6}$$

We now present a summary of the rest of the proof that is available in subsection 5.3. We use the following properties of bebc. Validity: If a correct node broadcasts a message, then the message is eventually delivered to every correct node. No-forge: Delivered messages are previously broadcast.

We know that a correct node is in the acknowledgement set. The component adds a node to the acknowledgement set only if a message is delivered from that node via bebc. By the no-forge property of bebc, the message should have been broadcast. Thus, a correct node has broadcast the message. By the validity property of bebc, the message is delivered to every correct node. When the component receives a message via bebc, if the message is not already in the pending set, it is rebroadcast. If it is already in the pending set, it can be shown that it is already broadcast. Thus, in any case, every correct node eventually (in the past or future) broadcasts the message. Thus, by the validity property of bebc, the message is delivered to every correct node from every correct node via bebc. When the component receives a message via bebc, it adds the sender to the acknowledgement set. The elements of the acknowledgement set always stay the same or increase. Thus, eventually forever, every correct node will have every correct node in its acknowledgement set for the message. As a majority of nodes is correct, the size of this set is more than half of the number of the nodes. As the periodic function is infinitely often called, it will be eventually called when the size of the acknowledgement set for the message is more than half of the number of the nodes. When the periodic function iterates the pending set, if the message is not in the delivered set, as its acknowledgement set is already large enough, it is delivered. On the other hand, if it is in the delivered set, it can be shown that it is already delivered. Thus, the message is eventually (in the past or future) delivered at every correct node.

## 3 Lowering Soundness

The following theorem states the soundness of the lowering transformation for compositional reasoning. If a top-level invariant $\mathcal{I}_i^{[]}$ is valid for the stack $\mathcal{S}_i$ and $\mathcal{S}_i$ is a substack of the stack $\mathcal{S}$, then the lowered invariant $\mathsf{lower}(i, \mathcal{I}_i^{[]})$ is valid for $\mathcal{S}$. We use the validity judgement $\vDash_\mathcal{S} \mathcal{A}$ that states that $\mathcal{A}$ is valid in every trace of $\mathcal{S}$. (Validity is defined more precisely in section 7, and the detailed proofs are available in subsection 5.2.)

**Theorem 3.** *For all $\mathcal{S}$, $c$, and $\overline{\mathcal{S}_i}$, such that $\mathcal{S} = \mathsf{stack}(c, \overline{\mathcal{S}_i})$, if $\vDash_{\mathcal{S}_i} \mathcal{I}_i^{[]}$ then $\vDash_\mathcal{S} \mathsf{lower}(i, \mathcal{I}_i^{[]})$.*

We now state the *compositional proof technique* and its soundness. The specifications of substacks can be lowered and used to derive the specification of the stack. Judgements of TLC are of the form $\Gamma \vdash_c \mathcal{A}$ where $\Gamma$ is the assumed assertions and $\mathcal{A}$ is the deduced assertion. Consider valid top-level invariants $\overline{\mathcal{I}_i^{[]}}$ for stacks $\overline{\mathcal{S}_i}$ and a stack $\mathcal{S}$ built by the component $c$ on top of $\overline{\mathcal{S}_i}$. The following theorem states that assuming the lowered invariants $\overline{\mathsf{lower}(i, \mathcal{I}_i^{[]})}$, any assertion that TLC deduces for $c$ is valid for $\mathcal{S}$.

**Corollary 1** (Composition Soundness). *For all $\mathcal{S}$, $c$, and $\overline{\mathcal{S}_i}$ such that $\mathcal{S} = \mathsf{stack}(c, \overline{\mathcal{S}_i})$, if $\overline{\vDash_{\mathcal{S}_i} \mathcal{I}_i^{[]}}$ and $\overline{\mathsf{lower}(i, \mathcal{I}_i^{[]})} \vdash_c \mathcal{A}$ then $\vDash_\mathcal{S} \mathcal{A}$.*

Let us present an overview of why the above theorem holds. We first define a few helper definitions and lemmas.

The transitions of the operational semantics generate labels $\ell$ that are tuples of ($n$, $d$, $o$, $e$, $\sigma$, $\sigma'$, $ors$, $ois$). A label represents the execution of an event and its components are the node identifier $n$, the location $d$, the orientation $o$, the user event $e$, the pre-state $\sigma$, the post-state $\sigma'$, issued output requests $ors$ and issued output indications $ois$ of the executed event. A trace $\tau$ is a sequence of labels. The set of traces of a stack $\mathcal{S}$ is denoted as $T(\mathcal{S})$. To determine satisfiability of temporal assertions, we define a model $m = (\tau, i, I)$ as a tuple of a trace $\tau$, a position $i$ in the trace and an interpretation $I$ (to evaluate rigid free variables). The set of models of a stack $\mathcal{S}$ is denoted as $M(\mathcal{S})$. (We elaborate these definitions in section 6 and section 7). A trace is pushed to a branch $i$ by appending $i$ to the location of its events. Pushing is naturally lifted to a model by pushing its trace.

**Definition 7** (Pushing a Trace and a Model).

$$\begin{aligned}
\mathsf{push}(i, \tau) &\triangleq \mathsf{map}\,(\lambda(d, e, n, s, s', ors, ois).\\
&\qquad\qquad (i ::: d, e, n, s, s', ors, ois), \tau) \\
\mathsf{push}(i, (\tau, j, I)) &\triangleq (\mathsf{push}(i, \tau), j, I)
\end{aligned}$$

The following lemma intuitively states that a substack generates a subset of the traces that it generates at the top level (modulo the location of events). More precisely, for every stack $\mathcal{S}$ with a substack $\mathcal{S}_i$, for any trace $\tau$ of $\mathcal{S}$, there exists a trace $\tau'$ of $\mathcal{S}_i$, such that subtrace of $\tau$ for $\mathcal{S}_i$ is equal to $\tau'$ pushed to branch $i$. We use $\tau|_{d\supseteq[i]}$ to denote the projection of $\tau$ over events whose location is an extension of $[i]$ i.e. the events that are from the stack at location $[i]$. This projection is simply lifted to models.

**Lemma 1.** *For all $c$, $\overline{\mathcal{S}_i}$ and $\tau$, if $\tau \in T(\mathsf{stack}(c, \overline{\mathcal{S}_i}))$, there exists $\tau' \in T(\mathcal{S}_i)$ such that $\tau|_{d\supseteq[i]} = \mathsf{push}(i, \tau')$.*

The above lemma is proved using a simulation from the semantics of $\mathsf{stack}(c, \overline{\mathcal{S}_i})$ to the semantics of $\mathcal{S}_i$. The following corollary for models is immediate from the definition of push on models (Definition 7).

**Corollary 2.** *For all $c$, $\overline{\mathcal{S}_i}$ and $m$, if $m \in M(\mathsf{stack}(c, \overline{\mathcal{S}_i}))$, there exists $m' \in M(\mathcal{S}_i)$ such that $m|_{d \supseteq [i]} = \mathsf{push}(i, m')$.*

The following lemma states that if a model $m$ satisfies a top-level assertion $\mathcal{A}^{[]}$, then the model resulted from pushing $m$ to branch $i$ satisfies the assertion resulted from pushing $\mathcal{A}^{[]}$ to branch $i$.

**Lemma 2.** *For all $m$, $\mathcal{A}^{[]}$ and $i$, if $m \vDash \mathcal{A}^{[]}$ then $\mathsf{push}(i, m) \vDash \mathsf{push}(i, \mathcal{A}^{[]})$.*

The above lemma is proved by induction on the structure of $\mathcal{A}^{[]}$. We now define extensions of a model that preserve the subtrace of branch $i$. A model is extended when its trace is extended. A model $(\tau'', 0, I)$ is an extension of the model $(\tau, 0, I)$ if $\tau''$ is an interleaving of $\tau$ with another trace $\tau'$; to preserve the events under branch $i$ in $\tau$, there should not be any event under branch $i$ in $\tau'$.

**Definition 8** (Extending a Model).

$$\mathsf{extend}((\tau, 0, I), i) \triangleq \{(\tau'', 0, I) \mid \exists \tau'.\ \tau'' \in \mathsf{interleave}(\tau, \tau') \land \forall j.\ d(\tau'(j)) \not\supseteq [i]\}$$

Based on the above definition, the following lemma states that if a model $m$ satisfies an invariant $\mathcal{I}^{[i]}$, then any extension of $m$ (that preserves the events of branch $i$) satisfies the restriction of $\mathcal{I}^{[i]}$ to branch $i$.

**Lemma 3.** *For all $m$, $i$, $\mathcal{I}^{[i]}$, $m'$, if $m \vDash \mathcal{I}^{[i]}$ and $m' \in \mathsf{extend}(m, i)$ then $m' \vDash \mathsf{restrict}(d \supseteq [i], \mathcal{I}^{[i]})$.*

A generalization of this lemma is proved by structural induction on the invariant $\mathcal{I}^{[i]}$. To prove this lemma for the not operator, the lemma is generalized to bi-implication. As we mentioned before, the next operator was excluded from the sublanguage of invariants. If we had mapped the next operator to the eventually operator, the forward implication would but the backward implication would not hold.

Now, use apply the above lemmas, to present a proof sketch for Theorem 3. From $\vDash_{\mathcal{S}_i} \mathcal{I}_i^{[]}$, we have that for every model $m'$ of $\mathcal{S}_i$, (1) $m' \vDash \mathcal{I}_i^{[]}$. By Lemma 2 on [1], both the model and the invariant can be pushed i.e. (2) $\mathsf{push}(i, m') \vDash \mathsf{push}(i, \mathcal{I}_i^{[]})$. By Corollary 2 on [2], for every model $m$ in $M(\mathcal{S})$, there exists a model $m'$ in $M(\mathcal{S}_i)$ such that (3) the projection $m|_{d \supseteq [i]}$ is equal to $\mathsf{push}(i, m')$. Thus, from [2] and [3], we have (4) $m|_{d \supseteq [i]} \vDash \mathsf{push}(i, \mathcal{I}_i^{[]})$. The whole model $m$ is an extension of the projection that is (5) $m \in \mathsf{extend}(i, m|_{d \supseteq [i]})$. Thus, by Lemma 3 on [4] and [5], we have that $m$ models the restriction of $\mathsf{push}(i, \mathcal{I}_i^{[]})$ i.e. $m \vDash \mathsf{restrict}(d \supseteq [i], \mathsf{push}(i, \mathcal{I}_i^{[]}))$ that by definition of $\mathsf{lower}$ (Definition 1) is $m \vDash \mathsf{lower}(i, \mathcal{I}_i^{[]})$. Therefore, as $m$ is an arbitrary model of $\mathcal{S}$, we have $\vDash_{\mathcal{S}} \mathsf{lower}(i, \mathcal{I}_i^{[]})$.

# 4 Components

## 4.1 Stubborn Links

Figure 14 shows the specification of stubborn links. It accepts requests $\mathsf{send}_{\mathsf{sl}}(n, m)$ to send the message $m$ to the node $n$. It issues indications $\mathsf{deliver}_{\mathsf{sl}}(n, m)$ to deliver the message $m$ sent by the node $n$. A stubborn link stubbornly retransmits messages. The stubborn delivery property states that once a message is sent, it is delivered infinitely often. More precisely, if a correct node $n$ sends a message $m$ to a correct node $n'$, then $n'$ delivers $m$ infinitely often. A stubborn link never forges a message. More precisely, if a node $n$ delivers a message $m$ with sender $n'$, then $m$ was in fact previously sent to $n$ by $n'$.

Figure 15 shows the component SLC that implements the stubborn link. It uses the underlying basic link to retransmit messages. Its state stores the set of sent messages as the set sent of pairs of destination and message with the initial state ∅. Upon receiving a request $\mathsf{send}_{\mathsf{sl}}(n', m)$, the pair $\langle n', m \rangle$ is added to the sent set and a $\mathsf{send}_{\mathsf{l}}(n', m)$ request is issued to the lower-level basic link. Upon receiving an indication $\mathsf{deliver}_{\mathsf{l}}(n, m)$ from the basic link, a $\mathsf{deliver}_{\mathsf{sl}}(n, m)$ indication is issued. The basic link transmits the message with losses. Therefore, the messages in the sent set are periodically resent by the basic link.

Stubborn Link Interface
$\text{Req}_\text{sl} \coloneqq \text{send}_\text{sl}(n, m)$   Requests to send message m to node n
$\text{Ind}_\text{sl} \coloneqq \text{deliver}_\text{sl}(n, m)$   Delivers message m sent by node n

$SL_1$ (Stubborn delivery):
  $n \in \text{Correct} \wedge n' \in \text{Correct} \rightarrow$
  $(n \bullet \top \downarrow \text{send}_\text{sl}(n', m)) \Rightarrow \Box \Diamond (n' \bullet \top \uparrow \text{deliver}_\text{sl}(n, m))$
  If a correct node $n$ sends a message $m$ to a correct node $n'$, then $n'$ delivers $m$ infinitely often.

$SL_2$ (No-forge):
  $(n \bullet \top \uparrow \text{deliver}_\text{sl}(n', m)) \leftharpoonup (n' \bullet \top \downarrow \text{send}_\text{sl}(n, m))$
  If a node $n$ delivers a message $m$ with sender $n'$, then $m$ was previously sent to $n$ by $n'$.

Figure 14: Stubborn Links Specification

```
SLC: Component Req_sl Ind_sl (Req_l, Ind_l) ≔
    let lc ≔ 0 in
    ⟨State ≔ ⟨sent: Set[⟨ℕ, M⟩]⟩,
     init ≔ λn. ∅,
     request ≔ λ n, s, ir.
         match ir with
         | send_sl(n', m) ⇒
             ⟨⟨s ∪ {⟨n', m⟩}⟩,
              [(lc, send_l(n', m))],
              []⟩
         end
     indication ≔ λ n', s, ii.
         match ii with
         | (lc, deliver_l(n, m) ⇒
             ⟨s, [], [deliver_sl(n, m)]⟩
         end
     periodic ≔ λ n, s.
         let ors ≔ map (λ⟨n, m⟩. (lc, send_l(n, m))) s in
         ⟨s, ors, []⟩⟩
```

Figure 15: Stubborn Link Component

## 4.2 Perfect Links

The specification of perfect links is presented in Figure 16. Perfect links accept requests $\mathsf{send}_{\mathsf{pl}}(n, m)$ to send the message $m$ to the node $n$ and issue indications $\mathsf{deliver}_{\mathsf{pl}}(n, m)$ to deliver a message $m$ sent by a node $n$. The reliable delivery property states that perfect links can reliably transmit messages between correct nodes. More precisely, it states that if a correct node $n$ sends a message $m$ to a correct node $n'$, then $n'$ will eventually deliver $m$. The no-duplication property states that perfect links do not redundantly deliver messages. More precisely, it states that if a message is sent at most once, it will be delivered at most once. The no-forge property states that perfect links do not forge messages. More precisely, it states that if a node $n$ delivers a message $m$ with sender $n'$, then $m$ was previously sent to $n$ by node $n'$.

Figure 17 presents the component PLC that implements the perfect link. It uses a stubborn link as the lower-level component. As the stubborn link delivers messages infinitely often, the component keeps track of delivered messages and ignores redelivered messages. It stores the number of messages sent by the current node counter initialized to zero and the set of received messages received initialized to empty. The counter is used to uniquely assign a number to each message sent by the same node. The set of received messages are the pairs of node identifier and the number of the message in that node. Upon a request to send a message, the counter is incremented and the message is sent using the stubborn link subcomponent with the new counter value. Upon an indication of delivery of a message from the stubborn link subcomponent, if the message is already received, it is ignored. Otherwise, the pair of the sending node and the message number are added to the received set and the component delivers the message.

Perfect Link Interface
$\mathsf{Req}_{\mathsf{pl}} \coloneqq \mathsf{send}_{\mathsf{pl}}(n,m)$    Requests to send message m to node n
$\mathsf{Ind}_{\mathsf{pl}} \coloneqq \mathsf{deliver}_{\mathsf{pl}}(n,m)$    Delivers message m sent by node n

$PL_1$ (Reliable delivery):
$n \in \mathsf{Correct} \wedge n' \in \mathsf{Correct} \rightarrow$
$(n \bullet \top \downarrow \mathsf{send}_{\mathsf{pl}}(n',m)) \rightsquigarrow (n' \bullet \top \uparrow \mathsf{deliver}_{\mathsf{pl}}(n,m))$
If a correct node $n$ sends a message $m$ to a correct node $n'$, then $n'$ will eventually deliver $m$.

$PL_2$ (No-duplication):
$[n' \bullet \top \downarrow \mathsf{send}_{\mathsf{pl}}(n,m) \Rightarrow$
$\quad \hat{\boxminus} \neg (n' \bullet \top \downarrow \mathsf{send}_{\mathsf{pl}}(n,m))] \rightarrow$
$[n \bullet \top \uparrow \mathsf{deliver}_{\mathsf{pl}}(n',m) \Rightarrow$
$\quad \hat{\boxminus} \neg (n \bullet \top \uparrow \mathsf{deliver}_{\mathsf{pl}}(n',m))]^*$
If a message is sent at most once, it will be delivered at most once.

$PL_3$ (No-forge):
$(n \bullet \top \uparrow \mathsf{deliver}_{\mathsf{pl}}(n',m)) \leftsquigarrow (n' \bullet \top \downarrow \mathsf{send}_{\mathsf{pl}}(n,m))$
If a node $n$ delivers a message $m$ with sender $n'$, then $m$ was previously sent to $n$ by node $n'$.

---

*It is notable that $(p \Rightarrow \hat{\boxminus} \neg p) \rightarrow (p \Rightarrow \hat{\Box} \neg p)$

Figure 16: Perfect Links Specification

```
PLC: Component Req_pl Ind_pl (Req_sl, Ind_sl) :=
    let slc := 0 in
    ⟨State := ⟨counter: Nat,
               received: Set[⟨ℕ, Nat⟩]⟩,
     init := λn. ⟨0, ∅⟩,

     request := λ n, s, ir.
         let ⟨c, r⟩ := s in
         match ir with
         | send_pl(n', m) ⇒
             let c' := c + 1 in
             let or := (slc, send_sl(n', ⟨c', m⟩)) in
             ⟨⟨c', r⟩, [or], []⟩,
         end

     indication := λ n', s, ii.
         let ⟨c, r⟩ =: s in
         match ii with
         | (slc, deliver_sl(n, ⟨c', m⟩)) ⇒
             if (⟨n, c'⟩ ∈ r)
                 ⟨s, [], []⟩
             else
                 let r' := r ∪ {⟨n, c'⟩} in
                 let oi := deliver_pl(n, m) in
                 ⟨⟨c, r'⟩, [], [oi]⟩
         end

     periodic := λ n, s. ⟨s, [], []⟩ ⟩
```

Figure 17: Perfect Link Component

## 4.3 Best-Effort Broadcast

The specification of best-effort broadcast is presented in Figure 18. Best-effort broadcast accepts requests $\mathsf{broadcast}_{\mathsf{beb}}(m)$ to broadcast the message $m$ and issues indications $\mathsf{deliver}_{\mathsf{beb}}(n, m)$ to deliver a message $m$ broadcast by a node $n$. The validity property states that the best-effort broadcast delivers the same set of messages from every correct node to every correct node. More precisely, it states that if a correct node broadcasts a message $m$, then every correct node eventually delivers $m$. The no-duplication property states that the best-effort broadcast does not redundantly deliver messages. More precisely, it states that if a message is broadcast at most once, it will be delivered at most once. The no-forge property states that the best-effort broadcast does not forge messages. More precisely, it states that if a correct node delivers a message $m$ with sender $n'$, then $m$ was previously broadcast by node $n'$.

Figure 19 presents the component BEBC that implements the best-effort broadcast. It uses a perfect link as the lower-level component. It does not store any state. Upon a request to broadcast a message, it sends the message to every node using the perfect-link subcomponent. Upon an indication of delivery of a message from the perfect link subcomponent, the component delivers the message.

**Best-Effort Broadcast**
$\text{Req}_{\text{beb}} \coloneqq \text{broadcast}_{\text{beb}}(m)$   Broadcast a message m to all nodes.
$\text{Ind}_{\text{beb}} \coloneqq \text{deliver}_{\text{beb}}(n, m)$   Delivers a message m broadcast by node n.

$\text{BEB}_1$ (Validity)
$n \in \text{Correct} \wedge n' \in \text{Correct} \rightarrow$
$(n' \bullet \top \downarrow \text{broadcast}_{\text{beb}}(m)) \rightsquigarrow$
$(n \bullet \top \uparrow \text{deliver}_{\text{beb}}(n', m))$
If a correct node broadcasts a message $m$, then every correct node eventually delivers $m$.

$\text{BEB}_2$ (No-duplication)
$[n' \bullet \top \downarrow \text{broadcast}_{\text{beb}}(m) \Rightarrow$
$\quad \hat{\boxminus} \neg (n' \bullet \top \downarrow \text{broadcast}_{\text{beb}}(m))] \rightarrow$
$[n \bullet \top \uparrow \text{deliver}_{\text{beb}}(n', m) \Rightarrow$
$\quad \hat{\boxminus} \neg (n \bullet \top \uparrow \text{deliver}_{\text{beb}}(n', m))]$
If a message is broadcast at most once, it will be delivered at most once.

$\text{BEB}_3$ (No-forge)
$(n \bullet \top \uparrow \text{deliver}_{\text{beb}}(n', m)) \leftsquigarrow$
$(n' \bullet \top \downarrow \text{broadcast}_{\text{beb}}(m))$
If a correct node delivers a message $m$ with sender $n'$, then $m$ was previously broadcast by node $n'$.

Figure 18: Best-Effort Broadcast Specification

$\text{BEBC}: \text{Component } \text{Req}_{\text{beb}} \text{ Ind}_{\text{beb}} \ (\text{Req}_{\text{pl}}, \text{Ind}_{\text{pl}}) \coloneqq$
$\quad \text{let plc} \coloneqq 0 \text{ in}$
$\quad \langle \text{State} \coloneqq \text{Unit}$
$\quad \ \ \text{init } \coloneqq \lambda n. \ \bot$
$\quad \ \ \text{request } \coloneqq \lambda \ n, s, ir.$
$\quad \quad \quad \text{match } ir \text{ with}$
$\quad \quad \quad | \ \text{broadcast}_{\text{beb}}(m) \Rightarrow$
$\quad \quad \quad \quad \text{let } ors \coloneqq \text{map } (\lambda n. \ (\text{plc}, \text{send}_{\text{pl}}(n, m))) \ \mathbb{N} \text{ in}$
$\quad \quad \quad \quad \langle s, ors, [] \rangle$
$\quad \quad \quad \text{end}$
$\quad \ \ \text{indication} \coloneqq \lambda \ n', s, ii.$
$\quad \quad \quad \text{match } ii \text{ with}$
$\quad \quad \quad | \ (\text{plc}, \text{deliver}_{\text{pl}}(n, m)) \Rightarrow$
$\quad \quad \quad \quad \langle s, [], [\text{deliver}_{\text{beb}}(n, m)] \rangle$
$\quad \quad \quad \text{end}$
$\quad \ \ \text{periodic} \coloneqq \lambda \ n, s.$
$\quad \quad \quad \langle s, [], [] \rangle \rangle$

Figure 19: Best-Effort Broadcast Component

## 4.4 Uniform Reliable Broadcast

Uniform reliable broadcast guarantees that if a node (even a faulty one) delivers a message, then every correct node also delivers it. It precludes the situation where a faulty node delivers a message and failes and a correct node never delivers the message. Uniform reliable broadcast guarantees that the set of messages delivered by correct nodes is always a superset of the messages delivered by faulty nodes; hence, it is called uniform.

Figure 20 presents the specification of uniform reliable broadcast. It accepts requests $\mathsf{broadcast}_{\mathsf{urb}}(m)$ to broadcast the message $m$ and issues indications $\mathsf{deliver}_{\mathsf{urb}}(n, m)$ to deliver a message $m$ broadcast by a node $n$. The validity property states that a correct node receives his own messages. More precisely, it states that if a correct node $n$ broadcasts a message $m$, then $n$ itself eventually delivers $m$. The no-duplication property states that messages are not redundantly delivered. More precisely, it states that if a message is broadcast at most once, it will be delivered at most once. The no-forge property states that messages are not forged. More precisely, it states that if a node delivers a message $m$ with sender $n'$, then $m$ was previously broadcast by node $n'$. The uniform agreement states that if a node delivers a message, a correct node does not miss it. More precisely, it states that if a message $m$ is delivered by some node (whether correct or faulty), then $m$ is eventually delivered by every correct node.

Uniform Reliable Broadcast
$\text{Req}_{\text{urb}} \coloneqq \text{broadcast}_{\text{urb}}(m)$
Broadcasts a message $m$ to all nodes.
$\text{Ind}_{\text{urb}} \coloneqq \text{deliver}_{\text{urb}}(n, m)$
Delivers a message $m$ broadcast by node $n$.

Assumption:
$|\text{Correct}| > |\mathbb{N}|/2$

$\text{URB}_1$ (Validity)
$n \in \text{Correct} \rightarrow$
$(n \bullet \top \downarrow \text{broadcast}_{\text{urb}}(m)) \rightsquigarrow$
$(n \bullet \top \uparrow \text{deliver}_{\text{urb}}(n, m))$
If a correct node $n$ broadcasts a message $m$, then $n$ itself eventually delivers $m$.

$\text{URB}_2$ (No-duplication)
$[n \bullet \top \downarrow \text{broadcast}_{\text{urb}}(m) \Rightarrow$
$\quad \hat{\boxminus} \neg (n \bullet \top \downarrow \text{broadcast}_{\text{urb}}(m))] \rightarrow$
$[(n' \bullet \top \uparrow \text{deliver}_{\text{urb}}(n, m)) \Rightarrow$
$\quad \hat{\boxminus} \neg (n' \bullet \top \uparrow \text{deliver}_{\text{urb}}(n, m))]$
If a message is broadcast at most once, it will be delivered at most once.

$\text{URB}_3$ (No-forge)
$(n \bullet \top \uparrow \text{deliver}_{\text{urb}}(n', m)) \leftsquigarrow$
$(n' \bullet \top \downarrow \text{broadcast}_{\text{urb}}(m))$
If a node delivers a message $m$ with sender $n'$, then $m$ was previously broadcast by node $n'$.

$\text{URB}_4$ (Uniform Agreement)
$n \in \text{Correct} \rightarrow$
$(n' \bullet \top \uparrow \text{deliver}_{\text{urb}}(n'', m)) \Rightarrow$
$\Diamond (n \bullet \top \uparrow \text{deliver}_{\text{urb}}(n'', m)) \vee$
$\Diamondleft (n \bullet \top \uparrow \text{deliver}_{\text{urb}}(n'', m))$
If a message $m$ is delivered by some node (whether correct or faulty), then $m$ is eventually delivered by every correct node.

Figure 20: Uniform Reliable Broadcast Specification

Figure 21 presents the uniform reliable broadcast component URBC that implements the uniform reliable broadcast. It uses a best-effort broadcast beb and assumes that a majority of processes are correct. It stores the number of messages sent by the current node *count*, the set of delivered messages *delivered*, the set of messages that are received but are pending for acknowledgement *pending*, and a mapping from each message to the set of nodes that have acknowledged the receipt of the message *ack*. Each message is uniquely identified by the identifier of the sender node and the number of the message in that node. The state *count* is initialized to zero, and the other sets are all initialized to ∅.

The high-level idea is that each node, before delivering a message, makes sure that a majority of nodes have acknowledged the receipt of the message. As a majority of nodes are correct, there is at least one correct node in the acknowledging nodes. Not only the sender but every receiver broadcasts the message. Thus, the correct acknowledging node broadcasts the message as well and its broadcast will be properly delivered by every correct node. The other correct nodes will in turn broadcast the message. Thus, every correct node eventually receives an acknowledgement from every correct node. There is a majority of correct nodes. Thus, each correct node will eventually receive acknowledgement from a majority of nodes and deliver the message.

Upon a broadcast request, the counter is incremented, the message is added to the pending set and is broadcast using beb together with the sender identifier and the new value of the counter. Upon a delivery indication by beb, it is recorded in the acknowledgement map that the sender has acknowledged the receipt of the message and if the message is not already in the pending set, it is added to the pending set and broadcast by beb. Every node broadcasts each message by beb only once when it receives it for the first time. In the periodic function, messages in the pending set are iterated, and if an acknowledgement for a message is received from a majority of processes, and it is not delivered before, then it is delivered.

URBC: Component Req$_{urb}$ Ind$_{urb}$ (Req$_{beb}$, Ind$_{beb}$) ≔
    let bebc ≔ 0 in
    ⟨State ≔
        ⟨$count$: Nat,
         $delivered$: Set[⟨$M$, $\mathbb{N}$, Nat⟩],
         $pending$: Set[⟨$M$, $\mathbb{N}$, Nat⟩],
         $ack$: Map[⟨$M$, $\mathbb{N}$, Nat⟩, Set[$\mathbb{N}$]]⟩
     init ≔ $\lambda n$. ⟨0, ∅, ∅, ∅⟩
     request ≔ $\lambda\ n, s, ir$.
        let ⟨$c, d, p, a$⟩ ≔ $s$ in
        match $ir$ with
        | broadcast$_{urb}(m)$ ⇒
           let $c' ≔ c + 1$ in
           let $p' ≔ p \cup \{⟨m, n, c'⟩\}$ in
           let $or$ ≔ (bebc, broadcast$_{beb}$⟨$m, n, c'$⟩) in
           ⟨⟨$c', d, p', a$⟩, [$or$], [ ]⟩
        end
     indication ≔ $\lambda\ n', s, ii$.
        let ⟨$c, d, p, a$⟩ ≔ $s$ in
        match $ii$ with
        | (bebc, deliver$_{beb}(n'', ⟨m, n, c'⟩)$) ⇒
           let $a' ≔ a[⟨m, n, c'⟩ \mapsto a(⟨m, n, c'⟩) \cup \{n''\}]$ in
           if ⟨$m, n, c'$⟩ ∉ $p$
              let $p' ≔ p \cup \{⟨m, n, c'⟩\}$ in
              let $or$ ≔ (bebc, broadcast$_{beb}$⟨$m, n, c'$⟩) in
              ⟨⟨$c, d, p', a'$⟩, [$or$], [ ]⟩
           else
              ⟨⟨$c, d, p, a'$⟩, [ ], [ ]⟩
        end
     periodic ≔ $\lambda\ n, s$.
        let ⟨$c, d, p, a$⟩ ≔ $s$ in
        let ⟨$d', ois$⟩ ≔ foldl (
           ($\lambda$ ⟨$d', ois$⟩, ⟨$m, n', c'$⟩.
              if $|a(⟨m, n', c'⟩)| > |\mathbb{N}|/2 \wedge ⟨m, n', c'⟩ \notin d$
                 ⟨$d' \cup \{⟨m, n', c'⟩\}, ois$ :: deliver$_{urb}(n', m)$⟩
              else
                 ⟨$d',\ ois$⟩),
           ⟨∅, [ ]⟩,
           $p$ ) in
        ⟨⟨$c, d \cup d', p, a$⟩, [ ], $ois$ ⟩ ⟩

Figure 21: Uniform Reliable Broadcast Component

## 4.5 Eventually Perfect Failure Detector

>Eventually Perfect Failure Detector
>$\mathsf{Req}_{\mathsf{epfd}} = \mathsf{Unit}$      None.
>$\mathsf{Ind}_{\mathsf{epfd}} \coloneqq \mathsf{suspect}(n)$    The indication that the node $n$ is suspected to have crashed.
>     $\mid \mathsf{restore}(n)$    The indication that the node $n$ is not suspected anymore.

$\mathsf{EPFD}_1$ (Strong Completeness):
   $\forall n, n'.\ n \in \mathsf{Correct} \wedge \neg n' \in \mathsf{Correct} \rightarrow$
   $\Diamond[(n \bullet \top \uparrow \mathsf{suspect}(n')) \wedge \Box\neg(n \bullet \top \uparrow \mathsf{restore}(n'))]$
   Every incorrect node is eventually permanently suspected by every correct node.

$\mathsf{EPFD}_2$ (Eventual Strong Accuracy):
   $\forall n, n'.\ n \in \mathsf{Correct} \wedge n' \in \mathsf{Correct} \rightarrow$
   $\Box\neg(n \bullet \top \uparrow \mathsf{suspect}(n')) \vee$
   $\Diamond[(n \bullet \top \uparrow \mathsf{restore}(n')) \wedge \Box\neg(n \bullet \top \uparrow \mathsf{suspect}(n'))]$
   Eventually no correct node is suspected by any correct node.

Figure 22: Eventual Perfect Failure Detector Specification

EPFDC: Component Req$_{epfd}$ Ind$_{epfd}$ (Req$_l$, Ind$_l$) ≔
    ⟨State ≔
        ⟨$alive$: Nat → Set[ℕ],
         $failed$: Set[ℕ],
         $r$: Nat⟩

    init ≔ ⟨ℕ, [ ], 0⟩

    request ≔ λ $n, s, ir$.
        ⟨$s$, [ ], [ ]⟩

    indication ≔ λ $n', s, ii$.
        let ⟨$a, f, r$⟩ ≔ $s$ in
        match $ii$ with
        | (0, deliver$_l$($n$, HB)) ⇒
            ⟨⟨$a[r ↦ a(r) ∪ \{n\}], f, r$⟩, [ ], [ ]⟩
        end

    periodic ≔ λ $n, s$.
        let ⟨$a, f, r$⟩ ≔ $s$ in
        let ⟨$f', ors, ois$⟩ ≔ foldl(
            ℕ
            (λ⟨$f', ors, ois$⟩, $n$.
                if ($n ∉ a(r) ∧ n ∉ f$)
                    ⟨$f' ∪ \{n\}$,
                        $ors ∪ \{$send$_l(n, $HB$)\}$
                        $ois ∪ \{$suspect$(n)\}$⟩
                else if ($n ∈ a(r) ∧ n ∈ f$)
                    ⟨$f' ∖ \{n\}$,
                        $ors ∪ \{$send$_l(n, $HB$)\}$
                        $ois ∪ \{$restore$(n)\}$⟩
                else
                    ⟨$f'$,
                        $ors ∪ \{$send$_l(n, $HB$)\}$
                        $ois$⟩)
            ⟨$f$, [ ], [ ]⟩ ) in
        ⟨⟨[ ], $f', r + 1$⟩, $ors, ois$⟩⟩

Figure 23: Eventual Perfect Failure Detector Component

## 4.6 Eventual Leader Elector

Eventual Leader Elector
$\mathsf{Req}_{\mathsf{eld}}$ = Unit     None.
$\mathsf{Ind}_{\mathsf{eld}} \coloneqq \mathsf{trust}(n)$     It indicates that the node $n$ is the leader.

$\mathsf{ELE}_1$(Eventual Leadership)
    Eventually every correct process trusts the same correct process.
    $\exists l.\ l \in \mathsf{Correct} \land$
      $[n \in \mathsf{Correct} \rightarrow$
      $\Diamond(n \bullet \top \uparrow \mathsf{trust}(l) \land \hat{\Box} \neg(n \bullet \top \uparrow \mathsf{trust}(l')))]$

Figure 24: Eventual Leader Elector Specification

ELEC: Component $\mathsf{Req}_{\mathsf{eld}}$ $\mathsf{Ind}_{\mathsf{eld}}$ ($\mathsf{Req}_{\mathsf{epfd}}, \mathsf{Ind}_{\mathsf{epfd}}$) ≔
    let epfd ≔ 0 in
    ⟨State ≔
        ⟨$suspected$: Set[ℕ],
         $leader$: ℕ⟩,

    init ≔ $\lambda n.\ \langle \varnothing, \bot \rangle$,

    request ≔ $\lambda\ n, s, ir.$
        $\langle s, [\,], [\,] \rangle$

    indication ≔ $\lambda\ n, s, ii.$
        let $\langle p, l \rangle \coloneqq s$ in
        match $ii$ with
        | (epfd, suspect($n'$)) ⇒
            let $p' \coloneqq p \cup \{n'\}$ in
            $\langle \langle p', l \rangle, [\,], [\,] \rangle$
        | (epfd, restore($n'$)) ⇒
            let $p' \coloneqq p \smallsetminus \{n'\}$ in
            $\langle \langle p', l \rangle, [\,], [\,] \rangle$
        end,

    periodic ≔ $\lambda\ n, s.$
        let $\langle p, l \rangle \coloneqq s$ in
        if $l \neq maxRank(\mathbb{N} \smallsetminus p)$ in
            let $l' = maxRank(\mathbb{N} \smallsetminus p)$ in
            $\langle \langle p, l' \rangle, [\,], [\,] \rangle$
        else
            $\langle s, [\,], [\,] \rangle \rangle$

Figure 25: Eventual Leader Elector Component

## 4.7 Epoch Consensus

Epoch Consensus
$\text{Req}_{ec} \coloneqq \text{propose}_{ec}(v) \mid \text{epoch}_{ec}(n, ts)$
    $\text{propose}_{ec}(v)$ proposes value $v$.
    $\text{epoch}_{ec}(n, ts)$ starts a new epoch with the leader $n$ and timestamp $ts$.
$\text{Ind}_{ec} \coloneqq \text{decide}_{ec}(v)$
    $\text{decide}_{ec}(v)$ outputs the decided value $v$.

$EC_1$ (Validity)
    $(n \bullet \top \uparrow \text{decide}_{ec}(v)) \Rightarrow$
    $\exists n'.\ \Diamond(n' \bullet \top \downarrow \text{propose}_{ec}(v))$
    If a node decides the value $v$, then $v$ was proposed by a node.

$EC_2$ (Uniform agreement)
    $n \bullet \top \uparrow \text{decide}_{ec}(v) \land$
    $\Diamond(n' \bullet \top \uparrow \text{decide}_{ec}(v')) \Rightarrow$
    $v = v'$
    No two nodes decide differently.

$EC_3$ (Integrity)
    $(n \bullet \top \uparrow \text{decide}_{ec}(v)) \Rightarrow$
    $\hat{\Box}\neg(n \bullet \top \uparrow \text{decide}_{ec}(v'))$
    Every node decides at most once.

$EC_4$ (Termination)
    $|\text{Correct}| > |\mathbb{N}|/2 \land n \in \text{Correct} \to$
    $(n \bullet \top \downarrow \text{propose}_{ec}(v)) \Rightarrow$
    $(n \bullet 1 \downarrow \text{epoch}_{ec}(n, ts)) \land \Box\neg(n \bullet 1 \downarrow \text{epoch}_{ec}(n', ts')) \Rightarrow$
    $\forall n'.\ n' \in \text{Correct} \to \Diamond\Diamond\exists v'.\ (n' \bullet 1 \uparrow \text{decide}_{ec}(v'))$
    If a correct node proposes and an epoch is started with that node as the leader, then every correct node eventually decides a value.

Figure 26: Epoch Consensus Specification

ECC: Component Req$_{ec}$ Ind$_{ec}$ (Req$_{beb}$, Ind$_{beb}$, Req$_{pl}$, Ind$_{pl}$) ≔
    let plc ≔ 0 in
    let bebc ≔ 1 in
    ⟨State ≔
        ⟨$ets$: Nat,
         $valts$: Nat, $val$: Nat,
         $rts$: Nat,
         $prop$: Nat,
         $wval$: Map[Nat, Nat]
         $states$: Map[$\mathbb{N}$, ⟨Nat, Nat⟩],
         $accepted$: Set[$\mathbb{N}$],
         $decided$: Bool⟩

    init ≔ $\lambda n.\ \langle -n, 0, \bot, 0, \bot, \lambda n.\ \bot, \varnothing, \varnothing,$ false⟩

    request ≔ $\lambda\ n, s, ir.$
        let $\langle ets, vts, val, rts, p, wv, st, ac, d \rangle \coloneqq s$ in
        match $ir$ with
        | propose$_{ec}(v)$ ⇒
            let $p' \coloneqq v$ in
            $\langle\langle ets, vts, val, rts, p', wv, st, ac, d\rangle, [\,], [\,]\rangle$
        | epoch$_{ec}(n_l, ts_l)\ \wedge\ n_l = n$ ⇒
            let $ets' \coloneqq ts_l$ in
            let $st' \coloneqq \varnothing$ in
            let $or \coloneqq$ (bebc, broadcast$_{beb}$(PREPARE$(ets')$)) in
            $\langle\langle ets', vts, val, rts, p, wv, st', ac, d\rangle, [or], [\,]\rangle$
        | _ ⇒
            $\langle s, [\,], [\,]\rangle$
        end

Figure 27: Epoch Consensus (part 1)

$\text{indication} := \lambda\, n', s, ii.$
  let $\langle ets, vts, val, rts, p, wv, st, ac, d\rangle := s$ in
  match $ii$ with
  | $(\text{bebc}, \text{deliver}_{\text{beb}}(n, m)) \Rightarrow$
      match $m$ with
      | $\text{Prepare}\,(ets') \;\wedge\; ets' > rts \;\Rightarrow$
          let $rts' := ets'$ in
          let $or := (\text{plc}, \text{send}_{\text{pl}}(n_l, \text{State}\,(ets', vts, val))$ in
          $\langle\langle ets, vts, val, rts', p, wv, st, ac, d\rangle, [or], [\,]\rangle$

      | $\text{Accept}\,(ets', v) \;\wedge\; ets' \geq rts \;\Rightarrow$
          let $\langle vts', val'\rangle := \langle ets', v\rangle$ in
          let $or := (\text{plc}, \text{send}_{\text{pl}}(n, \text{Accepted}(ets')))$ in
          $\langle\langle ets, vts', val', rts, p, wv, st, ac, d\rangle, [or], [\,]\rangle$

      | $\text{Decided}\,(ets', v) \;\wedge\; \neg d \;\Rightarrow$
          let $d' := \text{true}$ in
          let $oi := \text{decide}_{\text{ec}}(v)$
          $\langle\langle ets, vts, val, rts, p, wv, st, ac, d'\rangle, [\,], [oi]\rangle$

      | $\_ \Rightarrow \langle s, [\,], [\,]\rangle$
      end

  | $(\text{plc}, \text{deliver}_{\text{pl}}(n, m)) \Rightarrow$
      match $m$ with
      | $\text{State}\,(ets', ts, v) \;\wedge\; ets' = ets \;\Rightarrow$
          let $st' := st[n \mapsto \langle ts, v\rangle]$ in
          if $(|\text{dom}(st')| > |\mathbb{N}|/2 \;\wedge\; wv(ets) = \bot \;\wedge\; p \neq \bot)$
              let $st'' = \varnothing$
              let $\langle vts', val'\rangle := highest(st')$ in
              let $wv' := wv[ets \mapsto \text{if } (val' \neq \bot)\; val' \text{ else } p]$ in
              let $or := (\text{bebc}, \text{broadcast}_{\text{beb}}(\text{Accept}\,(ets, wv'(ets))))$ in
              $\langle\langle ets, vts', val', rts, p, wv', st'', ac, d\rangle, [or], [\,]\rangle$
          else
              $\langle\langle ets, vts, val, rts, p, wv, st', ac, d\rangle, [\,], [\,]\rangle$

      | $\text{Accepted}\,(ets') \;\wedge\; ets' = ets \;\Rightarrow$
          let $ac' := ac \cup \{n\}$ in
          if $(|ac'| > \mathbb{N}/2)$
              let $or := (\text{bebc}, \text{broadcast}_{\text{beb}}(\text{Decided}\,(ets, wv(ets))))$ in
              $\langle\langle ets, vts, val, rts, p, wv, st, ac', d\rangle, [or], [\,]\rangle$
          else
              $\langle\langle ets, vts, val, rts, p, wv, st, ac', d\rangle, [\,], [\,]\rangle$

      | $\_ \Rightarrow \langle s, [\,], [\,]\rangle$
      end
  end

$\text{periodic} := \lambda\, n, s.$
  $\langle s, [\,], [\,]\rangle\;\rangle$

Figure 28: Epoch Consensus (part 2)

## 4.8 Epoch Change

Epoch Change
$\mathsf{Ind}_\mathsf{ech} \coloneqq \mathsf{startEpoch}_\mathsf{ech}(ts, n_l)$
The event $\mathsf{startEpoch}_\mathsf{ech}(ts, n_l)$ starts the epoch identified by timestamp $ts$ with the leader $n_l$.

$\mathsf{ECH}_1$ (Monotonicity)
$\quad n \in \mathsf{Correct} \rightarrow$
$\quad (n \bullet \top \uparrow \mathsf{startEpoch}_\mathsf{ech}(ts, n_l)) \Rightarrow$
$\quad \hat{\Box}(n \bullet \top \uparrow \mathsf{startEpoch}_\mathsf{ech}(ts', n_l') \rightarrow ts' > ts)$
If a correct process starts an epoch $(ts, n_l)$ and later starts an epoch $(ts', n_l')$, then $ts' > ts$.

$\mathsf{ECH}_2$ (Consistency)
$\quad n \in \mathsf{Correct} \land n' \in \mathsf{Correct} \rightarrow$
$\quad (n \bullet \top \uparrow \mathsf{startEpoch}_\mathsf{ech}(ts, n_l)) \Rightarrow$
$\quad (n' \bullet \top \uparrow \mathsf{startEpoch}_\mathsf{ech}(ts, n_l')) \Rightarrow n_l = n_l'$
If a correct process starts an epoch $(ts, n_l)$ and another correct process starts an epoch $(ts, n_l')$, then $n_l = n_l'$.

$\mathsf{ECH}_3$ (Eventual leadership)
$\quad \exists ts, n_l.\ n_l \in \mathsf{Correct} \land$
$\quad [n \in \mathsf{Correct} \rightarrow$
$\quad\quad \Diamond[(n \bullet \top \uparrow \mathsf{startEpoch}_\mathsf{ech}(ts, n_l)) \land$
$\quad\quad\quad \hat{\Box}\neg(n \bullet \top \uparrow \mathsf{startEpoch}_\mathsf{ech}(ts', n_l'))]]$
There is a timestamp $ts$ and a correct process $n_l$ such that eventually every correct process starts an epoch with $ts$ and $n_l$ and does not start another epoch afterwards.

Figure 29: Epoch Change Specification

ECHC: Component $\mathsf{Ind}_{\mathsf{ech}}$ ($\mathsf{Req}_{\mathsf{eld}}, \mathsf{Ind}_{\mathsf{eld}}, \mathsf{Req}_{\mathsf{beb}}, \mathsf{Ind}_{\mathsf{beb}}, \mathsf{Req}_{\mathsf{pl}}, \mathsf{Ind}_{\mathsf{pl}}$) :=
    let plc := 0 in
    let bebc := 1 in
    let eldc := 2 in
    ⟨State :=
        ⟨$trusted$: ℕ,
        $lastts$: Int,
        $ts$: Int,
        $nk$: Set[Bool]⟩

    init :=
        ⟨$n_{l_0}, 0, rank(n), \varnothing$⟩

    request := $\lambda\, n, s, ir.\, \langle s, [\,], [\,]\rangle$

    indication := $\lambda\, n, s, ii.$
        let $\langle tr, lts, t, nk\rangle := s$ in
        match $ii$ with
        | (eldc, $\mathsf{trust}_{\mathsf{eld}}(n_l)$) ⇒
            let $tr' := n_l$ in
            let $nk[tr']$ := false in
            if ($tr' := n$)
                let $t' := t + N$ in
                let $or$ := (bebc, $\mathsf{broadcast}_{\mathsf{beb}}(\textsc{NewEpoch}(t'))$) in
                $\langle\langle tr', lts, t', nk\rangle, [or], [\,]\rangle$
            else
                $\langle\langle tr', lts, t, nk\rangle, [\,], [\,]\rangle$
        | (bebc, $\mathsf{deliver}_{\mathsf{beb}}(n_l, \textsc{NewEpoch}(newt))$) ⇒
            if ($tr = n_l \land nk[tr]$ = false)
                let $nk[tr]$ := true in
                let $or$ := (plc, $\mathsf{send}_{\mathsf{pl}}(n_l, \textsc{State}(lts))$) in
                $\langle\langle tr, lts', t, nk\rangle, [\,], [oi]\rangle$
            else if ($tr = n_l \land newt > lts$)
                let $lts' := newt$ in
                let $oi$ := (ech, $\mathsf{startEpoch}(lts', tr)$) in
                $\langle\langle tr, lts, t, nk\rangle, [or], [\,]\rangle$
            else if ($tr \neq n_l$)
                let $or$ := (plc, $\mathsf{send}_{\mathsf{pl}}(n_l, \textsc{Nack}(lts))$) in
                $\langle\langle tr, lts, t, nk\rangle, [or], [\,]\rangle$
            else
                $\langle s, [\,], [\,]\rangle$

Figure 30: Epoch Change Component (part 1)

```
        | (plc, deliver_pl(n', STATE(lts')))  ∧  tr = n  ⇒
            if (lts' > t)
                let t' := t + ((lts' − t)/N  +  1) * N in
                let or := (bebc, broadcast_beb(NEWEPOCH(t'))) in
                ⟨⟨tr, lts, t', nk⟩, [or], [ ]⟩
            else
                let or := (bebc, broadcast_beb(NEWEPOCH(t))) in
                ⟨⟨tr, lts, t, nk⟩, [or], [ ]⟩

        | (plc, deliver_pl(n', NACK(lts')))  ∧  tr = n  ⇒
            let or := (bebc, broadcast_beb(NEWEPOCH(t))) in
            ⟨⟨tr, lts, t, nk⟩, [or], [ ]⟩

        | _  ⇒
            ⟨s, [ ], [ ]⟩
      end

  periodic := λ n, s. ⟨s, [ ], [ ]⟩ ⟩
```

Figure 31: Epoch Change Component (part 2)

## 4.9 Uniform Consensus

Uniform Consensus
$\mathsf{Req}_{\mathsf{uc}} \coloneqq \mathsf{propose}_{\mathsf{uc}}(v)$
Propose value $v$ for consensus.
$\mathsf{Ind}_{\mathsf{uc}} \coloneqq \mathsf{decide}_{\mathsf{uc}}(v)$
Outputs a decided value $v$ of consensus.

$UC_1$ (Termination)
$|\mathsf{Correct}| > |\mathbb{N}|/2 \wedge n \in \mathsf{Correct} \to$
$[\forall n'.n' \in \mathsf{Correct} \to \exists v.\ v \neq \bot \wedge \Diamondminus(n' \bullet \top \downarrow \mathsf{propose}_{\mathsf{uc}}(v))] \Rightarrow$
$\exists v.\ \Diamondminus\Diamond(n \bullet \top \uparrow \mathsf{decide}_{\mathsf{uc}}(v))$
Every correct node eventually decides some value.

$UC_2$ (Validity)
$(n \bullet \top \uparrow \mathsf{decide}_{\mathsf{uc}}(v)) \leftsquigarrow$
$\exists n'.\ (n' \bullet \top \downarrow \mathsf{propose}_{\mathsf{uc}}(v))$
If a node decides $v$, then $v$ was proposed by some node.

$UC_3$ (Integrity)
$(n \bullet \top \uparrow \mathsf{decide}_{\mathsf{uc}}(v)) \Rightarrow$
$\hat{\Box}\neg(n \bullet \top \uparrow \mathsf{decide}_{\mathsf{uc}}(v'))$
No node decides twice.

$UC_4$ (Uniform agreement)
$(n \bullet \top \uparrow \mathsf{decide}_{\mathsf{uc}}(v)) \wedge \Diamond(n' \bullet \top \uparrow \mathsf{decide}_{\mathsf{uc}}(v')) \Rightarrow$
$v = v'$
No two nodes decide differently.

UCC: Component $\text{Req}_{uc}$ $\text{Ind}_{uc}$ $(\text{Req}_{ec}, \text{Ind}_{ec}, \text{Req}_{ech}, \text{Ind}_{ech})$ :=
    let echc := 0 in
    let ecc := 1 in
    ⟨State := ⟨$prop$ : Nat,
             $leader$ : $\mathbb{N}$,
             $ts$ : Nat
             $started$ : false⟩

    init := $\lambda n.\langle \bot, \bot, 0, \text{false}\rangle$

    request := $\lambda\ n, s, ir.$
        let $\langle p, l, ts, str\rangle$ := $s$ in
        match $ir$ with
        | $\text{propose}_{uc}(v) \Rightarrow$
            let $p' = v$ in
            let $or$ := $(\text{ecc}, \text{propose}_{ec}(v))$ in
            $\langle\langle p', l, ts, str\rangle, [or], []\rangle$
        end

    indication := $\lambda\ n', s, ii.$
        let $\langle p, l, ts, str\rangle$ := $s$ in
        match $ii$ with
        | $(\text{echc}, \text{startEpoch}_{ech}(ets, n_l)) \Rightarrow$
            let $l'$ := $n_l$ in
            let $ts'$ := $ets$ in
            let $str'$ := false
            if $(n' = n_l)$
                if $(p \neq \bot)$
                    let $str'$ := true
                    let $or$ := $(\text{ecc}, \text{epoch}_{ec}(n_l, ets))$ in
                    $\langle\langle p, l', st', str'\rangle, [or], []\rangle$
                else
                    $\langle\langle p, l', st', str'\rangle, [], []\rangle$
            else
                let $or$ := $(\text{ecc}, \text{epoch}_{ec}(n_l, ets))$ in
                $\langle\langle p, l', st', str'\rangle, [or], []\rangle$

        | $(\text{ecc}, \text{decide}_{ec}(v)) \Rightarrow$
            let $oi$ := $\text{decide}_{uc}(v)$ in
            $\langle s, [], [oi]\rangle$
        end

    periodic := $\lambda\ n, s.$
        let $\langle p, l, ts, str\rangle$ := $s$ in
        if $(l = n \wedge str = \text{false} \wedge p \neq \bot)$
            let $str'$ = true
            let $or$ := $(\text{ecc}, \text{epoch}_{ec}(n_l, ts))$ in
            $\langle\langle p, l, ts, str'\rangle, [or], []\rangle$
        else
            $\langle s, [], []\rangle\ \rangle$

Figure 32: Uniform Consensus Component

# 5 Proofs

## 5.1 Soundness

**Definition 9** (Transition).
$$\xrightarrow{\tau \cdot \tau'} \;=\; \xrightarrow{\tau}_t^* \xrightarrow{\tau'}_p$$

**Definition 10** (Initial World).
$w_0(\mathcal{S}) = (\lambda d, n.\ \text{let}\ (c, \_) = \mathcal{S}(d)\ \text{in}\ \text{init}_c(n), \varnothing, \varnothing, r_0)$

The trace semantics $T(\mathcal{S})$ of $\mathcal{S}$ is the set of traces $\tau$ of infinite transitions $\xrightarrow{\tau}^*$ starting from the initial state $w_0(\mathcal{S})$ (with any $r_{GST}$). We consider infinite traces to reason about liveness properties.

**Definition 11** (Traces of a stack).
$T(\mathcal{S}) = \{\tau \mid w_0(\mathcal{S}) \xrightarrow{\tau}^*\}$

**Definition 12** (Models of a stack).
$M(\mathcal{S}) = \{(\tau, 0, I_0) \mid \tau \in T(\mathcal{S})\}$

**Definition 13.** *Let*
$\mathsf{mself}(d, o) \triangleq$
$\quad (d = [\,] \land o = \downarrow) \lor$
$\quad (d = [\,] \land o = \updownarrow) \lor$
$\quad (d = [i] \land o = \uparrow)$
$\mathsf{mself}(\ell) \triangleq$
$\quad \mathsf{mself}(d(\ell), o(\ell))$

**Definition 14.** *Let*
$\mathsf{mcorrect}(\tau, n) =$
$\quad \forall j \geq 0.\ \exists k \geq j.$
$\quad \tau_k = (n, \_, \_, \_, \_, \_, \_, \_, \_)$

**Theorem 2.** (Soundness)
Let $\mathcal{S} = \mathsf{stack}(c, \overline{\mathcal{S}'})$
If $\Gamma \vdash_c \mathcal{A}$, then $\Gamma \vDash_\mathcal{S} \mathcal{A}$.

**Proof.**

Assumption:
    (1) $\Gamma \vdash_c \mathcal{A}$
Conclusion:
    $\Gamma \vDash_\mathcal{S} \mathcal{A}$
From Definition 5, the conclusion is
    $\forall m \in M(\mathcal{S}), m \vDash \Gamma \to m \vDash \mathcal{A}$
Assumption:
    (2) $m \in M(\mathcal{S})$
    (3) $m \vDash \Gamma$
Conclusion:
    (4) $m \vDash \mathcal{A}$

From Definition 12 and Definition 11 on [2], we have
    (5) $m = (\tau, 0, I_0)$
    (6) $w_0(\mathcal{S}) \xrightarrow{\tau} {}^*$

Induction on the derivation of [1]:

Case SEQ:
    (7) $\mathcal{A} \;=\; \mathsf{n} \neq n \Rightarrow \mathsf{s'}(\mathsf{n}) = \mathsf{s}(n)$
From Definition 6 on [4], [5] and [7], we need to show that for all $j \geq 0$:
    Let $\tau(j) = (n_j, r_j, d_j, o_j, e_j, \sigma_j, \sigma'_j, ors_j, ois_j)$,
    $d'_j = \begin{cases} d & \text{if } o = \downarrow \vee o = \updownarrow \\ \mathsf{tail}(d_j) & \text{else} \end{cases}$
    $n_j \neq n \to \sigma'(d'_j)(n) = \sigma(d'_j)(n)$
    Immediate from Lemma 4.

Case IR:
    (8) $\mathcal{A} \;=\; \forall e.\; \top \downarrow e \Rightarrow (\mathsf{s'}(\mathsf{n}), \mathsf{ois}, \mathsf{ors}) = \mathsf{request}(\mathsf{n}, \mathsf{s}(\mathsf{n}), e)$
From Definition 6 on [4], [5] and [8], we need to show that for all $j \geq 0$:
    Let $\tau(j) = (n_j, r_j, d_j, o_j, e_j, \sigma_j, \sigma'_j, ors_j, ois_j)$,
    $\forall e.\; d_j = [] \wedge o_j = \downarrow \wedge e_j = e \to$
        $(\sigma'(d_j)(n_j), ois_j, ors_j) = \mathsf{request}(n_j, \sigma(d_j)(n_j), e)$
Induction on the derivation of [6]:
    Cases except REQ: By induction hypothesis.
    Case REQ: Immediate from the assumption of REQ.
        $\sigma(d) = s$
        $\mathsf{request}(c, n, s(n), e) = (s'_n, (i, e_1), e_2)$
        $s' = s[n \mapsto s'_n]$
        $\sigma' = \sigma[d \mapsto s']$

Case II:
    Similar to the case IR.

Case PE:
    Similar to the case IR.

Case OI:

(9) $\mathcal{A} = \forall n, e.\ n \bullet e \in \mathsf{ois} \wedge \mathsf{self} \Rightarrow \hat{\Diamond}(n \bullet \top \uparrow e)$

From Definition 6 on [4], [5] and [9], we need to show that:

$\forall j \geq 0. \forall n, d, ois.$
$\quad \tau(j) = (n, \_, d, o, \_, \_, \_, \_, ois) \wedge e \in ois \wedge \mathsf{mself}(d, o) \rightarrow$
$\exists k > j.\ \tau_k = (n, \_, [], e, \uparrow, \_, \_, \_, \_)$

Case analysis on the disjuncts of Definition 13:

Case 1:

(10) $d = [] \wedge o = \downarrow$

We assume that

(11) $\tau(j) = (n, \_, [], \downarrow, \_, \_, \_, \_, ois)$

(12) $e \in ois$

We show that

$\exists k > j.\ \tau_k = (n, \_, [], \uparrow, e, \_, \_, \_, \_)$

Induction on the derivation of [6]:

Case REQ: From the assumption of REQ:

$\mathsf{request}(n, s(n), e') = (\_, \_, e)$
$\tau_2 = (n, \_, [], \uparrow, e, \_, \_, \_, \_) \cdot \tau_2'$
$\tau = (n, \_, [], \downarrow, e', \_, \_, \_, e) \cdot \_ \cdot \tau_2$

Thus

$\tau = (n, \_, [], \downarrow, e', \_, \_, \_, e) \cdot \_ \cdot (n, \_, [], \uparrow, e, \_, \_, \_, \_) \cdot \_$

After the event $(n, \_, [], \downarrow, e', \_, \_, \_, e)$,
the event $(n, \_, [], \uparrow, e, \_, \_, \_, \_)$ comes.
Also, the induction hypothesis is used for the
rest of events in $\tau$ such as the events in $\tau_2'$.

Cases except REQ:

By induction hypothesis.

Case 2:

(13) $d = [] \wedge o = \updownarrow$

Similar to case REQ.

Case 3:

(14) $d = [i] \wedge o = \uparrow$

Similar to Case 1.

Induction on the derivation of [6].

Rule IND is main case.

Case OI':

(15) $\mathcal{A} = \forall n, e.\ n \bullet \top \uparrow e \Rightarrow \hat{\boxminus}(n \bullet e \in \mathsf{ois} \wedge \mathsf{self})$

From Definition 6 on [4], [5] and [15], we need to show that:

$\forall j \geq 0.$ Let $\tau(j) = (n_j, \_, d_j, o_j, e_j, \_, \_, \_, \_)$,
$\forall n, e.\ n_j = n \wedge d_j = [] \wedge o_j = \uparrow \wedge e_j = e \rightarrow$
$\exists k < j.$ Let $\tau_k = (n_k, \_, d_k, o_k, \_, \_, \_, \_, ois_k)$,
$n_k = n \wedge e \in ois_k \wedge \mathsf{mself}(d_k, o_k)$

Induction on the derivation of [6]:

Cases except REQ, IND and PER: By induction hypothesis.

Case REQ: From the assumption of REQ:

$\mathsf{request}(c, n, s(n), e') = (\_, \_, e)$

$\tau_2 = (n, \_, [], \uparrow, e, \_, \_, \_, \_) \cdot \tau_2'$
$\tau = (n, \_, [], \downarrow, e', \_, \_, \_, e) \cdot \tau_1 \cdot \tau_2$
Before the event $(n, \_, [], \uparrow, e, \_, \_, \_, \_)$,
there is the event $(n, \_, [], \downarrow, e', \_, \_, \_, e)$.
$d = [] \wedge o = \downarrow$ satisfies the first disjunct of mself.
Also, the induction hypothesis is used for the
rest of events in $\tau$ such as the events in $\tau_2'$.

Case IND:
  Similar to Case REQ.
  The second disjunct of mself is satisfied.

Case PER:
  Similar to Case REQ.
  The first disjunct of mself is satisfied.

Case OR:
  Similar to the Case OI.

Case OR':
  Similar to the Case OI'.

Case INIT:
  (16) $\mathcal{A} = \text{\textcircled{S}} \; (\mathsf{s} = \lambda n. \; \mathsf{init}_c(n))$
  By [4], [5], and rule SELFM, we have
    $\tau_1' = \epsilon$
    $\tau_2' = \tau_{0..}|_\mathsf{mself}$
    $\tau' = \tau_1' \cdot \tau_2'$
    $i' = 0$
  We need to show that
    $(\tau_{0..}|_\mathsf{mself}, 0, I_0) \vDash (\mathsf{s} = (\lambda n. \; \mathsf{init}_c(n)))$
  By [6]
    $w_0(\mathcal{S}) \xrightarrow{\tau}^*$
  By Definition 10
    $w_0(\mathcal{S}) = (\lambda d, n. \; \mathsf{let} \; (c, \_) = \mathcal{S}(d) \; \mathsf{in} \; \mathsf{init}_c(n), \varnothing, \varnothing, r_0)$
  Thus, the state is equal to the user-defined init at the beginning.
    $\sigma(\tau_0')([])(n) = \mathsf{init}_c(n)$
  By Definition 9, $\rightarrow_t$ appears at the beginning of $\rightarrow$
    $\tau = \tau' \cdot \tau'' \cdot \tau'''$
    $w_0(\mathcal{S}) \xrightarrow{\tau'}{}_t^* \xrightarrow{\tau''}{}_p W' \xrightarrow{\tau'''}{}$
  There are two $\rightarrow_t$ transitions:
    The rule FAIL: It preserves the state $\sigma$.
    The rule REQUEST: It is a self transition.
  Thus, the init state is preserved to the first self step,
  that is either by the rule REQUEST or the rule PERIODIC.

Case ASELF:
  (17) $\mathcal{A} = \text{\textcircled{S}} \; \Box\mathsf{self}$
  From Definition 6 on [4], [5] and [17], we need to show that:

$\forall j \geq 0$. Let $(\tau|_{\mathsf{mself}})_j = (\_, \_, d_j, o_j, \_, \_, \_, \_, \_, \_)$,
   $\mathsf{mself}(d_j, o_j)$
That is
   $\forall j \geq 0$. Let $\tau(j) = (\_, \_, d_k, o_k, \_, \_, \_, \_, \_, \_)$,
   $\mathsf{mself}(d_j, o_j) \to \mathsf{mself}(d_j, o_j)$
That is trivial.

Case SINV:
   (18) $\mathcal{A}$ = ⓈI ↔ restrict(self, $\mathcal{I}$)

   Let $\mathcal{I} = \Box \mathcal{A}$
   where ○ and Ⓢ are not used in $\mathcal{A}$.

Forward Direction:
   We have
      (19) $m \models$ Ⓢ $\Box \mathcal{A}$
   We show that
      (20) $m \models$ restrict(self, $\Box \mathcal{A}$)
   Thus, from [5], we need to show that
      (21) $(\tau, 0, I_0) \models$ restrict(self, $\Box \mathcal{A}$)
   Thus, from Definition 3 on [21], we need to show that
      (22) $(\tau, 0, I_0) \models \Box(\mathsf{self} \to \mathsf{restrict}(\mathsf{self}, \mathcal{A}))$
   Thus, from Definition 6, we need to show that:
      $(\tau, k, I_0) \models \mathsf{self} \to \mathsf{restrict}(\mathsf{self}, \mathcal{A})$   forall $k \geq 0$

   From Definition 6 on [19] and [5], we have
      (23) $\tau' = \tau|_{\mathsf{mself}}$
      (24) $(\tau', 0, I_0) \models \Box \mathcal{A}$

   Let
      (25) $j$ be the location of the first mself event in $\tau$
   that is
      (26) $\tau'_0 = \tau(j)$
   We consider two cases:
      Case:
         (27) $k < j$
         From [25] and [27],
            (28) $\neg\mathsf{mself}(\tau_k)$
         Thus
            $(\tau, k, I_0) \models \mathsf{self} \to \mathsf{restrict}(\mathsf{self}, \mathcal{A})$
      Case:
         (29) $k \geq j$
         From Lemma 14 on [23], [26], [24],
            (30) $(\tau, j, I_0) \models$ restrict(self, $\Box \mathcal{A}$)
         From Definition 3 on [30],
            (31) $(\tau, j, I_0) \models \Box(\mathsf{self} \to \mathsf{restrict}(\mathsf{self}, \mathcal{A}))$
         From Definition 6 on [31], we have
            $(\tau, k, I_0) \models \mathsf{self} \to \mathsf{restrict}(\mathsf{self}, \mathcal{A})$   forall $k \geq j$

Backward Direction:
  We have
   (32)  $m \vDash \text{restrict}(\text{self}, \Box \mathcal{A})$
  We show that
   $m \vDash \text{\textcircled{S}} \Box \mathcal{A}$
  Let
   (33)  $\tau' = \tau|_{\text{mself}}$
  From [5] and Definition 6, we need to show that
   $(\tau', k, I_0) \vDash \mathcal{A}$   forall $k \geq 0$

  From [32] and [5], we have
   (34)  $(\tau, 0, I_0) \vDash \text{restrict}(\text{self}, \Box \mathcal{A})$
  Thus, from Definition 3 on [34], we have
   (35)  $(\tau, 0, I_0) \vDash \Box(\text{self} \rightarrow \text{restrict}(\text{self}, \mathcal{A}))$
  Thus, from Definition 6, we have
   (36)  $(\tau, j, I_0) \vDash \text{self} \rightarrow \text{restrict}(\text{self}, \mathcal{A})$   forall $j \geq 0$

  From [33], forall $k \geq 0$, there exists $j$ such that
   (37)  $\tau'_k = \tau(j)$
   (38)  $\text{mself}(\tau(j))$
  From [36] and [38],
   (39)  $(\tau, j, I_0) \vDash \text{restrict}(\text{self}, \mathcal{A})$
  From Lemma 14 on [33], [37], [39],
   $(\tau', k, I_0) \vDash \mathcal{A}$

Case POSTPRE:
  (40)  $\mathcal{A} \;=\; \text{\textcircled{S}} \, (\text{s}' = s \Leftrightarrow \bigcirc \text{s} = s)$
  From Definition 6 on [4], [5] and [40], we need to show that:
   $\forall j \geq 0. \;\; \sigma'((\tau|_{\text{mself}})_j)([\,]) = \sigma((\tau|_{\text{mself}})_{j+1})([\,])$
  That is straightforward from Lemma 5 and Lemma 6.

Case RSEQ:
Immediate from the periodic transition $\twoheadrightarrow_p$ that increments the round $r$.

Case RORD:
Immediate from the periodic transition $\twoheadrightarrow_p$ that increments the round $r$ for the transition $\twoheadrightarrow_{per}$ for the periodic events per.

Case GST:
Every message that is sent in round $r$ stays in $ms$ until the periodic transition $\twoheadrightarrow_p$ of round $r$ (Lemma 7).
After the GST round $r_{GST}$, the periodic transition $\twoheadrightarrow_p$ does not drop messages.
The periodic transition $\twoheadrightarrow_p$ of round $r$ includes the transition $\twoheadrightarrow_{msg}$ that delivers the messages.

Case FDUP:
  (41)  $\mathcal{A} =$
     $\Box \Diamond (n' \bullet d \uparrow \text{deliver}_{\text{l}}(n, m)) \rightarrow$

$$\square\diamond(n \bullet d \downarrow \mathsf{send}_\mathsf{l}(n', m))$$

We prove the contra-positive that is

(42) $\mathcal{A}$ =
$$\diamond\square\neg(n \bullet d \downarrow \mathsf{send}_\mathsf{l}(n', m)) \to$$
$$\diamond\square\neg(n' \bullet d \uparrow \mathsf{deliver}_\mathsf{l}(n, m))$$

From Definition 6 on [4], [5] and [42], we need to show that:

$(\exists j.\ \forall i \geq j.$
  $\neg(\mathsf{n}(\tau(i)) = n \land \mathsf{d}(\tau(i)) = d \land$
  $\mathsf{o}(\tau(i)) =\downarrow \land \mathsf{e}(\tau(i)) = \mathsf{send}_\mathsf{l}(n', m)))$
$\to$
$(\exists j.\ \forall i \geq j.$
  $\neg(\mathsf{n}(\tau(i)) = n' \land \mathsf{d}(\tau(i)) = d \land$
  $\mathsf{o}(\tau(i)) =\uparrow \land \mathsf{e}(\tau(i)) = \mathsf{deliver}_\mathsf{l}(n, m)))$

Immediate by Lemma 24.

Case NFORGE:

(43) $\mathcal{A}$ =
$$(n' \bullet d \uparrow \mathsf{deliver}_\mathsf{l}(n, m)) \Rightarrow$$
$$\ominus(n \bullet d \downarrow \mathsf{send}_\mathsf{l}(n', m))$$

From Definition 6 on [4], [5] and [43], we need to show that:

$\forall i.$
  $(\mathsf{n}(\tau(i)) = n' \land \mathsf{d}(\tau(i)) = d \land$
  $\mathsf{o}(\tau(i)) =\uparrow \land \mathsf{e}(\tau(i)) = \mathsf{deliver}_\mathsf{l}(n, m)) \to$
$\to$
$\exists j \leq i.$
  $(\mathsf{n}(\tau(j)) = n \land \mathsf{d}(\tau(j)) = d \land$
  $\mathsf{o}(\tau(j)) =\downarrow \land \mathsf{e}(\tau(j)) = \mathsf{send}_\mathsf{l}(n', m)$

Immediate by Lemma 27.

Case UNIOI:

(44) $\mathcal{A}$ =
$$(\mathsf{occ}(\mathsf{ois}, e) \leq 1 \land$$
$$\hat{\square}(\mathsf{n} = n \land \mathsf{self} \to e \notin \mathsf{ois}) \land$$
$$\hat{\boxminus}(\mathsf{n} = n \land \mathsf{self} \to e \notin \mathsf{ois})) \Rightarrow$$
$$(n \bullet \top \uparrow e) \Rightarrow$$
$$\hat{\square}\neg(n \bullet \top \uparrow e) \land \hat{\boxminus}\neg(n \bullet \top \uparrow e)$$

From Definition 6 on [4], [5] and [44], we need to show that:

$\forall \mathcal{S}, \tau, i, j, k, oi, n.$
$\mathsf{occ}(\mathsf{ois}(\tau(i)), e) \leq 1 \land$
$\forall k < i.\ (\mathsf{n}(\tau_k) = n \land \mathsf{mself}(\tau_k)) \to e \notin \mathsf{ois}(\tau_k) \land$
$\forall k > i.\ (\mathsf{n}(\tau_k) = n \land \mathsf{mself}(\tau_k)) \to e \notin \mathsf{ois}(\tau_k) \land$
$\mathsf{n}(\tau(j)) = n \land$
$\mathsf{d}(\tau(j)) = [] \land$
$\mathsf{o}(\tau(j)) = \uparrow \land$
$\mathsf{e}(\tau(j)) = e \to$
$k < j \to \neg[\mathsf{n}(\tau_k) = n \land \mathsf{d}(\tau_k) = [] \land \mathsf{o}(\tau_k) = \uparrow \land \mathsf{e}(\tau_k) = e]$
$k > j \to \neg[\mathsf{n}(\tau_k) = n \land \mathsf{d}(\tau_k) = [] \land \mathsf{o}(\tau_k) = \uparrow \land \mathsf{e}(\tau_k) = e]$

That is immediate from Lemma 32.

Case UniOR:
   Similar to rule UniOI

Case APer:
   (45) $\mathcal{A}$ =
      $n \in \mathsf{Correct} \to \square\Diamond(n \bullet \top \wr \mathsf{per})$
   From Definition 6 on [4], [5] and [45], we need to show that:
      $\forall n.\ n \in \mathbb{N} \wedge$
      $\forall i.\ \neg(\mathsf{n}(\tau(i)) = n \wedge \mathsf{d}(\tau(i)) = [] \wedge \mathsf{e}(\tau(i)) = \mathsf{fail}) \to$
      $\forall j.\ \exists k \geq j.$
         $\mathsf{n}(\tau_k) = n \wedge \mathsf{d}(\tau_k) = [] \wedge \mathsf{o}(\tau_k) = \wr \wedge \mathsf{e}(\tau_k) = \mathsf{per}$

   This is proved using Lemma 13 on [6].

Case Node:
   (46) $\mathcal{A}$ =
      $\square\ \mathsf{n} \in \mathbb{N}$
   From Definition 6 on [4], [5] and [46], we need to show that:
      $\forall i.\ \mathsf{n}(\tau(i)) \in \mathbb{N}$
   This is proved using Lemma 33.

**Corollary 2.**
For all $\Gamma$, $\mathcal{A}$, $\mathcal{S}$, $c$ and $\overline{\mathcal{S}'}$ such that $\mathcal{S} = \mathsf{stack}(c, \overline{\mathcal{S}'})$, if $\models_\mathcal{S} \Gamma$ and $\Gamma \vdash_c \mathcal{A}$, then $\models_\mathcal{S} \mathcal{A}$.

**Proof.**
Assumption:
   (1) $\models_\mathcal{S} \Gamma$
   (2) $\Gamma \vdash_c \mathcal{A}$
Conclusion:
   $\models_\mathcal{S} \mathcal{A}$

From Definition 4 on [1]
   (3) $\forall m \in M(\mathcal{S}).\ m \models \mathcal{A}$
From Theorem 2 on [2], we have
   (4) $\Gamma \models_\mathcal{S} \mathcal{A}$
From Definition 5 on [4], we have
   (5) $\forall m \in M(\mathcal{S}).\ m \models \Gamma \to m \models \mathcal{A}$
From [3] and [5], we have
   (6) $\forall m \in M(\mathcal{S}).\ m \models \mathcal{A}$
From Definition 4 on [6], we have
   $\models_\mathcal{S} \mathcal{A}$

**Lemma 4.**
*Only steps in a node change the state for that node.*

$\forall W_1, W_2, \tau, n, n', d, o, \sigma, \sigma'.$
$W_1 \xrightarrow{\tau}{}^* W_2 \wedge$
Let $(n, \_, d, o, \_, \sigma, \sigma', \_, \_) \in \tau$
$$d' = \begin{cases} d & \text{if } o = \downarrow \vee o = \updownarrow \\ \text{tail}(d) & \text{else} \end{cases}$$
$n' \neq n \rightarrow \sigma'(d')(n') = \sigma(d')(n')$

**Proof.**
Immediate from induction on steps.

**Lemma 5.**
*The post-state of every event is the same as the pre-state of its next event.*
$\forall W_1, W_2, \tau, j.$
$W_1 \xrightarrow{\tau}{}^* W_2$
$\rightarrow$
$\sigma'(\tau(j)) = \sigma(\tau(j+1))$

**Proof.**
Immediate from induction on steps.

**Lemma 6.**
*Only the self events change the top-level state.*
$\forall W_1, W_2, \tau, d, o, \sigma, \sigma'.$
$W_1 \xrightarrow{\tau}{}^* W_2 \wedge$
Let $(\_, \_, d, o, \_, \sigma, \sigma', \_, \_) \in \tau$
$\neg\textsf{mself}(d, o) \rightarrow$
$\sigma'([\,]) = \sigma([\,])$

**Proof.**
Immediate from induction on steps.

**Lemma 7.**
*The transitions $\xrightarrow{\tau}_t$, $\xrightarrow{\tau}_{req}$, $\xrightarrow{\tau}_{ind}$, and $\xrightarrow{\tau}_{per}$ preserve messages.*
$\forall \sigma, \sigma', ms, ms', f, f', r, r', \tau.$
$(\sigma, ms, f, r) \xrightarrow{\tau}{}_t^* (\sigma', ms', f', r') \vee$
$(\sigma, ms, f, r) \xrightarrow{\tau}{}_{req}^* (\sigma', ms') \vee$
$(\sigma, ms, f, r) \xrightarrow{\tau}{}_{ind}^* (\sigma', ms') \vee$
$(\sigma, ms, f, r) \xrightarrow{\tau}{}_{per}^* (\sigma', ms')$
$\rightarrow$
$ms \subseteq ms'$

**Proof.**
Immediate from induction on steps.

**Lemma 8.**
*If a node is failed, it remains failed.*
$\forall \sigma, \sigma', ms, ms', f, f', r, r', \tau, n.$
$[(\sigma, ms, f, r) \xrightarrow{\tau}{}^* (\sigma', ms', f', r') \vee$
$(\sigma, ms, f, r) \xrightarrow{\tau}{}_t^* (\sigma', ms', f', r') \vee$
$(\sigma, ms, f, r) \xrightarrow{\tau}{}_p (\sigma', ms', f', r')] \wedge$
$n \in f$
$\rightarrow$
$n \in f'$

**Proof.**
Immediate from Definition 9 and induction on steps.

**Lemma 9.**
*If a node is failed, it does not take any step.*
$\forall \sigma, \sigma', ms, ms', f, f', r, r', \tau, n.$
$[(\sigma, ms, f, r) \xrightarrow{\tau}{}^* (\sigma', ms', f', r') \vee$
$(\sigma, ms, f, r) \xrightarrow{\tau}{}_t^* (\sigma', ms', f', r') \vee$
$(\sigma, ms, f, r) \xrightarrow{\tau}{}_p (\sigma', ms', f', r') \vee$
$(\sigma, ms, f, r) \xrightarrow{\tau}{}_{req} (\sigma', ms') \vee$
$(\sigma, ms, f, r) \xrightarrow{\tau}{}_{ind} (\sigma', ms') \vee$
$(\sigma, ms, f, r) \xrightarrow{\tau}{}_{per} (\sigma', ms')] \wedge$
$n \notin \mathbb{N} \vee n \in f$
$\rightarrow$
$\nexists j. \ \mathsf{n}(\tau(j)) = n \wedge$

**Proof.**
Immediate from Definition 9, induction on steps and Lemma 8.

**Lemma 10.**
*If a node is not failed, it takes a periodic step on a p step.*
$\forall \sigma, \sigma', ms, ms', f, f', r, r', \tau, n.$
$(\sigma, ms, f, r) \xrightarrow{\tau}{}_p (\sigma', ms', f', r')$
$n \in \mathbb{N} \setminus f$
$\rightarrow$
$\exists i. \ \mathsf{n}(\tau(i)) = n \wedge \mathsf{d}(\tau(i)) = [\,] \wedge \mathsf{o}(\tau(i)) = \{\} \wedge \mathsf{e}(\tau(i)) = \mathsf{per}$

**Proof.**
Immediate from the definition of the rules rule PERIODIC and rule PER.

**Lemma 11.**
*If a node takes steps, it is not failed.*
$\forall \sigma, ms, f, r, \tau, n.$
$(\sigma, ms, f, r) \xrightarrow{\tau}{}^* \wedge$
$mcorrect(\tau, n)$
$\rightarrow$
$n \in \mathbb{N} \setminus f$

**Proof.**
Contra-positive is proved using induction on the steps and using Lemma 9 and Lemma 8.

**Lemma 12.**
*If a node is in the failed set, it has failed before.*
$\forall \sigma, ms, f, r, \tau, n.$
$w_0(\mathcal{S}) \xrightarrow{\tau}{}^* (\sigma, ms, f, r) \wedge$
$n \in f$
$\rightarrow$
$\exists i.\ n(\tau(i)) = n \wedge d(\tau(i)) = [\,] \wedge e(\tau(i)) = \textsf{fail}$

**Proof.**
By induction on the steps.

**Lemma 13.**
*For all $\mathcal{S}$ and $\tau$,*
$w_0(\mathcal{S}) \xrightarrow{\tau}{}^* \wedge n \in \mathbb{N}$
$\forall i.\ \neg(n(\tau(i)) = n \wedge d(\tau(i)) = [\,] \wedge e(\tau(i)) = \textsf{fail}) \rightarrow$
$\forall j.\ \exists k \geq j.$
$\quad n(\tau_k) = n \wedge d(\tau_k) = [\,] \wedge o(\tau_k) = \{\} \wedge\ e(\tau_k) = \textsf{per}$

**Proof.**
We assume
  (1) $w_0(\mathcal{S}) \xrightarrow{\tau}{}^*$
  (2) $n \in \mathbb{N}$
We prove the contra-positive that is
  (3) $\exists j.\ \nexists k \geq j.$
      $n(\tau_k) = n \wedge d(\tau_k) = [\,] \wedge o(\tau_k) = \{\} \wedge\ e(\tau_k) = \textsf{per} \rightarrow$
    $\exists i.\ n(\tau(i)) = n \wedge d(\tau(i)) = [\,] \wedge e(\tau(i)) = \textsf{fail}$
We assume
  (4) $\exists j.\ \nexists k \geq j.$
      $n(\tau_k) = n \wedge d(\tau_k) = [\,] \wedge o(\tau_k) = \{\} \wedge\ e(\tau_k) = \textsf{per} \rightarrow$
By Definition 9 on [1]
  (5) $\tau = \overline{\tau'_i \cdot \tau_i}$

(6) $W_1 = w_0(\mathcal{S})$

(7) $W_i \xrightarrow{\tau'_i}^*_t W'_i \xrightarrow{\tau_i}_p W_{i+1}$

From [4], [5],

(8) $\exists i.\ \forall \ell \in \tau_i.$
$\neg(\mathsf{n}(\ell) = n \wedge \mathsf{d}(\ell) = [] \wedge \mathsf{o}(\ell) = \zeta \wedge \mathsf{e}(\ell) = \mathsf{per})$

By the contra-positive of Lemma 10 on [7], [2] and [8]

(9) $n \in f(W'_i)$

By Lemma 12 on [7], [5], and [9]

$\exists i.\ \mathsf{n}(\tau(i)) = n \wedge \mathsf{d}(\tau(i)) = [] \wedge \mathsf{e}(\tau(i)) = \mathsf{fail}$

**Lemma 14.**
$\forall \tau, \tau', i, j.$
$\tau' = \tau|_{\mathit{mself}} \wedge$
$\tau'(i) = \tau(j) \rightarrow$
$(\tau', i, I) \models \mathcal{A} \leftrightarrow (\tau, j, I) \models \mathit{restrict}(\mathit{self}, \mathcal{A})$
where ○ and ⓢ are not used in $\mathcal{A}$.

**Proof.**
Similar to Lemma 36.

**Lemma 15.**
*A message is not added to the message set unless it is sent.*
$\forall \sigma, \sigma', ms, ms', f, f', r, r', \tau, n, n', d, m.$
$(n, n', d, m) \notin ms \wedge$
$[(\sigma, ms, f, r) \xrightarrow{\tau}^*_t (\sigma', ms', f', r') \vee$
$(\sigma, ms, f, r) \xrightarrow{\tau}^*_{\mathit{req}} (\sigma', ms') \vee$
$(\sigma, ms, f, r) \xrightarrow{\tau}^*_{\mathit{ind}} (\sigma', ms') \vee$
$(\sigma, ms, f, r) \xrightarrow{\tau}^*_{\mathit{per}} (\sigma', ms')] \wedge$
$(\forall i,\ \neg(\mathsf{n}(\tau(i)) = n \wedge \mathsf{d}(\tau(i)) = d \wedge \mathsf{o}(\tau(i)) = \downarrow \wedge \mathsf{e}(\tau(i)) = \mathsf{send}_l(n', m)))$
$\rightarrow$
$(n, n', d, m) \notin ms'$

**Proof.**
Immediate from induction on steps.

**Lemma 16.**
$\forall W_1, W_2, n, n', d, m.$
$W_1 \xrightarrow{\tau}_p W_2 \wedge$
$(\forall i. \neg(\mathsf{n}(\tau(i)) = n \wedge \mathsf{d}(\tau(i)) = d \wedge \mathsf{o}(\tau(i)) = \downarrow \wedge \mathsf{e}(\tau(i)) = \mathsf{send}_l(n', m)))$
$\rightarrow$
$(n, n', d, m) \notin \mathit{ms}(W_2)$

**Proof.**

By the definition of the transition $\to_p$ and Lemma 15.

**Lemma 17.**
$\forall W_1, W_2, \tau, \tau_1, \tau_2, n, n', d, m.$
$W_1 \xrightarrow{\tau_1}{}_t^* \xrightarrow{\tau_2}{}_p W_2 \wedge$
$\tau = \tau_1 \cdot \tau_2 \wedge$
$(n, n', d, m) \notin \mathsf{ms}(W_1) \wedge$
$(\forall i. \neg(\mathsf{n}(\tau(i)) = n \wedge \mathsf{d}(\tau(i)) = d \wedge \mathsf{o}(\tau(i)) = \downarrow \wedge \mathsf{e}(\tau(i)) = \mathsf{send}_l(n', m)))$
$\to$
$(n, n', d, m) \notin \mathsf{ms}(W_2)$

**Proof.**
By Lemma 15 and Lemma 16.

**Lemma 18.**
$\forall \sigma, \sigma', ms, ms', f, r, r', \tau, n, n', d, m.$
$(n, n', d, m) \notin ms \wedge$
$[(\sigma, ms, f, r) \xrightarrow{\tau}{}_t^* (\sigma', ms', f', r') \vee$
$(\sigma, ms, f, r) \xrightarrow{\tau}{}_{req}^* (\sigma', ms') \vee$
$(\sigma, ms, f, r) \xrightarrow{\tau}{}_{ind}^* (\sigma', ms') \vee$
$(\sigma, ms, f, r) \xrightarrow{\tau}{}_{per}^* (\sigma', ms')] \wedge$
$\to$
$(\forall i. \neg(\mathsf{n}(\tau(i)) = n' \wedge \mathsf{d}(\tau(i)) = d \wedge \mathsf{o}(\tau(i)) = \uparrow \wedge \mathsf{e}(\tau(i)) = \mathsf{deliver}_l(n, m)))$

**Proof.**
Immediate from induction on steps.

**Lemma 19.**
$\forall W_1, W_2, n, n', d, m.$
$W_1 \xrightarrow{\tau}{}_p W_2 \wedge$
$(n, n', d, m) \notin \mathsf{ms}(W_1) \wedge$
$\to$
$(\forall i. \neg(\mathsf{n}(\tau(i)) = n' \wedge \mathsf{d}(\tau(i)) = d \wedge \mathsf{o}(\tau(i)) = \uparrow \wedge \mathsf{e}(\tau(i)) = \mathsf{deliver}_l(n, m)))$

**Proof.**
By the definition of the transition $\to_p$ and Lemma 18.

**Lemma 20.**
$\forall W_1, W_2, \tau, \tau_1, \tau_2, n, n', d, m.$
$W_1 \xrightarrow{\tau_1}{}_t^* \xrightarrow{\tau_2}{}_p W_2 \wedge$
$\tau = \tau_1 \cdot \tau_2 \wedge$

$(n, n', d, m) \notin ms(W_1)$
$(\forall i. \neg (n(\tau(i)) = n \land d(\tau(i)) = d \land o(\tau(i)) = \downarrow \land e(\tau(i)) = send_l(n', m)))$
$\rightarrow$
$(\forall i. \neg (n(\tau(i)) = n' \land d(\tau(i)) = d \land o(\tau(i)) = \uparrow \land e(\tau(i)) = deliver_l(n, m)))$

**Proof.**
By Lemma 18 and then Lemma 15 and Lemma 19.

**Lemma 21.**
$\forall W_1, W_2, \tau, \tau_1, \tau_2, n, n', d, m.$
$W_1 \xrightarrow{\tau_1}{}^*_t \xrightarrow{\tau_2}_p W_2 \land$
$\tau = \tau_1 \cdot \tau_2 \land$
$(n, n', d, m) \notin ms(W_1) \land$
$(\forall i. \neg (n(\tau(i)) = n \land d(\tau(i)) = d \land o(\tau(i)) = \downarrow \land e(\tau(i)) = send_l(n', m)))$
$\rightarrow$
$(n, n', d, m) \notin ms(W_2) \land$
$(\forall i. \neg (n(\tau(i)) = n' \land d(\tau(i)) = d \land o(\tau(i)) = \uparrow \land e(\tau(i)) = deliver_l(n, m)))$

**Proof.**
By Lemma 17 and Lemma 20.

**Lemma 22.**
$\forall W_1, \tau, n, n', d, m.$
$W_1 \xrightarrow{\tau}{}^* \quad \land$
$(n, n', d, m) \notin ms(W_1) \land$
$(\forall i. \neg (n(\tau(i)) = n \land d(\tau(i)) = d \land o(\tau(i)) = \downarrow \land e(\tau(i)) = send_l(n', m)))$
$\rightarrow$
$(\forall i. \neg (n(\tau(i)) = n' \land d(\tau(i)) = d \land o(\tau(i)) = \uparrow \land e(\tau(i)) = deliver_l(n, m)))$

**Proof.**
By Definition 9, induction on the steps and Lemma 21.

**Lemma 23.**
$\forall W_1, W_2, \tau, \tau_1, \tau_2, n, n', d, m.$
$W_1 \xrightarrow{\tau_1}_p W_2 \xrightarrow{\tau_2}{}^* \quad \land$
$\tau = \tau_1 \cdot \tau_2 \land$
$(\forall i. \neg (n(\tau(i)) = n \land d(\tau(i)) = d \land o(\tau(i)) = \downarrow \land e(\tau(i)) = send_l(n', m)))$
$\rightarrow$
$(\forall i. \neg (n(\tau(i)) = n' \land d(\tau(i)) = d \land o(\tau(i)) = \uparrow \land e(\tau(i)) = deliver_l(n, m)))$

**Proof.**

By Lemma 16 and Lemma 22.

**Lemma 24.**
$\forall W, \tau, n, n', d, m.$
$W \xrightarrow{\tau}{}^{*} \;\land$
$(\exists j.\; \forall i \geq j.$
$\quad \neg(n(\tau(i)) = n \land d(\tau(i)) = d \land o(\tau(i)) = \downarrow \land e(\tau(i)) = \text{send}_l(n', m)))$
$\rightarrow$
$(\exists j.\; \forall i \geq j.$
$\quad \neg(n(\tau(i)) = n' \land d(\tau(i)) = d \land o(\tau(i)) = \uparrow \land e(\tau(i)) = \text{deliver}_l(n, m)))$

**Proof.**
By Definition 9 and Lemma 23.

**Lemma 25.**
$\forall \mathcal{S}, W, \tau, n, n', d, m, i.$
$w_0(\mathcal{S}) \xrightarrow{\tau}{}^{*} W \;\land$
$(\forall i. \neg(n(\tau(i)) = n \land d(\tau(i)) = d \land o(\tau(i)) = \downarrow \land e(\tau(i)) = \text{send}_l(n', m)))$
$\rightarrow$
$(n, n', d, m) \notin \text{ms}(W)$

**Proof.**
By Definition 9, induction on the steps and Lemma 21.

**Lemma 26.**
$\forall \mathcal{S}, W, W', \tau_1, \tau_2, \tau, n, n', d, m, i.$
$w_0(\mathcal{S}) \xrightarrow{\tau_1}{}^{*} W \xrightarrow{\tau_2}{}_{t}^{*} W' \;\land$
$\tau = \tau_1 \cdot \tau_2 \;\land$
$(\forall i. \neg(n(\tau(i)) = n \land d(\tau(i)) = d \land o(\tau(i)) = \downarrow \land e(\tau(i)) = \text{send}_l(n', m)))$
$\rightarrow$
$(n, n', d, m) \notin \text{ms}(W')$

**Proof.**
By Lemma 25 and Lemma 15.

**Lemma 27.**
$\forall \mathcal{S}, \tau, n, n', d, m, i.$
$w_0(\mathcal{S}) \xrightarrow{\tau}{}^{*} \;\land$
$(\forall j \leq i.$
$\quad \neg(n(\tau(j)) = n \land d(\tau(j)) = d \land o(\tau(j)) = \downarrow \land e(\tau(j)) = \text{send}_l(n', m))$
$\rightarrow$
$\neg(n(\tau(i)) = n' \land d(\tau(i)) = d \land o(\tau(i)) = \uparrow \land e(\tau(i)) = \text{deliver}_l(n, m)))$

**Proof.**
By Definition 9, we consider two cases:
Case 1: The index $i$ is in a $\rightarrow_t$ transition.
   (1) $w_0(\mathcal{S}) \xrightarrow{\tau_1}^* W' \xrightarrow{\tau_2}^*_t W''$
   (2) $\tau = \tau_1 \cdot \tau_2$
   (3) $|\tau_1| \leq i < |\tau_1| + |\tau_2|$
   Immediate from Lemma 18.
Case 2: The index $i$ is in a $\rightarrow_p$ transition.
   (4) $w_0(\mathcal{S}) \xrightarrow{\tau_1}^* W' \xrightarrow{\tau_2}^*_t W'' \xrightarrow{\tau_3}_p W'''$
   (5) $\tau = \tau_1 \cdot \tau_2$
   (6) $|\tau_1| + |\tau_2| \leq i < |\tau_1| + |\tau_2| + |\tau_3|$
   Immediate from Lemma 26 and Lemma 19.

**Definition 15.**
*We instrument the transition system with event ID (ei), parent ID (pi) and child index (ci).*
*The event ID ei uniquely identifies events in a trace.*
*The parent ID pi is the ID of the event that issued the event.*
*The child index ci for an event is the index of that event in the list of events that its parent event issued. For example, if a parent event issues the list of request events $[e_1, e_2, e_3]$, then the child index of $e_2$ is 1. Similarly, if a parent event issues the list of indication events $[e_3, e_4, e_5]$, then the child index of $e_4$ is 1.*
*To create unique IDs, a counter ei is added to the state of the transition system and the counter is weaved through the rules.*
*As an example, the rule* REQ *is updated as follows:*
*The state of the transition system includes the counter ei:*
$(s, ms, f, ei)$
*The events include the event ID ei, the parent ID pi and the child index ci:*
$(ei, pi, ci, \_, \_, \_, \_, \_, \_, \_, \_, \_)$
*We use the functions* ei, pi *and* ci *to extract the corresponding components of an event.*
*As the counter in the pre-state is ei, the event id of the main event is $ei + 1$ and the updated counter is passed in the pre-state to the next transition for the issued request event.*
*The parent Id of the issued indication and request events is $ei + 1$.*
*Let us recall that in the interest of simplicity, the rules show only one issued request event and one issued indication event.*
*The child index of the issued indication and request events is 0.*

REQ
$$\frac{\begin{array}{c} n \in \mathbb{N} \setminus f \quad \mathcal{S}(d) = (c, \_) \quad \sigma(d) = s \\ \text{request}_c(n, s(n), e) = (s'_n, [(i, e_1)], [e_2]) \\ s' = s[n \mapsto s'_n] \quad \sigma' = \sigma[d \mapsto s'] \\ (\sigma', ms, f, r, ei + 1) \xrightarrow{\tau_1}_{req} (\sigma_1, ms_1, ei') \quad \tau_1 = (\_, ei + 1, 0, n, r, i :: d, \downarrow, e_1, \_, \_, \_, \_) \cdot \tau'_1 \\ (\sigma_1, ms_1, f, r, ei') \xrightarrow{\tau_2}_{ind} (\sigma_2, ms_2, ei'') \quad \tau_2 = (\_, ei + 1, 0, n, r, d, \uparrow, e_2, \_, \_, \_, \_) \cdot \tau'_2 \\ \tau = (ei + 1, \_, \_, n, d, \downarrow, e, \sigma, \sigma', (i, e_1), e_2) \cdot \tau_1 \cdot \tau_2 \end{array}}{(\sigma, ms, f, r, ei) \xrightarrow{\tau}_{req} (\sigma_2, ms_2, ei'')}$$

*It is easy to show that there is a bi-simulation between the original and the instrumented transition system.*

**Lemma 28.**
*Based on Definition 15, the event ID uniquely identifies events.*
$\forall \mathcal{S}, \tau, i, j$
$w_0(\mathcal{S}) \xrightarrow{\tau}{}^* \wedge$
$ei(\tau(i)) = ei(\tau(j)) \rightarrow$
$i = j$

**Proof.**
By the invariants:
In every transition, the event ID in the post-state is greater than the event ID in the pre-state.
The ID of every event in a trace is greater than the ID passed in the pre-state of the transitions.
Straightforward induction on the steps.

**Lemma 29.**
*Based on Definition 15, the parent ID, child index and the event orientation uniquely identify events.*
$\forall \mathcal{S}, \tau, i, j.$
$w_0(\mathcal{S}) \xrightarrow{\tau}{}^* \wedge$
$pi(\tau(i)) = pi(\tau(j)) \wedge$
$ci(\tau(i)) = ci(\tau(j)) \wedge$
$o(\tau(i)) = o(\tau(j)) \rightarrow$
$i = j$

**Proof.**
By the invariants:
In every transition, the event ID in the post-state is greater than the event ID in the pre-state.
The ID of every event in a trace is greater than the ID passed in the pre-state of the transitions.
The event ID in the pre-state is greater than or equal to the child ID of the first event in the trace.
The event ID of every event is greater than its parent ID.
In the trace of every transition, the parent ID of the first event is less than the parent ID of later events.
Straightforward induction on the steps.

**Lemma 30.**
*Based on Definition 15, the parent ID and the child index of an event point back to the origin of that event.*
$\forall \mathcal{S}, \tau, i.$
$w_0(\mathcal{S}) \xrightarrow{\tau}{}^* \wedge$

$d(\tau(i)) = [] \wedge$
$o(\tau(i)) = \uparrow \rightarrow$
$ois(\tau(pi(\tau(i))))(ci(\tau(i))) = e(\tau(i)) \wedge$
$n(\tau(pi(\tau(i)))) = n(\tau(i)) \wedge$
$mself(\tau(pi(\tau(i))))$

**Proof.**
Straightforward induction on the steps.

**Lemma 31.**
*If an event is issued only once, every executed event that matches it has the same parent ID and child index.*
$\forall \mathcal{S}, \tau, i, j, e, n, d. \exists k.$
$w_0(\mathcal{S}) \xrightarrow{\tau}{}^* \wedge$
$occ(ois(\tau(i)), e) \leq 1 \wedge$
$\forall k \neq i. \ (n(\tau_k) = n \wedge mself(\tau_k)) \rightarrow e \notin ois(\tau_k) \wedge$
$n(\tau(j)) = n \wedge$
$d(\tau(j)) = [] \wedge$
$o(\tau(j)) = \uparrow \wedge$
$e(\tau(j)) = e \rightarrow$
$pi(\tau(j)) = i \wedge$
$ci(\tau(j)) = k$

**Proof.**
Immediate from Lemma 30.

**Lemma 32.**
$\forall \mathcal{S}, \tau, i, j, k, e, n, d.$
$w_0(\mathcal{S}) \xrightarrow{\tau}{}^* \wedge$
$occ(ois(\tau(i)), e) \leq 1 \wedge$
$\forall k \neq i. \ (n(\tau_k) = n \wedge mself(\tau_k)) \rightarrow e \notin ois(\tau_k) \wedge$
$n(\tau(j)) = n \wedge$
$d(\tau(j)) = [] \wedge$
$o(\tau(j)) = \uparrow \wedge$
$e(\tau(j)) = e \wedge$
$k \neq j \rightarrow$
$\neg [n(\tau_k) = n \ \wedge \ d(\tau_k) = [] \ \wedge \ o(\tau_k) = \uparrow \ \wedge \ e(\tau_k) = e]$

**Proof.**
Immediate from Lemma 31.

**Lemma 33.**
$\forall W, W', \tau, i.$

$W \xrightarrow{\tau}{}^* W'$

$\rightarrow$

$n(\tau(i)) \in \mathbb{N}$

**Proof.**
By induction on the steps.

**Lemma 34.**
*The derived rules are sound.*

Proof.
Case rule FLoss:
    Immediate from GST.

Case rule IRSe:
    By the rule IR, definition of self and rule SInv.
    $\textsf{self} \wedge a \Rightarrow b$
    $\textsf{self} \Rightarrow a \to b$
    $\Box(\textsf{self} \to a \to b)$
    Ⓢ $\Box(a \to b)$
    Ⓢ $(a \Rightarrow b)$

Case rule PeSe:
    Similar to rule IRSe.
    By the rule Pe, definition of self and rule SInv.

Case rule IISe:
    Similar to rule IRSe.
    By the rule II, definition of self and rule SInv.

Case rule ORSe:
    By the rule OR, and rule SInv.

Case rule OISe:
    By the rule OR, and rule SInv.

Case rule ORSe′:
    By the rule OR′, definition of self and rule SInv.

Case rule OISe′:
    By the rule OI′, definition of self and rule SInv.

Case rule IROI:
    By the rule IR and rule OI.

Case rule IIOI:
    By the rule II and rule OI.

Case rule PeOI:
    By the rule Pe and rule OI.

Case rule IROR:
    By the rule IR and rule OR.

Case rule IIOR:
   By the rule II and rule OR.

Case rule PeOR:
   By the rule Pe and rule OR.

Case rule IROISe to rule PeORSe:
   By the rule IROI and rule PeOR and rule SInv.

Case rule APerSe:
   By rule APer, rule InvS and rule ASelf.

Case rule UniORSe:
   By the rule UniOR and rule SInv.

Case rule UniOISe:
   By the rule UniOI and rule SInv.

Case rule CSelf:
   By the definition of self.

Case rule SEqSe:
   By rule SEq and rule SInv.

Case rule InvSe:
   First proving $\Box(\mathsf{self} \to \mathcal{A})$, by the definition of self,
   distributing $\vee$ over $\to$, using $\mathcal{A} \wedge \mathcal{A}' \to \mathcal{A} \vee \mathcal{A}'$, Lemma 84, and
   rule $\wedge r$.
   Then using rule ASelf and Lemma 79.

Case rule InvSe':
   Similar to the proof of rule InvSe'. Instead of $\Box(\mathsf{self} \to \mathcal{A})$,
   we first prove $\Box(\mathsf{self} \wedge \hat{\boxminus}\mathcal{A} \to \mathcal{A})$.
   At the end, we use Lemma 110.

Case rule InvUSe:
   Using rule InvSe' and rule IRSe, rule IISe, and rule PeSe.

Case rule InvMSe:
   By rule InvSe' by instantiating $\mathcal{A}$ to $\mathcal{A} \wedge \mathcal{A}'$ and Lemma 93.

Case rule InvMSe':
   By rule InvUSe by instantiating $\mathcal{A}$ to $\mathcal{A} \wedge \mathcal{A}'$ and Lemma 93.

Case rule InvLSe:
   By rule InvUSe and rule Gen.

Case rule InvL:

By rule InvLSe and rule SInv.

Case rule InvUSSe:
  By Lemma 82 with $p$ instantiated to $S(\mathsf{s}(n))$.
  Then using rule PostPre to convert $\circ \mathsf{s}(n)$ to $\mathsf{s}'(n)$.
  Then case analysis on whether $\mathsf{n} = n$.
  Then rule SEqSe for $\mathsf{n} \neq n$ and rule InvUSe for $\mathsf{n} = n$.

Case rule InvUSSe':
  By rule InvUSSe', Lemma 83 and rule Init.

Case rule InvSSe:
  By rule InvUSSe and rule Gen.

Case rule InvS:
  By rule InvSSe and rule SInv.

Case rule InvSSe':
  By rule InvSSe, Lemma 83 and rule Init.

Case rule InvS':
  By rule InvSSe', and rule SInv.

Case rule InvSSe'':
  By rule InvSSe, Lemma 81, rule PostPre.

Case rule InvS'':
  By rule InvSSe'', and rule SInv.

Case rule InvSASe:
  We have
      (1) $\neg S(\mathsf{init}_c(n))$
  By rule InvLSe with
  $\mathcal{A}$ instantiated to $n \bullet \neg S(\mathsf{s}(n)) \wedge S(\mathsf{s}'(n))) \to \mathcal{A}$,
  we have
      (2) $\vdash_c \circledS\, [(n \bullet \neg S(\mathsf{s}(n)) \wedge S(\mathsf{s}'(n))) \Rightarrow \mathcal{A}]$
  We need to show that
      $\vdash_c \circledS\, (S(\mathsf{s}(n)) \Rightarrow \hat{\Diamond}\mathcal{A})$
  We show the contra-positive that is
      $\vdash_c \circledS\, (\hat{\boxminus}\neg\mathcal{A} \Rightarrow \neg S(\mathsf{s}(n)))$
  By rule Init and [1]
      (3) $\vdash_c \circledS\, \neg S(\mathsf{s}(n))$
  The contra-positive of [2] is
      (4) $\vdash_c \circledS\, [(\neg\mathcal{A} \Rightarrow \mathsf{n} \neq n \vee S(\mathsf{s}(n)) \vee \neg S(\mathsf{s}'(n)))]$
  that is
      (5) $\vdash_c \circledS\, [(\neg\mathcal{A} \Rightarrow (\mathsf{n} \neq n \vee \neg S(\mathsf{s}(n)) \to \neg S(\mathsf{s}'(n))))]$
  By rule SEqSe and [5]
      (6) $\vdash_c \circledS\, [(\neg\mathcal{A} \Rightarrow$

$$(\neg S(\mathsf{s}(n))) \to \neg S(\mathsf{s}'(n))) \lor \neg S(\mathsf{s}(n))) \to \neg S(\mathsf{s}'(n))))]$$
that is
(7) $\vdash_c \circledS [(\neg \mathcal{A} \Rightarrow \neg S(\mathsf{s}(n)) \to \neg S(\mathsf{s}'(n))))]$
By rule POSTPRE and [7]
(8) $\vdash_c \circledS [(\neg \mathcal{A} \Rightarrow (\neg S(\mathsf{s}(n)) \to \circ \neg S(\mathsf{s}(n))))]$
By Lemma 103, we have
(9) $\vdash_c \circledS [(\neg S(\mathsf{s}(n))) \to$
$(\hat{\boxminus}((\neg S(\mathsf{s}(n))) \Rightarrow \circ(\neg S(\mathsf{s}(n)))) \Rightarrow \neg S(\mathsf{s}(n)))]$
By [9] and [3]
(10) $\vdash_c \circledS [(\hat{\boxminus}((\neg S(\mathsf{s}(n))) \Rightarrow \circ(\neg S(\mathsf{s}(n)))) \Rightarrow \neg S(\mathsf{s}(n)))]$
By [8] and [10]
$\vdash_c \circledS [(\hat{\boxminus}\neg \mathcal{A} \Rightarrow \neg S(\mathsf{s}(n)))]$

Case rule INVSA:
By rule INVSASE, and rule SINV.

Case rule INVMSIASE:
By rule INVSE on the assumptions, we have
$\vdash_c \circledS \hat{\boxminus}\mathcal{A} \land (\forall x.\ S(\mathsf{s}(n)) \to \hat{\diamond}\mathcal{A}')] \Rightarrow$
$\mathcal{A} \land (\forall x.\ S(\mathsf{s}'(n)) \to \diamond \mathcal{A}')$
By Lemma 110, we have
$\vdash_c \Box \mathcal{A}$
By rule POSTPRE, Lemma 114 and Axiom 18, we have
$\vdash_c \circledS (\forall x.\ S(\mathsf{s}(n)) \to \hat{\diamond}\mathcal{A}') \Rightarrow$
$\circ (\forall x.\ S(\mathsf{s}(n)) \to \hat{\diamond}\mathcal{A}')$
By Lemma 82
$\vdash_c \circledS (\forall x.\ S(\mathsf{s}(n)) \to \hat{\diamond}\mathcal{A}') \Rightarrow \Box(\forall x.\ S(\mathsf{s}(n)) \to \hat{\diamond}\mathcal{A}')$
From the assumption $\forall x.\neg S(\mathsf{init}_c(n))$ and rule INIT
$\vdash_c \circledS (\forall x.\ S(\mathsf{s}(n)) \to \hat{\diamond}\mathcal{A}')$
By Lemma 97
$\vdash_c \circledS \Box(\forall x.\ S(\mathsf{s}(n)) \to \hat{\diamond}\mathcal{A}')$

Case rule ASASE:
Assumption:
(11) $\vdash_c \circledS \mathcal{A} \Rightarrow S(\mathsf{s}'(n))$
(12) $\vdash_c \circledS S(\mathsf{s}(n)) \Rightarrow \Box \mathcal{A}'$
We prove
$\vdash_c \circledS \mathcal{A} \Rightarrow \hat{\Box}\mathcal{A}'$
By rule POSTPRE on [11],
(13) $\vdash_c \circledS \mathcal{A} \Rightarrow \circ S(\mathsf{s}(n))$
By Lemma 81 ($\circ$M) on [12]
(14) $\vdash_c \circledS \circ S(\mathsf{s}(n)) \Rightarrow \hat{\Box}\mathcal{A}'$
By Lemma 80 ($\Rightarrow$T) on [13] and [14],
$\vdash_c \circledS \mathcal{A} \Rightarrow \hat{\Box}\mathcal{A}'$

Case rule ASA:
By rule ASASE, and rule SINV.

Case rule APerSA:
  By rule APer, and rule Pe.

Case rule Quorum:
Let
  (15) $\Gamma = |\mathsf{Correct}| > t_1;\ N \subseteq \mathbb{N};\ |N| > t_2;\ t_1 + t_2 \geq |\mathbb{N}|$
We show that
  $\Gamma \vdash_c \exists n.\ n \in N \land n \in \mathsf{Correct}$

We have
  (16) $\Gamma \vdash_c \mathsf{Correct} \subseteq \mathbb{N}$
From [15]
  (17) $\Gamma \vdash_c |\mathsf{Correct}| > t_1$
  (18) $\Gamma \vdash_c N \subseteq \mathbb{N}$
  (19) $\Gamma \vdash_c |N| > t_2$
  (20) $\Gamma \vdash_c |t_1 + t_2| \geq |\mathbb{N}|$
From set theory
  (21) $\Gamma \vdash_c \forall S_1, S_2, S.\ \ S_1 \subseteq S \land S_1 \subseteq S \rightarrow S_1 \cup S_2 \subseteq S$
By [21] on [16] and [18]
  (22) $\Gamma \vdash_c \mathsf{Correct} \cup N \subseteq \mathbb{N}$
From set theory
  (23) $\Gamma \vdash_c \forall S_1, S_2.\ \ S_1 \subseteq S_2 \rightarrow |S_1| \leq |S_2|$
By [23] on [22]
  (24) $\Gamma \vdash_c |\mathsf{Correct} \cup N| \leq |\mathbb{N}|$
From [17], [19], [24], and [20]
  (25) $|\mathsf{Correct}| + |N| - |\mathsf{Correct} \cup N| > 0$
From set theory
  (26) $\Gamma \vdash_c \forall S_1, S_2.\ \ |S_1 \cup S_2| = |S_1| + |S_2| - |S_1 \cap S_2|$
that is
  (27) $\Gamma \vdash_c \forall S_1, S_2.\ \ |S_1 \cap S_2| = |S_1| + |S_2| - |S_1 \cup S_2|$
By [27] on [25]
  (28) $\Gamma \vdash_c |\mathsf{Correct} \cap N| > 0$
From set theory
  (29) $\Gamma \vdash_c \forall S.\ \ |S| > 0 \leftrightarrow S \neq \emptyset$
By [29] on [28]
  (30) $\Gamma \vdash_c \mathsf{Correct} \cap N \neq \emptyset$
From set theory
  (31) $\Gamma \vdash_c \forall S.\ \ S = \emptyset \leftrightarrow \nexists s.\ s \in S$
By contrapositive of [31] on [30]
  (32) $\Gamma \vdash_c \exists n.\ n \in \mathsf{Correct} \cap N$
From set theory
  (33) $\Gamma \vdash_c \forall S_1, S_2. \forall s.\ \ s \in S_1 \cap S_2 \leftrightarrow s \in S_1 \land s \in S_2$
By contrapositive of [33] on [32]
  (34) $\Gamma \vdash_c \exists n.\ n \in \mathsf{Correct} \land n \in N$

Thus, we showed that
  $|\mathsf{Correct}| > t_1;\ N \subseteq \mathbb{N};\ |N| > t_2;\ t_1 + t_2 \geq |\mathbb{N}| \vdash_c$

    $\exists n.\ n \in N \wedge n \in \mathsf{Correct}$
By rule $\rightarrow r$, we have
    $|\mathsf{Correct}| > t_1 \ \vdash_c$
    $N \subseteq \mathbb{N} \wedge |N| > t_2 \wedge t_1 + t_2 \geq |\mathbb{N}| \rightarrow \exists n.\ n \in N \wedge n \in \mathsf{Correct}$
As the assertion is non-temporal, we have
    $|\mathsf{Correct}| > t_1 \ \vdash_c$
    $N \subseteq \mathbb{N} \wedge |N| > t_2 \wedge t_1 + t_2 \geq |\mathbb{N}| \Rightarrow \exists n.\ n \in N \wedge n \in \mathsf{Correct}$

## 5.2 Composition

**Lemma 1.**
For all $\tau \in T(\mathsf{stack}(c, \overline{\mathcal{S}}))$ there exists $\tau' \in T(\mathcal{S}_i)$ such that $\tau|_{d \supseteq [i]} = \mathsf{push}(i, \tau')$

**Proof.** Induction on the transition steps of the network semantic (Figure 14) for $\tau$ :
The trace $\tau'$ on $\mathcal{S}_i$ with state $(\sigma', ms', f, r)$ is inductively build from the trace $\tau$ on $\mathsf{stack}(c, \overline{\mathcal{S}})$ with the state $(\sigma, ms, f, r)$.
$\sigma' = \lambda d.\ \sigma(i :: d)$ and
$ms' = \{(n, n', d, m) \mid (n, n', i :: d, m) \in ms\}$
The induction hypothesis is strengthened with the fact
$\forall d.\ \sigma(i :: d) = \sigma'(d)$
that is the state of the substack $i$ of $\mathcal{S}$ is the same as the state of $\mathcal{S}_i$.
On every step $j$ of $\tau$ on $(\sigma, ms, f, r)$:
(1) If the step is not in substack $i$ that is $d(\tau(j)) \not\supseteq [i]$, no step ($\epsilon$ step) is taken for $\tau'$.
(2) If the step is in substack $i$ that is $\tau(j) = (n, r, i :: d, o, e, s, s', ors, ois)$, a corresponding step $(n, r, d, o, e, s|_{d \supseteq [i]}, s'|_{d \supseteq [i]}, ors, ois)$ can be applied to $(\sigma', ms', f', r')$. Note that all the elements except $d$ and the states of the two events are the same. The state $s|_{d \supseteq [i]}$ is a projection over $s$ for locations that are an extension of $[i]$.

**Corollary 2.**
For all $c$, $\overline{\mathcal{S}_i}$ and $m \in M(\mathsf{stack}(c, \overline{\mathcal{S}_i}))$, there exists $m' \in M(\mathcal{S}_i)$ such that $m|_{d \supseteq [i]} = \mathsf{push}(i, m')$

**Proof.** Immediate from Definition 12 and Corollary 2.

**Lemma 2.** For all $m$, $\mathcal{A}^{[]}$ and $i$,
$m \vDash \mathcal{A}^{[]} \rightarrow$
$\mathsf{push}(i, m) \vDash \mathsf{push}(i, \mathcal{A}^{[]})$

**Proof.** Immediate from induction on the structure of $\mathcal{A}^{[]}$.

**Definition 16** (Expanding a Model).

$$\mathsf{ext}((\tau, j, I), i) \triangleq \{(\tau'', j'', I) \mid \\ \exists \tau'.\ \forall j.\ d(\tau'(j)) \not\supseteq [i] \land \tau'' \in \mathsf{interleave}(\tau, \tau') \land \tau''(j'') = \tau(j)\}$$

We assume that events of a trace are unique. Uniqueness of events can be simply provided by the semantics with separate counters in nodes.

**Lemma 3.**
For all $m$, $i$, $m'$ and $\mathcal{I}^{[i]}$,
$m \vDash \mathcal{I}^{[i]} \land$
$m' \in \mathsf{extend}(i, m) \rightarrow$
$m' \vDash \mathsf{restrict}(\mathsf{d} = [i], \mathcal{I}^{[i]})$.

**Proof.**
Let
  (1) $\mathcal{I}^{[i]} = \Box \mathcal{A}^{[i]}$
We assume that
  (2) $m \vDash \Box \mathcal{A}^{[i]}$
  (3) $m' \in \mathsf{extend}(i, m)$
We show that
  $m' \vDash \mathsf{restrict}(\mathsf{d} = [i], \Box \mathcal{A}^{[i]})$
From Definition 8 (the definition of extend), on [3], there exists $\tau, \tau'$ and $I$ such that
  (4) $m = (\tau, 0, I)$
  (5) $m' = (\tau'', 0, I)$
  (6) $\forall j.\ d(\tau'(j)) \not\sqsupseteq [i]$
  (7) $\tau'' \in \mathsf{interleave}(\tau, \tau')$
From [5], we need to show that
  $(\tau'', 0, I) \vDash \mathsf{restrict}(\mathsf{d} = [i], \Box \mathcal{A}^{[i]})$
From Definition 3 (layering of assertions), we need to show that
  $(\tau'', 0, I) \vDash \Box(\mathsf{d} = [i] \to \mathsf{restrict}(\mathsf{d} = [i], \mathcal{A}^{[i]}))$
From Definition 6 (the definition of the models relation), we need to show that
  $(\tau'', k'', I) \vDash (\mathsf{d} = [i] \to \mathsf{restrict}(\mathsf{d} = [i], \mathcal{A}^{[i]}))$ forall $k'' \geq 0$

By [7], let $j''$ be the index where the element at index 0 of $\tau$ appears i.e.
  (8) $\tau''(j'') = \tau(0)$
We consider two cases:

Case:
  (9) $k'' < j''$
  From [7], [9], [8], and [6],
    (10) $d(\tau''(k'')) \neq [i]$
  Therefore, the following assertion is trivially holds
    $(\tau'', k'', I) \vDash (\mathsf{d} = [i] \to \mathsf{restrict}(\mathsf{d} = [i], \mathcal{A}^{[i]}))$

Case:
  (11) $k'' \geq j''$
  From Definition 16 (the definition of ext), on [6], [7] and [8],
    (12) $(\tau'', j'', I) \in \mathsf{ext}(i, (\tau, 0, I))$
  From [12] and [4],
    (13) $(\tau'', j'', I) \in \mathsf{ext}(i, m)$
  By Lemma 35 on [2], [13]
    (14) $(\tau'', j'', I) \vDash \mathsf{restrict}(\mathsf{d} = [i], \Box \mathcal{A}^{[i]})$
  From Definition 3 (layering of assertions) on [14]
    (15) $(\tau'', j'', I) \vDash \Box(\mathsf{d} = [i] \to \mathsf{restrict}(\mathsf{d} = [i], \mathcal{A}^{[i]}))$
  From Definition 6 (the definition of the models relation) on [15]
    (16) $(\tau'', k'', I) \vDash (\mathsf{d} = [i] \to \mathsf{restrict}(\mathsf{d} = [i], \mathcal{A}^{[i]}))$
          forall $k'' \geq j''$
  From [16] and [11]

$$(\tau'', k'', I) \vDash (\mathsf{d} = [i] \to \mathsf{restrict}(\mathsf{d} = [i], \mathcal{A}^{[i]}))$$

**Lemma 35.**
*For all $m$, $i$, $m'$ and $\mathcal{A}^{[i]}$,*
$m \vDash \mathcal{A}^{[i]}\ \land$
$m' \in \mathsf{ext}(i, m)\ \to$
$m' \vDash \mathsf{restrict}(\mathsf{d} = [i], \mathcal{A}^{[i]})$

**Proof.**
Immediate from Lemma 36.

**Lemma 36.** *For all $m$, $i$, $\mathcal{A}^{[i]}$, and $m'$,*
*if $m' \in \mathsf{ext}(i, m)$, then*
$m \vDash \mathcal{A}^{[i]}\ \leftrightarrow\ m' \vDash \mathsf{restrict}(\mathsf{d} = [i], \mathcal{A}^{[i]})$

**Proof.**
We assume that
    (1) $m' \in \mathsf{ext}(i, m)$
From Definition 16 (the definition of ext), on [1], we have
    (2) $m = (\tau, j, I)$
    (3) $m' = (\tau'', j'', I)$
    (4) $\forall j.\ d(\tau'(j)) \not\supseteq [i]$
    (5) $\tau'' \in \mathsf{interleave}(\tau, \tau')$
    (6) $\tau''(j'') = \tau(j)$
We show that
    $m \vDash \mathcal{A}^{[i]}$
if and only if
    $m' \vDash \mathsf{restrict}(\mathsf{d} \supseteq [i], \mathcal{A}^{[i]})$

Induction on the structure of $\mathcal{A}^{[i]}$

Case $\mathsf{n} = t_1\ \land\ \mathsf{d} = d'\ \land\ \mathsf{o} = t_2\ \land\ \mathsf{e} = t_3\quad d' \supseteq [i]$:
    The forward direction:
    We assume that
        (7) $m \vDash \mathsf{n} = t_1\ \land\ \mathsf{d} = d'\ \land\ \mathsf{o} = t_2\ \land\ \mathsf{e} = t_3\quad d' \supseteq [i]$
    We show that
        $m' \vDash \mathsf{restrict}(\mathsf{d} \supseteq [i], \mathsf{n} = t_1\ \land\ \mathsf{d} = d'\ \land\ \mathsf{o} = t_2\ \land\ \mathsf{e} = t_3)$
    From Definition 6 (the definition of the models relation),
    on [2] and [7]
        (8) $m \vDash t_1 : v_1$
        (9) $m \vDash t_2 : v_2$
        (10) $m \vDash t_3 : v_3$
        (11) $n(\tau(j)) = v_1$
        (12) $d(\tau(j)) = d'\quad d' \supseteq [i]$
        (13) $o(\tau(j)) = v_2$

(14) $e(\tau(j)) = v_3$

From [2], [3] and [6] on [11], [12], [13] and [14]

(15) $m' \models t_1 : v_1$

(16) $m' \models t_2 : v_2$

(17) $m' \models t_3 : v_3$

(18) $n(\tau''(j'')) = v_1$

(19) $d(\tau''(j'')) = d' \quad d' \supseteq [i]$

(20) $o(\tau''(j'')) = v_2$

(21) $e(\tau''(j'')) = v_3$

From Definition 6 (the definition of the models relation), on [15] to [21]

(22) $m' \models (\mathsf{n} = n \,\wedge\, \mathsf{d} = d' \,\wedge\, \mathsf{o} = o \,\wedge\, \mathsf{e} = e) \quad d' \supseteq [i]$

From Definition 3, (layering of assertions)

(23) $\mathsf{restrict}(\mathsf{d} \supseteq [i], \mathsf{n} = n \,\wedge\, \mathsf{d} = d' \,\wedge\, \mathsf{o} = o \,\wedge\, \mathsf{e} = e) =$
$\mathsf{n} = n \,\wedge\, \mathsf{d} = d' \,\wedge\, \mathsf{o} = o \,\wedge\, \mathsf{e} = e$

From [22], [23]

$m' \models \mathsf{restrict}(\mathsf{d} = [i], \mathsf{n} = n \,\wedge\, \mathsf{d} = d' \,\wedge\, \mathsf{o} = o \,\wedge\, \mathsf{e} = e)$

The backward direction is similar.

Case $\mathcal{A}_1^{[i]} \wedge \mathcal{A}_2^{[i]}$:

Immediate from the induction hypothesis.

Case $\neg \mathcal{A}^{[i]}$:

Forward direction Immediate from the backward direction of the induction hypothesis and vice versa.

Case $\Box \mathcal{A}^{[i]}$:

The forward direction:

We assume that

(24) $m \models \Box \mathcal{A}^{[i]}$

We show that

$m' \models \mathsf{restrict}(\mathsf{d} = [i], \Box \mathcal{A}^{[i]})$

From Definition 3 (layering of assertions), we show that

$m' \models \Box(\mathsf{d} = [i] \rightarrow \mathsf{restrict}(\mathsf{d} = [i], \mathcal{A}^{[i]}))$

From Definition 6 (the definition of the models relation), on [3], we show that

$(\tau'', k'', I) \models (\mathsf{d} = [i] \rightarrow \mathsf{restrict}(\mathsf{d} = [i], \mathcal{A}^{[i]}))$
forall $k'' \geq j''$

From Definition 6 (the definition of the models relation), on [2] and [24]

(25) $(\tau, k, I) \models \mathcal{A}^{[i]} \quad$ forall $k \geq j$

For all $k''$, such that

(26) $k'' \geq j''$

We consider two cases:

Case:

(27) $\mathsf{d}(\tau''(k'')) \neq [i]$

The assertion is trivially true.

$(\tau'', k'', I) \models (\mathsf{d} = [i] \to \mathsf{restrict}(\mathsf{d} = [i], \mathcal{A}^{[i]}))$

Case:

(28) $\mathsf{d}(\tau''_( k'')) = [i]$

We need to show that
$(\tau'', k'', I) \models \mathsf{restrict}(\mathsf{d} = [i], \mathcal{A}^{[i]})$

From [5], [4], and [6] on [28] and [26], there exists $k$, such that

(29) $k \geq j$

(30) $\tau''(k'') = \tau_k$

From [25] and [29]

(31) $(\tau, k, I) \models \mathcal{A}^{[i]}$

From [4], [5] and [30],

(32) $(\tau'', k'', I) \in \mathsf{ext}(i, (\tau, k, I))$

By the induction hypothesis on [31] and [32],
$(\tau'', k'', I) \models \mathsf{restrict}(\mathsf{d} = [i], \mathcal{A}^{[i]})$

The backward direction is similar.

Other cases are similar.

**Theorem 1 (Composition).**
For all $\mathcal{S}$, $c$, and $\overline{\mathcal{S}_i}$,
$\mathcal{S} = \mathsf{stack}(c, \overline{\mathcal{S}_i}) \wedge$
$\models_{\mathcal{S}_i} \mathcal{I}_i^{[]}$
$\to$
$\models_{\mathcal{S}} \mathsf{lower}(i, \mathcal{I}_i^{[]})$

**Proof.**
We assume that

(1) $\mathcal{S} = \mathsf{stack}(c, \overline{\mathcal{S}_i})$

(2) $\models_{\mathcal{S}_i} \mathcal{I}_i^{[]}$

We show that
$\mathcal{S} \models \mathsf{lower}(i, \mathcal{I}_i^{[]})$

From Definition 4 on [2]

(3) $\forall m' \in M(\mathcal{S}_i).\ m' \models \mathcal{I}_i^{[]}$

By Lemma 2 on [3]

(4) $\forall m' \in M(\mathcal{S}_i).\ \mathsf{push}(i, m') \models \mathsf{push}(i, \mathcal{I}_i^{[]})$

By Corollary 2

(5) $\forall m \in M(\mathcal{S}).\ \exists m' \in M(\mathcal{S}_i).$
$m|_{\mathsf{d} \supseteq [i]} = \mathsf{push}(i, m')$

From [5] and [4]

(6) $\forall m \in M(\mathcal{S}).\ m|_{\mathsf{d} \supseteq [i]} \models \mathsf{push}(i, \mathcal{I}_i^{[]})$

We have

(7) $\forall m \in M(\mathcal{S}).\ m \in \mathsf{extend}(i, m|_{\mathsf{d} \supseteq [i]})$

By Lemma 3 on [6] and [7], we have

(8) $\forall m \in M(\mathcal{S}).\ m \models \mathsf{restrict}(\mathsf{d} = [i], \mathsf{push}(i, \mathcal{I}_i^{[]}))$

From Definition 1 on [8], we have

(9) $\forall m \in M(\mathcal{S}).\ m \vDash \mathsf{lower}(i, \mathcal{I}_i^{[]})$

From Definition 4 on [8], we have

(10) $\vDash_{\mathcal{S}} \mathsf{lower}(i, \mathcal{I}_i^{[]})$

**Corollary 1 (Composition Soundness).**
For all $\mathcal{S}$, $c$, and $\overline{\mathcal{S}_i}$,
$\mathcal{S} = \mathsf{stack}(c, \overline{\mathcal{S}_i}) \wedge$
$\overline{\vDash_{\mathcal{S}_i} \mathcal{I}_i^{[]}} \wedge$
$\overline{\mathsf{lower}(i, \mathcal{I}_i^{[]})} \vdash_c \mathcal{A}$
$\rightarrow$
$\vDash_{\mathcal{S}} \mathcal{A}$.

**Proof.**
We assume that

(1) $\mathcal{S} = \mathsf{stack}(c, \overline{\mathcal{S}_i})$

(2) $\overline{\vDash_{\mathcal{S}_i} \mathcal{I}_i^{[]}}$

(3) $\overline{\mathsf{lower}(i, \mathcal{I}_i^{[]})} \vdash_c \mathcal{A}$

We show that

$\vDash_{\mathcal{S}} \mathcal{A}$

From Theorem 1 on [2], we have

(4) $\overline{\vDash_{\mathcal{S}} \mathsf{lower}(i, \mathcal{I}_i^{[]})}$

From Corollary 2 on [4] and [3], we have

$\vDash_{\mathcal{S}} \mathcal{A}$

## 5.3 Component Verification

### 5.3.1 Stubborn Links

**Theorem 4.** *($SL_1$: Stubborn delivery)*
If a correct node $n$ sends a message $m$ to a correct node $n'$, then $n'$ delivers $m$ infinitely often.

$\vdash_{\mathsf{SLC}} n \in \mathsf{Correct} \wedge n' \in \mathsf{Correct} \rightarrow$
$\quad (n \bullet \top \downarrow \mathsf{send}_{\mathsf{sl}}(n', m)) \Rightarrow \Box \Diamond (n' \bullet \top \uparrow \mathsf{deliver}_{\mathsf{sl}}(n, m))$

**Proof.**

The proof idea: upon a stubborn link send request for a message $m$, $m$ is added to the sent set. The periodic section of SLC, reissues a send request for every message in the sent set. The periodic function executes infinitely often. Thus, send requests are infinitely often issued for every messages in the send set. By the fair-loss property of the basic link, if a message is infinitely sent, then it will be infinitely often delivered. If a basic deliver indication is issued, then a stubborn link deliver indication is issued and then eventually executed.

We assume
  (1) $\Gamma = n \in \mathsf{Correct}; n' \in \mathsf{Correct}$
We prove that
  $\Gamma \vdash_{\mathsf{SLC}} (n \bullet \top \downarrow \mathsf{send}_{\mathsf{sl}}(n', m)) \Rightarrow$
  $\qquad \Box \Diamond (n' \bullet \top \uparrow \mathsf{deliver}_{\mathsf{sl}}(n, m))$

By rule IR,
  (2) $\Gamma \vdash_{\mathsf{SLC}} (n \bullet \top \downarrow \mathsf{send}_{\mathsf{sl}}(n', m)) \Rightarrow$
  $\qquad (\mathsf{self} \wedge \langle n', m \rangle \in \mathsf{s}'(n))$
By rule INVS″ with $S = \lambda s.\ \langle n', m \rangle \in s$,
  (3) $\Gamma \vdash_{\mathsf{SLC}} (\mathsf{self} \wedge \langle n', m \rangle \in \mathsf{s}'(n)) \Rightarrow$
  $\qquad \hat{\Box}(\mathsf{self} \Rightarrow \langle n', m \rangle \in \mathsf{s}(n))$
From [2] and [3],
  (4) $\Gamma \vdash_{\mathsf{SLC}} (n \bullet \top \downarrow \mathsf{send}_{\mathsf{sl}}(n', m)) \Rightarrow$
  $\qquad \hat{\Box}(\mathsf{self} \Rightarrow \langle n', m \rangle \in \mathsf{s}(n))$

By rule APERSA with:
  $S = \lambda s.\ \langle n', m \rangle \in s$ and
  $\mathcal{A} = n \bullet (0, \mathsf{send}_{\mathsf{l}}(n', m)) \in \mathsf{ors} \wedge \mathsf{self}$ on [1], we have
  (5) $\Gamma \vdash_{\mathsf{SLC}} n \in \mathsf{Correct} \rightarrow$
  $\qquad (\mathsf{self} \Rightarrow \langle n', m \rangle \in \mathsf{s}(n)) \Rightarrow$
  $\qquad \Box \Diamond (n \bullet (0, \mathsf{send}_{\mathsf{l}}(n', m)) \in \mathsf{ors} \wedge \mathsf{self})$
From [4] and [5], we have
  (6) $\Gamma \vdash_{\mathsf{SLC}} (n \bullet \top \downarrow \mathsf{send}_{\mathsf{sl}}(n', m)) \Rightarrow$

$$\Box\Diamond(n \bullet (0, \mathsf{send}_\mathsf{l}(n', m)) \in \mathsf{ors} \wedge \mathsf{self})$$

From rule OR,
    (7) $\Gamma \vdash_\mathsf{SLC} (n \bullet (0, \mathsf{send}_\mathsf{l}(n, m)) \in \mathsf{ors} \wedge \mathsf{self}) \Rightarrow$
            $\Diamond(n \bullet 0 \downarrow \mathsf{send}_\mathsf{l}(n, m))$

From [6] and [7],
    (8) $\Gamma \vdash_\mathsf{SLC} (n \bullet \top \downarrow \mathsf{send}_\mathsf{sl}(n', m)) \Rightarrow$
            $\Box\Diamond\Diamond(n \bullet 0 \downarrow \mathsf{send}_\mathsf{l}(n, m))$

From Lemma 87 on [8],
    (9) $\Gamma \vdash_\mathsf{SLC} (n \bullet \top \downarrow \mathsf{send}_\mathsf{sl}(n', m)) \Rightarrow$
            $\Box\Diamond(n \bullet 0 \downarrow \mathsf{send}_\mathsf{l}(n, m))$

By rule FLoss on [1] and [9],
    (10) $\Gamma \vdash_\mathsf{SLC} (n \bullet \top \downarrow \mathsf{send}_\mathsf{sl}(n', m)) \Rightarrow$
            $\Box\Diamond(n' \bullet 0 \downarrow \mathsf{deliver}_\mathsf{l}(n, m))$

From rule IIOI,
    (11) $\Gamma \vdash_\mathsf{SLC} (n' \bullet 0 \downarrow \mathsf{deliver}_\mathsf{l}(n, m)) \Rightarrow$
            $\Diamond(n' \bullet \top \downarrow \mathsf{deliver}_\mathsf{sl}(n, m))$

From [10] and [11],
    (12) $\Gamma \vdash_\mathsf{SLC} (n \bullet \top \downarrow \mathsf{send}_\mathsf{sl}(n', m)) \Rightarrow$
            $\Box\Diamond\Diamond(n' \bullet \top \downarrow \mathsf{deliver}_\mathsf{sl}(n, m))$

From Lemma 87 and [12]
  $\Gamma \vdash_\mathsf{SLC} (n \bullet \top \downarrow \mathsf{send}_\mathsf{sl}(n', m)) \Rightarrow$
            $\Box\Diamond(n' \bullet \top \downarrow \mathsf{deliver}_\mathsf{sl}(n, m))$

**Theorem 5.** *($SL_2$: No-forge)*
If a node $n$ delivers a message $m$ with sender $n'$, then $m$ was previously sent to $n$ by $n'$.

$$\vdash_{\mathsf{SLC}} (n \bullet \top \uparrow \mathsf{deliver}_{\mathsf{sl}}(n', m)) \rightsquigarrow (n' \bullet \top \downarrow \mathsf{send}_{\mathsf{sl}}(n, m))$$

**Proof.**

The proof idea: If a stubborn link delivery event is executed, it is issued by a previous basic link delivery event. By the no-forge property of basic links, we know that any basic link delivery event is preceded by a basic link send event. A basic link send event is only issued by a previous stubborn link send event.

By rule OI',
(1) $\vdash_{\mathsf{SLC}} (n \bullet \top \uparrow \mathsf{deliver}_{\mathsf{sl}}(n', m)) \Rightarrow$
$\quad\quad \diamondsuit(n \bullet \mathsf{deliver}_{\mathsf{sl}}(n', m) \in \mathsf{ois} \land \mathsf{self})$

By rule INVL,
(2) $\vdash_{\mathsf{SLC}} (n \bullet \mathsf{deliver}_{\mathsf{sl}}(n', m) \in \mathsf{ois} \land \mathsf{self}) \Rightarrow$
$\quad\quad (n \bullet 0 \uparrow \mathsf{deliver}_{\mathsf{l}}(n', m))$

By using [1] and [2],
$\vdash_{\mathsf{SLC}} (n \bullet \top \uparrow \mathsf{deliver}_{\mathsf{sl}}(n', m)) \Rightarrow$
$\quad\quad \diamondsuit(n \bullet 0 \uparrow \mathsf{deliver}_{\mathsf{l}}(n', m))$

That is,
(3) $\vdash_{\mathsf{SLC}} (n \bullet \top \uparrow \mathsf{deliver}_{\mathsf{sl}}(n', m)) \rightsquigarrow$
$\quad\quad (n \bullet 0 \uparrow \mathsf{deliver}_{\mathsf{l}}(n', m))$

By rule NFORGE,
(4) $\vdash_{\mathsf{SLC}} (n \bullet 0 \uparrow \mathsf{deliver}_{\mathsf{l}}(n', m)) \rightsquigarrow$
$\quad\quad (n' \bullet 0 \downarrow \mathsf{send}_{\mathsf{l}}(n, m))$

By Lemma 88 on [3], [4],
(5) $\vdash_{\mathsf{SLC}} (n \bullet \top \uparrow \mathsf{deliver}_{\mathsf{sl}}(n', m)) \rightsquigarrow$
$\quad\quad (n' \bullet 0 \downarrow \mathsf{send}_{\mathsf{l}}(n, m))$

By rule OR',
(6) $\vdash_{\mathsf{SLC}} (n' \bullet 0 \downarrow \mathsf{send}_{\mathsf{l}}(n, m)) \Rightarrow$
$\quad\quad \diamondsuit(n' \bullet (0, \mathsf{send}_{\mathsf{l}}(n, m)) \in \mathsf{ors} \land \mathsf{self})$

By rule INVL,
(7) $\vdash_{\mathsf{SLC}} (n' \bullet (0, \mathsf{send}_{\mathsf{l}}(n, m)) \in \mathsf{ors} \land \mathsf{self}) \Rightarrow$
$\quad\quad (n' \bullet \top \downarrow \mathsf{send}_{\mathsf{sl}}(n, m))$

From [6] and [7],
$\vdash_{\mathsf{SLC}} (n' \bullet 0 \downarrow \mathsf{send}_{\mathsf{l}}(n, m)) \Rightarrow$
$\quad\quad \diamondsuit(n' \bullet \top \downarrow \mathsf{send}_{\mathsf{sl}}(n, m))$

That is,
(8) $\vdash_{\mathsf{SLC}} (n' \bullet 0 \downarrow \mathsf{send}_{\mathsf{l}}(n, m)) \rightsquigarrow$
$\quad\quad (n' \bullet \top \downarrow \mathsf{send}_{\mathsf{sl}}(n, m))$

From [5] and [8],
$\vdash_{\mathsf{SLC}} (n \bullet \top \uparrow \mathsf{deliver}_{\mathsf{sl}}(n', m)) \rightsquigarrow$

$(n' \bullet \top \downarrow \mathsf{send}_{\mathsf{sl}}(n, m))$

### 5.3.2 Perfect Links

**Definition 17.**
Lowering the properties of the stubborn link.
$\Gamma = \mathsf{SL}'_1, \mathsf{SL}'_2$

$\mathsf{SL}'_1 = \mathsf{lower}(0, \mathsf{SL}_1) =$
   $n \in \mathsf{Correct} \wedge n' \in \mathsf{Correct} \rightarrow$
   $(n \bullet 0 \downarrow \mathsf{send}_{\mathsf{sl}}(n', m)) \Rightarrow$
   $\Box \Diamond (n' \bullet 0 \uparrow \mathsf{deliver}_{\mathsf{sl}}(n, m))$

$\mathsf{SL}'_2 = \mathsf{lower}(0, \mathsf{SL}_2) =$
   $(n \bullet 0 \uparrow \mathsf{deliver}_{\mathsf{sl}}(n', m)) \rightsquigarrow$
   $(n' \bullet 0 \downarrow \mathsf{send}_{\mathsf{sl}}(n, m))$

**Theorem 6.** *(PL$_1$: Reliable delivery)*
If a correct node $n$ sends a message $m$ to a correct node $n'$, then $n'$ will eventually deliver $m$.

$\Gamma \vdash_{\mathsf{PLC}}$
$\quad n \in \mathsf{Correct} \land n' \in \mathsf{Correct} \rightarrow$
$\quad (n \bullet \top \downarrow \mathsf{send}_{\mathsf{pl}}(n', m) \rightsquigarrow$
$\quad (n' \bullet \top \uparrow \mathsf{deliver}_{\mathsf{pl}}(n, m))$
where
$\Gamma$ is defined in Definition 17.

**Proof.**

The proof idea: As a $\mathsf{send}_{\mathsf{pl}}$ event is executed at the sender, a $\mathsf{send}_{\mathsf{sl}}$ event is issued and eventually executes. Thus, by the stubborn link, a $\mathsf{deliver}_{\mathsf{sl}}$ event eventually executes at the receiver. At the $\mathsf{deliver}_{\mathsf{sl}}$ event, either the sender and counter pair is already in the current state or not. (1) If it is not, a $\mathsf{deliver}_{\mathsf{pl}}$ event is issued that eventually executes. This is the desired conclusion. (2) If the counter is already in the current state, a $\mathsf{deliver}_{\mathsf{sl}}$ event with the same sender and counter (with potentially a different message) has executed in the past that has issued a $\mathsf{deliver}_{\mathsf{pl}}$ event that will eventually execute. There cannot be two $\mathsf{deliver}_{\mathsf{sl}}$ events from the same sender with the same counter. Thus, the two $\mathsf{deliver}_{\mathsf{sl}}$ events are the same and carry the same message. Thus, the $\mathsf{deliver}_{\mathsf{pl}}$ event that will be executed is with the sent message.

We assume
$\quad \mathcal{A}_1 = n \in \mathsf{Correct} \land n' \in \mathsf{Correct}$
$\quad \Gamma' = \Gamma; \mathcal{A}_1$
We prove that
$\quad \Gamma' \vdash_{\mathsf{PLC}} (n \bullet \top \downarrow \mathsf{send}_{\mathsf{pl}}(n', m)) \rightsquigarrow (n' \bullet \top \uparrow \mathsf{deliver}_{\mathsf{pl}}(n, m))$

By rule IR, the definition of request and rule OR, there exists $c$ such that
$\quad$ (1) $\Gamma' \vdash_{\mathsf{PLC}} n \bullet \top \downarrow \mathsf{send}_{\mathsf{pl}}(n', m) \Rightarrow$
$\qquad \mathsf{self} \land n \bullet (0, \mathsf{send}_{\mathsf{sl}}(n', \langle c, m \rangle)) \in \mathsf{ors} \land \Diamond(n \bullet 0 \downarrow \mathsf{send}_{\mathsf{sl}}(n', \langle c, m \rangle))$

From $SL'_1$ and Axiom 1,
$\quad$ (2) $\Gamma' \vdash_{\mathsf{PLC}} n \bullet 0 \downarrow \mathsf{send}_{\mathsf{sl}}(n', \langle c, m \rangle) \Rightarrow$
$\qquad \Diamond(n' \bullet 0 \uparrow \mathsf{deliver}_{\mathsf{sl}}(n, \langle c, m \rangle))$

By rule Lemma 89 on [1] and [2],
$\quad$ (3) $\Gamma' \vdash_{\mathsf{PLC}} (n \bullet \top \downarrow \mathsf{send}_{\mathsf{pl}}(n', m)) \Rightarrow$

$$\text{self} \land n \bullet (0, \text{send}_\text{sl}(n', \langle c, m \rangle)) \in \text{ors} \land \Diamond(n' \bullet 0 \uparrow \text{deliver}_\text{sl}(n, \langle c, m \rangle))$$

By rule Lemma 99 on [3] and Lemma 38,
$$\Gamma' \vdash_{\mathsf{PLC}} n \bullet \top \downarrow \text{send}_\text{pl}(n', m) \Rightarrow$$
$$\text{self} \land n \bullet (0, \text{send}_\text{sl}(n', \langle c, m \rangle)) \in \text{ors} \land$$
$$\Diamond[\Diamond(n' \bullet \top \uparrow \text{deliver}_\text{pl}(n, m)) \lor$$
$$\exists m.\ \hat{\Diamond}[\text{self} \land (0, \text{send}_\text{sl}(n', \langle c, m \rangle)) \in \text{ors} \land \Diamond(n' \bullet \top \uparrow \text{deliver}_\text{pl}(n, m))]]$$

that is
$$\Gamma' \vdash_{\mathsf{PLC}} n \bullet \top \downarrow \text{send}_\text{pl}(n', m) \Rightarrow$$
$$\text{self} \land n \bullet (0, \text{send}_\text{sl}(n', \langle c, m \rangle)) \in \text{ors} \land$$
$$(\Diamond(n' \bullet \top \uparrow \text{deliver}_\text{pl}(n, m)) \lor$$
$$\exists m.\ \hat{\Diamond}[\text{self} \land n \bullet (0, \text{send}_\text{sl}(n', \langle c, m \rangle)) \in \text{ors} \land \Diamond(n' \bullet \top \uparrow \text{deliver}_\text{pl}(n, m))] \lor$$
$$\exists m.\ [\text{self} \land n \bullet (0, \text{send}_\text{sl}(n', \langle c, m \rangle)) \in \text{ors} \land \Diamond(n' \bullet \top \uparrow \text{deliver}_\text{pl}(n, m))] \lor$$
$$\exists m.\ \hat{\Diamond}[\text{self} \land n \bullet (0, \text{send}_\text{sl}(n', \langle c, m \rangle)) \in \text{ors} \land \Diamond(n' \bullet \top \uparrow \text{deliver}_\text{pl}(n, m))])$$

By Lemma 37
$$\Gamma' \vdash_{\mathsf{PLC}} n \bullet \top \downarrow \text{send}_\text{pl}(n', m) \Rightarrow$$
$$\text{self} \land n \bullet (0, \text{send}_\text{sl}(n', \langle c, m \rangle)) \in \text{ors} \land$$
$$(\Diamond(n' \bullet \top \uparrow \text{deliver}_\text{pl}(n, m)) \lor$$
$$\exists m.\ [\text{self} \land n \bullet (0, \text{send}_\text{sl}(n', \langle c, m \rangle)) \in \text{ors} \land \Diamond(n' \bullet \top \uparrow \text{deliver}_\text{pl}(n, m))])$$

Thus,
$$\Gamma' \vdash_{\mathsf{PLC}} n \bullet \top \downarrow \text{send}_\text{pl}(n', m) \Rightarrow$$
$$\Diamond(n' \bullet \top \uparrow \text{deliver}_\text{pl}(n, m))$$

**Lemma 37.**
$$\Gamma' \vdash_{\mathsf{PLC}} \text{self} \land n \bullet (0, \text{send}_\text{sl}(n', \langle c, m \rangle)) \in \text{ors} \Rightarrow$$
$$\hat{\Box}\neg[\text{self} \land n \bullet (0, \text{send}_\text{sl}(n', \langle c, m' \rangle)) \in \text{ors}]$$
$$\hat{\boxminus}\neg[\text{self} \land n \bullet (0, \text{send}_\text{sl}(n', \langle c, m' \rangle)) \in \text{ors}]$$

**Proof.**
By the rule INVL
(1) $\Gamma' \vdash_{\mathsf{PLC}} \text{self} \land n \bullet (0, \text{send}_\text{sl}(n', \langle c, m \rangle)) \in \text{ors} \Rightarrow$
$\text{counter}(\text{s}'(n)) = c$

By the rule INVS
(2) $\Gamma' \vdash_{\mathsf{PLC}} \text{self} \land \text{counter}(\text{s}(n)) \geq c \Rightarrow$
$(\text{self} \Rightarrow \text{counter}(\text{s}(n)) \geq c)$

By the rule ASA on [1] and [2],
(3) $\Gamma' \vdash_{\mathsf{PLC}} n \bullet \text{self} \land (0, \text{send}_\text{sl}(n', \langle c, m \rangle)) \in \text{ors} \Rightarrow$
$\hat{\Box}(\text{self} \rightarrow \text{counter}(\text{s}(n)) \geq c)$

By the rule INVL
(4) $\Gamma' \vdash_{\mathsf{PLC}} \text{self} \land \text{counter}(\text{s}(n)) \geq c \Rightarrow$
$\neg(n \bullet (0, \text{send}_\text{sl}(n', \langle c, m' \rangle)) \in \text{ors})$

From [3] and [4]
$\Gamma' \vdash_{\mathsf{PLC}} \text{self} \land n \bullet (0, \text{send}_\text{sl}(n', \langle c, m \rangle)) \in \text{ors} \Rightarrow$
$\hat{\Box}(\text{self} \rightarrow \neg(n \bullet (0, \text{send}_\text{sl}(n', \langle c, m' \rangle)) \in \text{ors}))$

that is
$\Gamma' \vdash_{\mathsf{PLC}} \text{self} \land n \bullet (0, \text{send}_\text{sl}(n', \langle c, m \rangle)) \in \text{ors} \Rightarrow$

$$\hat{\Box}\neg(\text{self} \land n \bullet (0, \text{send}_{\text{sl}}(n', \langle c, m' \rangle)) \in \text{ors})$$

The proof of the second conjunct is similar.

**Lemma 38.**
$\Gamma' \vdash_{\text{PLC}}$
$\quad (n' \bullet 0 \uparrow \text{deliver}_{\text{sl}}(n, \langle c, m \rangle)) \Rightarrow$
$\quad \Diamond(n' \bullet \top \uparrow \text{deliver}_{\text{pl}}(n, m)) \lor$
$\quad \exists m. \Diamondleft[(n \bullet \top \downarrow \text{send}_{\text{pl}}(n, m)) \land \Diamond(n' \bullet \top \uparrow \text{deliver}_{\text{pl}}(n, m))]$

**Proof.**
Immediate from considering two cases $(n, m) \notin \text{received}(\text{s}(n'))$ and $(n, m) \in \text{received}(\text{s}(n'))$ and Lemma 39 and Lemma 40.

**Lemma 39.**
$\Gamma' \vdash_{\text{PLC}} (n' \bullet 0 \uparrow \text{deliver}_{\text{sl}}(n, \langle c, m \rangle)) \land \langle n, c \rangle \notin \text{received}(\text{s}(n')) \Rightarrow$
$\quad \Diamond(n' \bullet \top \uparrow \text{deliver}_{\text{pl}}(n, m))$

**Proof.**
Immediate from rule II and rule OI.

**Lemma 40.**
$\Gamma' \vdash_{\text{PLC}} \text{self} \land \text{n} = n' \land \langle n, c \rangle \in \text{received}(\text{s}(n')) \Rightarrow$
$\quad \exists m. \Diamondleft[n \bullet (0, \text{send}_{\text{sl}}(n', \langle c, m \rangle)) \in \text{ors} \land \text{self} \land \Diamond(n' \bullet \top \uparrow \text{deliver}_{\text{pl}}(n, m))]$

**Proof.**
By Lemma 41
(1) $\Gamma' \vdash_{\text{PLC}} \langle n, c \rangle \in \text{received}(\text{s}(n')) \land \text{self} \Rightarrow$
$\quad \exists m. \Diamondleft[n' \bullet 0 \uparrow \text{deliver}_{\text{sl}}(n, \langle c, m \rangle) \land \text{deliver}_{\text{pl}}(n, m) \in \text{ois}]$

By $SL_2'$
(2) $\Gamma' \vdash_{\text{PLC}} n' \bullet 0 \uparrow \text{deliver}_{\text{sl}}(n, \langle c, m \rangle) \Rightarrow \Diamondleft(n \bullet 0 \downarrow \text{send}_{\text{sl}}(n', \langle c, m \rangle))$
By rule OR',
(3) $\Gamma' \vdash_{\text{PLC}} (n \bullet 0 \downarrow \text{send}_{\text{sl}}(n', \langle c, m \rangle)) \Rightarrow \Diamondleft(n \bullet (0, \text{send}_{\text{sl}}(n', \langle c, m \rangle)) \in \text{ors} \land \text{self})$
By the Lemma 89 on [2], and [3],
(4) $\Gamma' \vdash_{\text{PLC}} n' \bullet 0 \uparrow \text{deliver}_{\text{sl}}(n, \langle c, m \rangle) \Rightarrow \Diamondleft(n \bullet (0, \text{send}_{\text{sl}}(n', \langle c, m \rangle)) \in \text{ors} \land \text{self})$

By rule OI,
(5) $\Gamma' \vdash_{\text{PLC}} (n' \bullet \top \uparrow \text{deliver}_{\text{pl}}(n, m) \in \text{ois}) \Rightarrow \Diamond(n' \bullet \top \uparrow \text{deliver}_{\text{pl}}(n, m))$

By [1], [4], [5],
$\quad \Gamma' \vdash_{\text{PLC}} \text{self} \land \langle n, m \rangle \in \text{received}(\text{s}(n')) \Rightarrow \exists m. \Diamondleft[\Diamondleft(n \bullet (0, \text{send}_{\text{sl}}(n', \langle c, m \rangle)) \in \text{ors} \land \text{self}) \land \Diamond(n' \bullet \top \uparrow \text{deliver}_{\text{pl}}(n, m))]$
that is

$\Gamma' \vdash_{\mathsf{PLC}} \mathsf{self} \wedge (n,m) \in \mathsf{received}(\mathsf{s}(n')) \Rightarrow \exists m.\ \ominus[n \bullet (0, \mathsf{send}_{\mathsf{sl}}(n', \langle c,m \rangle)) \in \mathsf{ors} \wedge \mathsf{self} \wedge \Diamond(n' \bullet \top \uparrow \mathsf{deliver}_{\mathsf{pl}}(n,m))]$

**Lemma 41.**
$\Gamma' \vdash_{\mathsf{PLC}} \mathsf{self} \wedge \langle n,c \rangle \in \mathsf{received}(\mathsf{s}(n')) \Rightarrow$
$\quad \exists m.\ \ominus[n' \bullet 0 \uparrow \mathsf{deliver}_{\mathsf{sl}}(n, \langle c,m \rangle) \wedge \mathsf{deliver}_{\mathsf{pl}}(n,m) \in \mathsf{ois}]$

**Proof.**
We use rule INVSA with
$\quad n$ instantiated to $n'$,
$\quad S$ instantiated to $\lambda s.\ \langle n,c \rangle \in \mathsf{received}(s)$ and
$\quad \mathcal{A}$ instantiated to $\exists m.\ n' \bullet 0 \uparrow \mathsf{deliver}_{\mathsf{sl}}(n, \langle c,m \rangle) \wedge \mathsf{deliver}_{\mathsf{pl}}(n,m) \in \mathsf{ois}$
From the definition of PLC
$\quad \mathsf{init}_{\mathsf{PLC}} = \lambda n.\ \langle 0, \varnothing \rangle$
Thus
$\quad (n,m) \notin \mathsf{received}(\mathsf{init}_{\mathsf{PLC}}(n))$
Thus
$\quad (1) \quad \neg S(\mathsf{init}_{\mathsf{PLC}}(n))$

From the definition of request, we have
$\quad (2) \quad \mathsf{o} = \downarrow \wedge \mathsf{d} = \top \wedge$
$\qquad \mathsf{request}_c(\mathsf{n}, \mathsf{s}(\mathsf{n}), \mathsf{e}) = (\mathsf{s}'(\mathsf{n}), \mathsf{ois}, \mathsf{ors}) \wedge$
$\qquad \langle n,c \rangle \notin \mathsf{received}(\mathsf{s}(\mathsf{n})) \wedge \langle n,c \rangle \in \mathsf{received}(\mathsf{s}'(\mathsf{n})) \rightarrow$
$\qquad \exists m.\ n' \bullet 0 \uparrow \mathsf{deliver}_{\mathsf{sl}}(n, \langle c,m \rangle) \wedge \mathsf{deliver}_{\mathsf{pl}}(n,m) \in \mathsf{ois}$
and similarly for indication and periodic.

By rule INVSA on [1] and [2],
$\quad \Gamma' \vdash_{\mathsf{PLC}} \mathsf{self} \wedge (n,m) \in \mathsf{received}(\mathsf{s}(n')) \Rightarrow$
$\qquad \exists m.\ \ominus[n' \bullet 0 \uparrow \mathsf{deliver}_{\mathsf{sl}}(n, \langle c,m \rangle) \wedge \mathsf{deliver}_{\mathsf{pl}}(n,m) \in \mathsf{ois}]$

**Theorem 7.** *(PL$_2$: No-duplication)*
If a message is sent at most once, it will be delivered at most once.

$\Gamma \vdash_{\mathsf{PLC}} [n' \bullet \top \downarrow \mathsf{send}_{\mathsf{pl}}(n,m) \Rightarrow \hat{\boxminus}\neg(n' \bullet \top \downarrow \mathsf{send}_{\mathsf{pl}}(n,m))] \rightarrow$
$\qquad [n \bullet \top \uparrow \mathsf{deliver}_{\mathsf{pl}}(n',m) \Rightarrow \hat{\boxminus}\neg(n \bullet \top \uparrow \mathsf{deliver}_{\mathsf{pl}}(n',m))]$

where
$\Gamma$ is defined in Definition 17.

**Proof.**

Proof idea: If there is a perfect-link delivery event, there is a stubborn link delivery event before it. That stubborn link delivery event is the event that issues one perfect-link delivery event and there is no event before or after it that issues a perfect-link delivery event (Lemma 42). Therefore, no perfect-link delivery event other than the current one can be executed.

We assume
$\quad \mathcal{A}_1 = n' \bullet \top \downarrow \mathsf{send}_{\mathsf{pl}}(n,m) \Rightarrow \hat{\boxminus}\neg(n' \bullet \top \downarrow \mathsf{send}_{\mathsf{pl}}(n,m))$
$\quad \Gamma' = \Gamma; \mathcal{A}_1$
and prove
$\quad \Gamma' \vdash_{\mathsf{PLC}} n \bullet \top \uparrow \mathsf{deliver}_{\mathsf{pl}}(n',m) \Rightarrow \hat{\boxminus}\neg(n \bullet \top \uparrow \mathsf{deliver}_{\mathsf{pl}}(n',m))$

By rule OI',
$\quad$(1) $\Gamma' \vdash_{\mathsf{PLC}} n \bullet \top \uparrow \mathsf{deliver}_{\mathsf{pl}}(n',m) \Rightarrow$
$\qquad\qquad \Diamondminus(\mathsf{n} = n \wedge \mathsf{deliver}_{\mathsf{pl}}(n',m) \in \mathsf{ois} \wedge \mathsf{self})$
By rule INVL,
$\quad$(2) $\Gamma' \vdash_{\mathsf{PLC}} \mathsf{self} \wedge \mathsf{n} = n \wedge \mathsf{deliver}_{\mathsf{pl}}(n',m) \in \mathsf{ois} \Rightarrow$
$\qquad\qquad \mathsf{self} \wedge \mathsf{n} = n \wedge \mathsf{occ}(\mathsf{ois}, \mathsf{deliver}_{\mathsf{pl}}(n',m)) = 1]$
By Lemma 99 on [1] and [2],
$\quad$(3) $\Gamma' \vdash_{\mathsf{PLC}} n \bullet \top \uparrow \mathsf{deliver}_{\mathsf{pl}}(n',m) \Rightarrow$
$\qquad\qquad \Diamondminus(\mathsf{self} \wedge \mathsf{n} = n \wedge \mathsf{occ}(\mathsf{ois}, \mathsf{deliver}_{\mathsf{pl}}(n',m)) = 1])$

From Lemma 42,
$\quad$(4) $\Gamma' \vdash_{\mathsf{PLC}} \circledS [\mathsf{n} = n \wedge \mathsf{deliver}_{\mathsf{pl}}(n',m) \in \mathsf{ois} \Rightarrow$
$\qquad\qquad \hat{\Box}(\mathsf{n} = n \rightarrow \mathsf{deliver}_{\mathsf{pl}}(n',m) \notin \mathsf{ois}) \wedge$
$\qquad\qquad \hat{\boxminus}(\mathsf{n} = n \rightarrow \mathsf{deliver}_{\mathsf{pl}}(n',m) \notin \mathsf{ois})]$
By rule SINV on [4],
$\quad$(5) $\Gamma' \vdash_{\mathsf{PLC}} \mathsf{self} \wedge \mathsf{n} = n \wedge \mathsf{deliver}_{\mathsf{pl}}(n',m) \in \mathsf{ois} \Rightarrow$
$\qquad\qquad \hat{\Box}(\mathsf{self} \wedge \mathsf{n} = n \rightarrow \mathsf{deliver}_{\mathsf{pl}}(n',m) \notin \mathsf{ois}) \wedge$
$\qquad\qquad \hat{\boxminus}(\mathsf{self} \wedge \mathsf{n} = n \rightarrow \mathsf{deliver}_{\mathsf{pl}}(n',m) \notin \mathsf{ois})$

From [3] and [5],
$\quad$(6) $\Gamma' \vdash_{\mathsf{PLC}} (n \bullet \top \uparrow \mathsf{deliver}_{\mathsf{pl}}(n',m)) \Rightarrow$
$\qquad\qquad \Diamondminus(\mathsf{occ}(\mathsf{ois}, \mathsf{deliver}_{\mathsf{pl}}(n',m)) = 1] \wedge$
$\qquad\qquad \hat{\Box}(\mathsf{self} \wedge \mathsf{n} = n \rightarrow \mathsf{deliver}_{\mathsf{pl}}(n',m) \notin \mathsf{ois}) \wedge$

$$\hat{\boxminus}(\mathsf{self} \land \mathsf{n} = n \to \mathsf{deliver}_{\mathsf{pl}}(n', m) \notin \mathsf{ois}))$$
From rule UniOI,
(7) $\Gamma' \vdash_{\mathsf{PLC}} (\mathsf{occ}(\mathsf{ois}, \mathsf{deliver}_{\mathsf{pl}}(n', m)) \leq 1 \land$
$\hat{\boxdot}(\mathsf{self} \land \mathsf{n} = n \to \mathsf{deliver}_{\mathsf{pl}}(n', m) \notin \mathsf{ois}) \land$
$\hat{\boxminus}(\mathsf{self} \land \mathsf{n} = n \to \mathsf{deliver}_{\mathsf{pl}}(n', m) \notin \mathsf{ois})) \Rightarrow$
$n \bullet \top \uparrow \mathsf{deliver}_{\mathsf{pl}}(n', m) \Rightarrow \hat{\boxminus}\neg(n \bullet \top \uparrow \mathsf{deliver}_{\mathsf{pl}}(n', m))$

From [6] and [7],
$\Gamma' \vdash_{\mathsf{PLC}} n \bullet \top \uparrow \mathsf{deliver}_{\mathsf{pl}}(n', m) \Rightarrow$
$\diamondsuit[n \bullet \top \uparrow \mathsf{deliver}_{\mathsf{pl}}(n', m) \Rightarrow \hat{\boxminus}\neg(n \bullet \top \uparrow \mathsf{deliver}_{\mathsf{pl}}(n', m))]$
which leads to
$\Gamma' \vdash_{\mathsf{PLC}} n \bullet \top \uparrow \mathsf{deliver}_{\mathsf{pl}}(n', m) \Rightarrow$
$n \bullet \top \uparrow \mathsf{deliver}_{\mathsf{pl}}(n', m) \Rightarrow \hat{\boxminus}\neg(n \bullet \top \uparrow \mathsf{deliver}_{\mathsf{pl}}(n', m))$
which leads to
$\Gamma' \vdash_{\mathsf{PLC}} n \bullet \top \uparrow \mathsf{deliver}_{\mathsf{pl}}(n', m) \Rightarrow$
$\hat{\boxminus}\neg(n \bullet \top \uparrow \mathsf{deliver}_{\mathsf{pl}}(n', m))$

**Lemma 42.**
At every nodes, a perfect-link delivery event for a message is issued at most once.
$\Gamma' \vdash_{\mathsf{PLC}} \circledS [(\mathsf{n} = n \land \mathsf{deliver}_{\mathsf{pl}}(n', m) \in \mathsf{ois}) \Rightarrow$
$\hat{\boxdot}(\mathsf{n} = n \to \mathsf{deliver}_{\mathsf{pl}}(n', m) \notin \mathsf{ois}) \land$
$\hat{\boxminus}(\mathsf{n} = n \to \mathsf{deliver}_{\mathsf{pl}}(n', m) \notin \mathsf{ois})]$

**Proof.**

Proof idea: A perfect-link delivery event is only issued in the execution of a stubborn link delivery event. In the execution of this event, the sender and counter pair is recorded and kept in the received set. In the future events, the received set and the past execution of the stubborn link delivery event prevents issuance of any perfect-link delivery event (Lemma 43).

By rule InvLSe, there exists $c$ such that
$\Gamma' \vdash_{\mathsf{PLC}} \circledS [(\mathsf{n} = n \land \mathsf{deliver}_{\mathsf{pl}}(n', m) \in \mathsf{ois}) \Rightarrow$
$(n', c) \in \mathsf{received}(\mathsf{s}'(n)) \land n \bullet 0 \uparrow \mathsf{deliver}_{\mathsf{sl}}(n', \langle c, m \rangle)]$
Thus, by Lemma 90,
(1) $\Gamma' \vdash_{\mathsf{PLC}} \circledS [(\mathsf{n} = n \land \mathsf{deliver}_{\mathsf{pl}}(n', m) \in \mathsf{ois}) \Rightarrow$
$(n', c) \in \mathsf{received}(\mathsf{s}'(n)) \land \diamondsuit(n \bullet 0 \uparrow \mathsf{deliver}_{\mathsf{sl}}(n', \langle c, m \rangle))]$

By Lemma 43,
(2) $\Gamma' \vdash_{\mathsf{PLC}} \circledS [(n', c) \in \mathsf{received}(\mathsf{s}(n)) \land \diamondsuit(n \bullet 0 \uparrow \mathsf{deliver}_{\mathsf{sl}}(n', \langle c, m \rangle)) \Rightarrow$
$\mathsf{n} = n \to \mathsf{deliver}_{\mathsf{pl}}(n', m) \notin \mathsf{ois}]$
By rule InvSSe,
(3) $\Gamma' \vdash_{\mathsf{PLC}} \circledS [(n', c) \in \mathsf{received}(\mathsf{s}(n)) \Rightarrow$

$$\Box(n', c) \in \mathsf{received}(\mathsf{s}(n))]$$

By Lemma 107,

(4) $\Gamma' \vdash_{\mathsf{PLC}} \circledS [\Diamondblack(n \bullet 0 \uparrow \mathsf{deliver}_{\mathsf{sl}}(n', \langle c, m \rangle)) \Rightarrow$
$\Box(\Diamondblack(n \bullet 0 \uparrow \mathsf{deliver}_{\mathsf{sl}}(n', \langle c, m \rangle)))]$

From [3] an [4]

(5) $\Gamma' \vdash_{\mathsf{PLC}} \circledS [(n', c) \in \mathsf{received}(\mathsf{s}(n)) \wedge \Diamondblack(n \bullet 0 \uparrow \mathsf{deliver}_{\mathsf{sl}}(n', \langle c, m \rangle)) \Rightarrow$
$\Box((n', c) \in \mathsf{received}(\mathsf{s}(n)) \wedge \Diamondblack(n \bullet 0 \uparrow \mathsf{deliver}_{\mathsf{sl}}(n', \langle c, m \rangle)))]$

By Lemma 100 on [5] and [2],

(6) $\Gamma' \vdash_{\mathsf{PLC}} \circledS [(n', c) \in \mathsf{received}(\mathsf{s}(n)) \wedge \Diamondblack(n \bullet 0 \uparrow \mathsf{deliver}_{\mathsf{sl}}(n', \langle c, m \rangle)) \Rightarrow$
$\Box(\mathsf{n} = n \rightarrow \mathsf{deliver}_{\mathsf{pl}}(n', m) \notin \mathsf{ois})]$

By rule ASASE on [1] and [6],

(7) $\Gamma' \vdash_{\mathsf{PLC}} \circledS [\mathsf{n} = n \wedge \mathsf{deliver}_{\mathsf{pl}}(n', m) \in \mathsf{ois} \Rightarrow$
$\hat{\Box}(\Box(\mathsf{n} = n \rightarrow \mathsf{deliver}_{\mathsf{pl}}(n', m) \notin \mathsf{ois}))]$

that is

(8) $\Gamma' \vdash_{\mathsf{PLC}} \circledS [\mathsf{n} = n \wedge \mathsf{deliver}_{\mathsf{pl}}(n', m) \in \mathsf{ois} \Rightarrow$
$\hat{\Box}(\mathsf{n} = n \rightarrow \mathsf{deliver}_{\mathsf{pl}}(n', m) \notin \mathsf{ois})]$

By Lemma 108 on [8],

(9) $\Gamma' \vdash_{\mathsf{PLC}} \circledS [\mathsf{n} = n \wedge \mathsf{deliver}_{\mathsf{pl}}(n', m) \in \mathsf{ois} \Rightarrow$
$\hat{\Box}(\mathsf{n} = n \rightarrow \mathsf{deliver}_{\mathsf{pl}}(n', m) \notin \mathsf{ois}) \wedge$
$\hat{\boxminus}(\mathsf{n} = n \rightarrow \mathsf{deliver}_{\mathsf{pl}}(n', m) \notin \mathsf{ois})]$

**Lemma 43.**
$\Gamma' \vdash_{\mathsf{PLC}} \circledS [(n', c) \in \mathsf{received}(\mathsf{s}(n)) \wedge \Diamondblack(n \bullet 0 \uparrow \mathsf{deliver}_{\mathsf{sl}}(n', \langle c, m \rangle)) \Rightarrow$
$\mathsf{n} = n \rightarrow \mathsf{deliver}_{\mathsf{pl}}(n', m) \notin \mathsf{ois}]$

**Proof.**

Proof idea: A perfect-link delivery event is only issued in the execution of a stubborn link delivery event. A stubborn link delivery event is executed in the past. Any other stubborn link delivery event with the same sender and message has the same counter (Lemma 44). On any stubborn link delivery event, if the pair of the sender and the counter are already in the delivered set, the indication function of the protocol does not issue a perfect-link delivery event. Thus, no perfect-link delivery event with the same sender and message is issued.

We use rule INVSE.
The request and periodic obligations are trivial by rule IR and rule PE.
For the indication obligation, we have to prove that
$\Gamma' \vdash_{\mathsf{PLC}} \circledS [\forall i, e.\ i \uparrow e \Rightarrow$
$(n', c, m) \in \mathsf{received}(\mathsf{s}(n)) \rightarrow$

$$\mathsf{n} = n \to \mathsf{deliver}_{\mathsf{pl}}(n', m) \notin \mathsf{ois}$$

That is
$$\Gamma' \vdash_{\mathsf{PLC}} \circledS \ [\forall i, e.\ n \bullet i \uparrow e\ \wedge\ (n', c) \in \mathsf{received}(\mathsf{s}(n))\ \wedge\ \Diamond(n \bullet 0 \uparrow \mathsf{deliver}_{\mathsf{sl}}(n', \langle c, m \rangle)) \Rightarrow$$
$$\mathsf{deliver}_{\mathsf{pl}}(n', m) \notin \mathsf{ois}$$

By rule IISE,
$$\Gamma' \vdash_{\mathsf{PLC}} \circledS \ [n \bullet i \uparrow e \Rightarrow$$
$$\exists n', c, m.\ i = 0\ \wedge\ e = \mathsf{deliver}_{\mathsf{sl}}(n', \langle c, m \rangle)$$

Thus, we show that
$$\Gamma' \vdash_{\mathsf{PLC}} \circledS \ [\forall n'', c', m'.$$
$$n \bullet 0 \uparrow \mathsf{deliver}_{\mathsf{sl}}(n'', (c', m')) \wedge (n', c) \in \mathsf{received}(\mathsf{s}(n)) \wedge \Diamond(n \bullet 0 \uparrow \mathsf{deliver}_{\mathsf{sl}}(n', \langle c, m \rangle)) \Rightarrow$$

$$\mathsf{deliver}_{\mathsf{pl}}(n', m) \notin \mathsf{ois}$$

We consider two cases whether $(n'' = n' \wedge m' = m)$:

If $\neg(n'' = n' \wedge m' = m)$, the result is immediate from the rule IISE.

Thus, we show that
$$\Gamma' \vdash_{\mathsf{PLC}} \circledS \ [n \bullet 0 \uparrow \mathsf{deliver}_{\mathsf{sl}}(n', \langle c', m \rangle) \wedge (n', c) \in \mathsf{received}(\mathsf{s}(n)) \wedge \Diamond(n \bullet 0 \uparrow \mathsf{deliver}_{\mathsf{sl}}(n', \langle c, m \rangle)) \Rightarrow$$

$$\mathsf{deliver}_{\mathsf{pl}}(n', m) \notin \mathsf{ois}$$

By Lemma 44, we need to show that
$$\Gamma' \vdash_{\mathsf{PLC}} \circledS \ [n \bullet 0 \uparrow \mathsf{deliver}_{\mathsf{sl}}(n', \langle c, m \rangle)\ \wedge\ (n', c) \in \mathsf{received}(\mathsf{s}(n)) \Rightarrow$$
$$\mathsf{deliver}_{\mathsf{pl}}(n', m) \notin \mathsf{ois}$$

That is immediate from rule IISE.

**Lemma 44.**
Every two stubborn link delivery events with the same sender and message have the same counter.
$$\Gamma' \vdash_{\mathsf{PLC}} \circledS \ [(n \bullet 0 \uparrow \mathsf{deliver}_{\mathsf{sl}}(n', \langle c, m \rangle)\ \wedge$$
$$\Diamond(n \bullet 0 \uparrow \mathsf{deliver}_{\mathsf{sl}}(n', \langle c', m \rangle))) \Rightarrow$$
$$c = c']$$

**Proof.**

Proof idea: The execution of any stubborn link delivery event is preceded by a stubborn link send event. The send event is preceded by its issuance. A stubborn link send event is only issued by a perfect-link send event. Therefore, from the fact that there are two stubborn link delivery events, we have that in the past there has been two perfect-link send events. However, the assumption is that there is at most one perfect-link send event. Therefore, the two events are the same and they send the same counter. Thus, the counters of the two stubborn link delivery events are the same.

By $SL'_2$,

(1) $\Gamma' \vdash_{\mathsf{PLC}} n \bullet 0 \uparrow \mathsf{deliver}_{\mathsf{sl}}(n', \langle c, m \rangle) \Rightarrow$
$\diamondsuit(n' \bullet 0 \downarrow \mathsf{send}_{\mathsf{sl}}(n, \langle c, m \rangle))$

By rule OR$'$

(2) $\Gamma' \vdash_{\mathsf{PLC}} n' \bullet 0 \downarrow \mathsf{send}_{\mathsf{sl}}(n, \langle c, m \rangle) \Rightarrow$
$\diamondsuit(n' \bullet (0, \mathsf{send}_{\mathsf{sl}}(n, \langle c, m \rangle)) \in \mathsf{ors} \wedge \mathsf{self})$

By the rule INVL,

(3) $\Gamma' \vdash_{\mathsf{PLC}} (n' \bullet (0, \mathsf{send}_{\mathsf{sl}}(n, \langle c, m \rangle)) \in \mathsf{ors} \wedge \mathsf{self}) \Rightarrow$
$(n' \bullet \top \downarrow \mathsf{send}_{\mathsf{pl}}(n, m)) \wedge \mathsf{counter}(\mathsf{s}'(n)) = c$

By Lemma 89 and Lemma 99 on [1], [2] and [3],

(4) $\Gamma' \vdash_{\mathsf{PLC}} n \bullet 0 \uparrow \mathsf{deliver}_{\mathsf{sl}}(n', \langle c, m \rangle) \Rightarrow$
$\diamondsuit(n' \bullet \top \downarrow \mathsf{send}_{\mathsf{pl}}(n, m)) \wedge \mathsf{counter}(\mathsf{s}'(n)) = c$

By rule SINV on [4],

(5) $\Gamma' \vdash_{\mathsf{PLC}} \circledS \, [n \bullet 0 \uparrow \mathsf{deliver}_{\mathsf{sl}}(n', \langle c, m \rangle) \Rightarrow$
$\diamondsuit(n' \bullet \top \downarrow \mathsf{send}_{\mathsf{pl}}(n, m)) \wedge \mathsf{counter}(\mathsf{s}'(n)) = c]$

Similarly, we can derive

(6) $\Gamma' \vdash_{\mathsf{PLC}} \circledS \, [n \bullet 0 \uparrow \mathsf{deliver}_{\mathsf{sl}}(n', \langle c', m \rangle) \Rightarrow$
$\diamondsuit(n' \bullet \top \downarrow \mathsf{send}_{\mathsf{pl}}(n, m)) \wedge \mathsf{counter}(\mathsf{s}'(n)) = c']$

Thus, from [5] and [6], we need to show that

$\Gamma' \vdash_{\mathsf{PLC}} \circledS \, [(\diamondsuit((n' \bullet \top \downarrow \mathsf{send}_{\mathsf{pl}}(n, m)) \wedge \mathsf{counter}(\mathsf{s}'(n)) = c) \wedge$
$\diamondsuit\diamondsuit((n' \bullet \top \downarrow \mathsf{send}_{\mathsf{pl}}(n, m)) \wedge \mathsf{counter}(\mathsf{s}'(n)) = c')) \Rightarrow$
$c = c']$

That is

$\Gamma' \vdash_{\mathsf{PLC}} \circledS \, [(\diamondsuit((n' \bullet \top \downarrow \mathsf{send}_{\mathsf{pl}}(n, m)) \wedge \mathsf{counter}(\mathsf{s}'(n)) = c) \wedge$
$\diamondsuit((n' \bullet \top \downarrow \mathsf{send}_{\mathsf{pl}}(n, m)) \wedge \mathsf{counter}(\mathsf{s}'(n)) = c')) \Rightarrow$
$c = c']$

By Lemma 105, there are two cases that are similar.

$\Gamma' \vdash_{\mathsf{PLC}} \circledS \, [\diamondsuit((n' \bullet \top \downarrow \mathsf{send}_{\mathsf{pl}}(n, m)) \wedge \mathsf{counter}(\mathsf{s}'(n)) = c \wedge$
$\diamondsuit((n' \bullet \top \downarrow \mathsf{send}_{\mathsf{pl}}(n, m)) \wedge \mathsf{counter}(\mathsf{s}'(n)) = c')) \Rightarrow$
$c = c']$

By assumption $\mathcal{A}_1$, we need to prove

$\Gamma' \vdash_{\mathsf{PLC}} \circledS \, [\diamondsuit(\hat{\boxminus}\neg(n' \bullet \top \downarrow \mathsf{send}_{\mathsf{pl}}(n, m)) \wedge \mathsf{counter}(\mathsf{s}'(n)) = c \wedge$
$\diamondsuit((n' \bullet \top \downarrow \mathsf{send}_{\mathsf{pl}}(n, m)) \wedge \mathsf{counter}(\mathsf{s}'(n)) = c')) \Rightarrow$
$c = c']$

By Lemma 106, we need to prove

$\Gamma' \vdash_{\mathsf{PLC}} \circledS \, [\diamondsuit(\mathsf{counter}(\mathsf{s}'(n)) = c \wedge$
$(n' \bullet \top \downarrow \mathsf{send}_{\mathsf{pl}}(n, m)) \wedge \mathsf{counter}(\mathsf{s}'(n)) = c')] \Rightarrow$
$c = c']$

that is trivial.

**Theorem 8.** *($PL_3$: No-forge)*
If a node $n$ delivers a message $m$ with sender $n'$, then $m$ was previously sent to $n$ by node $n'$.

$\Gamma \vdash_{\mathsf{PLC}} (n \bullet \top \uparrow \mathsf{deliver}_{\mathsf{pl}}(n', m)) \leftsquigarrow (n' \bullet \top \downarrow \mathsf{send}_{\mathsf{pl}}(n, m))$
where
$\Gamma$ is defined in Definition 17.

**Proof.**
By rule OI′,
  (1) $\Gamma \vdash_{\mathsf{PLC}} (n \bullet \top \uparrow \mathsf{deliver}_{\mathsf{pl}}(n', m)) \Rightarrow$
      $\Diamond\!\!\!\text{-}(n \bullet \mathsf{deliver}_{\mathsf{pl}}(n', m) \in \mathsf{ois} \land \mathsf{self})$
By rule INVL,
  (2) $\Gamma \vdash_{\mathsf{PLC}} (n \bullet \mathsf{deliver}_{\mathsf{pl}}(n', m) \in \mathsf{ois} \land \mathsf{self}) \Rightarrow$
      $\exists c.\ (n \bullet 0 \uparrow \mathsf{deliver}_{\mathsf{sl}}(n', \langle c, m \rangle))$
By Lemma 99 on [1] and [2],
  $\Gamma \vdash_{\mathsf{PLC}} (n \bullet \top \uparrow \mathsf{deliver}_{\mathsf{pl}}(n', m)) \Rightarrow$
      $\exists c.\ \Diamond\!\!\!\text{-}(n \bullet 0 \uparrow \mathsf{deliver}_{\mathsf{sl}}(n', \langle c, m \rangle))$
that is
  (3) $\Gamma \vdash_{\mathsf{PLC}} n \bullet \top \uparrow \mathsf{deliver}_{\mathsf{pl}}(n', m) \leftsquigarrow$
      $\exists c.\ (n \bullet 0 \uparrow \mathsf{deliver}_{\mathsf{sl}}(n', \langle c, m \rangle))$

By rule OR′,
  (4) $\Gamma \vdash_{\mathsf{PLC}} \forall c.\ (n' \bullet 0 \downarrow \mathsf{send}_{\mathsf{sl}}(n, \langle c, m \rangle)) \Rightarrow$
      $\Diamond\!\!\!\text{-}(n' \bullet (0, \mathsf{send}_{\mathsf{sl}}(n, \langle c, m \rangle)) \in \mathsf{ors} \land \mathsf{self})$
By rule INVL
  (5) $\Gamma \vdash_{\mathsf{PLC}} \forall c.\ n' \bullet (0, \mathsf{send}_{\mathsf{sl}}(n, \langle c, m \rangle)) \in \mathsf{ors} \land \mathsf{self} \Rightarrow$
      $(n' \bullet \top \downarrow \mathsf{send}_{\mathsf{pl}}(n, \langle c, m \rangle))$
By Lemma 99 on [4] and [5]
  (6) $\Gamma \vdash_{\mathsf{PLC}} \forall c.\ (n' \bullet 0 \downarrow \mathsf{send}_{\mathsf{sl}}(n, \langle c, m \rangle)) \leftsquigarrow$
      $(n' \bullet \top \downarrow \mathsf{send}_{\mathsf{pl}}(n, m))$

From Lemma 88 on [3], $SL'_2$ and [6],
  $\Gamma \vdash_{\mathsf{PLC}} (n \bullet \top \uparrow \mathsf{deliver}_{\mathsf{pl}}(n', m)) \leftsquigarrow$
      $(n' \bullet \top \downarrow \mathsf{send}_{\mathsf{pl}}(n, m))$

### 5.3.3 Best-Effort Broadcast

**Definition 18.**
$\Gamma = \mathsf{PL}'_1; \mathsf{PL}'_2; \mathsf{PL}'_3$

$\mathsf{PL}'_1 = \mathsf{lower}(0, \mathsf{PL}_1) =$
$\quad \forall n, n', m.$
$\quad n \in \mathsf{Correct} \wedge n' \in \mathsf{Correct} \to$
$\quad (n \bullet 0 \downarrow \mathsf{send}_{\mathsf{pl}}(n', m) \rightsquigarrow$
$\quad (n' \bullet 0 \uparrow \mathsf{deliver}_{\mathsf{pl}}(n, m))$

$\mathsf{PL}'_2 = \mathsf{lower}(0, \mathsf{PL}_2) =$
$\quad\quad [n' \bullet 0 \downarrow \mathsf{send}_{\mathsf{pl}}(n, m) \Rightarrow \hat{\boxminus}\neg(n' \bullet 0 \downarrow \mathsf{send}_{\mathsf{pl}}(n, m))] \to$
$\quad\quad [n \bullet 0 \uparrow \mathsf{deliver}_{\mathsf{pl}}(n', m) \Rightarrow \hat{\boxminus}\neg(n \bullet 0 \uparrow \mathsf{deliver}_{\mathsf{pl}}(n', m))]]$

$\mathsf{PL}'_3 = \mathsf{lower}(0, \mathsf{PL}_3) =$
$\quad (n \bullet 0 \uparrow \mathsf{deliver}_{\mathsf{pl}}(n', m)) \leftsquigarrow$
$\quad (n' \bullet 0 \downarrow \mathsf{send}_{\mathsf{pl}}(n, m))$

**Theorem 9.** *(BEB$_1$: Validity)*
If a correct node broadcasts a message $m$, then every correct node eventually delivers $m$.

$\Gamma \vdash_{\mathsf{BEBC}} \forall n, n', m.$
$\quad n \in \mathsf{Correct} \wedge n' \in \mathsf{Correct} \rightarrow$
$\quad [(n' \bullet \top \downarrow \mathsf{broadcast}_{\mathsf{beb}}(m)) \leadsto$
$\quad (n \bullet \top \uparrow \mathsf{deliver}_{\mathsf{beb}}(n', m))]$
where
$\Gamma$ is defined in Definition 18.

**Proof.**

The proof idea: The assumption is that the nodes $n$ and $n'$ are both correct. Upon the execution of a $\mathsf{broadcast}_{\mathsf{beb}}$ of a message $m$ from $n$ to $n'$, the component sends $m$ to every node using the perfect link subcomponent. By the reliable delivery property of the perfect link, as both the sender $n$ and receiver $n'$ are correct, $m$ is eventually delivered to $n'$. Upon the delivery of $m$ by the perfect link, the component issues the delivery of $m$ that eventually executes.

We assume
$\quad$ (1) $\Gamma' = \Gamma; n \in \mathsf{Correct}; n' \in \mathsf{Correct}$
We prove
$\quad \Gamma' \vdash (n' \bullet \top \downarrow \mathsf{broadcast}_{\mathsf{beb}}(m)) \Rightarrow \Diamond(n \bullet \top \uparrow \mathsf{deliver}_{\mathsf{beb}}(n', m))$

By rule IROR, and the definition of request,
$\quad$ (2) $\Gamma' \vdash_{\mathsf{BEBC}} (n \bullet \top \downarrow \mathsf{broadcast}_{\mathsf{beb}}(m)) \Rightarrow \Diamond(n \bullet 0 \downarrow \mathsf{send}_{\mathsf{pl}}(n', m))$

From $PL_1$ and [1],
$\quad \Gamma' \vdash_{\mathsf{BEBC}} (n \bullet 0 \downarrow \mathsf{send}_{\mathsf{pl}}(n', m)) \leadsto (n' \bullet 0 \uparrow \mathsf{deliver}_{\mathsf{pl}}(n, m))$
That is,
$\quad$ (3) $\Gamma' \vdash_{\mathsf{BEBC}} (n \bullet 0 \downarrow \mathsf{send}_{\mathsf{pl}}(n', m)) \Rightarrow \Diamond(n' \bullet 0 \uparrow \mathsf{deliver}_{\mathsf{pl}}(n, m))$

By rule IIOI,
$\quad$ (4) $\Gamma' \vdash_{\mathsf{BEBC}} (n' \bullet 0 \uparrow \mathsf{deliver}_{\mathsf{pl}}(n, m)) \Rightarrow \Diamond(n' \bullet \top \uparrow \mathsf{deliver}_{\mathsf{beb}}(n, m))$

By Lemma 89 on [2], [3] and [4],
$\quad \Gamma' \vdash_{\mathsf{BEBC}} (n \bullet \top \downarrow \mathsf{broadcast}_{\mathsf{beb}}(m) \Rightarrow \Diamond(n' \bullet \top \uparrow \mathsf{deliver}_{\mathsf{beb}}(n, m))$
That is,
$\quad \Gamma' \vdash_{\mathsf{BEBC}} (n \bullet \top \downarrow \mathsf{broadcast}_{\mathsf{beb}}(m) \leadsto (n' \bullet \top \uparrow \mathsf{deliver}_{\mathsf{beb}}(n, m))$

**Theorem 10.** *(BEB$_2$: No-duplication)*
If a message is broadcast at most once, it will be delivered at most once.

$\Gamma \vdash_{\mathsf{BEBC}} [(n' \bullet \top \downarrow \mathsf{broadcast}_{\mathsf{beb}}(m)) \Rightarrow \hat{\boxminus} \neg (n' \bullet \top \downarrow \mathsf{broadcast}_{\mathsf{beb}}(m))] \to$
$\qquad [(n \bullet \top \uparrow \mathsf{deliver}_{\mathsf{beb}}(n', m)) \Rightarrow \hat{\boxminus} \neg (n \bullet \top \uparrow \mathsf{deliver}_{\mathsf{beb}}(n', m))]$

where
$\Gamma$ is defined in Definition 18.

**Proof.**

The proof idea: We assume that a message is broadcast at most once. By the definition of BEBC, a message is sent by the perfect link subcomponent plc once, only when a broadcast request is processed. Thus, BEBC sends the message to every node at most once by plc. Thus, by the no-duplication property of plc, the message is delivered to every node at most once by plc. By the definition of BEBC, a delivery is issued only when a delivery indication is received from plc. Thus, BEBC issues delivery at every node at most once.

We assume
(1) $\Gamma' = \Gamma; [n' \bullet \top \downarrow \mathsf{broadcast}_{\mathsf{beb}}(m) \Rightarrow \hat{\boxminus} \neg (n' \bullet \top \downarrow \mathsf{broadcast}_{\mathsf{beb}}(m))]$

We want to prove
$\Gamma' \vdash_{\mathsf{BEBC}} (n \bullet \top \uparrow \mathsf{deliver}_{\mathsf{beb}}(n', m)) \Rightarrow \hat{\boxminus} \neg (n \bullet \top \uparrow \mathsf{deliver}_{\mathsf{beb}}(n', m))$

From [1],
(2) $\Gamma' \vdash_{\mathsf{BEBC}} n' \bullet \top \downarrow \mathsf{broadcast}_{\mathsf{beb}}(m) \Rightarrow$
$\qquad \hat{\boxminus} \neg (n' \bullet \top \downarrow \mathsf{broadcast}_{\mathsf{beb}}(m))$

By Lemma 108 on [2],
(3) $\Gamma' \vdash_{\mathsf{BEBC}} n' \bullet \top \downarrow \mathsf{broadcast}_{\mathsf{beb}}(m) \Rightarrow$
$\qquad \hat{\boxminus} \neg (n' \bullet \top \downarrow \mathsf{broadcast}_{\mathsf{beb}}(m)) \wedge$
$\qquad \hat{\square} \neg (n' \bullet \top \downarrow \mathsf{broadcast}_{\mathsf{beb}}(m))$

By rule INVL,
(4) $\Gamma' \vdash_{\mathsf{BEBC}} (n' \bullet (0, \mathsf{send}_p(n, m)) \in \mathsf{ors} \wedge \mathsf{self}) \Rightarrow$
$\qquad (n' \bullet \top \downarrow \mathsf{broadcast}_{\mathsf{beb}}(m))$

The contra-positive of [4] is
(5) $\Gamma' \vdash_{\mathsf{BEBC}} \neg (n' \bullet \top \downarrow \mathsf{broadcast}_{\mathsf{beb}}(m)) \Rightarrow$
$\qquad \neg (n' \bullet (0, \mathsf{send}_p(n, m)) \in \mathsf{ors} \wedge \mathsf{self})$

From [3], and [5]
(6) $\Gamma' \vdash_{\mathsf{BEBC}} n' \bullet \top \downarrow \mathsf{broadcast}_{\mathsf{beb}}(m) \Rightarrow$
$\qquad \hat{\boxminus} \neg (n' \bullet (0, \mathsf{send}_p(n, m)) \in \mathsf{ors} \wedge \mathsf{self}) \wedge$

$$\hat{\Box} \neg (n' \bullet (0, \mathsf{send}_p(n,m)) \in \mathsf{ors} \wedge \mathsf{self})$$
that is
(7) $\Gamma' \vdash_{\mathsf{BEBC}} n' \bullet \top \downarrow \mathsf{broadcast}_{\mathsf{beb}}(m) \Rightarrow$
$$\hat{\ominus}(n' \bullet \mathsf{self} \to (0, \mathsf{send}_p(n,m)) \notin \mathsf{ors}) \wedge$$
$$\hat{\Box}(n' \bullet \mathsf{self} \to (0, \mathsf{send}_p(n,m)) \notin \mathsf{ors})$$

By rule OR$'$
(8) $\Gamma' \vdash_{\mathsf{BEBC}} (n' \bullet 0 \downarrow \mathsf{send}_{\mathsf{pl}}(n,m)) \Rightarrow \Diamond(n' \bullet (0, \mathsf{send}_{\mathsf{pl}}(n,m)) \in \mathsf{ors} \wedge \mathsf{self})$

By rule INVL with $\mathcal{A} =$
$(n' \bullet (0, \mathsf{send}_{\mathsf{pl}}(n,m)) \in \mathsf{ors}) \to (n' \bullet \top \downarrow \mathsf{broadcast}_{\mathsf{beb}}(m)) \wedge \mathsf{occ}(\mathsf{ors}, (0, \mathsf{send}_p(n,m))) = 1$

(9) $\Gamma' \vdash_{\mathsf{BEBC}} \mathsf{self} \wedge (n' \bullet (0, \mathsf{send}_{\mathsf{pl}}(n,m)) \in \mathsf{ors}) \Rightarrow$
$$(n' \bullet \top \downarrow \mathsf{broadcast}_{\mathsf{beb}}(m)) \wedge \mathsf{occ}(\mathsf{ors}, (0, \mathsf{send}_p(n,m))) = 1$$

From [8] and [9],
(10) $\Gamma' \vdash_{\mathsf{BEBC}} (n' \bullet 0 \downarrow \mathsf{send}_{\mathsf{pl}}(n,m)) \Rightarrow$
$$\Diamond[(n' \bullet \top \downarrow \mathsf{broadcast}_{\mathsf{beb}}(m)) \wedge \mathsf{occ}(\mathsf{ors}, (0, \mathsf{send}_p(n,m))) = 1]$$

From [10] and [7],
(11) $\Gamma' \vdash_{\mathsf{BEBC}} (n' \bullet 0 \downarrow \mathsf{send}_{\mathsf{pl}}(n,m)) \Rightarrow \Diamond[$
$$\mathsf{occ}(\mathsf{ors}, (0, \mathsf{send}_p(n,m))) = 1 \wedge$$
$$\hat{\ominus}(n' \bullet \mathsf{self} \to (0, \mathsf{send}_p(n,m)) \notin \mathsf{ors}) \wedge$$
$$\hat{\Box}(n' \bullet \mathsf{self} \to (0, \mathsf{send}_p(n,m)) \notin \mathsf{ors})]$$

By rule UNIOR on [11]
(12) $\Gamma' \vdash_{\mathsf{BEBC}} (n' \bullet 0 \downarrow \mathsf{send}_{\mathsf{pl}}(n,m)) \Rightarrow \Diamond[$
$$(n' \bullet 0 \downarrow \mathsf{send}_{\mathsf{pl}}(n,m)) \Rightarrow$$
$$\hat{\ominus} \neg(n' \bullet 0 \downarrow \mathsf{send}_{\mathsf{pl}}(n,m))]$$

By Lemma 109 on [12]
(13) $\Gamma' \vdash_{\mathsf{BEBC}} (n' \bullet 0 \downarrow \mathsf{send}_{\mathsf{pl}}(n,m)) \Rightarrow$
$$\hat{\ominus} \neg(n' \bullet 0 \downarrow \mathsf{send}_{\mathsf{pl}}(n,m))$$

By $PL'_2$ on [13],
(14) $\Gamma' \vdash_{\mathsf{BEB}} n \bullet \top \uparrow \mathsf{deliver}_{\mathsf{pl}}(n',m) \Rightarrow$
$$\hat{\ominus} \neg(n \bullet \top \uparrow \mathsf{deliver}_{\mathsf{pl}}(n',m))$$

By Lemma 108 on [14],
(15) $\Gamma' \vdash_{\mathsf{BEB}} (n \bullet \top \uparrow \mathsf{deliver}_{\mathsf{pl}}(n',m)) \Rightarrow$
$$\hat{\ominus} \neg(n \bullet \top \uparrow \mathsf{deliver}_{\mathsf{pl}}(n',m)) \wedge$$
$$\hat{\Box} \neg(n \bullet \top \uparrow \mathsf{deliver}_{\mathsf{pl}}(n',m))$$

By rule INVL,
(16) $\Gamma' \vdash_{\mathsf{BEBC}} (n \bullet (\mathsf{deliver}_{\mathsf{beb}}(n',m)) \in \mathsf{ois} \wedge \mathsf{self}) \Rightarrow (n \bullet 0 \uparrow \mathsf{deliver}_{\mathsf{pl}}(n',m))$

The contra-positive of [16] is
(17) $\Gamma' \vdash_{\mathsf{BEBC}} \neg(n \bullet 0 \uparrow \mathsf{deliver}_{\mathsf{pl}}(n',m)) \Rightarrow \neg(n \bullet (\mathsf{deliver}_{\mathsf{beb}}(n',m)) \in \mathsf{ois} \wedge \mathsf{self})$

From [15], and [17],
(18) $\Gamma' \vdash_{\mathsf{BEB}} (n \bullet \top \uparrow \mathsf{deliver}_{\mathsf{pl}}(n',m)) \Rightarrow$
$$\hat{\ominus} \neg(n \bullet (\mathsf{deliver}_{\mathsf{beb}}(n',m)) \in \mathsf{ois} \wedge \mathsf{self}) \wedge$$
$$\hat{\Box} \neg(n \bullet (\mathsf{deliver}_{\mathsf{beb}}(n',m)) \in \mathsf{ois} \wedge \mathsf{self})$$

that is
(19) $\Gamma' \vdash_{\mathsf{BEB}} (n \bullet \top \uparrow \mathsf{deliver}_{\mathsf{pl}}(n',m)) \Rightarrow$

$$\hat{\boxminus}(n \bullet \text{self} \to (\text{deliver}_{\text{beb}}(n', m) \notin \text{ois}) \land$$
$$\hat{\boxdot}(n \bullet \text{self} \to (\text{deliver}_{\text{beb}}(n', m) \notin \text{ois})$$

By rule OR$'$
  (20) $\Gamma' \vdash_{\text{BEBC}} (n \bullet \top \uparrow \text{deliver}_{\text{beb}}(n', m)) \Rightarrow \Diamond(n \bullet (0, \text{deliver}_{\text{beb}}(n', m)) \in \text{ois} \land \text{self})$
By rule INVL with $\mathcal{A} =$
  $(n \bullet (0, \text{deliver}_{\text{beb}}(n', m)) \in \text{ois}) \to (n' \bullet 0 \uparrow \text{deliver}_{\text{pl}}(n', m)) \land \text{occ}(\text{ois}, (0, \text{deliver}_{\text{beb}}(n', m))) = 1$
  (21) $\Gamma' \vdash_{\text{BEBC}} \text{self} \land (n \bullet (0, \text{deliver}_{\text{beb}}(n', m)) \in \text{ois}) \Rightarrow$
              $(n' \bullet 0 \uparrow \text{deliver}_{\text{pl}}(n', m)) \land \text{occ}(\text{ois}, (0, \text{deliver}_{\text{beb}}(n', m))) = 1$
From [20] and [21],
  (22) $\Gamma' \vdash_{\text{BEBC}} (n \bullet \top \uparrow \text{deliver}_{\text{beb}}(n', m)) \Rightarrow$
              $\Diamond[(n' \bullet 0 \uparrow \text{deliver}_{\text{pl}}(n', m)) \land \text{occ}(\text{ois}, (0, \text{deliver}_{\text{beb}}(n', m))) = 1]$
From [22] and [19],
  (23) $\Gamma' \vdash_{\text{BEBC}} (n \bullet \top \uparrow \text{deliver}_{\text{beb}}(n', m)) \Rightarrow \Diamond[$
              $\text{occ}(\text{ois}, (0, \text{deliver}_{\text{beb}}(n', m))) = 1 \land$
              $\hat{\boxminus}(n \bullet \text{self} \to (\text{deliver}_{\text{beb}}(n', m) \notin \text{ois}) \land$
              $\hat{\boxdot}(n \bullet \text{self} \to (\text{deliver}_{\text{beb}}(n', m) \notin \text{ois})]$
By rule UNIOI on [23],
  (24) $\Gamma' \vdash_{\text{BEBC}} (n \bullet \top \uparrow \text{deliver}_{\text{beb}}(n', m)) \Rightarrow \Diamond[$
              $(n \bullet \top \uparrow \text{deliver}_{\text{beb}}(n', m)) \Rightarrow$
              $\hat{\boxminus}\neg(n \bullet \top \uparrow \text{deliver}_{\text{beb}}(n', m))]$
By Lemma 109 on [24]
  $\Gamma' \vdash_{\text{BEBC}} (n \bullet \top \uparrow \text{deliver}_{\text{beb}}(n', m)) \Rightarrow$
              $\hat{\boxminus}\neg(n \bullet \top \uparrow \text{deliver}_{\text{beb}}(n', m))$

**Theorem 11.** *(BEB$_3$: No-forge)*
If a node delivers a message $m$ with sender $n'$, then $m$ was previously broadcast by node $n'$.

$\Gamma \vdash_{\mathsf{BEBC}} (n \bullet \top \uparrow \mathsf{deliver}_{\mathsf{beb}}(n', m)) \leftsquigarrow (n' \bullet \top \downarrow \mathsf{broadcast}_{\mathsf{beb}}(m))$
where
$\Gamma$ is defined in Definition 18.

**Proof.**

The proof idea: If a best-effort broadcast delivery event is executed, it is previously issued. The component issues a best-effort broadcast delivery event only when a delivery event from the perfect link is processed. By the no-forge property of the perfect link, every delivery indication is preceded by a corresponding send event. The component issues a perfect link send request only when a broadcast request is processed.

By rule OI',
    (1) $\Gamma \vdash_{\mathsf{BEBC}} (n \bullet \top \uparrow \mathsf{deliver}_{\mathsf{beb}}(n', m)) \Rightarrow \diamondsuit(n \bullet \mathsf{deliver}_{\mathsf{beb}}(n', m) \in \mathsf{ois} \wedge \mathsf{self})$
By rule INVL
    (2) $\Gamma \vdash_{\mathsf{BEBC}} (n \bullet \mathsf{deliver}_{\mathsf{beb}}(n', m) \in \mathsf{ois} \wedge \mathsf{self}) \Rightarrow (n \bullet 0 \uparrow \mathsf{deliver}_{\mathsf{pl}}(n', m))$
From [1] and [2]
    $\Gamma \vdash_{\mathsf{BEBC}} (n \bullet \top \uparrow \mathsf{deliver}_{\mathsf{beb}}(n', m)) \Rightarrow \diamondsuit(n \bullet 0 \uparrow \mathsf{deliver}_{\mathsf{pl}}(n', m))$
that is,
    (3) $\Gamma \vdash_{\mathsf{BEBC}} (n \bullet \top \uparrow \mathsf{deliver}_{\mathsf{beb}}(n', m)) \leftsquigarrow (n \bullet 0 \uparrow \mathsf{deliver}_{\mathsf{pl}}(n', m))$

From $PL_3$
    (4) $\Gamma \vdash_{\mathsf{BEBC}} (n \bullet 0 \uparrow \mathsf{deliver}_{\mathsf{pl}}(n', m)) \leftsquigarrow (n' \bullet 0 \downarrow \mathsf{send}_{\mathsf{pl}}(n, m))$

By rule OR',
    (5) $\Gamma \vdash_{\mathsf{BEBC}} (n' \bullet 0 \downarrow \mathsf{send}_{\mathsf{pl}}(n, m)) \Rightarrow \diamondsuit(n' \bullet (0, \mathsf{send}_{\mathsf{pl}}(n, m)) \in \mathsf{ors} \wedge \mathsf{self})$
By rule INVL
    (6) $\Gamma \vdash_{\mathsf{BEBC}} (n' \bullet (0, \mathsf{send}_{\mathsf{pl}}(n, m)) \in \mathsf{ors} \wedge \mathsf{self}) \Rightarrow (n' \bullet \top \downarrow \mathsf{broadcast}_{\mathsf{beb}}(m))$
From [5] and [6]
    $\Gamma \vdash_{\mathsf{BEBC}} (n' \bullet 0 \downarrow \mathsf{send}_{\mathsf{pl}}(n, m)) \Rightarrow \diamondsuit(n' \bullet \top \downarrow \mathsf{broadcast}_{\mathsf{beb}}(m))$
that is
    (7) $\Gamma \vdash_{\mathsf{BEBC}} n' \bullet 0 \downarrow \mathsf{send}_{\mathsf{pl}}(n, m)) \leftsquigarrow (n' \bullet \top \downarrow \mathsf{broadcast}_{\mathsf{beb}}(m))$

From Lemma 88 on [3], [4] and [7],
    $\Gamma \vdash_{\mathsf{BEBC}} (n \bullet \top \uparrow \mathsf{deliver}_{\mathsf{beb}}(n', m)) \leftsquigarrow (n' \bullet \top \downarrow \mathsf{broadcast}_{\mathsf{beb}}(m))$

### 5.3.4 Uniform Reliable Broadcast

**Definition 19.**
$\Gamma =$
  $|\mathsf{Correct}| > \mathbb{N}/2;$
  $\mathsf{BEB}'_1; \mathsf{BEB}'_2; \mathsf{BEB}'_3$

$\mathsf{BEB}'_1 = \mathsf{lower}(0, \mathsf{BEB}_1) =$
  $n \in \mathsf{Correct} \wedge n' \in \mathsf{Correct} \to$
  $(n' \bullet 0 \downarrow \mathsf{broadcast}_{\mathsf{beb}}(m)) \rightsquigarrow$
  $(n \bullet 0 \uparrow \mathsf{deliver}_{\mathsf{beb}}(n', m))$

$\mathsf{BEB}'_2 = \mathsf{lower}(0, \mathsf{BEB}_2) =$
  $[n' \bullet 0 \downarrow \mathsf{broadcast}_{\mathsf{beb}}(n, m) \Rightarrow$
    $\hat{\boxminus} \neg (n' \bullet 0 \downarrow \mathsf{broadcast}_{\mathsf{beb}}(n, m))] \to$
  $[n \bullet 0 \uparrow \mathsf{deliver}_{\mathsf{beb}}(n', m) \Rightarrow$
    $\hat{\boxminus} \neg (n \bullet 0 \uparrow \mathsf{deliver}_{\mathsf{beb}}(n', m))]$

$\mathsf{BEB}'_3 = \mathsf{lower}(0, \mathsf{BEB}_3) =$
  $(n \bullet 0 \uparrow \mathsf{deliver}_{\mathsf{beb}}(n', m)) \leftarrowtail$
  $(n' \bullet 0 \downarrow \mathsf{broadcast}_{\mathsf{beb}}(m))$

**Theorem 12.** *($URB_1$: Validity)*
If a correct node $n$ broadcasts a message $m$, then $n$ itself eventually delivers $m$.

$\Gamma \vdash_{\mathsf{URBC}} \forall n.\ n \in \mathsf{Correct} \rightarrow$
$\qquad (n \bullet \top \downarrow \mathsf{broadcast}_{\mathsf{urb}}(m)) \rightsquigarrow (n \bullet \top \uparrow \mathsf{deliver}_{\mathsf{urb}}(n, m))$

where
$\Gamma$ is defined in Definition 19.

**Proof.**

The proof idea: Upon a broadcast request for a message $m$ by a node $n$, the request function broadcasts $m$ using the best-effort broadcast subcomponent bebc. By the validity property of bebc, as $n$ is correct, bebc will eventually deliver $m$ to every correct node. Upon delivery of a message from bebc, the indication function checks whether the message is in the pending set. If it does not find the message in the pending set, it rebroadcasts the message using bebc. Otherwise, the node has already broadcast the message. Therefore, every correct node broadcasts $m$ using bebc. Therefore, by the validity of bebc, bebc eventually delivers $m$ from all correct nodes to the node $n$. Upon delivery of a message from bebc, the indication function adds the sender to the set of nodes that have acknowledged the message. Thus, at node $n$, the identifiers of all the correct nodes will eventually be in the acknowledged set for $m$. This acknowledgement set never shrinks. On the next execution of the periodic function either $m$ is in the delivered set or not. If it is not in the delivered set, since acknowledgement from all the correct nodes has been received and the correct nodes are a majority, the delivery condition is satisfied and the delivery event for $m$ is issued.

Otherwise, if $m$ is not in the delivered set, the delivery of $m$ is previously issued. This delivery cannot be before the original request, based on the following two invariants. First, the counter of a node monotonically increases. Second, the counter of every message in the pending set is less than or equal to the counter of the node. Based on the two invariants, if the delivery of $m$ happens before its broadcast request, the counter of $n$ should have decreased from the former to the latter.

We assume
 (1) $\Gamma' = \Gamma; n \in \mathsf{Correct}$
We prove that
 $\Gamma \vdash_{\mathsf{URBC}} (n \bullet \top \downarrow \mathsf{broadcast}_{\mathsf{urb}}(m)) \rightsquigarrow (n \bullet \top \uparrow \mathsf{deliver}_{\mathsf{urb}}(n, m))$

By rule IRSE and ORSE,
 (2) $\Gamma \vdash_{\mathsf{URBC}} \textcircled{S}\ (n \bullet \top \downarrow \mathsf{broadcast}_{\mathsf{urb}}(m)) \Rightarrow$
  $\exists c.\ count(\mathsf{s}(n)) = c \wedge \Diamond (n \bullet 0 \downarrow \mathsf{broadcast}_{\mathsf{beb}}(\langle m, n, c+1 \rangle))$
By Lemma 45 on [2] and [1]
 (3) $\Gamma \vdash_{\mathsf{URBC}} \textcircled{S}\ (n \bullet \top \downarrow \mathsf{broadcast}_{\mathsf{urb}}(m)) \Rightarrow$
  $\exists c.\ count(\mathsf{s}(n)) = c \wedge \Diamond [$
   $\diamondsuit (n \bullet \top \uparrow \mathsf{deliver}_{\mathsf{urb}}(n, m)) \wedge \Box\, c + 1 \leq count(\mathsf{s}(n))\ \vee$
   $\Diamond (n \bullet \top \uparrow \mathsf{deliver}_{\mathsf{urb}}(n, m))]$
By Lemma 112 on [3] and simplification
 $\Gamma \vdash_{\mathsf{URBC}} \textcircled{S}\ (n \bullet \top \downarrow \mathsf{broadcast}_{\mathsf{urb}}(m)) \Rightarrow$

$$\diamond(n \bullet \top \uparrow \mathsf{deliver}_{\mathsf{urb}}(n, m)) \vee$$
$$[\exists c.\ count(\mathsf{s}(n)) = c\ \wedge \diamond(\square c + 1 \leq count(\mathsf{s}(n)))]$$

that is

$$\Gamma \vdash_{\mathsf{URBC}} \circledS\ (n \bullet \top \downarrow \mathsf{broadcast}_{\mathsf{urb}}(m)) \Rightarrow$$
$$\diamond(n \bullet \top \uparrow \mathsf{deliver}_{\mathsf{urb}}(n, m)) \vee$$
$$[\exists c.\ count(\mathsf{s}(n)) = c\ \wedge c + 1 \leq count(\mathsf{s}(n))]$$

that is

$$\Gamma \vdash_{\mathsf{URBC}} \circledS\ (n \bullet \top \downarrow \mathsf{broadcast}_{\mathsf{urb}}(m)) \Rightarrow$$
$$\diamond(n \bullet \top \uparrow \mathsf{deliver}_{\mathsf{urb}}(n, m))$$

that by rule SInv is

$$\Gamma \vdash_{\mathsf{URBC}} (n \bullet \top \downarrow \mathsf{broadcast}_{\mathsf{urb}}(m)) \leadsto (n \bullet \top \uparrow \mathsf{deliver}_{\mathsf{urb}}(n, m))$$

**Lemma 45.**
$$\Gamma \vdash_{\mathsf{URBC}} \circledS\ n \in \mathsf{Correct} \wedge n' \in \mathsf{Correct} \rightarrow$$
$$n \bullet 0 \downarrow \mathsf{broadcast}_{\mathsf{beb}}(\langle m, n_s, c \rangle) \Rightarrow$$
$$\Leftrightarrow (n' \bullet \top \uparrow \mathsf{deliver}_{\mathsf{urb}}(n_s, m)) \wedge \square c \leq count(\mathsf{s}(n')) \vee$$
$$\diamond(n' \bullet \top \uparrow \mathsf{deliver}_{\mathsf{urb}}(n_s, m))$$

**Proof.**
Immediate from Lemma 46 and Lemma 48.

**Lemma 46.**
$$\Gamma \vdash_{\mathsf{URBC}} \circledS\ n \in \mathsf{Correct} \wedge n' \in \mathsf{Correct} \rightarrow$$
$$n \bullet 0 \downarrow \mathsf{broadcast}_{\mathsf{beb}}(\langle m, n_s, c \rangle) \Rightarrow$$
$$\diamond \square |ack(\mathsf{s}(n'))(\langle m, n_s, c \rangle)| > |\mathbb{N}|/2$$

**Proof.**
We assume that
  (1) $\Gamma' = \Gamma$; $n \in \mathsf{Correct}$; $n' \in \mathsf{Correct}$
We show that
$$\Gamma \vdash_{\mathsf{URBC}} \circledS\ n \bullet 0 \downarrow \mathsf{broadcast}_{\mathsf{beb}}(\langle m, n_s, c \rangle) \Rightarrow$$
$$\diamond \square |ack(\mathsf{s}(n'))(\langle m, n_s, c \rangle)| > |\mathbb{N}|/2$$

By Definition 19 ($\mathsf{BEB}'_1$) on [1] and rule InvS,
  (2) $\Gamma' \vdash_{\mathsf{URBC}} \circledS\ \forall n'.\ n' \in \mathsf{Correct} \rightarrow$
$$(n \bullet 0 \downarrow \mathsf{broadcast}_{\mathsf{beb}}(\langle m, n_s, c \rangle)) \Rightarrow$$
$$\diamond(n' \bullet 0 \uparrow \mathsf{deliver}_{\mathsf{beb}}(n, \langle m, n_s, c \rangle))$$

By Lemma 47
  (3) $\Gamma' \vdash_{\mathsf{URBC}} \circledS\ \forall n', n''.$
$$(n' \bullet 0 \uparrow \mathsf{deliver}_{\mathsf{beb}}(n'', \langle m, n_s, c \rangle)) \Rightarrow$$
$$\Leftrightarrow (n' \bullet 0 \downarrow \mathsf{broadcast}_{\mathsf{beb}}(\langle m, n_s, c \rangle)) \vee$$
$$\diamond(n' \bullet 0 \downarrow \mathsf{broadcast}_{\mathsf{beb}}(\langle m, n_s, c \rangle))$$

By Definition 19 ($\mathsf{BEB}'_1$) and rule InvS,
  (4) $\Gamma' \vdash_{\mathsf{URBC}} \circledS\ \forall n, n'.\ n \in \mathsf{Correct} \wedge n' \in \mathsf{Correct} \rightarrow$

$$(n \bullet 0 \downarrow \mathsf{broadcast}_{\mathsf{beb}}(\langle m, n_s, c\rangle)) \Rightarrow$$
$$\Diamond(n' \bullet 0 \uparrow \mathsf{deliver}_{\mathsf{beb}}(n, \langle m, n_s, c\rangle))$$

From [3] and [4]
$$\Gamma' \vdash_{\mathsf{URBC}} ⑤ \ \forall n, n', n''. \ n \in \mathsf{Correct} \wedge n' \in \mathsf{Correct} \to$$
$$(n \bullet 0 \uparrow \mathsf{deliver}_{\mathsf{beb}}(n'', \langle m, n_s, c\rangle)) \Rightarrow$$
$$\hat{\Diamond}\Diamond(n' \bullet 0 \uparrow \mathsf{deliver}_{\mathsf{beb}}(n, \langle m, n_s, c\rangle)) \vee$$
$$\Diamond\Diamond(n' \bullet 0 \uparrow \mathsf{deliver}_{\mathsf{beb}}(n, \langle m, n_s, c\rangle))$$

that by Lemma 112, Lemma 86 and Lemma 87 is
(5) $\Gamma' \vdash_{\mathsf{URBC}} ⑤ \ \forall n, n', n''. \ n \in \mathsf{Correct} \wedge n' \in \mathsf{Correct} \to$
$$(n \bullet 0 \uparrow \mathsf{deliver}_{\mathsf{beb}}(n'', \langle m, n_s, c\rangle)) \Rightarrow$$
$$\hat{\Diamond}(n' \bullet 0 \uparrow \mathsf{deliver}_{\mathsf{beb}}(n, \langle m, n_s, c\rangle)) \vee$$
$$\Diamond(n' \bullet 0 \uparrow \mathsf{deliver}_{\mathsf{beb}}(n, \langle m, n_s, c\rangle))$$

From [2] and [5]
(6) $\Gamma' \vdash_{\mathsf{URBC}} ⑤ \ \forall n'. \ n' \in \mathsf{Correct} \to$
$$(n \bullet 0 \downarrow \mathsf{broadcast}_{\mathsf{beb}}(\langle m, n_s, c\rangle)) \Rightarrow \Diamond[$$
$$\forall n''. \ n'' \in \mathsf{Correct} \to$$
$$\hat{\Diamond}(n'' \bullet 0 \uparrow \mathsf{deliver}_{\mathsf{beb}}(n', \langle m, n_s, c\rangle)) \vee$$
$$\Diamond(n'' \bullet 0 \uparrow \mathsf{deliver}_{\mathsf{beb}}(n', \langle m, n_s, c\rangle))]$$

that is
(7) $\Gamma' \vdash_{\mathsf{URBC}} ⑤ \ \forall n'. \ n' \in \mathsf{Correct} \wedge n'' \in \mathsf{Correct} \to$
$$(n \bullet 0 \downarrow \mathsf{broadcast}_{\mathsf{beb}}(\langle m, n_s, c\rangle)) \Rightarrow \Diamond[$$
$$\hat{\Diamond}(n'' \bullet 0 \uparrow \mathsf{deliver}_{\mathsf{beb}}(n', \langle m, n_s, c\rangle)) \vee$$
$$\Diamond(n'' \bullet 0 \uparrow \mathsf{deliver}_{\mathsf{beb}}(n', \langle m, n_s, c\rangle))]$$

that by Lemma 112, Lemma 86 and Lemma 87 is
(8) $\Gamma' \vdash_{\mathsf{URBC}} ⑤ \ \forall n'. \ n' \in \mathsf{Correct} \wedge n'' \in \mathsf{Correct} \to$
$$(n \bullet 0 \downarrow \mathsf{broadcast}_{\mathsf{beb}}(\langle m, n_s, c\rangle)) \Rightarrow$$
$$\hat{\Diamond}(n'' \bullet 0 \uparrow \mathsf{deliver}_{\mathsf{beb}}(n', \langle m, n_s, c\rangle)) \vee$$
$$\Diamond(n'' \bullet 0 \uparrow \mathsf{deliver}_{\mathsf{beb}}(n', \langle m, n_s, c\rangle))$$

By rule IISE,
(9) $\Gamma' \vdash_{\mathsf{URBC}} ⑤ \ \forall n, n'.$
$$(n \bullet 0 \uparrow \mathsf{deliver}_{\mathsf{beb}}(n', \langle m, n_s, c\rangle)) \Rightarrow$$
$$n' \in ack(\mathsf{s}'(n))(\langle m, n_s, c\rangle)$$

By rule INVSSE″,
(10) $\Gamma' \vdash_{\mathsf{URBC}} ⑤ \ \forall n, n'.$
$$n' \in ack(\mathsf{s}'(n))(\langle m, n_s, c\rangle) \Rightarrow$$
$$\hat{\Box} n' \in ack(\mathsf{s}(n))(\langle m, n_s, c\rangle)$$

From [9] and [10]
(11) $\Gamma' \vdash_{\mathsf{URBC}} ⑤ \ \forall n, n'.$
$$(n \bullet 0 \uparrow \mathsf{deliver}_{\mathsf{beb}}(n', \langle m, n_s, c\rangle)) \Rightarrow$$
$$\hat{\Box} n' \in ack(\mathsf{s}(n))(\langle m, n_s, c\rangle)$$

From [8] and [11]
$\Gamma' \vdash_{\mathsf{URBC}} ⑤ \ \forall n', n''. \ n' \in \mathsf{Correct} \wedge n'' \in \mathsf{Correct} \to$
$$(n \bullet 0 \downarrow \mathsf{broadcast}_{\mathsf{beb}}(\langle m, n_s, c\rangle)) \Rightarrow$$
$$\hat{\Diamond}\hat{\Box} n' \in ack(\mathsf{s}(n''))(\langle m, n_s, c\rangle) \vee$$
$$\Diamond\hat{\Box} n' \in ack(\mathsf{s}(n''))(\langle m, n_s, c\rangle)$$

that is
(12) $\Gamma' \vdash_{\mathsf{URBC}} \circledS\ \forall n', n''.\ n' \in \mathsf{Correct} \wedge n'' \in \mathsf{Correct} \rightarrow$
$(n \bullet 0 \downarrow \mathsf{broadcast}_{\mathsf{beb}}(\langle m, n_s, c \rangle)) \Rightarrow$
$\Diamond \Box n' \in ack(\mathsf{s}(n''))(\langle m, n_s, c \rangle)$

From [12] and [1] (instantiating $n''$ with $n'$ and renaming $n'$ to $n''$),
(13) $\Gamma' \vdash_{\mathsf{URBC}} \circledS\ \forall n''.\ n'' \in \mathsf{Correct} \rightarrow$
$(n \bullet 0 \downarrow \mathsf{broadcast}_{\mathsf{beb}}(\langle m, n_s, c \rangle)) \Rightarrow$
$\Diamond \Box n'' \in ack(\mathsf{s}(n'))(\langle m, n_s, c \rangle)$

that is
(14) $\Gamma' \vdash_{\mathsf{URBC}} \circledS\ (n \bullet 0 \downarrow \mathsf{broadcast}_{\mathsf{beb}}(\langle m, n_s, c \rangle)) \Rightarrow$
$\Diamond \Box \forall n''.\ n'' \in \mathsf{Correct} \rightarrow n'' \in ack(\mathsf{s}(n'))(\langle m, n_s, c \rangle)$

that is
(15) $\Gamma' \vdash_{\mathsf{URBC}} \circledS\ (n \bullet 0 \downarrow \mathsf{broadcast}_{\mathsf{beb}}(\langle m, n_s, c \rangle)) \Rightarrow$
$\Diamond \Box \mathsf{Correct} \subseteq ack(\mathsf{s}(n'))(\langle m, n_s, c \rangle)$

By Definition 19 (a majority of correct nodes) on [1]
$\Gamma' \vdash_{\mathsf{URBC}} \circledS\ (n \bullet 0 \downarrow \mathsf{broadcast}_{\mathsf{beb}}(\langle m, n_s, c \rangle)) \Rightarrow$
$\Diamond \Box |ack(\mathsf{s}(n'))(\langle m, n_s, c \rangle)| > |\mathbb{N}|/2$

**Lemma 47.**
$\Gamma \vdash_{\mathsf{URBC}} \circledS$
$(n \bullet 0 \uparrow \mathsf{deliver}_{\mathsf{beb}}(n'', \langle m, n', c \rangle)) \Rightarrow$
$\hat{\Diamond}(n \bullet 0 \downarrow \mathsf{broadcast}_{\mathsf{beb}}(\langle m, n', c \rangle)) \vee$
$\Diamond(n \bullet 0 \downarrow \mathsf{broadcast}_{\mathsf{beb}}(\langle m, n', c \rangle))$

**Proof.**
By rule INVSA with
$S = \lambda s.\ \langle m, n', c \rangle \in pending(s)$
$\mathcal{A} = (0, \mathsf{broadcast}_{\mathsf{beb}}(\langle m, n', c \rangle)) \in \mathsf{ors}$
(1) $\Gamma \vdash_{\mathsf{URBC}} \circledS\ \langle m, n', c \rangle \in pending(\mathsf{s}(n)) \Rightarrow$
$\hat{\Diamond}(n \bullet (0, \mathsf{broadcast}_{\mathsf{beb}}(\langle m, n', c \rangle)) \in \mathsf{ors})$

By rule ORSE
(2) $\Gamma \vdash_{\mathsf{URBC}} \circledS\ n \bullet (0, \mathsf{broadcast}_{\mathsf{beb}}(\langle m, n', c \rangle)) \in \mathsf{ors} \Rightarrow$
$\Diamond(n \bullet 0 \downarrow \mathsf{broadcast}_{\mathsf{beb}}(\langle m, n', c \rangle))$

By Lemma 80 and Lemma 112 on [1] and [2]
(3) $\Gamma \vdash_{\mathsf{URBC}} \circledS\ \langle m, n', c \rangle \in pending(\mathsf{s}(n)) \Rightarrow$
$\hat{\Diamond}(n \bullet 0 \downarrow \mathsf{broadcast}_{\mathsf{beb}}(\langle m, n', c \rangle)) \vee$
$\Diamond(n \bullet 0 \downarrow \mathsf{broadcast}_{\mathsf{beb}}(\langle m, n', c \rangle))$

By rule IIORSE,
(4) $\Gamma \vdash_{\mathsf{URBC}} (n \bullet 0 \uparrow \mathsf{deliver}_{\mathsf{beb}}(n, \langle m, n', c \rangle)) \wedge \langle m, n', c \rangle \notin pending(\mathsf{s}(n)) \Rightarrow$
$\Diamond(n \bullet 0 \downarrow \mathsf{broadcast}_{\mathsf{beb}}(\langle m, n', c \rangle))$

From [3] and [4]
(5) $\Gamma \vdash_{\mathsf{URBC}} \circledS\ [\langle m, n', c \rangle \in pending(\mathsf{s}(n))] \vee$
$[(n \bullet 0 \uparrow \mathsf{deliver}_{\mathsf{beb}}(n, \langle m, n', c \rangle)) \wedge \langle m, n', c \rangle \notin pending(\mathsf{s}(n))] \Rightarrow$
$\hat{\Diamond}(n \bullet 0 \downarrow \mathsf{broadcast}_{\mathsf{beb}}(\langle m, n', c \rangle)) \vee$

$\diamond(n \bullet 0 \downarrow \mathsf{broadcast}_{\mathsf{beb}}(\langle m, n', c \rangle))$

that (after adding a conjunct) is

(6) $\Gamma \vdash_{\mathsf{URBC}} \circledS [(n \bullet 0 \uparrow \mathsf{deliver}_{\mathsf{beb}}(n, \langle m, n', c \rangle)) \wedge \langle m, n', c \rangle \in pending(\mathsf{s}(n))] \vee$
$[(n \bullet 0 \uparrow \mathsf{deliver}_{\mathsf{beb}}(n, \langle m, n', c \rangle)) \wedge \langle m, n', c \rangle \notin pending(\mathsf{s}(n))] \Rightarrow$
$\diamondsuit(n \bullet 0 \downarrow \mathsf{broadcast}_{\mathsf{beb}}(\langle m, n', c \rangle)) \vee$
$\diamond(n \bullet 0 \downarrow \mathsf{broadcast}_{\mathsf{beb}}(\langle m, n', c \rangle))$

that (after folding distribution of or over and) is

(7) $\Gamma \vdash_{\mathsf{URBC}} \circledS [(n \bullet 0 \uparrow \mathsf{deliver}_{\mathsf{beb}}(n, \langle m, n', c \rangle))] \wedge$
$[\langle m, n', c \rangle \in pending(\mathsf{s}(n)) \vee \langle m, n', c \rangle \notin pending(\mathsf{s}(n))] \Rightarrow$
$\diamondsuit(n \bullet 0 \downarrow \mathsf{broadcast}_{\mathsf{beb}}(\langle m, n', c \rangle)) \vee$
$\diamond(n \bullet 0 \downarrow \mathsf{broadcast}_{\mathsf{beb}}(\langle m, n', c \rangle))$

that is

(8) $\Gamma \vdash_{\mathsf{URBC}} \circledS (n \bullet 0 \uparrow \mathsf{deliver}_{\mathsf{beb}}(n, \langle m, n', c \rangle)) \Rightarrow$
$\diamondsuit(n \bullet 0 \downarrow \mathsf{broadcast}_{\mathsf{beb}}(\langle m, n', c \rangle)) \vee$
$\diamond(n \bullet 0 \downarrow \mathsf{broadcast}_{\mathsf{beb}}(\langle m, n', c \rangle))$

**Lemma 48.**
$\Gamma \vdash_{\mathsf{URBC}} \circledS \; n \in \mathsf{Correct} \rightarrow$
$\diamond\square |ack(\mathsf{s}(n))(\langle m, n', c \rangle)| > |\mathbb{N}|/2 \Rightarrow$
$\diamondsuit(n \bullet \top \uparrow \mathsf{deliver}_{\mathsf{urb}}(n', m)) \wedge \square c \le count(\mathsf{s}(n)) \vee$
$\diamond(n \bullet \top \uparrow \mathsf{deliver}_{\mathsf{urb}}(n', m))$

**Proof.**
We assume that

(1) $\Gamma' = \Gamma; \; n \in \mathsf{Correct}$

We show that

$\Gamma' \vdash_{\mathsf{URBC}} \circledS \diamond\square |ack(\mathsf{s}(n))(\langle m, n', c \rangle)| > |\mathbb{N}|/2 \Rightarrow$
$\diamondsuit(n \bullet \top \uparrow \mathsf{deliver}_{\mathsf{urb}}(n', m)) \vee$
$\diamond(n \bullet \top \uparrow \mathsf{deliver}_{\mathsf{urb}}(n', m))$

We consider two cases: $\langle m, n_s, c \rangle \in delivered(\mathsf{s}(n))$ and $\langle m, n_s, c \rangle \notin delivered(\mathsf{s}(n))$.
By Lemma 49 on

(2) $\Gamma \vdash_{\mathsf{URBC}} \circledS \; \langle m, n', c \rangle \in delivered(\mathsf{s}(n)) \Rightarrow$
$\diamondsuit(n \bullet \top \uparrow \mathsf{deliver}_{\mathsf{urb}}(n', m) \wedge \square c \le count(\mathsf{s}(n))) \vee$
$\diamond(n \bullet \top \uparrow \mathsf{deliver}_{\mathsf{urb}}(n', m))$

By rule PEOISE

(3) $\Gamma \vdash_{\mathsf{URBC}} \circledS \; n \bullet \top \wr \mathsf{per} \wedge$
$\langle m, n_s, c \rangle \in pending(\mathsf{s}(n)) \wedge |ack(\mathsf{s}(n))(\langle m, n_s, c \rangle)| > |\mathbb{N}|/2 \wedge \langle m, n_s, c \rangle \notin delivered(\mathsf{s}(n)) \Rightarrow$

$\diamond(n \bullet \top \uparrow \mathsf{deliver}_{\mathsf{urb}}(n, m))$

By rule InvSSE'

(4) $\Gamma \vdash_{\mathsf{URBC}} \circledS \; ack(\mathsf{s}(n))(\langle m, n_s, c \rangle \ne \varnothing \Rightarrow \langle m, n_s, c \rangle \in pending(\mathsf{s}(n))$

From [3] and [4]

(5) $\Gamma \vdash_{\mathsf{URBC}} \circledS \; n \bullet \top \wr \mathsf{per} \wedge$
$|ack(\mathsf{s}(n))(\langle m, n_s, c \rangle)| > |\mathbb{N}|/2 \wedge \langle m, n_s, c \rangle \notin delivered(\mathsf{s}(n)) \Rightarrow$

$$\diamond(n \bullet \top \uparrow \mathsf{deliver_{urb}}(n,m))$$

By rule APERSE on [1]

(6) $\Gamma \vdash_{\mathsf{URBC}} \circledS \square\diamond(n \bullet \top \wr \mathsf{per})$

From [6]

(7) $\Gamma' \vdash_{\mathsf{URBC}} \circledS \square|ack(\mathsf{s}(n))(\langle m,n',c\rangle)| > |\mathbb{N}|/2 \Rightarrow$
$$\diamond[n \bullet \top \wr \mathsf{per} \wedge |ack(\mathsf{s}(n))(\langle m,n',c\rangle)| > |\mathbb{N}|/2]$$

By Lemma 87 on [7]

(8) $\Gamma' \vdash_{\mathsf{URBC}} \circledS \diamond\square|ack(\mathsf{s}(n))(\langle m,n',c\rangle)| > |\mathbb{N}|/2 \Rightarrow$
$$\diamond[n \bullet \top \wr \mathsf{per} \wedge |ack(\mathsf{s}(n))(\langle m,n',c\rangle)| > |\mathbb{N}|/2]$$

From [8]

(9) $\Gamma' \vdash_{\mathsf{URBC}} \circledS \diamond\square|ack(\mathsf{s}(n))(\langle m,n',c\rangle)| > |\mathbb{N}|/2 \Rightarrow \diamond$
$$\langle m,n',c\rangle \in delivered(\mathsf{s}(n)) \vee$$
$$[n \bullet \top \wr \mathsf{per} \wedge |ack(\mathsf{s}(n))(\langle m,n',c\rangle)| > |\mathbb{N}|/2 \wedge \langle m,n',c\rangle \notin delivered(\mathsf{s}(n))]$$

From [9], [2] and [5]

(10) $\Gamma' \vdash_{\mathsf{URBC}} \circledS \diamond\square|ack(\mathsf{s}(n))(\langle m,n',c\rangle)| > |\mathbb{N}|/2 \Rightarrow \diamond$
$$\Diamondleft(n \bullet \top \uparrow \mathsf{deliver_{urb}}(n',m) \wedge \square c \leq count(\mathsf{s}(n))) \vee$$
$$\diamond(n \bullet \top \uparrow \mathsf{deliver_{urb}}(n,m))$$

By Lemma 112 and Lemma 87 on [10]

$\Gamma' \vdash_{\mathsf{URBC}} \circledS \diamond\square|ack(\mathsf{s}(n))(\langle m,n',c\rangle)| > |\mathbb{N}|/2 \Rightarrow$
$$\Diamondleft(n \bullet \top \uparrow \mathsf{deliver_{urb}}(n',m)) \wedge \square c \leq count(\mathsf{s}(n)) \vee$$
$$\diamond(n \bullet \top \uparrow \mathsf{deliver_{urb}}(n,m))$$

**Lemma 49.**
$\Gamma \vdash_{\mathsf{URBC}} \circledS \langle m,n_s,c\rangle \in delivered(\mathsf{s}(n)) \Rightarrow$
$$[\diamond(n \bullet \top \uparrow \mathsf{deliver_{urb}}(n_s,m)) \vee$$
$$\Diamondleft(n \bullet \top \uparrow \mathsf{deliver_{urb}}(n_s,m) \wedge \square c \leq count(\mathsf{s}(n)))]$$

**Proof.**

By rule INVSASE with
  $n$ instantiated to $n$,
  $S$ instantiated to $\lambda s.\ \langle m,n_s,c\rangle \in delivered(s)$ and
  $\mathcal{A}$ instantiated to $\mathsf{deliver_{urb}}(\langle n,m\rangle)) \in \mathsf{ois} \wedge \langle m,n_s,c\rangle \in pending(\mathsf{s}(n))$

(1) $\Gamma \vdash_{\mathsf{URBC}} \circledS \langle m,n_s,c\rangle \in delivered(\mathsf{s}(n)) \Rightarrow$
$$\Diamondleft[n \bullet \mathsf{deliver_{urb}}(\langle n_s,m\rangle)) \in \mathsf{ois} \wedge \langle m,n_s,c\rangle \in pending(\mathsf{s}(n))]$$

By rule OISE,

(2) $\Gamma' \vdash_{\mathsf{URBC}} \circledS (n \bullet \mathsf{deliver_{urb}}(n_s,m) \in \mathsf{ois}) \Rightarrow \diamond(n \bullet \top \uparrow \mathsf{deliver_{urb}}(n_s,m))$

From [1] and [2],

(3) $\Gamma \vdash_{\mathsf{URBC}} \circledS \langle m,n_s,c\rangle \in delivered(\mathsf{s}(n)) \Rightarrow$
$$\Diamondleft[\langle m,n_s,c\rangle \in pending(\mathsf{s}(n)) \wedge \diamond(n \bullet \top \uparrow \mathsf{deliver_{urb}}(n_s,m))]$$

By rule INVSSE′

(4) $\Gamma \vdash_{\mathsf{URBC}} \circledS \ \langle m,n_s,c\rangle \in pending(\mathsf{s}(n)) \Rightarrow c \leq count(\mathsf{s}(n))$

From [3] and [4],
  (5) $\Gamma \vdash_{\mathsf{URBC}} \circledS \ \langle m, n_s, c\rangle \in delivered(\mathsf{s}(n)) \Rightarrow$
      $\diamondsuit[c \leq count(\mathsf{s}(n)) \wedge \diamondsuit(n \bullet \top \uparrow \mathsf{deliver}_{\mathsf{urb}}(n_s, m))]$

By rule INVSSE
  (6) $\Gamma \vdash_{\mathsf{URBC}} \circledS \ c \leq count(\mathsf{s}(n)) \Rightarrow (c \leq count(\mathsf{s}(n)))$

From [5] and [6],
  (7) $\Gamma \vdash_{\mathsf{URBC}} \circledS \ \langle m, n_s, c\rangle \in delivered(\mathsf{s}(n)) \Rightarrow$
      $\diamondsuit[(\square c \leq count(\mathsf{s}(n))) \wedge \diamondsuit(n \bullet \top \uparrow \mathsf{deliver}_{\mathsf{urb}}(n_s, m))]$
By Lemma 115 and Lemma 112 on [7],
  $\Gamma \vdash_{\mathsf{URBC}} \circledS \ \langle m, n_s, c\rangle \in delivered(\mathsf{s}(n)) \Rightarrow$
      $[\diamondsuit(n \bullet \top \uparrow \mathsf{deliver}_{\mathsf{urb}}(n_s, m) \wedge \square c \leq count(\mathsf{s}(n))) \vee$
      $\diamondsuit(n \bullet \top \uparrow \mathsf{deliver}_{\mathsf{urb}}(n_s, m) \wedge \square c \leq count(\mathsf{s}(n)))]$
that is
  $\Gamma \vdash_{\mathsf{URBC}} \circledS \ \langle m, n_s, c\rangle \in delivered(\mathsf{s}(n)) \Rightarrow$
      $[\diamondsuit(n \bullet \top \uparrow \mathsf{deliver}_{\mathsf{urb}}(n_s, m)) \vee$
      $\diamondsuit(n \bullet \top \uparrow \mathsf{deliver}_{\mathsf{urb}}(n_s, m) \wedge \square c \leq count(\mathsf{s}(n)))]$

**Theorem 13.** *(URB$_2$: No-duplication)*
If a message is broadcast at most once, it will be delivered at most once.

$\Gamma \vdash_{\mathsf{URBC}} [n' \bullet \top \downarrow \mathsf{broadcast}_{\mathsf{urb}}(m) \Rightarrow \hat{\boxminus}\neg(n' \bullet \top \downarrow \mathsf{broadcast}_{\mathsf{urb}}(m))] \rightarrow$
$\qquad [(n \bullet \top \uparrow \mathsf{deliver}_{\mathsf{urb}}(n', m)) \Rightarrow \hat{\boxminus}\neg(n \bullet \top \uparrow \mathsf{deliver}_{\mathsf{urb}}(n', m))]$
where
$\Gamma$ is defined in Definition 19.

**Proof.**

The proof idea: The uniform reliable broadcast for a message $m$ by a node $n$ is executed at most once. When a uniform reliable broadcast is executed, a best-effort broadcast request is issued. As $m$ is broadcast from $n$ at most once, it is broadcast with a unique counter $c$. By the definition of the component, a best-effort broadcast is issued only when a uniform reliable broadcast or a best-effort delivery is executed. By the no-forge property of best-effort broadcast, a best-effort delivery is executed only if the execution of a corresponding best-effort broadcast precedes it. Thus by a mutual induction, every best-effort delivery for $m$ is executed with the initial unique counter $c$. A message is added to the pending set only when a best-effort delivery is executed. Thus, the message $m$ from $n$ is added to the pending set only with the initial unique counter $c$. The periodic function issues a uniform reliable delivery for a message only when it is in the pending set and not delivered before. When it issues a uniform reliable delivery for a message, it is added to the delivered set. Thus, if a uniform reliable delivery for $m$ from $n$ is issued at a node, it will not be issued again. Thus, a uniform reliable delivery for $m$ from $n$ is executed at a node at most once.

We assume
  (1) $\Gamma' = \Gamma$;
      $n' \bullet \top \downarrow \mathsf{broadcast}_{\mathsf{urb}}(m) \Rightarrow \hat{\boxminus}\neg(n' \bullet \top \downarrow \mathsf{broadcast}_{\mathsf{urb}}(m))$
and prove
  $\Gamma' \vdash_{\mathsf{URBC}} n \bullet \top \uparrow \mathsf{deliver}_{\mathsf{urb}}(n', m) \Rightarrow \hat{\boxminus}\neg(n \bullet \top \uparrow \mathsf{deliver}_{\mathsf{urb}}(n', m))$

By rule OI$'$
  (2) $\Gamma \vdash_{\mathsf{URBC}} (n \bullet \top \uparrow \mathsf{deliver}_{\mathsf{urb}}(n', m)) \Rightarrow$
      $\qquad \Diamondsuit(n \bullet \mathsf{deliver}_{\mathsf{urb}}(n', m) \in \mathsf{ois} \wedge \mathsf{self})$

From Lemma 57 and Lemma 108,
(3) $\Gamma' \vdash_{\mathsf{URBC}} \circledS \, [n \bullet \mathsf{deliver}_{\mathsf{urb}}(n', m) \in \mathsf{ois} \Rightarrow$
$\qquad \mathsf{occ}(\mathsf{ois}, \mathsf{deliver}_{\mathsf{urb}}(n', m)) \leq 1 \wedge$
$\qquad \hat{\Box} \neg (n \bullet \mathsf{deliver}_{\mathsf{urb}}(n', m) \in \mathsf{ois}) \wedge$
$\qquad \hat{\boxminus} \neg (n \bullet \mathsf{deliver}_{\mathsf{urb}}(n', m) \in \mathsf{ois})]$

From rule UniOISe,
(4) $\Gamma' \vdash_{\mathsf{URBC}} \circledS \, (\mathsf{occ}(\mathsf{ois}, \mathsf{deliver}_{\mathsf{urb}}(n', m)) \leq 1 \wedge$
$\qquad \hat{\Box}(\mathsf{n} = n \to \mathsf{deliver}_{\mathsf{urb}}(n', m) \notin \mathsf{ois}) \wedge$
$\qquad \hat{\boxminus}(\mathsf{n} = n \to \mathsf{deliver}_{\mathsf{urb}}(n', m) \notin \mathsf{ois})) \Rightarrow$
$\qquad n \bullet \top \uparrow \mathsf{deliver}_{\mathsf{urb}}(n', m) \Rightarrow$
$\qquad \hat{\boxminus} \neg (n \bullet \top \uparrow \mathsf{deliver}_{\mathsf{urb}}(n', m))$

From [3] and [4]
$\Gamma' \vdash_{\mathsf{URBC}} \circledS \, n \bullet \top \uparrow \mathsf{deliver}_{\mathsf{urb}}(n', m) \Rightarrow \Diamond[$
$\qquad n \bullet \top \uparrow \mathsf{deliver}_{\mathsf{urb}}(n', m) \Rightarrow$
$\qquad \hat{\boxminus} \neg (n \bullet \top \uparrow \mathsf{deliver}_{\mathsf{urb}}(n', m))]$

That is,
$\Gamma' \vdash_{\mathsf{URBC}} \circledS \, n \bullet \top \uparrow \mathsf{deliver}_{\mathsf{urb}}(n', m) \Rightarrow$
$\qquad \hat{\boxminus} \neg (n \bullet \top \uparrow \mathsf{deliver}_{\mathsf{urb}}(n', m)$

By rule SInv
$\Gamma' \vdash_{\mathsf{URBC}} n \bullet \top \uparrow \mathsf{deliver}_{\mathsf{urb}}(n', m) \Rightarrow$
$\qquad \hat{\boxminus} \neg (n \bullet \top \uparrow \mathsf{deliver}_{\mathsf{urb}}(n', m)$

**Lemma 50.**
$\Gamma' \vdash_{\mathsf{URBC}} \circledS \, [\forall n, n', n'', m, c.$
$\qquad \Box(n \bullet 0 \uparrow \mathsf{deliver}_{\mathsf{beb}}(n'', \langle m, n', c \rangle)) \to$
$\qquad \Diamond[(n' \bullet \top \downarrow \mathsf{broadcast}_{\mathsf{urb}}(m)) \wedge count(\mathsf{s}(n')) = c - 1]]$

**Proof.**
Using Lemma 111 with
$\mathcal{A} = \circledS \, [\forall n', n'', m, c.$
$\qquad (n'' \bullet 0 \downarrow \mathsf{broadcast}_{\mathsf{beb}}(\langle m, n', c \rangle)) \to$
$\qquad \Diamond[(n' \bullet \top \downarrow \mathsf{broadcast}_{\mathsf{urb}}(m)) \wedge count(\mathsf{s}(n')) = c - 1]$
$\mathcal{A}' = \circledS \, [\forall n, n', n'', m, c.$
$\qquad (n \bullet 0 \uparrow \mathsf{deliver}_{\mathsf{beb}}(n'', \langle m, n', c \rangle)) \to$
$\qquad \Diamond[(n' \bullet \top \downarrow \mathsf{broadcast}_{\mathsf{urb}}(m)) \wedge count(\mathsf{s}(n')) = c - 1]]$
and then Lemma 51 and Lemma 52.

**Lemma 51.**
$\Gamma' \vdash_{\mathsf{URBC}} \circledS \, [\hat{\boxminus}[\forall n, n', n'', m, c.$
$\qquad (n \bullet 0 \uparrow \mathsf{deliver}_{\mathsf{beb}}(n'', \langle m, n', c \rangle)) \to$
$\qquad \Diamond[(n' \bullet \top \downarrow \mathsf{broadcast}_{\mathsf{urb}}(m)) \wedge count(\mathsf{s}(n')) = c - 1]]$

$$\overrightarrow{[\forall n', n'', m, c.}$$
$$(n'' \bullet 0 \downarrow \mathsf{broadcast}_{\mathsf{beb}}(\langle m, n', c \rangle)) \to$$
$$\diamondsuit[(n' \bullet \top \downarrow \mathsf{broadcast}_{\mathsf{urb}}(m)) \land count(\mathsf{s}(n')) = c - 1]]]$$

**Proof.**
Let
  (1) $\Gamma'' = \Gamma'$;
    $\circledS \, [\hat{\boxminus}[\forall n, n', n'', m, c.$
      $(n \bullet 0 \uparrow \mathsf{deliver}_{\mathsf{beb}}(n'', \langle m, n', c \rangle)) \to$
        $\diamondsuit[(n' \bullet \top \downarrow \mathsf{broadcast}_{\mathsf{urb}}(m)) \land count(\mathsf{s}(n')) = c - 1]]]$
We prove that
  $\Gamma'' \vdash_{\mathsf{URBC}} \circledS \, [\forall n', n'', m, c.$
    $(n'' \bullet 0 \downarrow \mathsf{broadcast}_{\mathsf{beb}}(\langle m, n', c \rangle)) \to$
      $\diamondsuit[(n' \bullet \top \downarrow \mathsf{broadcast}_{\mathsf{urb}}(m)) \land count(\mathsf{s}(n')) = c - 1]]$

From [1]
  (2) $\Gamma'' \vdash_{\mathsf{URBC}} \circledS \, [\hat{\boxminus}[\forall n, n', n'', m, c.$
      $(n \bullet 0 \uparrow \mathsf{deliver}_{\mathsf{beb}}(n'', \langle m, n', c \rangle)) \to$
        $\diamondsuit[(n' \bullet \top \downarrow \mathsf{broadcast}_{\mathsf{urb}}(m)) \land count(\mathsf{s}(n')) = c - 1]]]$

By rule OISE',
  (3) $\Gamma'' \vdash_{\mathsf{URBC}} \circledS \, [(n'' \bullet 0 \downarrow \mathsf{broadcast}_{\mathsf{beb}}(\langle m, n', c \rangle)) \to$
      $\hat{\diamondsuit}(n'' \bullet (0, \mathsf{broadcast}_{\mathsf{beb}}(\langle m, n', c \rangle)) \in \mathsf{ors})]$

By rule INVLSE,
  (4) $\Gamma'' \vdash_{\mathsf{URBC}} \circledS \, [(n'' \bullet \mathsf{broadcast}_{\mathsf{beb}}(\langle m, n', c \rangle) \in \mathsf{ois}) \Rightarrow$
      $[(n'' \bullet \top \downarrow \mathsf{broadcast}_{\mathsf{urb}}(m)) \land count(\mathsf{s}(n'')) = c - 1 \land n'' = n'] \lor$
      $\exists n'''. \, (n'' \bullet 0 \uparrow \mathsf{deliver}_{\mathsf{beb}}(n''', \langle m, n', c \rangle))]$

From [4],
  (5) $\Gamma'' \vdash_{\mathsf{URBC}} \circledS \, [\hat{\diamondsuit}(n'' \bullet \mathsf{broadcast}_{\mathsf{beb}}(\langle m, n', c \rangle) \in \mathsf{ois}) \Rightarrow$
      $\hat{\diamondsuit}[(n' \bullet \top \downarrow \mathsf{broadcast}_{\mathsf{urb}}(m)) \land count(\mathsf{s}(n')) = c - 1] \lor$
      $\hat{\diamondsuit}\exists n'''. \, (n'' \bullet 0 \uparrow \mathsf{deliver}_{\mathsf{beb}}(n''', \langle m, n', c \rangle))]$

From [5] and [2]
  (6) $\Gamma'' \vdash_{\mathsf{URBC}} \circledS \, [\hat{\diamondsuit}(n'' \bullet \mathsf{broadcast}_{\mathsf{beb}}(\langle m, n', c \rangle) \in \mathsf{ois}) \Rightarrow$
      $\hat{\diamondsuit}[(n' \bullet \top \downarrow \mathsf{broadcast}_{\mathsf{urb}}(m)) \land count(\mathsf{s}(n')) = c - 1] \lor$
      $\hat{\diamondsuit}\diamondsuit[(n' \bullet \top \downarrow \mathsf{broadcast}_{\mathsf{urb}}(m)) \land count(\mathsf{s}(n')) = c - 1]]$
that is
  (7) $\Gamma'' \vdash_{\mathsf{URBC}} \circledS \, [\hat{\diamondsuit}(n'' \bullet \mathsf{broadcast}_{\mathsf{beb}}(\langle m, n', c \rangle) \in \mathsf{ois}) \Rightarrow$
      $\hat{\diamondsuit}[(n' \bullet \top \downarrow \mathsf{broadcast}_{\mathsf{urb}}(m)) \land count(\mathsf{s}(n')) = c - 1]$

From [3] and [7]
  (8) $\Gamma'' \vdash_{\mathsf{URBC}} \circledS \, [(n'' \bullet 0 \downarrow \mathsf{broadcast}_{\mathsf{beb}}(\langle m, n', c \rangle)) \to$
      $\hat{\diamondsuit}[(n' \bullet \top \downarrow \mathsf{broadcast}_{\mathsf{urb}}(m)) \land count(\mathsf{s}(n')) = c - 1]$

Thus,
$$\Gamma'' \vdash_{\text{URBC}} \text{\textcircled{S}} \, [(n'' \bullet 0 \downarrow \text{broadcast}_{\text{beb}}(\langle m, n', c \rangle)) \rightarrow$$
$$\Diamond [(n' \bullet \top \downarrow \text{broadcast}_{\text{urb}}(m)) \land count(\text{s}(n')) = c - 1]$$

**Lemma 52.**
$$\Gamma' \vdash_{\text{URBC}} \text{\textcircled{S}} \, [\hat{\boxminus}[\forall n', n'', m, c.$$
$$(n'' \bullet 0 \downarrow \text{broadcast}_{\text{beb}}(\langle m, n', c \rangle)) \rightarrow$$
$$\Diamond [(n' \bullet \top \downarrow \text{broadcast}_{\text{urb}}(m)) \land count(\text{s}(n')) = c - 1]]$$
$$\rightarrow$$
$$[\forall n, n', n'', m, c.$$
$$(n \bullet 0 \uparrow \text{deliver}_{\text{beb}}(n'', \langle m, n', c \rangle)) \rightarrow$$
$$\Diamond [(n' \bullet \top \downarrow \text{broadcast}_{\text{urb}}(m)) \land count(\text{s}(n')) = c - 1]]]$$

**Proof.**
Let

(1) $\Gamma'' = \Gamma';$
$\text{\textcircled{S}} \, [\hat{\boxminus}[\forall n', n'', m, c.$
$(n'' \bullet 0 \downarrow \text{broadcast}_{\text{beb}}(\langle m, n', c \rangle)) \rightarrow$
$\Diamond [(n' \bullet \top \downarrow \text{broadcast}_{\text{urb}}(m)) \land count(\text{s}(n')) = c - 1]]]$

We prove that
$$\Gamma'' \vdash_{\text{URBC}} \text{\textcircled{S}} \, [\forall n, n', n'', m, c.$$
$$(n \bullet 0 \uparrow \text{deliver}_{\text{beb}}(n'', \langle m, n', c \rangle)) \rightarrow$$
$$\Diamond [(n' \bullet \top \downarrow \text{broadcast}_{\text{urb}}(m)) \land count(\text{s}(n')) = c - 1]]$$

From [1]

(2) $\Gamma'' \vdash_{\text{URBC}} \text{\textcircled{S}} \, [\forall n', n'', m, c.$
$\hat{\boxminus}(n'' \bullet 0 \downarrow \text{broadcast}_{\text{beb}}(\langle m, n', c \rangle)) \rightarrow$
$\Diamond [(n' \bullet \top \downarrow \text{broadcast}_{\text{urb}}(m)) \land count(\text{s}(n')) = c - 1]]$

From $\Gamma$, $\text{BEB}_3$ and rule SINV, we have

(3) $\Gamma'' \vdash_{\text{URBC}} \text{\textcircled{S}} \, [\forall n, n', n'', m, c.$
$(n \bullet 0 \uparrow \text{deliver}_{\text{beb}}(n'', \langle m, n', c \rangle)) \Rightarrow$
$\hat{\Diamond}(n'' \bullet 0 \downarrow \text{broadcast}_{\text{beb}}(\langle m, n', c \rangle))$

From [3] and [2]

(4) $\Gamma'' \vdash_{\text{URBC}} \text{\textcircled{S}} \, [\forall n, n', n'', m, c.$
$(n \bullet 0 \uparrow \text{deliver}_{\text{beb}}(n'', \langle m, n', c \rangle)) \rightarrow$
$\hat{\Diamond}\Diamond [(n' \bullet \top \downarrow \text{broadcast}_{\text{urb}}(m)) \land count(\text{s}(n')) = c - 1]]$

that is

(5) $\Gamma'' \vdash_{\text{URBC}} \text{\textcircled{S}} \, [\forall n, n', n'', m, c.$
$(n \bullet 0 \uparrow \text{deliver}_{\text{beb}}(n'', \langle m, n', c \rangle)) \rightarrow$
$\Diamond [(n' \bullet \top \downarrow \text{broadcast}_{\text{urb}}(m)) \land count(\text{s}(n')) = c - 1]]$

**Lemma 53.**
$\Gamma' \vdash_{\mathsf{URBC}} \text{\textcircled{S}}\, [\forall n, n', m, c.$
$\quad \langle m, n', c \rangle \in pending(\mathsf{s}(n)) \Rightarrow$
$\quad \Diamond[(n' \bullet \top \downarrow \mathsf{broadcast}_{\mathsf{urb}}(m)) \wedge count(\mathsf{s}(n')) = c - 1]]$

**Proof.**
By rule INVSASE
(1) $\Gamma' \vdash_{\mathsf{URBC}} \text{\textcircled{S}}\, [\langle m, n', c \rangle \in pending(\mathsf{s}(n)) \Rightarrow \hat{\Diamond}[$
$\quad ((n' \bullet \top \downarrow \mathsf{broadcast}_{\mathsf{urb}}(m)) \wedge count(\mathsf{s}(n')) = c - 1 \wedge n' = n) \vee$
$\quad (n \bullet 0 \uparrow \mathsf{deliver}_{\mathsf{beb}}(n'', \langle m, n', c \rangle))]]$

By Lemma 50,
(2) $\Gamma' \vdash_{\mathsf{URBC}} \text{\textcircled{S}}\, [\forall n, n', n'', m, c.$
$\quad (n \bullet 0 \uparrow \mathsf{deliver}_{\mathsf{beb}}(n'', \langle m, n', c \rangle)) \Rightarrow$
$\quad \Diamond[(n' \bullet \top \downarrow \mathsf{broadcast}_{\mathsf{urb}}(m)) \wedge count(\mathsf{s}(n')) = c - 1]]$

From [1] and [2]
(3) $\Gamma' \vdash_{\mathsf{URBC}} \text{\textcircled{S}}\, [\langle m, n', c \rangle \in pending(\mathsf{s}(n)) \Rightarrow \hat{\Diamond}[$
$\quad ((n' \bullet \top \downarrow \mathsf{broadcast}_{\mathsf{urb}}(m)) \wedge count(\mathsf{s}(n')) = c - 1) \vee$
$\quad \Diamond[(n' \bullet \top \downarrow \mathsf{broadcast}_{\mathsf{urb}}(m)) \wedge count(\mathsf{s}(n')) = c - 1]]]$

thus
(4) $\Gamma' \vdash_{\mathsf{URBC}} \text{\textcircled{S}}\, [\langle m, n', c \rangle \in pending(\mathsf{s}(n)) \Rightarrow$
$\quad \Diamond[(n' \bullet \top \downarrow \mathsf{broadcast}_{\mathsf{urb}}(m)) \wedge count(\mathsf{s}(n')) = c - 1]]$

**Lemma 54.**
$\Gamma' \vdash_{\mathsf{URBC}} \text{\textcircled{S}}\, [\forall n, c, c'. \; c \neq c' \rightarrow$
$\quad \langle m, n', c \rangle \in pending(\mathsf{s}(n)) \Rightarrow$
$\quad \boxminus \langle m, n', c' \rangle \notin pending(\mathsf{s}(n)) \; \wedge$
$\quad \Box \langle m, n', c' \rangle \notin pending(\mathsf{s}(n))]$

**Proof.**
We show the contra-positive of the first conjunct.
$\Gamma' \vdash_{\mathsf{URBC}} \text{\textcircled{S}}\, [c \neq c' \rightarrow$
$\quad \Diamondminus \langle m, n', c' \rangle \in pending(\mathsf{s}(n)) \Rightarrow$
$\quad \neg \langle m, n', c \rangle \notin pending(\mathsf{s}(n))]$

By Lemma 53,
(1) $\Gamma' \vdash_{\mathsf{URBC}} \text{\textcircled{S}}\, [\forall n, n', n'', m, c.$
$\quad \langle m, n', c \rangle \in pending(\mathsf{s}(n)) \Rightarrow$
$\quad \Diamond[(n' \bullet \top \downarrow \mathsf{broadcast}_{\mathsf{urb}}(m)) \wedge count(\mathsf{s}(n')) = c - 1]]$

From [1],
(2) $\Gamma' \vdash_{\mathsf{URBC}} \text{\textcircled{S}}\, [\Diamondminus \langle m, n', c' \rangle \in pending(\mathsf{s}(n)) \Rightarrow$
$\quad \Diamondminus \Diamond[(n' \bullet \top \downarrow \mathsf{broadcast}_{\mathsf{urb}}(m)) \wedge count(\mathsf{s}(n')) = c' - 1]]$

that is
(3) $\Gamma' \vdash_{\mathsf{URBC}} \text{\textcircled{S}}\, [\Diamondminus \langle m, n', c' \rangle \in pending(\mathsf{s}(n)) \Rightarrow$
$\quad \Diamond[(n' \bullet \top \downarrow \mathsf{broadcast}_{\mathsf{urb}}(m)) \wedge count(\mathsf{s}(n')) = c' - 1]]$

From the definition of $\Gamma'$,
- (4) $\Gamma' \vdash_{\mathsf{URBC}} \circledS\ [(n' \bullet \top \downarrow \mathsf{broadcast}_{\mathsf{urb}}(m)) \Rightarrow$
  $\hat{\boxminus} \neg (n' \bullet \top \downarrow \mathsf{broadcast}_{\mathsf{urb}}(m))]$

From [4] and Lemma 108
- (5) $\Gamma' \vdash_{\mathsf{URBC}} \circledS\ [(n' \bullet \top \downarrow \mathsf{broadcast}_{\mathsf{urb}}(m)) \wedge count(\mathsf{s}(n')) = c' - 1 \Rightarrow$
  $\hat{\boxminus} \neg (n' \bullet \top \downarrow \mathsf{broadcast}_{\mathsf{urb}}(m)) \wedge$
  $\hat{\square} \neg (n' \bullet \top \downarrow \mathsf{broadcast}_{\mathsf{urb}}(m))]$

thus
- (6) $\Gamma' \vdash_{\mathsf{URBC}} \circledS\ [c \neq c' \rightarrow$
  $(n' \bullet \top \downarrow \mathsf{broadcast}_{\mathsf{urb}}(m)) \wedge count(\mathsf{s}(n')) = c' - 1 \Rightarrow$
  $\boxminus \neg [(n' \bullet \top \downarrow \mathsf{broadcast}_{\mathsf{urb}}(m)) \wedge count(\mathsf{s}(n')) = c - 1] \wedge$
  $\square \neg [(n' \bullet \top \downarrow \mathsf{broadcast}_{\mathsf{urb}}(m)) \wedge count(\mathsf{s}(n')) = c - 1]]$

From [3] and [6]
- (7) $\Gamma' \vdash_{\mathsf{URBC}} \circledS\ [c \neq c' \rightarrow$
  $\Diamond \langle m, n', c' \rangle \in pending(\mathsf{s}(n)) \Rightarrow$
  $\Diamond [\boxminus \neg [(n' \bullet \top \downarrow \mathsf{broadcast}_{\mathsf{urb}}(m)) \wedge count(\mathsf{s}(n')) = c - 1] \wedge$
  $\square \neg [(n' \bullet \top \downarrow \mathsf{broadcast}_{\mathsf{urb}}(m)) \wedge count(\mathsf{s}(n')) = c - 1]]]$

thus
- (8) $\Gamma' \vdash_{\mathsf{URBC}} \circledS\ [c \neq c' \rightarrow$
  $\Diamond \langle m, n', c' \rangle \in pending(\mathsf{s}(n)) \Rightarrow$
  $\boxminus \neg [(n' \bullet \top \downarrow \mathsf{broadcast}_{\mathsf{urb}}(m)) \wedge count(\mathsf{s}(n')) = c - 1]]$

From the contra-positive of [1]
- (9) $\Gamma' \vdash_{\mathsf{URBC}} \circledS\ [\boxminus \neg [(n' \bullet \top \downarrow \mathsf{broadcast}_{\mathsf{urb}}(m)) \wedge count(\mathsf{s}(n')) = c - 1] \Rightarrow$
  $\neg \langle m, n', c \rangle \in pending(\mathsf{s}(n))]$

From [8] and [9]
- (10) $\Gamma' \vdash_{\mathsf{URBC}} \circledS\ [c \neq c' \rightarrow$
  $\Diamond \langle m, n', c' \rangle \in pending(\mathsf{s}(n)) \Rightarrow$
  $\neg \langle m, n', c \rangle \notin pending(\mathsf{s}(n))]$

Proof of the second conjunct is similar.

**Lemma 55.**
$\Gamma' \vdash_{\mathsf{URBC}} \circledS\ [\forall n, n', m, c.$
  $\square \mathsf{occ}(pending(\mathsf{s}(n)), \langle m, n', c \rangle) \leq 1]$

**Proof.**
By rule INVSSE',
- (1) $\Gamma' \vdash_{\mathsf{URBC}} \circledS\ \forall n, m, c.$
  $\langle m, n, c \rangle \in pending(\mathsf{s}(n)) \Rightarrow$
  $c \leq count(\mathsf{s}(n))$

From the definition of request and [1]
- (2) $\Gamma' \vdash_{\mathsf{URBC}} \circledS\ [\forall n', m, c.$

$$(\mathsf{s'(n), ois, ors}) = \mathsf{request}_c(\mathsf{n}, \mathsf{s(n)}, e) \wedge \mathsf{occ}(pending(\mathsf{s(n)}, \langle m, n', c \rangle)) \leq 1 \Rightarrow$$
$$\mathsf{occ}(pending(\mathsf{s'(n)}, \langle m, n', c \rangle)) \leq 1]$$

By rule INVUSSE'
$$\Gamma' \vdash_{\mathsf{URBC}} \circledS \; [\forall n, n', m, c.$$
$$\Box \mathsf{occ}(pending(\mathsf{s}(n), \langle m, n', c \rangle)) \leq 1]$$

The request case is from [2]. The indication and periodic cases are trivial.

**Lemma 56.**
$$\Gamma' \vdash_{\mathsf{URBC}} \circledS \; [\forall n, n', m$$
$$\Box \; |\{\langle m, n' \rangle \mid \exists c. \; \langle m, n', c \rangle \in pending(\mathsf{s}(n))\}| \leq 1]$$

**Proof.**
Immediate from Lemma 54 and Lemma 55.

**Lemma 57.**
The uniform-reliable broadcast delivery event is issued at most once at every node.
$$\Gamma' \vdash_{\mathsf{URBC}} \circledS \; [n \bullet \mathsf{deliver}_{\mathsf{urb}}(n', m) \in \mathsf{ois} \Rightarrow$$
$$\mathsf{occ}(\mathsf{ois}, \mathsf{deliver}_{\mathsf{urb}}(n', m)) \leq 1 \wedge$$
$$\hat{\boxminus} \neg (n \bullet \mathsf{deliver}_{\mathsf{urb}}(n', m) \in \mathsf{ois})]$$

**Proof.**

By rule INVLSE
(1) $\Gamma' \vdash_{\mathsf{URBC}} \circledS \; [(n \bullet \mathsf{deliver}_{\mathsf{urb}}(n', m) \in \mathsf{ois}) \Rightarrow$
$(\mathsf{s'}(n), \mathsf{ois}, \mathsf{ors}) = \mathsf{periodic}_c(\mathsf{n}, \mathsf{s}(n))]$

From the definition of periodic and Lemma 56
(2) $\Gamma' \vdash_{\mathsf{URBC}} \circledS \; [(\mathsf{s'}(n), \mathsf{ois}, \mathsf{ors}) = \mathsf{periodic}_c(\mathsf{n}, \mathsf{s}(n)) \Rightarrow$
$\mathsf{occ}(\mathsf{ois}, \mathsf{deliver}_{\mathsf{urb}}(n', m)) \leq 1]$

From [1] and [2]
$\Gamma' \vdash_{\mathsf{URBC}} \circledS \; [(n \bullet \mathsf{deliver}_{\mathsf{urb}}(n', m) \in \mathsf{ois}) \Rightarrow$
$\mathsf{occ}(\mathsf{ois}, \mathsf{deliver}_{\mathsf{urb}}(n', m)) \leq 1]$

We show the contra-positive of
(3) $\Gamma' \vdash_{\mathsf{URBC}} \circledS \; [n \bullet \mathsf{deliver}_{\mathsf{urb}}(n', m) \in \mathsf{ois} \Rightarrow$
$\hat{\boxminus} \neg (n \bullet \mathsf{deliver}_{\mathsf{urb}}(n', m) \in \mathsf{ois})]$

that is
(4) $\Gamma' \vdash_{\mathsf{URBC}} \circledS \; [\hat{\diamondsuit}(n \bullet \mathsf{deliver}_{\mathsf{urb}}(n', m) \in \mathsf{ois}) \Rightarrow$
$\neg (n \bullet \mathsf{deliver}_{\mathsf{urb}}(n', m) \in \mathsf{ois})]$

By rule INVLSE

(5) $\Gamma' \vdash_{\mathsf{URBC}} \circledS \,[(n \bullet \mathsf{deliver}_{\mathsf{urb}}(n', m) \in \mathsf{ois}) \Rightarrow$
$\exists c.\ \langle m, n', c \rangle \in pending(\mathsf{s}(n)) \wedge$
$\langle m, n', c \rangle \notin delivered(\mathsf{s}(n)) \wedge$
$\langle m, n', c \rangle \in delivered(\mathsf{s}'(n))]$

From the contra-positive of [5] is

(6) $\Gamma' \vdash_{\mathsf{URBC}} \circledS \,[\forall c.\ \langle m, n', c \rangle \notin pending(\mathsf{s}(n)) \vee$
$\langle m, n', c \rangle \in delivered(\mathsf{s}(n))] \Rightarrow$
$\neg(n \bullet \mathsf{deliver}_{\mathsf{urb}}(n', m) \in \mathsf{ois})]$

By rule INVSSE″

(7) $\Gamma' \vdash_{\mathsf{URBC}} \circledS \,[\langle m, n', c \rangle \in delivered(\mathsf{s}'(n)) \Rightarrow$
$\hat{\Box} \langle m, n', c \rangle \in delivered(\mathsf{s}(n))]$

From [5] and [7]

(8) $\Gamma' \vdash_{\mathsf{URBC}} \circledS \,[(n \bullet \mathsf{deliver}_{\mathsf{urb}}(n', m) \in \mathsf{ois})] \Rightarrow$
$\exists c.\ \langle m, n', c \rangle \in pending(\mathsf{s}(n)) \wedge$
$\hat{\Box} \langle m, n', c \rangle \in delivered(\mathsf{s}(n))]$

From [8]

(9) $\Gamma' \vdash_{\mathsf{URBC}} \circledS \,[\hat{\Diamond}(n \bullet \mathsf{deliver}_{\mathsf{urb}}(n', m) \in \mathsf{ois}) \Rightarrow$
$\hat{\Diamond}[\exists c.\ \langle m, n', c \rangle \in pending(\mathsf{s}(n)) \wedge$
$\hat{\Box} \langle m, n', c \rangle \in delivered(\mathsf{s}(n))]]$

From [9] and Lemma 54

(10) $\Gamma' \vdash_{\mathsf{URBC}} \circledS \,[\hat{\Diamond}(n \bullet \mathsf{deliver}_{\mathsf{urb}}(n', m) \in \mathsf{ois}) \Rightarrow$
$\hat{\Diamond}[\exists c.\ \forall c'.\ c \neq c' \rightarrow \Box \langle m, n', c' \rangle \notin pending(\mathsf{s}(n)) \wedge$
$\hat{\Box} \langle m, n', c \rangle \in delivered(\mathsf{s}(n))]]$

thus

(11) $\Gamma' \vdash_{\mathsf{URBC}} \circledS \,[\hat{\Diamond}(n \bullet \mathsf{deliver}_{\mathsf{urb}}(n', m) \in \mathsf{ois}) \Rightarrow$
$\exists c.\ \forall c'.\ c \neq c' \rightarrow$
$\langle m, n', c' \rangle \notin pending(\mathsf{s}(n)) \wedge \langle m, n', c \rangle \in delivered(\mathsf{s}(n))]$

From [11] and [6]

(12) $\Gamma' \vdash_{\mathsf{URBC}} \circledS \,[\hat{\Diamond}(n \bullet \mathsf{deliver}_{\mathsf{urb}}(n', m) \in \mathsf{ois}) \Rightarrow$
$\neg(n \bullet \mathsf{deliver}_{\mathsf{urb}}(n', m) \in \mathsf{ois})]$

**Theorem 14.** *(URB$_3$: No-forge)*
If a node delivers a message $m$ with sender $n'$, then $m$ was previously broadcast by node $n'$.

$\Gamma \vdash_{\mathsf{URBC}} (n \bullet \top \uparrow \mathsf{deliver}_{\mathsf{urb}}(n', m)) \leftsquigarrow (n' \bullet \top \downarrow \mathsf{broadcast}_{\mathsf{urb}}(m))$
where
$\Gamma$ is defined in Definition 19.

**Proof.**

The proof idea: A uniform-reliable delivery event is issued only in the periodic function. The periodic function only issues messages from the pending set. As shown in the proof of the no-forge property, any message that is put in the pending set is previously broadcast by the uniform-reliable broadcast.

By rule OISE',
 (1) $\Gamma \vdash_{\mathsf{URBC}} ⓢ (n \bullet \top \uparrow \mathsf{deliver}_{\mathsf{urb}}(n', m)) \Rightarrow \diamondsuit(n \bullet \mathsf{deliver}_{\mathsf{urb}}(n', m) \in \mathsf{ois})$

By rule InvLSe
 (2) $\Gamma \vdash_{\mathsf{URBC}} ⓢ (n \bullet \mathsf{deliver}_{\mathsf{urb}}(n', m) \in \mathsf{ois}) \Rightarrow \exists c.\ \langle m, n', c \rangle \in pending(\mathsf{s}(n))$
By Lemma 53
 (3) $\Gamma' \vdash_{\mathsf{URBC}} ⓢ \langle m, n', c \rangle \in pending(\mathsf{s}(n)) \Rightarrow \diamondsuit(n' \bullet \top \downarrow \mathsf{broadcast}_{\mathsf{urb}}(m))$
From [1], [2] and [3]
 (4) $\Gamma \vdash_{\mathsf{URBC}} ⓢ (n \bullet \top \uparrow \mathsf{deliver}_{\mathsf{urb}}(n', m)) \Rightarrow \diamondsuit(n' \bullet \top \downarrow \mathsf{broadcast}_{\mathsf{urb}}(m))$
From [4], and rule SInv,
 $\Gamma \vdash_{\mathsf{URBC}} (n \bullet \top \uparrow \mathsf{deliver}_{\mathsf{urb}}(n', m)) \Rightarrow \diamondsuit(n' \bullet \top \downarrow \mathsf{broadcast}_{\mathsf{urb}}(m))$

**Theorem 15.** *(URB$_4$: Uniform Agreement)*
If a message $m$ is delivered by some node (whether correct or faulty),
then $m$ is eventually (in the past or future) delivered by every correct node.

$\Gamma \vdash_{\mathsf{URBC}} \forall n, n', n_s.\ n \in \mathsf{Correct} \to$
$\quad (n' \bullet \top \uparrow \mathsf{deliver}_{\mathsf{urb}}(n_s, m)) \Rightarrow$
$\quad \Diamond(n \bullet \top \uparrow \mathsf{deliver}_{\mathsf{urb}}(n_s, m)) \lor$
$\quad \Diamondleft(n \bullet \top \uparrow \mathsf{deliver}_{\mathsf{urb}}(n_s, m))$

where
$\Gamma$ is defined in Definition 19.

**Proof.**

The proof idea: A node has delivered the message. The protocol delivers a message only in the periodic function when acknowledgement from a majority of nodes is received for the message. As a majority of nodes are correct, there is at least one correct node in the acknowledging set. A node is added to the acknowledging set only if a message is received (via best-effort broadcast) from it. By the no-forge property of the best-effort broadcast, the node should have broadcast the message. Thus, a correct node has broadcast the message. By the validity property of the best-effort broadcast, every correct node delivers the message. When a message is delivered (via best-effort broadcast), if it is not already in the pending set, it is rebroadcast. If it is already in the pending set, it is already broadcast. Thus, every correct node eventually (in the past or future) broadcasts the message. Thus, by the validity property of the best-effort broadcast, the message is delivered to every correct node from every correct node (via best-effort broadcast). When a message is delivered (via the best-effort broadcast), the sender is added to the acknowledgement set and always remains in it. Thus, eventually forever, every correct node will have every correct node in its acknowledgement set for the message. As at least a majority of nodes are correct, the size of this set is more than half of the number of the nodes. As the periodic function is infinitely often called, it will be eventually called when the size of the acknowledgement set for the message is more than half of the number of the nodes. The protocol maintains the invariant that every message that is in an acknowledgement set is in the pending set. When the periodic function iterates the pending set, if the message is not in the delivered set, as its acknowledgement set is already large enough, it is delivered. On the other hand, if it is in the delivered set, it is already delivered. Thus, the message is eventually (in the past or future) delivered at every correct node.

We assume that
(1) $\Gamma' = \Gamma; n \in \mathsf{Correct}$
We show that
$\Gamma \vdash_{\mathsf{URBC}} (n' \bullet \top \uparrow \mathsf{deliver}_{\mathsf{urb}}(n_s, m)) \Rightarrow$
$\quad \Diamond(n \bullet \top \uparrow \mathsf{deliver}_{\mathsf{urb}}(n_s, m)) \lor$
$\quad \Diamondleft(n \bullet \top \uparrow \mathsf{deliver}_{\mathsf{urb}}(n_s, m))$

By rule OISE',
(2) $\Gamma \vdash_{\mathsf{URBC}} \circledS (n' \bullet \top \uparrow \mathsf{deliver}_{\mathsf{urb}}(n_s, m)) \Rightarrow$
$\quad \Diamondleft(n' \bullet \mathsf{deliver}_{\mathsf{urb}}(n_s, m) \in \mathsf{ois})$
By rule INVLSE
(3) $\Gamma \vdash_{\mathsf{URBC}} \circledS (n' \bullet \mathsf{deliver}_{\mathsf{urb}}(n_s, m) \in \mathsf{ois}) \Rightarrow$
$\quad \exists c.\ |ack(\mathsf{s}(n'))(\langle m, n_s, c\rangle)| > |\mathbb{N}|/2$

By Lemma 58
    (4)  $\Gamma \vdash_{\mathsf{URBC}} \circledS \ \square ack(\mathsf{s}(n'))(\langle m, n_s, c\rangle) \subseteq \mathbb{N}$

It is obvious that
    (5)  $\Gamma \vdash_{\mathsf{URBC}} \circledS \ \square |\mathbb{N}|/2 + |\mathbb{N}|/2 \geq |\mathbb{N}|$

By rule QUORUM on Definition 19 (correct majority), [4] and [5] on [3]
    (6)  $\Gamma \vdash_{\mathsf{URBC}} \circledS \ (n' \bullet \mathsf{deliver}_{\mathsf{urb}}(n_s, m) \in \mathsf{ois}) \Rightarrow$
          $\exists c, n''.\ n'' \in \mathsf{Correct} \wedge n'' \in ack(\mathsf{s}(n'))(\langle m, n_s, c\rangle)$

By Lemma 99 on [2] and [6]
    (7)  $\Gamma \vdash_{\mathsf{URBC}} \circledS \ (n' \bullet \top \uparrow \mathsf{deliver}_{\mathsf{urb}}(n_s, m)) \Rightarrow$
          $\Diamond \exists c, n''.\ n'' \in \mathsf{Correct} \wedge n'' \in ack(\mathsf{s}(n'))(\langle m, n_s, c\rangle)$

By rule INVSASE with
    $S = n'' \in ack(\mathsf{s}(n'))(\langle m, n_s, c\rangle)$
    $\mathcal{A} = n' \bullet 0 \uparrow \mathsf{deliver}_{\mathsf{beb}}(n'', \langle m, n_s, c\rangle)$
    (8)  $\Gamma \vdash_{\mathsf{URBC}} \circledS \ (n'' \in ack(\mathsf{s}(n'))(\langle m, n_s, c\rangle)) \Rightarrow$
          $\Diamond(n' \bullet 0 \uparrow \mathsf{deliver}_{\mathsf{beb}}(n'', \langle m, n_s, c\rangle))$

By Definition 19 ($\mathsf{BEB}_3'$) and rule SINV,
    (9)  $\Gamma \vdash_{\mathsf{URBC}} \circledS \ (n' \bullet 0 \uparrow \mathsf{deliver}_{\mathsf{beb}}(n'', \langle m, n_s, c\rangle)) \Rightarrow$
          $\Diamond(n'' \bullet 0 \downarrow \mathsf{broadcast}_{\mathsf{beb}}(\langle m, n_s, c\rangle))$

By Lemma 88 on [8] and [9],
    (10)  $\Gamma \vdash_{\mathsf{URBC}} \circledS \ (n'' \in ack(\mathsf{s}(n'))(\langle m, n_s, c\rangle)) \Rightarrow$
          $\Diamond(n'' \bullet 0 \downarrow \mathsf{broadcast}_{\mathsf{beb}}(\langle m, n_s, c\rangle))$

By Lemma 88 on [7] and [10],
    (11)  $\Gamma \vdash_{\mathsf{URBC}} \circledS \ (n' \bullet \top \uparrow \mathsf{deliver}_{\mathsf{urb}}(n_s, m)) \Rightarrow$
          $\Diamond \exists c, n''.\ n'' \in \mathsf{Correct} \wedge (n'' \bullet 0 \downarrow \mathsf{broadcast}_{\mathsf{beb}}(\langle m, n_s, c\rangle))$

By Lemma 45 on [11] and [1], and then Lemma 112
    (12)  $\Gamma \vdash_{\mathsf{URBC}} \circledS \ (n' \bullet \top \uparrow \mathsf{deliver}_{\mathsf{urb}}(n_s, m)) \Rightarrow$
          $\Diamond(n \bullet \top \uparrow \mathsf{deliver}_{\mathsf{urb}}(n_s, m)) \vee$
          $\Diamond(n \bullet \top \uparrow \mathsf{deliver}_{\mathsf{urb}}(n_s, m))$

By rule SINV on [12]
    $\Gamma \vdash_{\mathsf{URBC}} (n' \bullet \top \uparrow \mathsf{deliver}_{\mathsf{urb}}(n_s, m)) \Rightarrow$
          $\Diamond(n \bullet \top \uparrow \mathsf{deliver}_{\mathsf{urb}}(n_s, m)) \vee$
          $\Diamond(n \bullet \top \uparrow \mathsf{deliver}_{\mathsf{urb}}(n_s, m))$

**Lemma 58.**
$\Gamma \vdash_{\mathsf{URBC}} \circledS \ \square ack(\mathsf{s}(n))(\langle m, n_s, c\rangle) \subseteq \mathbb{N}$

**Proof.**
We show that
    $\Gamma \vdash_{\mathsf{URBC}} \circledS \ \forall n.\ n \in ack(\mathsf{s}(n))(\langle m, n_s, c\rangle) \Rightarrow n \in \mathbb{N}$

By rule INVSASE
    (1)  $\Gamma \vdash_{\mathsf{URBC}} \circledS \ n \in ack(\mathsf{s}(n))(\langle m, n_s, c\rangle) \Rightarrow \Diamond(n \bullet 0 \uparrow \mathsf{deliver}_{\mathsf{beb}}(n_s, \langle m, n', c\rangle))$

By Definition 19 ($\mathsf{BEB}_3'$)

(2) $\Gamma \vdash_{\mathsf{URBC}} \circledS\ (n \bullet 0 \uparrow \mathsf{deliver}_{\mathsf{beb}}(n_s, \langle m, n', c \rangle)) \Rightarrow \Diamond (n_s \bullet 0 \downarrow \mathsf{broadcast}_{\mathsf{beb}}(\langle m, n', c \rangle))$

By Lemma 88 on [1], [2] and rule NODE,

(3) $\Gamma \vdash_{\mathsf{URBC}} \circledS\ n \in ack(\mathsf{s}(n))(\langle m, n_s, c \rangle) \Rightarrow \Diamond n \in \mathbb{N}$

that is ($n$ is rigid.)

$\Gamma \vdash_{\mathsf{URBC}} \circledS\ n \in ack(\mathsf{s}(n))(\langle m, n_s, c \rangle) \Rightarrow n \in \mathbb{N}$

### 5.3.5 Eventually Perfect Failure Detector

**Theorem 16.** *(EPFD$_1$: Strong Completeness)*
Every incorrect node is eventually permanently suspected by every correct node.
$\forall n, n'.\ n \in \mathsf{Correct} \land \neg n' \in \mathsf{Correct} \rightarrow$
$\quad \Diamond[(n \bullet \top \uparrow \mathsf{suspect}(n')) \land \Box \neg(n \bullet \top \uparrow \mathsf{restore}(n'))].$

**Proof.**

By the rule APER, the periodic handler of an incorrect node $n'$ is executed only finitely often. So the periodic handler of $n'$ is eventually never executed. By the rule INVL, only the periodic handler sends heartbeat messages. Therefore, $n'$ will eventually never send heartbeat messages. By the rule NFORGE, a message is received only if it is sent. Thus, no node will eventually receive heartbeat messages from $n'$. By the rule INVL, the node $n'$ is added to the *active* set of the node $n$ only when $n$ receives a heartbeat message from $n'$. Therefore, eventually after a round $r$, the node $n'$ will be absent in $active(r)$ of the node $n$. Therefore, in round $r + 1$, the periodic handler of $n$ issues a suspect event for $n'$, if $n'$ is not in the $failed$ set. If it is in the $failed$ set, it is already suspected and not restored since. From round $r + 1$, the node $n'$ will be never restored. This is because by the rule INVL, the restore indication for $n'$ is issued only when it is found in the *active* set.

**Theorem 17.** *(EPFD$_2$: Eventual Strong Accuracy)*
Eventually no correct node is suspected by any correct node.
$\forall n, n'.\ n \in \mathsf{Correct} \land n' \in \mathsf{Correct} \rightarrow$
$\quad \Box \neg(n \bullet \top \uparrow \mathsf{suspect}(n')) \lor$
$\quad \Diamond[(n \bullet \top \uparrow \mathsf{restore}(n')) \land \Box \neg(n \bullet \top \uparrow \mathsf{suspect}(n'))].$

**Proof.**

In the periodic handler, every correct node $n'$ sends a heartbeat message to every node in every round $r$. By the rule GST, after the round global stabilization time (GST), every correct node $n$ delivers every message set to it in the same round. On the receipt of the heartbeat message, the indication handler adds the sender $n'$ to the *active* set of $n$. By the rule RSEQ, the periodic event per of round $r + 1$ is executed after the indication events $\downarrow$ of round $r$. By the rule INVS, the node $n'$ remains in the *active* set of $n$. When the periodic event per of round $r + 1$ executes, the node $n$ finds $n'$ in *active* set. The node $n$ restores $n'$ if it is suspected before and does not suspect $n'$. By the rule INVL, suspecting nodes is only issued in the periodic handler. In the periodic handler, the node $n'$ is always found in the *active* set and hence is never suspected again.

### 5.3.6 Eventual Leader Elector

**Theorem 18.** *(ELE$_1$ (Eventual Leadership))*
Eventually every correct process trusts the same correct process.
$\exists l.\ l \in \mathsf{Correct} \wedge$
  $[n \in \mathsf{Correct} \rightarrow$
  $\Diamond(n \bullet \top \uparrow \mathsf{trust}(l)) \wedge \hat{\Box} \neg (n \bullet \top \uparrow \mathsf{trust}(l')))].$

**Proof.**

Immediate from the two properties EPFD$_1$ and EPFD$_2$. Eventually every node will have every incorrect node in the suspected set and every correct node not in the suspected set: the suspected set will be the set of incorrect nodes. The periodic event after the latest such event applies the *maxRank* funciton to the set of correct nodes and deterministically chooses the leader.

### 5.3.7 Epoch Consensus

**Definition 20.**

$\Gamma =$
   $|\mathsf{Correct}| > \mathbb{N}/2;$
   $\mathsf{PL}'_1, \mathsf{PL}'_2, \mathsf{PL}'_3, \mathsf{BEB}'_1; \mathsf{BEB}'_2; \mathsf{BEB}'_3$

$\mathsf{PL}'_1 = \mathsf{lower}(0, \mathsf{PL}_1) =$
   $n \in \mathsf{Correct} \land n' \in \mathsf{Correct} \rightarrow$
   $(n \bullet 0 \downarrow \mathsf{send}_{\mathsf{pl}}(n', m) \rightsquigarrow$
   $n \bullet 0 \uparrow \mathsf{deliver}_{\mathsf{pl}}(n, m))$

$\mathsf{PL}'_2 = \mathsf{lower}(0, \mathsf{PL}_2) =$
   $[n' \bullet 0 \downarrow \mathsf{send}_{\mathsf{pl}}(n, m) \Rightarrow$
      $\hat{\boxminus}\neg(n' \bullet 0 \downarrow \mathsf{send}_{\mathsf{pl}}(n, m))] \rightarrow$
   $[n \bullet 0 \uparrow \mathsf{deliver}_{\mathsf{pl}}(n', m) \Rightarrow$
      $\hat{\boxminus}\neg(n \bullet 0 \uparrow \mathsf{deliver}_{\mathsf{pl}}(n', m))]$

$\mathsf{PL}'_3 = \mathsf{lower}(0, \mathsf{PL}_3) =$
   $(n \bullet 0 \uparrow \mathsf{deliver}_{\mathsf{pl}}(n', m)) \leftsquigarrow$
   $(n' \bullet 0 \downarrow \mathsf{send}_{\mathsf{pl}}(n, m))$

$\mathsf{BEB}'_1 = \mathsf{lower}(1, \mathsf{BEB}_1) =$
   $n \in \mathsf{Correct} \land n' \in \mathsf{Correct} \rightarrow$
   $(n' \bullet 1 \downarrow \mathsf{broadcast}_{\mathsf{beb}}(m)) \rightsquigarrow$
   $(n \bullet 1 \uparrow \mathsf{deliver}_{\mathsf{beb}}(n', m))$

$\mathsf{BEB}'_2 = \mathsf{lower}(1, \mathsf{BEB}_2) =$
   $[n' \bullet 1 \downarrow \mathsf{broadcast}_{\mathsf{beb}}(m) \Rightarrow$
      $\hat{\boxminus}\neg(n' \bullet 1 \downarrow \mathsf{broadcast}_{\mathsf{beb}}(m))] \rightarrow$
   $[n \bullet 1 \uparrow \mathsf{deliver}_{\mathsf{beb}}(n', m) \Rightarrow$
      $\hat{\boxminus}\neg(n \bullet 1 \uparrow \mathsf{deliver}_{\mathsf{beb}}(n', m))]$

$\mathsf{BEB}'_3 = \mathsf{lower}(1, \mathsf{BEB}_3) =$
   $(n \bullet 1 \uparrow \mathsf{deliver}_{\mathsf{beb}}(n', m)) \leftsquigarrow$
   $(n' \bullet 1 \downarrow \mathsf{broadcast}_{\mathsf{beb}}(m))$

$\mathsf{Cons} =$
   $(n \bullet \top \downarrow \mathsf{epoch}_{\mathsf{ec}}(ts, n_l)) \Rightarrow$
   $(n' \bullet \top \downarrow \mathsf{epoch}_{\mathsf{ec}}(ts, n'_l)) \Rightarrow$
   $n_l = n'_l$

**Theorem 19.** *(EC$_1$: Validity)*
If a node decides the value $v$, then $v$ was proposed by the current leader $n_l$ or is passed to a node during initialization.
$\Gamma \vdash_{\mathsf{ECC}}$
$\quad (n \bullet \top \uparrow \mathsf{decide}_{\mathsf{ec}}(v)) \Rightarrow$
$\quad \Diamond (n_l \bullet \top \downarrow \mathsf{propose}_{\mathsf{ec}}(v))$
where
$\Gamma$ is defined in Definition 20.

**Proof.**
We assume
$\quad$ (1) $\Gamma' = \Gamma; v \neq \bot$
We show
$\quad \Gamma' \vdash_{\mathsf{ECC}} (n \bullet \top \uparrow \mathsf{decide}_{\mathsf{ec}}(v)) \Rightarrow$
$\qquad\qquad \Diamond (n_l \bullet \top \downarrow \mathsf{propose}_{\mathsf{ec}}(v))$
By Lemma 59,
$\quad$ (2) $\Gamma' \vdash_{\mathsf{ECC}} \textcircled{S} \ (n \bullet \top \uparrow \mathsf{decide}_{\mathsf{ec}}(v)) \Rightarrow$
$\qquad\qquad \Diamond (wval(\mathsf{s}(n_l)) = v \wedge v \neq \bot)$
By INVSASE with,
$\quad n$ instantiated to $n_l$,
$\quad S$ instantiated to $\lambda s.\ wval(s) \neq \bot$ and
$\quad \mathcal{A}$ instantiated to $wval(\mathsf{s}'(n_l)) = prop(\mathsf{s}(n_l)) \vee \exists n'.\ (\_, wval(\mathsf{s}'(n_l))) = states(\mathsf{s}'(n_l))(n')$
$\quad$ (3) $\Gamma \vdash_{\mathsf{ECC}} \textcircled{S} \ wval(\mathsf{s}(n_l)) \neq \bot \Rightarrow$
$\qquad\qquad \hat{\Diamond}(wval(\mathsf{s}'(n_l)) = prop(\mathsf{s}(n_l)) \vee \exists n'.\ (\_, wval(\mathsf{s}'(n_l))) = states(\mathsf{s}'(n_l))(n'))$
From [2] and [3] and Lemma 99 and Lemma 86
$\quad$ (4) $\Gamma' \vdash_{\mathsf{ECC}} \textcircled{S} \ (n \bullet \top \uparrow \mathsf{decide}_{\mathsf{ec}}(v)) \Rightarrow$
$\qquad\qquad \hat{\Diamond}(v = prop(\mathsf{s}(n_l)) \vee \exists n'.\ (\_, v) = states(\mathsf{s}'(n_l))(n')) \wedge v \neq \bot)$
By POSTPRE on [5]
$\quad$ (5) $\Gamma' \vdash_{\mathsf{ECC}} \textcircled{S} \ (n \bullet \top \uparrow \mathsf{decide}_{\mathsf{ec}}(v)) \Rightarrow$
$\qquad\qquad \hat{\Diamond}(v = prop(\mathsf{s}(n_l)) \vee \bigcirc \exists n'.\ (\_, v) = states(\mathsf{s}(n_l))(n')) \wedge v \neq \bot)$

By INVSASE with,
$\quad n$ instantiated to $n_l$,
$\quad S$ instantiated to $\lambda s.\ prop(s) = v \wedge prop(s) \neq \bot$ and
$\quad \mathcal{A}$ instantiated to $\mathsf{propose}_{\mathsf{ec}}(v)$
$\quad$ (6) $\Gamma \vdash_{\mathsf{ECC}} \textcircled{S} \ prop(\mathsf{s}(n_l)) = v \wedge prop(\mathsf{s}(n_l)) \neq \bot \Rightarrow$
$\qquad\qquad \hat{\Diamond}(n_l \bullet \top \downarrow \mathsf{propose}_{\mathsf{ec}}(v))$

From [5], [6] and Lemma 60,
$\quad$ (7) $\Gamma' \vdash_{\mathsf{ECC}} \textcircled{S} \ (n \bullet \top \uparrow \mathsf{decide}_{\mathsf{ec}}(v)) \Rightarrow$
$\qquad\qquad \hat{\Diamond}(\Diamond(n_l \bullet \top \downarrow \mathsf{propose}_{\mathsf{ec}}(v)) \vee \bigcirc [\exists n'.\ \Diamond(n_l \bullet \top \downarrow \mathsf{propose}_{\mathsf{ec}}(v)))]$

By Lemma 86 and Lemma 121 on [7]
$\quad$ (8) $\Gamma' \vdash_{\mathsf{ECC}} \textcircled{S} \ (n \bullet \top \uparrow \mathsf{decide}_{\mathsf{ec}}(v)) \Rightarrow$

$$\diamond(n_l \bullet \top \downarrow \mathsf{propose}_{\mathsf{ec}}(v))$$

**Lemma 59.**
$\Gamma \vdash_{\mathsf{ECC}} \circledS (n \bullet \top \uparrow \mathsf{decide}_{\mathsf{ec}}(v)) \Rightarrow$
$\quad \diamond[|states(\mathsf{s}(n_l))| > |\mathbb{N}|/2 \land wval(\mathsf{s}(n_l)) = v \land v \neq \bot \land$
$\quad highest(states(\mathsf{s}(n_l))) \neq \bot \to wval(\mathsf{s}(n_l)) = highest(states(\mathsf{s}(n_l)))]$

By OISE',
(9) $\Gamma' \vdash_{\mathsf{ECC}} \circledS (n \bullet \top \uparrow \mathsf{decide}_{\mathsf{ec}}(v)) \Rightarrow$
$\quad \hat{\diamond}(n \bullet \top \uparrow \mathsf{decide}_{\mathsf{ec}}(v) \in \mathsf{ois})$

By INVLSE,
(10) $\Gamma' \vdash_{\mathsf{ECC}} \circledS (n \bullet \top \uparrow \mathsf{decide}_{\mathsf{ec}}(v) \in \mathsf{ois}) \Rightarrow$
$\quad (n \bullet 1 \uparrow \mathsf{deliver}_{\mathsf{beb}}(n_l, \mathrm{DECIDED}(\langle ts, v \rangle)))$

From $\Gamma''$ and $\mathrm{BEB}'_3$, we have,
(11) $\Gamma' \vdash_{\mathsf{ECC}} (n \bullet 1 \uparrow \mathsf{deliver}_{\mathsf{beb}}(n_l, \mathrm{DECIDED}(\langle ts, v \rangle))) \Rightarrow$
$\quad \diamond(n_l \bullet 1 \downarrow \mathsf{broadcast}_{\mathsf{beb}}(\mathrm{DECIDED}(\langle ts, v \rangle)))$

By ORSE',
(12) $\Gamma' \vdash_{\mathsf{ECC}} \circledS (n_l \bullet 1 \downarrow \mathsf{broadcast}_{\mathsf{beb}}(\mathrm{DECIDED}(\langle ts, v \rangle))) \Rightarrow$
$\quad \hat{\diamond}(n_l \bullet (1, \mathsf{broadcast}_{\mathsf{beb}}(\mathrm{DECIDED}(\langle ts, v \rangle))) \in \mathsf{ors})$

From [10], [11], and [12] and SINV
(13) $\Gamma' \vdash_{\mathsf{ECC}} \circledS (n \bullet \top \uparrow \mathsf{decide}_{\mathsf{ec}}(v) \in \mathsf{ois}) \Rightarrow$
$\quad \hat{\diamond}(n_l \bullet (1, \mathsf{broadcast}_{\mathsf{beb}}(\mathrm{DECIDED}(\langle ts, v \rangle))) \in \mathsf{ors})$

By INVLSE
(14) $\Gamma' \vdash_{\mathsf{ECC}} \circledS (n_l \bullet (1, \mathsf{broadcast}_{\mathsf{beb}}(\mathrm{DECIDED}(\langle ts, v \rangle))) \in \mathsf{ors}) \Rightarrow$
$\quad (n_l \bullet 0 \uparrow \mathsf{deliver}_{\mathsf{pl}}(n, \mathrm{ACCEPTED}(ts))) \land$
$\quad (wval(\mathsf{s}(n_l)) = v)$

From $\Gamma'$, $\mathrm{PL}'_3$ and SINV, we have,
(15) $\Gamma' \vdash_{\mathsf{ECC}} (n_l \bullet 0 \uparrow \mathsf{deliver}_{\mathsf{pl}}(n, \mathrm{ACCEPTED}(ts))) \Rightarrow$
$\quad \diamond(n \bullet 0 \downarrow \mathsf{send}_{\mathsf{pl}}(n_l, \mathrm{ACCEPTED}(ts)))$

By ORSE',
(16) $\Gamma' \vdash_{\mathsf{ECC}} \circledS (n \bullet 0 \downarrow \mathsf{send}_{\mathsf{pl}}(n_l, \mathrm{ACCEPTED}(ts))) \Rightarrow$
$\quad \hat{\diamond}(n_l \bullet (0, \mathsf{send}_{\mathsf{pl}}(n_l, \mathrm{ACCEPTED}(ts))) \in \mathsf{ors})$

By INVLSE
(17) $\Gamma' \vdash_{\mathsf{ECC}} (n \bullet 0 \downarrow \mathsf{send}_{\mathsf{pl}}(n_l, \mathrm{ACCEPTED}(ts)) \in \mathsf{ors}) \Rightarrow$
$\quad \exists v'.\ (n \bullet 1 \uparrow \mathsf{deliver}_{\mathsf{beb}}(n_l, \mathrm{ACCEPT}(\langle ts, v' \rangle)))$

From $\Gamma'$, $\mathrm{BEB}'_3$ and SINV, we have,
(18) $\Gamma' \vdash_{\mathsf{ECC}} (n \bullet 1 \uparrow \mathsf{deliver}_{\mathsf{beb}}(n_l, \mathrm{ACCEPT}(\langle ts, v' \rangle))) \Rightarrow$
$\quad \diamond(n_l \bullet 1 \downarrow \mathsf{broadcast}_{\mathsf{beb}}(\mathrm{ACCEPT}(\langle ts, v' \rangle)))$

By ORSE',
(19) $\Gamma' \vdash_{\mathsf{ECC}} \circledS (n_l \bullet 1 \downarrow \mathsf{broadcast}_{\mathsf{beb}}(\mathrm{ACCEPT}(\langle ts, v' \rangle))) \Rightarrow$
$\quad \hat{\diamond}(n_l \bullet (1, \mathsf{broadcast}_{\mathsf{beb}}(\mathrm{ACCEPT}(\langle ts, v' \rangle))) \in \mathsf{ors})$

By INVLSE
(20) $\Gamma' \vdash_{\mathsf{ECC}} \circledS (n_l \bullet (1, \mathsf{broadcast}_{\mathsf{beb}}(\mathrm{ACCEPT}(\langle ts, v' \rangle))) \in \mathsf{ors}) \Rightarrow$
$\quad |states(\mathsf{s}'(n_l))| > |\mathbb{N}|/2 \land wval(\mathsf{s}'(n_l)) = v' \land v' \neq \bot \land$
$\quad highest(states(\mathsf{s}'(n_l))) \neq \bot \to wval(\mathsf{s}'(n_l)) = highest(states(\mathsf{s}'(n_l)))$

By Lemma 88, Lemma 99 and Lemma 86 on [14] and [13] to [20]
(21) $\Gamma' \vdash_{\mathsf{ECC}} \circledS (n \bullet \top \uparrow \mathsf{decide}_{\mathsf{ec}}(v)) \Rightarrow$
$\quad \hat{\diamond}[(wval(\mathsf{s}'(n_l)) = v) \land$

$$\hat{\diamond}[|states(\mathsf{s}'(n_l))| > |\mathbb{N}|/2 \wedge wval(\mathsf{s}'(n_l)) = v' \wedge v' \neq \bot \wedge$$
$$highest(states(\mathsf{s}'(n_l))) \neq \bot \rightarrow wval(\mathsf{s}'(n_l)) = highest(states(\mathsf{s}'(n_l)))]]$$

By INVSSE″

(22) $\Gamma' \vdash_{\mathsf{ECC}} wval(\mathsf{s}'(n_l)) = v' \wedge v' \neq \bot \Rightarrow$
$$\hat{\Box} wval(\mathsf{s}(n_l)) = v'$$

From [21] and [22]

(23) $\Gamma' \vdash_{\mathsf{ECC}} \circledS (n \bullet \top \uparrow \mathsf{decide}_{\mathsf{ec}}(v)) \Rightarrow$
$$\hat{\diamond}[(wval(\mathsf{s}(n_l)) = v) \wedge$$
$$\hat{\diamond}[|states(\mathsf{s}'(n_l))| > |\mathbb{N}|/2 \wedge wval(\mathsf{s}'(n_l)) = v' \wedge v' \neq \bot \wedge$$
$$highest(states(\mathsf{s}'(n_l))) \neq \bot \rightarrow wval(\mathsf{s}'(n_l)) = highest(states(\mathsf{s}'(n_l)))] \wedge$$
$$\hat{\diamond}\hat{\Box} wval(\mathsf{s}(n_l)) = v']$$

that is

$\Gamma' \vdash_{\mathsf{ECC}} \circledS (n \bullet \top \uparrow \mathsf{decide}_{\mathsf{ec}}(v)) \Rightarrow$
$$\hat{\diamond}[(wval(\mathsf{s}(n_l)) = v) \wedge$$
$$\hat{\diamond}[|states(\mathsf{s}'(n_l))| > |\mathbb{N}|/2 \wedge wval(\mathsf{s}'(n_l)) = v' \wedge v' \neq \bot \wedge$$
$$highest(states(\mathsf{s}'(n_l))) \neq \bot \rightarrow wval(\mathsf{s}'(n_l)) = highest(states(\mathsf{s}'(n_l)))] \wedge$$
$$wval(\mathsf{s}(n_l)) = v']$$

that is

$\Gamma' \vdash_{\mathsf{ECC}} \circledS (n \bullet \top \uparrow \mathsf{decide}_{\mathsf{ec}}(v)) \Rightarrow$
$$\hat{\diamond}[v = v' \wedge$$
$$\hat{\diamond}[|states(\mathsf{s}'(n_l))| > |\mathbb{N}|/2 \wedge wval(\mathsf{s}'(n_l)) = v' \wedge v' \neq \bot \wedge$$
$$highest(states(\mathsf{s}'(n_l))) \neq \bot \rightarrow wval(\mathsf{s}'(n_l)) = highest(states(\mathsf{s}'(n_l)))]]$$

that is

$\Gamma' \vdash_{\mathsf{ECC}} \circledS (n \bullet \top \uparrow \mathsf{decide}_{\mathsf{ec}}(v)) \Rightarrow$
$$\hat{\diamond}[$$
$$\hat{\diamond}[|states(\mathsf{s}'(n_l))| > |\mathbb{N}|/2 \wedge wval(\mathsf{s}'(n_l)) = v \wedge v \neq \bot \wedge$$
$$highest(states(\mathsf{s}'(n_l))) \neq \bot \rightarrow wval(\mathsf{s}'(n_l)) = highest(states(\mathsf{s}'(n_l)))]]$$

that by Lemma 86 is

(24) $\Gamma' \vdash_{\mathsf{ECC}} \circledS (n \bullet \top \uparrow \mathsf{decide}_{\mathsf{ec}}(v)) \Rightarrow$
$$\hat{\diamond}[|states(\mathsf{s}'(n_l))| > |\mathbb{N}|/2 \wedge wval(\mathsf{s}'(n_l)) = v \wedge v \neq \bot \wedge$$
$$highest(states(\mathsf{s}'(n_l))) \neq \bot \rightarrow wval(\mathsf{s}'(n_l)) = highest(states(\mathsf{s}'(n_l)))]$$

By POSTPRE and Lemma 121

(25) $\Gamma' \vdash_{\mathsf{ECC}} \circledS (n \bullet \top \uparrow \mathsf{decide}_{\mathsf{ec}}(v)) \Rightarrow$
$$\diamond[|states(\mathsf{s}(n_l))| > |\mathbb{N}|/2 \wedge wval(\mathsf{s}(n_l)) = v \wedge v \neq \bot \wedge$$
$$highest(states(\mathsf{s}'(n_l))) \neq \bot \rightarrow wval(\mathsf{s}(n_l)) = highest(states(\mathsf{s}(n_l)))]$$

**Lemma 60.**
$\Gamma \vdash_{\mathsf{ECC}} \circledS [(val(\mathsf{s}(n)) = v) \vee$
$\quad (states(\mathsf{s}(n_l))(n') = (\_, v))] \Rightarrow$
$\quad \diamond(n_l \bullet \top \downarrow \mathsf{propose}_{\mathsf{ec}}(v))$

**Proof.**

By Lemma 110 on Lemma 61 and Lemma 62.

**Lemma 61.**
$\Gamma \vdash_{\mathsf{ECC}} \circledS$
$\quad \hat{\boxminus}[(states(\mathsf{s}(n_l))(n') = (\_, v)) \to$
$\quad\quad \diamondsuit(n_l \bullet \top \downarrow \mathsf{propose}_{\mathsf{ec}}(v))]$
$\quad \Rightarrow$
$\quad (val(\mathsf{s}(n)) = v) \Rightarrow$
$\quad\quad \diamondsuit(n_l \bullet \top \downarrow \mathsf{propose}_{\mathsf{ec}}(v))$

**Proof.**
By INVSASE with,
$\quad n$ instantiated to $n$,
$\quad S$ instantiated to $\lambda s.\ val(s) = v$ and
$\quad \mathcal{A}$ instantiated to $\mathsf{propose}_{\mathsf{ec}}(v)$
$\quad$(1) $\Gamma' \vdash_{\mathsf{ECC}} \circledS\ (val(\mathsf{s}(n)) = v) \Rightarrow$
$\quad\quad\quad \hat{\diamondsuit}(n \bullet 1 \uparrow \mathsf{deliver}_{\mathsf{beb}}(n_l, \textsc{Accept}(\langle ts, v \rangle)))$
From $\Gamma'$ and $\mathsf{BEB}'_3$, we have,
$\quad$(2) $\Gamma' \vdash_{\mathsf{ECC}} (n \bullet 1 \uparrow \mathsf{deliver}_{\mathsf{beb}}(n_l, \textsc{Accept}(\langle ts, v \rangle))) \Rightarrow$
$\quad\quad\quad \diamondsuit(n_l \bullet 1 \downarrow \mathsf{broadcast}_{\mathsf{beb}}(\textsc{Accept}(\langle ts, v \rangle)))$
By Lemma 99 and Lemma 86 on [1] and [2] and SINV
$\quad$(3) $\Gamma' \vdash_{\mathsf{ECC}} \circledS\ (val(\mathsf{s}(n)) = v) \Rightarrow$
$\quad\quad\quad \diamondsuit(n_l \bullet 1 \downarrow \mathsf{broadcast}_{\mathsf{beb}}(\textsc{Accept}(\langle ts, v \rangle)))$
By ORSE',
$\quad$(4) $\Gamma' \vdash_{\mathsf{ECC}} \circledS\ (n_l \bullet 1 \downarrow \mathsf{broadcast}_{\mathsf{beb}}(\textsc{Accept}(\langle ts, v \rangle))) \Rightarrow$
$\quad\quad\quad \hat{\diamondsuit}(n_l \bullet (1, \mathsf{broadcast}_{\mathsf{beb}}(\textsc{Accept}(\langle ts, v \rangle))) \in \mathsf{ors})$
By INVLSE,
$\quad$(5) $\Gamma' \vdash_{\mathsf{ECC}} \circledS\ (n_l \bullet (1, \mathsf{broadcast}_{\mathsf{beb}}(\textsc{Accept}(\langle ts, v \rangle))) \in \mathsf{ors}) \Rightarrow$
$\quad\quad\quad (v = prop(\mathsf{s}(n_l)) \land v \neq \bot) \lor$
$\quad\quad\quad \exists n'.\ (\_, v) = states(\mathsf{s}(n_l))(n')$
By INVSASE with,
$\quad n$ instantiated to $n_l$,
$\quad S$ instantiated to $\lambda s.\ prop(s) = v \land prop(s) \neq \bot$ and
$\quad \mathcal{A}$ instantiated to $\mathsf{propose}_{\mathsf{ec}}(v)$
$\quad$(6) $\Gamma \vdash_{\mathsf{ECC}} \circledS\ prop(\mathsf{s}(n_l)) = v \land prop(\mathsf{s}(n_l)) \neq \bot \Rightarrow$
$\quad\quad\quad \hat{\diamondsuit}(n_l \bullet \top \downarrow \mathsf{propose}_{\mathsf{ec}}(v))$
By Lemma 88, Lemma 99 and Lemma 86 on [3] to [6]
$\quad$(7) $\Gamma' \vdash_{\mathsf{ECC}} \circledS\ (val(\mathsf{s}(n)) = v) \Rightarrow \hat{\diamondsuit}$
$\quad\quad\quad [(n_l \bullet \top \downarrow \mathsf{propose}_{\mathsf{ec}}(v)) \lor$
$\quad\quad\quad \exists n'.\ (\_, v) = states(\mathsf{s}(n_l))(n')]$
After adding a premise,
$\quad$(8) $\Gamma' \vdash_{\mathsf{ECC}} \circledS\ \hat{\boxminus}[(states(\mathsf{s}(n_l))(n') = (\_, v)) \to$
$\quad\quad\quad \diamondsuit(n_l \bullet \top \downarrow \mathsf{propose}_{\mathsf{ec}}(v))] \Rightarrow$
$\quad\quad\quad (val(\mathsf{s}(n)) = v) \Rightarrow$
$\quad\quad\quad\quad \hat{\diamondsuit}(n_l \bullet \top \downarrow \mathsf{propose}_{\mathsf{ec}}(v)) \lor$
$\quad\quad\quad\quad \hat{\diamondsuit} \exists n'.\ (\_, v) = states(\mathsf{s}(n_l))(n')$

After applying the premise in the conclusion,
that is replacing
$$(states(\mathsf{s}(n_l)))(n') = (\_, v))$$
with
$$\Diamond(n_l \bullet \top \downarrow \mathsf{propose}_{\mathsf{ec}}(v))$$
we have

(9) $\Gamma' \vdash_{\mathsf{ECC}} \circledS \;\; \hat{\boxminus}[(states(\mathsf{s}(n_l)))(n') = (\_, v)) \rightarrow$
$\Diamond(n_l \bullet \top \downarrow \mathsf{propose}_{\mathsf{ec}}(v))] \Rightarrow$
$(val(\mathsf{s}(n)) = v) \Rightarrow$
$\hat{\Diamond}(n_l \bullet \top \downarrow \mathsf{propose}_{\mathsf{ec}}(v)) \vee$
$\hat{\Diamond}\Diamond(n_l \bullet \top \downarrow \mathsf{propose}_{\mathsf{ec}}(v))$

that is

(10) $\Gamma' \vdash_{\mathsf{ECC}} \circledS \;\; \hat{\boxminus}[(states(\mathsf{s}(n_l)))(n') = (\_, v)) \rightarrow$
$\Diamond(n_l \bullet \top \downarrow \mathsf{propose}_{\mathsf{ec}}(v))] \Rightarrow$
$(val(\mathsf{s}(n)) = v) \Rightarrow$
$\Diamond(n_l \bullet \top \downarrow \mathsf{propose}_{\mathsf{ec}}(v))$

**Lemma 62.**
$\Gamma \vdash_{\mathsf{ECC}} \circledS$
$\hat{\boxminus}[(val(\mathsf{s}(n)) = v) \rightarrow$
$\Diamond(n_l \bullet \top \downarrow \mathsf{propose}_{\mathsf{ec}}(v))]$
$\Rightarrow$
$(states(\mathsf{s}(n_l)))(n') = (\_, v)) \Rightarrow$
$\Diamond(n_l \bullet \top \downarrow \mathsf{propose}_{\mathsf{ec}}(v))$

**Proof.**
By INVSASE with,
$n$ instantiated to $n_l$,
$S$ instantiated to $\lambda s. \; states(s)(n') = (vts, v)$ and
$\mathcal{A}$ instantiated to $(n_l \bullet 0 \uparrow \mathsf{deliver}_{\mathsf{pl}}(n, \text{STATE}\langle vts, v \rangle)$

(1) $\Gamma \vdash_{\mathsf{ECC}} \circledS \; states(\mathsf{s}(n_l))(n') = (vts, v) \Rightarrow$
$\hat{\Diamond}(n_l \bullet 0 \uparrow \mathsf{deliver}_{\mathsf{pl}}(n', \text{STATE}\langle vts, v \rangle))$

From $\Gamma'$ and $\mathsf{PL}_3$ we have,

(2) $\Gamma \vdash_{\mathsf{ECC}} \circledS \; (n_l \bullet 0 \uparrow \mathsf{deliver}_{\mathsf{pl}}(n', \text{STATE}\langle vts, v \rangle)) \Rightarrow$
$\Diamond(n' \bullet 0 \downarrow \mathsf{send}_{\mathsf{pl}}(n_l, \text{STATE}\langle vts, v \rangle))$

By Lemma 99 and Lemma 86 on [1] and [2] and SINV

(3) $\Gamma \vdash_{\mathsf{ECC}} \circledS \; states(\mathsf{s}(n_l))(n') = (vts, v) \Rightarrow$
$\Diamond(n' \bullet 0 \downarrow \mathsf{send}_{\mathsf{pl}}(n_l, \text{STATE}\langle vts, v \rangle))$

By ORSE' on [3],

(4) $\Gamma \vdash_{\mathsf{ECC}} \circledS \;\; (n' \bullet 0 \downarrow \mathsf{send}_{\mathsf{pl}}(n_l, \text{STATE}\langle vts, v \rangle)) \Rightarrow$
$\hat{\Diamond}(n' \bullet (0, \mathsf{send}_{\mathsf{pl}}(n_l, \text{STATE}\langle vts, v \rangle)) \in \mathsf{ors})$

By INVLSE,

(5) $\Gamma \vdash_{\mathsf{ECC}} \circledS \; (n' \bullet (0, \mathsf{send}_{\mathsf{pl}}(n_l, \text{STATE}\langle vts, v \rangle)) \in \mathsf{ors}) \Rightarrow$
$val(\mathsf{s}(n')) = v$

By Lemma 88, Lemma 99 and Lemma 86 on [3] to [5]

(6) $\Gamma \vdash_{\mathsf{ECC}} \circledS \; states(\mathsf{s}(n_l))(n') = (vts, v) \Rightarrow$

$$\hat{\diamond} val(\mathsf{s}(n')) = v$$

After adding a premise,

(7) $\Gamma' \vdash_{\mathsf{ECC}} \circledS\ \hat{\boxminus}[(val(\mathsf{s}(n)) = v) \to$
$\diamond(n_l \bullet \top \downarrow \mathsf{propose}_{\mathsf{ec}}(v))]$
$\Rightarrow$
$states(\mathsf{s}(n_l))(n') = (vts, v) \Rightarrow$
$\hat{\diamond} val(\mathsf{s}(n')) = v$

After applying the premise in the conclusion,

(8) $\Gamma' \vdash_{\mathsf{ECC}} \circledS\ \hat{\boxminus}[(val(\mathsf{s}(n)) = v) \to$
$\diamond(n_l \bullet \top \downarrow \mathsf{propose}_{\mathsf{ec}}(v))]$
$\Rightarrow$
$states(\mathsf{s}(n_l))(n') = (vts, v) \Rightarrow$
$\diamond(n_l \bullet \top \downarrow \mathsf{propose}_{\mathsf{ec}}(v))$

**Theorem 20.** *($EC_2$: Uniform Agreement)*
No two nodes decide differently.
$\Gamma \vdash_{\mathsf{ECC}}$
$\quad (n \bullet \top \uparrow \mathsf{decide}_{\mathsf{ec}}(v)) \Rightarrow$
$\quad (n' \bullet \top \uparrow \mathsf{decide}_{\mathsf{ec}}(v')) \Rightarrow$
$\quad v = v'$
where
$\Gamma$ is defined in Definition 20.

**Proof.**
We prove the equivalent formula
(1) $\Gamma \vdash_{\mathsf{ECC}}$
$\quad \Diamond(n \bullet \top \uparrow \mathsf{decide}_{\mathsf{ec}}(v)) \wedge$
$\quad (n' \bullet \top \uparrow \mathsf{decide}_{\mathsf{ec}}(v')) \Rightarrow$
$\quad v = v'$
By OI',
(2) $\Gamma' \vdash_{\mathsf{ECC}} (n \bullet \top \uparrow \mathsf{decide}_{\mathsf{ec}}(v)) \Rightarrow$
$\quad \hat{\Diamond}(n \bullet \mathsf{decide}_{\mathsf{ec}}(v) \in \mathsf{ois} \wedge \mathsf{self})$
By INVL,
(3) $\Gamma' \vdash_{\mathsf{ECC}} (\mathsf{self} \wedge n \bullet \mathsf{decide}_{\mathsf{ec}}(v) \in \mathsf{ois}) \Rightarrow$
$\quad \exists ts, n_l.\ (n \bullet 1 \uparrow \mathsf{deliver}_{\mathsf{beb}}(n_l, \mathrm{DECIDED}(ts, v)))$
From [2] and [3], we have
(4) $\Gamma \vdash_{\mathsf{ECC}} [\Diamond(n \bullet \top \uparrow \mathsf{decide}_{\mathsf{ec}}(v)) \wedge$
$\quad (n' \bullet \top \uparrow \mathsf{decide}_{\mathsf{ec}}(v'))] \Rightarrow$
$\quad \exists ts_1, n_{l1}, ts_2, n_{l2}.$
$\quad \Diamond(n \bullet 1 \uparrow \mathsf{deliver}_{\mathsf{beb}}(n_{l1}, \mathrm{DECIDED}(ts_1, v))) \wedge$
$\quad \Diamond(n' \bullet 1 \uparrow \mathsf{deliver}_{\mathsf{beb}}(n_{l2}, \mathrm{DECIDED}(ts_2, v')))$
Without loss of generality, we assume that
(5) $\Gamma \vdash_{\mathsf{ECC}} ts_1 \leq ts_2$
From Lemma 63 on [5], we have
(6) $\Gamma \vdash_{\mathsf{ECC}} (n \bullet 1 \uparrow \mathsf{deliver}_{\mathsf{beb}}(n_{l1}, \mathrm{DECIDED}(ts_1, v))) \Rightarrow$
$\quad \Box[(n' \bullet 1 \uparrow \mathsf{deliver}_{\mathsf{beb}}(n_{l2}, \mathrm{DECIDED}(ts_2, v'))) \rightarrow v' = v] \wedge$
$\quad \boxminus[(n' \bullet 1 \uparrow \mathsf{deliver}_{\mathsf{beb}}(n_{l2}, \mathrm{DECIDED}(ts_2, v'))) \rightarrow v' = v]$
From [4] and [6], we have
(7) $\Gamma \vdash_{\mathsf{ECC}} [\Diamond(n \bullet \top \uparrow \mathsf{decide}_{\mathsf{ec}}(v)) \wedge$
$\quad (n' \bullet \top \uparrow \mathsf{decide}_{\mathsf{ec}}(v'))] \Rightarrow$
$\quad v = v'$

**Lemma 63.**
$\Gamma \vdash_{\mathsf{ECC}}\ ts_2 \geq ts_1 \rightarrow$
$\quad (n \bullet 1 \uparrow \mathsf{deliver}_{\mathsf{beb}}(n_l, \mathrm{DECIDED}(ts_1, v))) \Rightarrow$
$\quad \Box[(n' \bullet 1 \uparrow \mathsf{deliver}_{\mathsf{beb}}(n_l, \mathrm{DECIDED}(ts_2, v'))) \rightarrow v' = v] \wedge$
$\quad \boxminus[(n' \bullet 1 \uparrow \mathsf{deliver}_{\mathsf{beb}}(n_l, \mathrm{DECIDED}(ts_2, v'))) \rightarrow v' = v]$
where
$\Gamma$ is defined in Definition 20.

**Proof.**

Immediate from Lemma 64 and Lemma 68.

**Lemma 64.**
$\Gamma \vdash_{\mathsf{ECC}}$
$\quad (n \bullet 1 \uparrow \mathsf{deliver}_{\mathsf{beb}}(n_l, \text{Decided}(ts, v))) \leftsquigarrow$
$\quad \exists n'. \ (n' \bullet 1 \uparrow \mathsf{deliver}_{\mathsf{beb}}(n_l, \text{Accept}(ts, v)))$
where
$\Gamma$ is defined in Definition 20.

**Proof.**
By $\mathsf{BEB}'_3$,
$\quad$ (1) $\ \Gamma \vdash_{\mathsf{ECC}} (n \bullet 1 \uparrow \mathsf{deliver}_{\mathsf{beb}}(n_l, \text{Decided}(ts, v))) \Rightarrow$
$\qquad\qquad \diamondsuit(n_l \bullet 1 \downarrow \mathsf{broadcast}_{\mathsf{beb}}(\text{Decided}(ts, v)))$
By $\mathsf{OR}'$,
$\quad$ (2) $\ \Gamma \vdash_{\mathsf{ECC}} (n_l \bullet 1 \downarrow \mathsf{broadcast}_{\mathsf{beb}}(\text{Decided}(ts, v))) \Rightarrow$
$\qquad\qquad \diamondsuit(n_l \bullet (1, \mathsf{broadcast}_{\mathsf{beb}}(\text{Decided}(ts, v))) \in \mathsf{ors} \wedge \mathsf{self})$
By $\mathsf{InvL}$,
$\quad$ (3) $\ \Gamma \vdash_{\mathsf{ECC}} (n_l \bullet (1, \mathsf{broadcast}_{\mathsf{beb}}(\text{Decided}(ts, v))) \in \mathsf{ors} \wedge \mathsf{self}) \Rightarrow$
$\qquad\qquad \exists n'. \ (n_l \bullet 0 \uparrow \mathsf{deliver}_{\mathsf{pl}}(n', \text{Accepted}(ts)))$
By $\mathsf{PL}'_3$,
$\quad$ (4) $\ \Gamma \vdash_{\mathsf{ECC}} (n_l \bullet 0 \uparrow \mathsf{deliver}_{\mathsf{pl}}(n', \text{Accepted}(ts))) \Rightarrow$
$\qquad\qquad \diamondsuit(n' \bullet 0 \downarrow \mathsf{send}_{\mathsf{pl}}(n_l, \text{Accepted}(ts)))$
By $\mathsf{OR}'$,
$\quad$ (5) $\ \Gamma \vdash_{\mathsf{ECC}} (n' \bullet 0 \downarrow \mathsf{send}_{\mathsf{pl}}(n_l, \text{Accepted}(ts))) \Rightarrow$
$\qquad\qquad \diamondsuit(n' \bullet (0, \mathsf{send}_{\mathsf{pl}}(n_l, \text{Accepted}(ts))) \in \mathsf{ors} \wedge \mathsf{self})$
By $\mathsf{InvL}$,
$\quad$ (6) $\ \Gamma \vdash_{\mathsf{ECC}} (n' \bullet (0, \mathsf{send}_{\mathsf{pl}}(n_l, \text{Accepted}(ts, v))) \in \mathsf{ors} \wedge \mathsf{self}) \Rightarrow$
$\qquad\qquad \exists v', n'_l. \ (n' \bullet 1 \uparrow \mathsf{deliver}_{\mathsf{beb}}(n'_l, \text{Accept}(ts, v')))$
From [1]-[6]
$\quad$ (7) $\ \Gamma \vdash_{\mathsf{ECC}} (n \bullet 1 \uparrow \mathsf{deliver}_{\mathsf{beb}}(n_l, \text{Decided}(ts, v))) \Rightarrow$
$\qquad\qquad \exists v', n', n'_l. \ \diamondsuit(n' \bullet 1 \uparrow \mathsf{deliver}_{\mathsf{beb}}(n'_l, \text{Accept}(ts, v')))$
From [7], Lemma 65 and Lemma 66
$\quad$ (8) $\ \Gamma \vdash_{\mathsf{ECC}} (n \bullet 1 \uparrow \mathsf{deliver}_{\mathsf{beb}}(n_l, \text{Decided}(ts, v))) \leftsquigarrow$
$\qquad\qquad \exists n'. \ (n' \bullet 1 \uparrow \mathsf{deliver}_{\mathsf{beb}}(n_l, \text{Accept}(ts, v)))$

**Lemma 65.**
$\Gamma \vdash_{\mathsf{ECC}}$
$\quad (n_l \bullet 1 \downarrow \mathsf{broadcast}_{\mathsf{beb}}(\text{Decided}(ts, v))) \Rightarrow$
$\quad (n'_l \bullet 1 \downarrow \mathsf{broadcast}_{\mathsf{beb}}(\text{Accept}(ts, v'))) \Rightarrow$
$\quad n_l = n'_l$
where
$\Gamma$ is defined in Definition 20.

**Proof.**
By a chain use of $\mathsf{OR}'$, $\mathsf{InvL}$, $\mathsf{BEB}'_3$, and $\mathsf{PL}'_3$,

(1) $\Gamma \vdash_{\mathsf{ECC}}$
$(n_l \bullet 1 \downarrow \mathsf{broadcast}_{\mathsf{beb}}(\textsc{Decided}(ts, v))) \Rightarrow$
$\exists n. \Diamond(n \bullet \top \downarrow \mathsf{epoch}_{\mathsf{ec}}(ts, n))$

(2) $\Gamma \vdash_{\mathsf{ECC}}$
$(n'_l \bullet 1 \downarrow \mathsf{broadcast}_{\mathsf{beb}}(\textsc{Accept}(ts, v'))) \Rightarrow$
$\exists n. \Diamond(n \bullet \top \downarrow \mathsf{epoch}_{\mathsf{ec}}(ts, n))$

From Definition 20,

(3) $\Gamma \vdash_{\mathsf{ECC}} (n \bullet \top \downarrow \mathsf{epoch}_{\mathsf{ec}}(ts, n_l)) \Rightarrow$
$ts \neq \bot$

From [1], [2], [3]

(4) $\Gamma \vdash_{\mathsf{ECC}} (n_l \bullet 1 \downarrow \mathsf{broadcast}_{\mathsf{beb}}(\textsc{Decided}(ts, v))) \Rightarrow$
$ts \neq \bot$

(5) $\Gamma \vdash_{\mathsf{ECC}} (n'_l \bullet 1 \downarrow \mathsf{broadcast}_{\mathsf{beb}}(\textsc{Accept}(ts, v'))) \Rightarrow$
$ts \neq \bot$

By IR,

(6) $\Gamma \vdash_{\mathsf{ECC}}$
$(n_l \bullet 1 \downarrow \mathsf{broadcast}_{\mathsf{beb}}(\textsc{Decided}(ts, v))) \Rightarrow$
$ts = ets(\mathsf{s}(n_l))$

(7) $\Gamma \vdash_{\mathsf{ECC}}$
$(n'_l \bullet 1 \downarrow \mathsf{broadcast}_{\mathsf{beb}}(\textsc{Accept}(ts, v'))) \Rightarrow$
$ts = ets(\mathsf{s}(n'_l))$

By InvSA,

(8) $\Gamma \vdash_{\mathsf{ECC}}$
$ets(\mathsf{s}(n_l)) = ts \wedge ts \neq \bot \wedge \mathsf{self} \Rightarrow$
$\Diamond(n_l \bullet \top \downarrow \mathsf{epoch}(n_l, ts))$

From Definition 20,

(9) $\Gamma \vdash_{\mathsf{ECC}}$
$(n \bullet \top \downarrow \mathsf{epoch}_{\mathsf{ec}}(ts, n_l)) \Rightarrow$
$(n' \bullet \top \downarrow \mathsf{epoch}_{\mathsf{ec}}(ts, n'_l)) \Rightarrow$
$n_l = n'_l$

From [4]-[9]

$\Gamma \vdash_{\mathsf{ECC}}$
$(n_l \bullet 1 \downarrow \mathsf{broadcast}_{\mathsf{beb}}(\textsc{Decided}(ts, v))) \Rightarrow$
$(n'_l \bullet 1 \downarrow \mathsf{broadcast}_{\mathsf{beb}}(\textsc{Accept}(ts, v'))) \Rightarrow$
$n_l = n'_l$

**Lemma 66.**
$\Gamma \vdash_{\mathsf{ECC}}$
$(n_l \bullet 1 \downarrow \mathsf{broadcast}_{\mathsf{beb}}(\textsc{Decided}(ts, v))) \Rightarrow$
$(n_l \bullet 1 \downarrow \mathsf{broadcast}_{\mathsf{beb}}(\textsc{Accept}(ts, v'))) \Rightarrow$
$v = v'$
where
$\Gamma$ is defined in Definition 20.

**Proof.**
By InvL,

(1) $\Gamma \vdash_{\mathsf{ECC}}$
$(n_l \bullet (1, \mathsf{broadcast}_{\mathsf{beb}}(\textsc{Accept}(ts, v))) \in \mathsf{ors}) \wedge \mathsf{self} \Rightarrow$
$v = wval(\mathsf{s}'(n_l))(ts) \wedge v \neq \bot$

By INVS,

(2) $\Gamma \vdash_{\mathsf{ECC}}$
$wval(\mathsf{s}(n))(ts) = v \wedge v \neq \bot \wedge \mathsf{self} \Rightarrow$
$\Box(\mathsf{self} \Rightarrow wval(\mathsf{s}(n))(ts) = v)$

By ASA on [1] and [2],

(3) $\Gamma \vdash_{\mathsf{ECC}}$
$(n_l \bullet (1, \mathsf{broadcast}_{\mathsf{beb}}(\textsc{Accept}(ts, v))) \in \mathsf{ors}) \wedge \mathsf{self} \Rightarrow$
$\hat{\Box}(\mathsf{self} \to wval(\mathsf{s}(n_l))(ts) = v)$

By INVL,

(4) $\Gamma \vdash_{\mathsf{ECC}}$
$(n_l \bullet (1, \mathsf{broadcast}_{\mathsf{beb}}(\textsc{Decided}(ts, v'))) \in \mathsf{ors}) \wedge \mathsf{self} \Rightarrow$
$v' = wval(\mathsf{s}(n_l))(ts)$

From [3] and [4], we have

(5) $\Gamma \vdash_{\mathsf{ECC}}$
$(n_l \bullet (1, \mathsf{broadcast}_{\mathsf{beb}}(\textsc{Accept}(ts, v))) \in \mathsf{ors}) \wedge \mathsf{self} \Rightarrow$
$(n_l \bullet (1, \mathsf{broadcast}_{\mathsf{beb}}(\textsc{Decided}(ts, v'))) \in \mathsf{ors}) \wedge \mathsf{self} \Rightarrow$
$v = v'$

By OR',

(6) $\Gamma \vdash_{\mathsf{ECC}} (n_l \bullet 1 \downarrow \mathsf{broadcast}_{\mathsf{beb}}(\textsc{Decided}(ts, v))) \Rightarrow$
$\Diamond(n_l \bullet (1, \mathsf{broadcast}_{\mathsf{beb}}(\textsc{Decided}(ts, v))) \in \mathsf{ors} \wedge \mathsf{self})$

By OR',

(7) $\Gamma \vdash_{\mathsf{ECC}} (n_l \bullet 1 \downarrow \mathsf{broadcast}_{\mathsf{beb}}(\textsc{Accept}(ts, v'))) \Rightarrow$
$\Diamond(n_l \bullet (1, \mathsf{broadcast}_{\mathsf{beb}}(\textsc{Accept}(ts, v))) \in \mathsf{ors} \wedge \mathsf{self})$

From [5], [6], and [7], we have

(8) $\Gamma \vdash_{\mathsf{ECC}}$
$(n_l \bullet 1 \downarrow \mathsf{broadcast}_{\mathsf{beb}}(\textsc{Decided}(ts, v))) \Rightarrow$
$(n_l \bullet 1 \downarrow \mathsf{broadcast}_{\mathsf{beb}}(\textsc{Accept}(ts, v'))) \Rightarrow$
$v = v'$

**Lemma 67.**
$\Gamma \vdash_{\mathsf{ECC}}$
$(n \bullet 1 \uparrow \mathsf{deliver}_{\mathsf{beb}}(n_l, \textsc{Decided}(ts, v))) \Rightarrow$
$\Box[(n_l \bullet (1, \mathsf{broadcast}_{\mathsf{beb}}(\textsc{Accept}(ts, v'))) \in \mathsf{ors} \wedge \mathsf{self}) \to v' = v] \wedge$
$\boxminus[(n_l \bullet (1, \mathsf{broadcast}_{\mathsf{beb}}(\textsc{Accept}(ts, v'))) \in \mathsf{ors} \wedge \mathsf{self}) \to v' = v]$

where
$\Gamma$ is defined in Definition 20.

**Proof.**
By Lemma 64,
$\Gamma \vdash_{\mathsf{ECC}}$
$(n \bullet 1 \uparrow \mathsf{deliver}_{\mathsf{beb}}(n_l, \textsc{Decided}(ts, v))) \leftsquigarrow$
$\exists n'.\ (n_l \bullet (1, \mathsf{broadcast}_{\mathsf{beb}}(\textsc{Accept}(ts, v))) \in \mathsf{ors} \wedge \mathsf{self})$

Thus, we show the equivalent formula
$\Gamma \vdash_{\mathsf{ECC}}$

$$(n \bullet 1 \uparrow \mathsf{deliver}_{\mathsf{beb}}(n_l, \textsc{Accept}(ts, v))) \Rightarrow$$
$$(n' \bullet 1 \uparrow \mathsf{deliver}_{\mathsf{beb}}(n_l, \textsc{Accept}(ts, v'))) \Rightarrow$$
$$v' = v$$

The rest of the proof is similar to the proof of Lemma 65 and Lemma 66.

**Lemma 68.**
$\Gamma \vdash_{\mathsf{ECC}} \ ts_2 \geq ts_1 \to$
  $(n \bullet 1 \uparrow \mathsf{deliver}_{\mathsf{beb}}(n_l, \textsc{Decided}(ts_1, v))) \Rightarrow$
  $\Box[(n' \bullet 1 \uparrow \mathsf{deliver}_{\mathsf{beb}}(n_l, \textsc{Accept}(ts_2, v'))) \to v' = v] \land$
  $\boxminus[(n' \bullet 1 \uparrow \mathsf{deliver}_{\mathsf{beb}}(n_l, \textsc{Accept}(ts_2, v'))) \to v' = v]$
where
$\Gamma$ is defined in Definition 20.

**Proof.**
Proof by induction on $ts_2$ starting from $ts_1$.
(1) Base Case: $ts_2 = ts_1$
  Immediate from Lemma 67.
(2) Inductive Case:
  We assume that
  (3) $\Gamma \vdash_{\mathsf{ECC}} \forall ts. \ ts_1 \leq ts < ts_2 \to$
    $(n \bullet 1 \uparrow \mathsf{deliver}_{\mathsf{beb}}(n_l, \textsc{Decided}(ts_1, v))) \Rightarrow$
    $\Box[(n' \bullet 1 \uparrow \mathsf{deliver}_{\mathsf{beb}}(n_l, \textsc{Accept}(ts, v'))) \to v' = v] \land$
    $\boxminus[(n' \bullet 1 \uparrow \mathsf{deliver}_{\mathsf{beb}}(n_l, \textsc{Accept}(ts, v'))) \to v' = v]$
  We show that
    $\Gamma \vdash_{\mathsf{ECC}}$
      $(n \bullet 1 \uparrow \mathsf{deliver}_{\mathsf{beb}}(n_l, \textsc{Decided}(ts_1, v))) \Rightarrow$
      $\Box[(n' \bullet 1 \uparrow \mathsf{deliver}_{\mathsf{beb}}(n_l, \textsc{Accept}(ts_2, v'))) \to v' = v] \land$
      $\boxminus[(n' \bullet 1 \uparrow \mathsf{deliver}_{\mathsf{beb}}(n_l, \textsc{Accept}(ts_2, v'))) \to v' = v]$
  By $\mathsf{BEB}'_3$,
  (4) $\Gamma \vdash_{\mathsf{ECC}} (n \bullet 1 \uparrow \mathsf{deliver}_{\mathsf{beb}}(n_l, \textsc{Accept}(ts_2, v'))) \Rightarrow$
      $\Diamondsuit(n_l \bullet 1 \downarrow \mathsf{broadcast}_{\mathsf{beb}}(\textsc{Accept}(ts_2, v')))$
  By $\mathsf{OR}'$,
  (5) $\Gamma \vdash_{\mathsf{ECC}} (n_l \bullet 1 \downarrow \mathsf{broadcast}_{\mathsf{beb}}(\textsc{Accept}(ts_2, v'))) \Rightarrow$
      $\Diamondsuit(n_l \bullet (1, \mathsf{broadcast}_{\mathsf{beb}}(\textsc{Accept}(ts_2, v'))) \in \mathsf{ors} \land \mathsf{self})$
  From [4] and [5], we need to show that
    $\Gamma \vdash_{\mathsf{ECC}}$
      $(n \bullet 1 \uparrow \mathsf{deliver}_{\mathsf{beb}}(n_l, \textsc{Decided}(ts_1, v))) \Rightarrow$
      $\Box[(n_l \bullet (1, \mathsf{broadcast}_{\mathsf{beb}}(\textsc{Accept}(ts_2, v'))) \in \mathsf{ors} \land \mathsf{self}) \to v' = v] \land$
      $\boxminus[(n_l \bullet (1, \mathsf{broadcast}_{\mathsf{beb}}(\textsc{Accept}(ts_2, v'))) \in \mathsf{ors} \land \mathsf{self}) \to v' = v]$
  From [3], we need to show that
    $\Gamma \vdash_{\mathsf{ECC}}$
      $(n \bullet 1 \uparrow \mathsf{deliver}_{\mathsf{beb}}(n_l, \textsc{Decided}(ts_1, v))) \land$
      $\forall ts. \ ts_1 \leq ts < ts_2 \to$
      $\Box[(n' \bullet 1 \uparrow \mathsf{deliver}_{\mathsf{beb}}(n_l, \textsc{Accept}(ts, v'))) \to v' = v] \land$
      $\boxminus[(n' \bullet 1 \uparrow \mathsf{deliver}_{\mathsf{beb}}(n_l, \textsc{Accept}(ts, v'))) \to v' = v]$
      $\Rightarrow$

$\Box[(n_l \bullet (1, \mathsf{broadcast}_{\mathsf{beb}}(\mathrm{Accept}(ts_2, v'))) \in \mathsf{ors} \land \mathsf{self}) \to v' = v] \land$
$\boxminus[(n_l \bullet (1, \mathsf{broadcast}_{\mathsf{beb}}(\mathrm{Accept}(ts_2, v'))) \in \mathsf{ors} \land \mathsf{self}) \to v' = v]$

We can show that

(6) $\Gamma \vdash_{\mathsf{ECC}}$
$(n \bullet 1 \uparrow \mathsf{deliver}_{\mathsf{beb}}(n_l, \mathrm{Decided}(ts, v))) \Rightarrow$
$\exists N.\ |N| > |\mathbb{N}|/2 \land \forall n'.\ n' \in N \to \Diamond(valts(\mathsf{s}'(n')) = ts \land val(\mathsf{s}'(n')) = v \land ts > rts(\mathsf{s}(n')))$

By INVLSE, we have

(7) $\Gamma \vdash_{\mathsf{ECC}}$
$(n_l \bullet (1, \mathsf{broadcast}_{\mathsf{beb}}(\mathrm{Accept}(ts_2, v'))) \in \mathsf{ors} \land \mathsf{self}) \Rightarrow$
$|states(\mathsf{s}'(n_l))| > |\mathbb{N}|/2 \land wval(\mathsf{s}'(n_l))(ts_2) = v' \land v' \neq \bot \land$
$highest(states(\mathsf{s}'(n_l))) \neq \bot \to wval(\mathsf{s}'(n_l))(ts_2) = highest(states(\mathsf{s}'(n_l)))$

We have (from the two intersecting sets)

(8) $\Gamma \vdash_{\mathsf{ECC}}$
$\forall N.\ |N| > |\mathbb{N}|/2 \land$
$\forall n'.\ n' \in N \to \Diamond(valts(\mathsf{s}'(n')) = ts_1 \land val(\mathsf{s}'(n')) = v \land ts > rts(\mathsf{s}(n'))) \land$
$|states(\mathsf{s}'(n_l))| > |\mathbb{N}|/2$
$\Rightarrow$
$\exists n'.\ n' \in dom(states(\mathsf{s}'(n_l))) \land \Diamond(valts(\mathsf{s}'(n')) = ts_1 \land val(\mathsf{s}'(n')) = v \land ts_1 > rts(\mathsf{s}(n')))$

By INVSA, then OI$'$ and INVL

(9) $\Gamma \vdash_{\mathsf{ECC}}$
$\mathsf{self} \land states(\mathsf{s}'(n_l))(n') = \langle ts_{n'}, v_{n'} \rangle \land ets(\mathsf{s}(n_l)) = ts_2 \Rightarrow$
$\Diamond[(n' \bullet 1 \uparrow \mathsf{deliver}_{\mathsf{beb}}(n_l, \mathrm{Prepare}(ts_2))) \land$
$valts(\mathsf{s}(n')) = ts_{n'} \land val(\mathsf{s}(n')) = v_{n'} \land rts(\mathsf{s}'(n')) = ts_2]$

By INVS$''$

(10) $\Gamma \vdash_{\mathsf{ECC}}$
$(\mathsf{self} \land rts(\mathsf{s}'(n')) \geq ts_2) \Rightarrow \hat{\Box}(\mathsf{self} \to rts(\mathsf{s}(n')) \geq ts_2)$

From [10], we have

(11) $\Gamma \vdash_{\mathsf{ECC}}$
$ts_1 \leq ts_2 \to$
$\Diamond(valts(\mathsf{s}'(n')) = ts_1 \land val(\mathsf{s}'(n')) = v \land ts_1 > rts(\mathsf{s}(n'))) \land$
$\Diamond(valts(\mathsf{s}(n')) = ts_{n'} \land val(\mathsf{s}(n')) = v_{n'} \land rts(\mathsf{s}'(n')) = ts_2) \Rightarrow$
$\Diamond[(valts(\mathsf{s}(n')) = ts_{n'} \land val(\mathsf{s}(n')) = v_{n'} \land rts(\mathsf{s}'(n')) = ts_2)$
$\Diamond(valts(\mathsf{s}'(n')) = ts_1 \land val(\mathsf{s}'(n')) = v \land ts_1 > rts(\mathsf{s}(n'))]$

Thus, we have

(12) $\Gamma \vdash_{\mathsf{ECC}}$
$(n \bullet 1 \uparrow \mathsf{deliver}_{\mathsf{beb}}(n_l, \mathrm{Decided}(ts, v))) \land$
$\forall ts.\ ts_1 \leq ts < ts_2 \to$
$\Box[(n' \bullet 1 \uparrow \mathsf{deliver}_{\mathsf{beb}}(n_l, \mathrm{Accept}(ts, v'))) \to v' = v] \land$
$\boxminus[(n' \bullet 1 \uparrow \mathsf{deliver}_{\mathsf{beb}}(n_l, \mathrm{Accept}(ts, v'))) \to v' = v]$
$\Rightarrow$
$(n_l \bullet (1, \mathsf{broadcast}_{\mathsf{beb}}(\mathrm{Accept}(ts_2, v'))) \in \mathsf{ors} \land \mathsf{self}) \Rightarrow$
$\exists ts, n'.\ states(\mathsf{s}'(n_l))(n') = \langle ts, v \rangle \land ts_1 \leq ts < ts_2 \land$
$highest(states(\mathsf{s}'(n_l))) = v' = v$

**Theorem 21.** *(EC$_3$: Integrity)*
Every node decides at most once.
$\Gamma \vdash_{\mathsf{ECC}} (n \bullet \top \uparrow \mathsf{decide}_{\mathsf{ec}}(v)) \Rightarrow$
$\quad\quad \hat{\Box}\neg(n \bullet \top \uparrow \mathsf{decide}_{\mathsf{ec}}(v'))$
where
$\Gamma$ is defined in Definition 20.

**Proof.**
By INVL,
 (1)   $\Gamma' \vdash_{\mathsf{ECC}} n \bullet \top \uparrow \mathsf{decide}_{\mathsf{ec}}(v) \Rightarrow$
     $deciced(\mathsf{s}(n)) = \mathsf{false} \wedge decided(\mathsf{s}'(n)) = \mathsf{true}$
By INVSSE,
 (2)   $\Gamma' \vdash_{\mathsf{ECC}} decided(\mathsf{s}(n_l)) = \mathsf{true} \Rightarrow$
     $\Box(decided(\mathsf{s}(n_l)) = \mathsf{true})$
By [1] and [2],
 (3)   $\Gamma' \vdash_{\mathsf{ECC}} n \bullet \top \uparrow \mathsf{decide}_{\mathsf{ec}}(v) \Rightarrow$
     $\Box decided(\mathsf{s}'(n)) = \mathsf{true}$
By POSTPRE on [3],
 (4)   $\Gamma' \vdash_{\mathsf{ECC}} n \bullet \top \uparrow \mathsf{decide}_{\mathsf{ec}}(v) \Rightarrow$
     $\Box \circ decided(\mathsf{s}(n)) = \mathsf{true}$
That is,
 (5)   $\Gamma' \vdash_{\mathsf{ECC}} n \bullet \top \uparrow \mathsf{decide}_{\mathsf{ec}}(v) \Rightarrow$
     $\hat{\Box} decided(\mathsf{s}(n)) = \mathsf{true}$
The contra-positive of [1] and changing $v$ to $v'$,
 (6)   $\Gamma' \vdash_{\mathsf{ECC}} \neg(deciced(\mathsf{s}(n)) = \mathsf{false} \wedge decided(\mathsf{s}'(n)) = \mathsf{true}) \Rightarrow$
     $\neg(n \bullet \top \uparrow \mathsf{decide}_{\mathsf{ec}}(v'))$
That is,
 (7)   $\Gamma' \vdash_{\mathsf{ECC}} (deciced(\mathsf{s}(n)) = true) \Rightarrow$
     $\neg(n \bullet \top \uparrow \mathsf{decide}_{\mathsf{ec}}(v'))$
From [5] and [7]
 (8)   $\Gamma' \vdash_{\mathsf{ECC}} n \bullet \top \uparrow \mathsf{decide}_{\mathsf{ec}}(v) \Rightarrow$
     $\hat{\Box}\neg(n \bullet \top \uparrow \mathsf{decide}_{\mathsf{ec}}(v'))$

**Theorem 22.** *($EC_4$: Termination)*
If a correct node proposes and an epoch is started with that node as the leader, then every correct node eventually decides a value.

$\Gamma \vdash_{\mathsf{ECC}} n_l \in \mathsf{Correct} \to$
$\quad |\mathsf{Correct}| > |\mathbb{N}|/2 \wedge n \in \mathsf{Correct} \to \Diamond(n \bullet \top \downarrow \mathsf{propose}_{ec}(v)) \Rightarrow$
$\quad\quad (n \bullet \top \downarrow \mathsf{epoch}_{ec}(n, ts)) \Rightarrow$
$\quad\quad \forall n'.\ n' \in \mathsf{Correct} \to \Diamond\Diamond \exists v'.\ (n' \bullet \top \uparrow \mathsf{decide}_{ec}(v'))$

where
$\Gamma$ is defined in Definition 20.

**Proof.**
By IROR,
(1) $\Gamma \vdash_{\mathsf{ECC}} (n_l \bullet \top \downarrow \mathsf{epoch}_{ec}(n, ts)) \wedge n = n_l \Rightarrow$
$\quad\quad \Diamond(n_l \bullet 1 \downarrow \mathsf{broadcast}_{beb}(\textsc{Prepare}(ts)))$
By $\mathsf{BEB}_1'$ on [1],
(2) $\Gamma \vdash_{\mathsf{ECC}} \forall n.\ n \in \mathsf{Correct} \to (n_l \bullet 1 \downarrow \mathsf{broadcast}_{beb}(\textsc{Prepare}(ts))) \Rightarrow$
$\quad\quad \Diamond(n \bullet 1 \uparrow \mathsf{deliver}_{beb}(n_l, \textsc{Prepare}(ts)))$
By IIOR,
(3) $\Gamma \vdash_{\mathsf{ECC}} \forall n.\ n \in \mathsf{Correct} \to (n \bullet 1 \uparrow \mathsf{deliver}_{beb}(n_l, \textsc{Prepare}(ts)) \wedge ts > rts(\mathsf{s}(n))) \Rightarrow$
$\quad\quad \Diamond(n \bullet 0 \downarrow \mathsf{send}_{pl}(n_l, \textsc{State}\langle ts, vts, val \rangle))$
By $\mathsf{PL}_1'$ on [3] and using the Lemma 89,
(4) $\Gamma \vdash_{\mathsf{ECC}} \forall n.\ n \in \mathsf{Correct} \to (n \bullet 1 \uparrow \mathsf{deliver}_{beb}(n_l, \textsc{Prepare}(ts)) \wedge ts > rts(\mathsf{s}(n))) \Rightarrow$
$\quad\quad \Diamond(n_l \bullet 0 \downarrow \mathsf{deliver}_{pl}(n_l, \textsc{State}\langle ts, vts, val \rangle))$
By II,
(5) $\Gamma \vdash_{\mathsf{ECC}} (n_l \bullet 0 \downarrow \mathsf{deliver}_{pl}(n_l, \textsc{State}\langle ts, vts, val \rangle) \wedge ets(\mathsf{s}(n_l)) = ts) \Rightarrow$
$\quad\quad \mathsf{self} \wedge \langle vts, val \rangle \in states(\mathsf{s}'(n_l))(n)$
By INVS,
(6) $\Gamma \vdash_{\mathsf{ECC}} \mathsf{self} \wedge \langle vts, val \rangle \in states(\mathsf{s}'(n_l))(n) \Rightarrow$
$\quad\quad (\mathsf{self} \Rightarrow \langle vts, val \rangle \in states(\mathsf{s}'(n_l))(n))$
From [2] and [6] to [4],
(7) $\Gamma \vdash_{\mathsf{ECC}} \forall n.\ n \in \mathsf{Correct} \to (n_l \bullet 1 \downarrow \mathsf{broadcast}_{beb}(\textsc{Prepare}(ts))) \Rightarrow$
$\quad\quad \Diamond[(n_l \bullet 0 \downarrow \mathsf{deliver}_{pl}(n, \textsc{State}\langle ts, vts, val \rangle) \wedge) \wedge$
$\quad\quad\quad \mathsf{self} \Rightarrow \langle vts, val \rangle \in states(\mathsf{s}'(n_l))(n)]$
Thus, there exists $n'$ such that
(8) $\Gamma \vdash_{\mathsf{ECC}} (n_l \bullet 1 \downarrow \mathsf{broadcast}_{beb}(\textsc{Prepare}(ts))) \Rightarrow$
$\quad\quad \Diamond[(n_l \bullet 0 \downarrow \mathsf{deliver}_{pl}(n', \textsc{State}\langle ts, vts, val \rangle)) \wedge$
$\quad\quad\quad \mathsf{self} \Rightarrow \forall n.\ n \in \mathsf{Correct} \to \langle vts, val \rangle \in states(\mathsf{s}'(n_l))(n)]$
That is,
(9) $\Gamma \vdash_{\mathsf{ECC}} (n_l \bullet 1 \downarrow \mathsf{broadcast}_{beb}(\textsc{Prepare}(ts))) \Rightarrow$
$\quad\quad \Diamond[(n_l \bullet 0 \downarrow \mathsf{deliver}_{pl}(n', \textsc{State}\langle ts, vts, val \rangle)) \wedge$
$\quad\quad\quad \mathsf{self} \Rightarrow (|states(\mathsf{s}'(n_l))| \geq |\mathsf{Correct}|)]$
By Definition 20 (a majority of correct nodes) on [9],
(10) $\Gamma \vdash_{\mathsf{ECC}} (n_l \bullet 1 \downarrow \mathsf{broadcast}_{beb}(\textsc{Prepare}(ts))) \Rightarrow$
$\quad\quad \Diamond[(n_l \bullet 0 \downarrow \mathsf{deliver}_{pl}(n', \textsc{State}\langle ts, vts, val \rangle)) \wedge$
$\quad\quad\quad \mathsf{self} \Rightarrow (|states(\mathsf{s}'(n_l))| \geq |\mathbb{N}|/2)]$
Thus, as the deliver event is a self event

(11) $\Gamma \vdash_{\mathsf{ECC}} (n_l \bullet 1 \downarrow \mathsf{broadcast_{beb}}(\textsc{Prepare}(ts))) \Rightarrow$
$\Diamond[(n_l \bullet 0 \downarrow \mathsf{deliver_{pl}}(n', \textsc{State}\langle ts, vts, val\rangle)) \wedge$
$|states(\mathsf{s}'(n_l))| \geq |\mathbb{N}|/2]$

By Lemma 69 on [11]

(12) $\Gamma \vdash_{\mathsf{ECC}} (n_l \bullet 1 \downarrow \mathsf{broadcast_{beb}}(\textsc{Prepare}(ts))) \Rightarrow$
$\Diamond[\Diamond(n_l \bullet 1 \downarrow \mathsf{broadcast_{beb}}(\textsc{Accept}\langle t, v\rangle)) \vee$
$\Diamondminus(n_l \bullet 1 \downarrow \mathsf{broadcast_{beb}}(\textsc{Accept}\langle t, v\rangle))]$

By Lemma 112 on [12]

(13) $\Gamma \vdash_{\mathsf{ECC}} (n_l \bullet 1 \downarrow \mathsf{broadcast_{beb}}(\textsc{Prepare}(ts))) \Rightarrow$
$\Diamondminus(n_l \bullet 1 \downarrow \mathsf{broadcast_{beb}}(\textsc{Accept}\langle t, v\rangle)) \vee$
$\Diamond(n_l \bullet 1 \downarrow \mathsf{broadcast_{beb}}(\textsc{Accept}\langle t, v\rangle))$

From [1] to [3] and [13],

(14) $\Gamma \vdash_{\mathsf{ECC}} (n_l \bullet \top \downarrow \mathsf{epoch_{ec}}(n, ts)) \wedge n = n_l \Rightarrow$
$\Diamond[\Diamondminus(n_l \bullet 1 \downarrow \mathsf{broadcast_{beb}}(\textsc{Accept}\langle t, v\rangle)) \vee$
$\Diamond(n_l \bullet 1 \downarrow \mathsf{broadcast_{beb}}(\textsc{Accept}\langle t, v\rangle))]$

By Lemma 112 on [15],

(15) $\Gamma \vdash_{\mathsf{ECC}} (n_l \bullet \top \downarrow \mathsf{epoch_{ec}}(n, ts)) \wedge n = n_l \Rightarrow$
$\Diamondminus(n_l \bullet 1 \downarrow \mathsf{broadcast_{beb}}(\textsc{Accept}\langle t, v\rangle)) \vee$
$\Diamond(n_l \bullet 1 \downarrow \mathsf{broadcast_{beb}}(\textsc{Accept}\langle t, v\rangle))$

By $\mathsf{BEB}'_1$ on [15],

(16) $\Gamma \vdash_{\mathsf{ECC}} \forall n.\ n \in \mathsf{Correct} \rightarrow (n_l \bullet 1 \downarrow \mathsf{broadcast_{beb}}(\textsc{Accept}\langle t, v\rangle)) \Rightarrow$
$\Diamondminus\Diamond(n \bullet 1 \uparrow \mathsf{deliver_{beb}}(\textsc{Accept}\langle t, v\rangle)) \vee$
$\Diamond\Diamond(n \bullet 1 \uparrow \mathsf{deliver_{beb}}(\textsc{Accept}\langle t, v\rangle))$

By Lemma 112 and Lemma 87 on [16],

(17) $\Gamma \vdash_{\mathsf{ECC}} \forall n.\ n \in \mathsf{Correct} \rightarrow (n_l \bullet 1 \downarrow \mathsf{broadcast_{beb}}(\textsc{Accept}\langle t, v\rangle)) \Rightarrow$
$\Diamondminus(n \bullet 1 \uparrow \mathsf{deliver_{beb}}(\textsc{Accept}\langle t, v\rangle)) \vee$
$\Diamond(n \bullet 1 \uparrow \mathsf{deliver_{beb}}(\textsc{Accept}\langle t, v\rangle))$

By IIOR,

(18) $\Gamma \vdash_{\mathsf{ECC}} (n \bullet 1 \uparrow \mathsf{deliver_{beb}}(\textsc{Accept}\langle t, v\rangle)) \wedge t > rts(\mathsf{s}(n)) \Rightarrow$
$\Diamond(n \bullet 0 \downarrow \mathsf{send_{pl}}(n_l, \textsc{Accepted}\langle t\rangle))$

From [17], [18] and $\mathsf{PL}'_1$,

(19) $\Gamma \vdash_{\mathsf{ECC}} \forall n.\ n \in \mathsf{Correct} \rightarrow (n_l \bullet 1 \downarrow \mathsf{broadcast_{beb}}(\textsc{Accept}\langle t, v\rangle)) \Rightarrow$
$\Diamondminus\Diamond(n_l \bullet 0 \uparrow \mathsf{deliver_{pl}}(n, \textsc{Accepted}\langle t\rangle)) \vee$
$\Diamond\Diamond(n_l \bullet 0 \uparrow \mathsf{deliver_{pl}}(n, \textsc{Accepted}\langle t\rangle))$

By Lemma 112, Lemma 87 on [19] and [1],

(20) $\Gamma \vdash_{\mathsf{ECC}} \forall n.\ n \in \mathsf{Correct} \rightarrow (n_l \bullet 1 \downarrow \mathsf{broadcast_{beb}}(\textsc{Accept}\langle t, v\rangle)) \Rightarrow$
$\Diamondminus(n_l \bullet 0 \uparrow \mathsf{deliver_{pl}}(n, \textsc{Accepted}\langle t\rangle)) \vee$
$\Diamond(n_l \bullet 0 \uparrow \mathsf{deliver_{pl}}(n, \textsc{Accepted}\langle t\rangle))$

By II,

(21) $\Gamma \vdash_{\mathsf{ECC}} (n_l \bullet 0 \uparrow \mathsf{deliver_{pl}}(n, \textsc{Accepted}\langle t\rangle)) \wedge ets(\mathsf{s}(n_l)) = t \Rightarrow$
$\mathsf{self} \wedge (n \in accepted(\mathsf{s}'(n_l)))$

By INVS,

(22) $\Gamma \vdash_{\mathsf{ECC}} \mathsf{self} \wedge (n \in accepted(\mathsf{s}'(n_l))) \Rightarrow$
$\mathsf{self} \Rightarrow (n \in accepted(\mathsf{s}'(n_l)))$

From [20] to [22], and Lemma 112,

(23) $\Gamma \vdash_{\mathsf{ECC}} \forall n.\ n \in \mathsf{Correct} \rightarrow (n_l \bullet 1 \downarrow \mathsf{broadcast_{beb}}(\textsc{Accept}\langle t, v\rangle)) \Rightarrow$
$\Diamondminus\Diamond[(n_l \bullet 0 \uparrow \mathsf{deliver_{pl}}(n, \textsc{Accepted}\langle t\rangle)) \wedge$
$\mathsf{self} \Rightarrow (n \in accepted(\mathsf{s}'(n_l)))]$

There exists $n'$ such that

(24) $\Gamma \vdash_{\mathsf{ECC}} (n_l \bullet 1 \downarrow \mathsf{broadcast}_{\mathsf{beb}}(\textsc{Accept}\langle t, v\rangle)) \Rightarrow$
$\Leftrightarrow\Diamond[(n_l \bullet 0 \uparrow \mathsf{deliver}_{\mathsf{pl}}(n', \textsc{Accepted}\langle t\rangle)) \wedge$
$\mathsf{self} \Rightarrow \forall n.\ n \in \mathsf{Correct} \rightarrow n \in accepted(\mathsf{s}'(n_l))]$

That is,

(25) $\Gamma \vdash_{\mathsf{ECC}} (n_l \bullet 1 \downarrow \mathsf{broadcast}_{\mathsf{beb}}(\textsc{Accept}\langle t, v\rangle)) \Rightarrow$
$\Leftrightarrow\Diamond[(n_l \bullet 0 \uparrow \mathsf{deliver}_{\mathsf{pl}}(n', \textsc{Accepted}\langle t\rangle)) \wedge$
$\mathsf{self} \Rightarrow \forall n.\ \mathsf{Correct} \subseteq accepted(\mathsf{s}'(n_l))]$

By Definition 20 (a majority of correct nodes) on [25],

(26) $\Gamma \vdash_{\mathsf{ECC}} (n_l \bullet 1 \downarrow \mathsf{broadcast}_{\mathsf{beb}}(\textsc{Accept}\langle t, v\rangle)) \Rightarrow$
$\Leftrightarrow\Diamond[(n_l \bullet 0 \uparrow \mathsf{deliver}_{\mathsf{pl}}(n', \textsc{Accepted}\langle t\rangle)) \wedge$
$\mathsf{self} \Rightarrow |accepted(\mathsf{s}'(n_l))| > |\mathbb{N}|/2$

Thus, as the deliver event is a self event

(27) $\Gamma \vdash_{\mathsf{ECC}} (n_l \bullet 1 \downarrow \mathsf{broadcast}_{\mathsf{beb}}(\textsc{Accept}\langle t, v\rangle)) \Rightarrow$
$\Leftrightarrow\Diamond[(n_l \bullet 0 \uparrow \mathsf{deliver}_{\mathsf{pl}}(n', \textsc{Accepted}\langle t\rangle)) \wedge$
$|accepted(\mathsf{s}'(n_l))| > |\mathbb{N}|/2$

By InvL,

(28) $\Gamma \vdash_{\mathsf{ECC}} (n_l \bullet 0 \uparrow \mathsf{deliver}_{\mathsf{pl}}(n', \textsc{Accepted}\langle t\rangle)) \wedge$
$|accepted(\mathsf{s}'(n_l))| > |\mathbb{N}|/2\ \wedge\ ets(\mathsf{s}(n_l)) = t \Rightarrow$
$\Diamond(n_l \bullet 1 \downarrow \mathsf{broadcast}_{\mathsf{beb}}(\textsc{Decided}\langle v\rangle))]$

From [27], [28] and Lemma 112

(29) $\Gamma \vdash_{\mathsf{ECC}} (n_l \bullet 1 \downarrow \mathsf{broadcast}_{\mathsf{beb}}(\textsc{Accept}\langle t, v\rangle)) \Rightarrow$
$\Leftrightarrow(n_l \bullet 1 \downarrow \mathsf{broadcast}_{\mathsf{beb}}(\textsc{Decided}\langle v\rangle)) \vee$
$\Diamond(n_l \bullet 1 \downarrow \mathsf{broadcast}_{\mathsf{beb}}(\textsc{Decided}\langle v\rangle))$

By $\mathsf{BEB}'_1$,

(30) $\Gamma \vdash_{\mathsf{ECC}} \forall n.\ n \in \mathsf{Correct} \rightarrow (n_l \bullet 1 \downarrow \mathsf{broadcast}_{\mathsf{beb}}(\textsc{Decided}\langle v\rangle)) \Rightarrow$
$\Diamond(n \bullet 1 \uparrow \mathsf{deliver}_{\mathsf{beb}}(n_l, \textsc{Decided}\langle v\rangle))$

From [29], [30], Lemma 87, and Lemma 112

(31) $\Gamma \vdash_{\mathsf{ECC}} \forall n.\ n \in \mathsf{Correct} \rightarrow (n_l \bullet 1 \downarrow \mathsf{broadcast}_{\mathsf{beb}}(\textsc{Accept}\langle t, v\rangle)) \Rightarrow$
$\Leftrightarrow(n_l \bullet 1 \uparrow \mathsf{deliver}_{\mathsf{beb}}(n_l, \textsc{Decided}\langle v\rangle)) \vee$
$\Diamond(n_l \bullet 1 \uparrow \mathsf{deliver}_{\mathsf{beb}}(n_l, \textsc{Decided}\langle v\rangle))$

By IIOI

(32) $\Gamma \vdash_{\mathsf{ECC}} (n \bullet 1 \uparrow \mathsf{deliver}_{\mathsf{beb}}(n_l, \textsc{Decided}\langle v\rangle)) \wedge decided(\mathsf{s}(n)) = \mathsf{false} \Rightarrow$
$\Diamond(n \bullet \top \uparrow \mathsf{decide}_{\mathsf{ec}}(v))$

From [31] and [32],

(33) $\Gamma \vdash_{\mathsf{ECC}} \forall n.\ n \in \mathsf{Correct} \rightarrow (n_l \bullet 1 \downarrow \mathsf{broadcast}_{\mathsf{beb}}(\textsc{Accept}\langle t, v\rangle)) \Rightarrow$
$\Leftrightarrow\Diamond(n \bullet \top \uparrow \mathsf{decide}_{\mathsf{ec}}(v)) \vee$
$\Diamond\Diamond(n \bullet \top \uparrow \mathsf{decide}_{\mathsf{ec}}(v))$

By Lemma 112 and Lemma 87 on [33],

(34) $\Gamma \vdash_{\mathsf{ECC}} \forall n.\ n \in \mathsf{Correct} \rightarrow (n_l \bullet 1 \downarrow \mathsf{broadcast}_{\mathsf{beb}}(\textsc{Accept}\langle t, v\rangle)) \Rightarrow$
$\Leftrightarrow(n \bullet \top \uparrow \mathsf{decide}_{\mathsf{ec}}(v)) \vee$
$\Diamond(n \bullet \top \uparrow \mathsf{decide}_{\mathsf{ec}}(v))$

From [15] and [34],

(35) $\Gamma \vdash_{\mathsf{ECC}} \forall n.\ n \in \mathsf{Correct} \rightarrow (n_l \bullet \top \downarrow \mathsf{epoch}_{\mathsf{ec}}(n, ts)) \wedge n = n_l \Rightarrow$
$\Leftrightarrow[\Leftrightarrow(n \bullet \top \uparrow \mathsf{decide}_{\mathsf{ec}}(v)) \vee$
$\Diamond(n \bullet \top \uparrow \mathsf{decide}_{\mathsf{ec}}(v))] \vee$
$\Diamond[\Leftrightarrow(n \bullet \top \uparrow \mathsf{decide}_{\mathsf{ec}}(v)) \vee$
$\Diamond(n \bullet \top \uparrow \mathsf{decide}_{\mathsf{ec}}(v))]$

By Lemma 112, Lemma 87 and Lemma 86 on [15],
- (36) $\Gamma \vdash_{\mathsf{ECC}} \forall n.\ n \in \mathsf{Correct} \to (n_l \bullet \top \downarrow \mathsf{epoch}_{\mathsf{ec}}(n, ts)) \wedge n = n_l \Rightarrow$
  $\hat{\Diamond}(n \bullet \top \uparrow \mathsf{decide}_{\mathsf{ec}}(v)) \vee$
  $\Diamond(n \bullet \top \uparrow \mathsf{decide}_{\mathsf{ec}}(v))$

That is,
- (37) $\Gamma \vdash_{\mathsf{ECC}} n_l \in \mathsf{Correct} \to$
  $|\mathsf{Correct}| > |\mathbb{N}|/2 \wedge n \in \mathsf{Correct} \to$
  $(n_l \bullet \top \downarrow \mathsf{epoch}_{\mathsf{ec}}(n_l, ts)) \Rightarrow$
  $\forall n'.\ n' \in \mathsf{Correct} \to \hat{\Diamond}\Diamond \exists v'.\ (n' \bullet \top \uparrow \mathsf{decide}_{\mathsf{ec}}(v'))$

**Lemma 69.**
$\Gamma \vdash_{\mathsf{ECC}}$
  $[(n_l \bullet 0 \downarrow \mathsf{deliver}_{\mathsf{pl}}(n, \mathrm{STATE}\langle ts, vts, val \rangle)) \wedge$
  $(|states(\mathsf{s}'(n_l))| \geq |\mathbb{N}|/2)] \Rightarrow$
  $\hat{\Diamond}(n_l \bullet 1 \downarrow \mathsf{broadcast}_{\mathsf{beb}}(\mathrm{ACCEPT}\langle t, v \rangle)) \vee$
  $\Diamond(n_l \bullet 1 \downarrow \mathsf{broadcast}_{\mathsf{beb}}(\mathrm{ACCEPT}\langle t, v \rangle))$

**Proof.**
There are two cases, the first case is $wval(\mathsf{s}(n_l)) = \bot$.

By INVL
- (1) $\Gamma \vdash_{\mathsf{ECC}} [(n_l \bullet 0 \downarrow \mathsf{deliver}_{\mathsf{pl}}(n, \mathrm{STATE}\langle ts, vts, val \rangle)) \wedge$
  $(|states(\mathsf{s}'(n_l))| \geq |\mathbb{N}|/2) \wedge wval(\mathsf{s}(n_l)) = \bot] \Rightarrow$
  $\mathsf{self} \wedge (n_l \bullet (1, \mathsf{broadcast}_{\mathsf{beb}}(\mathrm{ACCEPT}\langle ts, wval(\mathsf{s}'(n_l))(ts)\rangle)) \in \mathsf{ors}$

By OR
- (2) $\Gamma \vdash_{\mathsf{ECC}} \mathsf{self} \wedge (n_l \bullet (1, \mathsf{broadcast}_{\mathsf{beb}}(\mathrm{ACCEPT}\langle v \rangle)) \in \mathsf{ors}) \Rightarrow$
  $\hat{\Diamond}(n_l \bullet 1 \downarrow \mathsf{broadcast}_{\mathsf{beb}}(\mathrm{ACCEPT}\langle ts, wval(\mathsf{s}'(n_l))(ts)\rangle))$

From [1] and [2],
- (3) $\Gamma \vdash_{\mathsf{ECC}} [(n_l \bullet 0 \downarrow \mathsf{deliver}_{\mathsf{pl}}(n, \mathrm{STATE}\langle ts, vts, val \rangle)) \wedge$
  $(|states(\mathsf{s}'(n_l))| \geq |\mathbb{N}|/2) \wedge wval(\mathsf{s}(n_l)) = \bot] \Rightarrow$
  $\hat{\Diamond}(n_l \bullet 1 \downarrow \mathsf{broadcast}_{\mathsf{beb}}(\mathrm{ACCEPT}\langle ts, wval(\mathsf{s}'(n_l))(ts)\rangle))$

That is,
- (4) $\Gamma \vdash_{\mathsf{ECC}} [(n_l \bullet 0 \downarrow \mathsf{deliver}_{\mathsf{pl}}(n, \mathrm{STATE}\langle ts, vts, val \rangle)) \wedge$
  $(|states(\mathsf{s}'(n_l))| \geq |\mathbb{N}|/2) \wedge wval(\mathsf{s}(n_l)) = \bot] \Rightarrow$
  $\Diamond(n_l \bullet 1 \downarrow \mathsf{broadcast}_{\mathsf{beb}}(\mathrm{ACCEPT}\langle ts, wval(\mathsf{s}'(n_l))(ts)\rangle))$

The second case is $wval(\mathsf{s}(n_l))(ts) = v \wedge v \neq \bot$, by INVSA with,
  $S$ instantiated to $wval(\mathsf{s}(n_l))(ts) \neq \bot$
  $A$ instantiated to $(n_l \bullet 1 \downarrow \mathsf{broadcast}_{\mathsf{beb}}(\mathrm{ACCEPT}\langle ts, wval(\mathsf{s}'(n_l))(ts)\rangle))$
- (5) $\Gamma \vdash_{\mathsf{ECC}} \mathsf{self} \wedge wval(\mathsf{s}(n_l))(ts) \wedge v \neq \bot \Rightarrow$
  $\hat{\Diamond}(n_l \bullet 1 \downarrow \mathsf{broadcast}_{\mathsf{beb}}(\mathrm{ACCEPT}\langle ts, wval(\mathsf{s}'(n_l))(ts)\rangle))$

That is,
- (6) $\Gamma \vdash_{\mathsf{ECC}} [(n_l \bullet 0 \downarrow \mathsf{deliver}_{\mathsf{pl}}(n, \mathrm{STATE}\langle ts, vts, val \rangle)) \wedge$
  $(|states(\mathsf{s}'(n_l))| \geq |\mathbb{N}|/2) \wedge wval(\mathsf{s}(n_l)) = v \wedge v \neq \bot] \Rightarrow$
  $\hat{\Diamond}(n_l \bullet 1 \downarrow \mathsf{broadcast}_{\mathsf{beb}}(\mathrm{ACCEPT}\langle ts, wval(\mathsf{s}'(n_l))(ts)\rangle))$

From [4] and [6],

(7) $\Gamma \vdash_{\mathsf{ECC}} [(n_l \bullet 0 \downarrow \mathsf{deliver}_{\mathsf{pl}}(n, \textsc{State}\langle ts, vts, val\rangle)) \wedge$
$\quad\quad\quad (|states(\mathsf{s}'(n_l))| \geq |\mathbb{N}|/2) \wedge wval(\mathsf{s}(n_l)) = \bot] \vee$
$\quad\quad\quad [(n_l \bullet 0 \downarrow \mathsf{deliver}_{\mathsf{pl}}(n, \textsc{State}\langle ts, vts, val\rangle)) \wedge$
$\quad\quad\quad (|states(\mathsf{s}'(n_l))| \geq |\mathbb{N}|/2) \wedge wval(\mathsf{s}(n_l)) = v \wedge v \neq \bot] \Rightarrow$
$\quad\quad\quad \Diamond(n_l \bullet 1 \downarrow \mathsf{broadcast}_{\mathsf{beb}}(\textsc{Accept}\langle ts, wval(\mathsf{s}'(n_l))(ts)\rangle)) \vee$
$\quad\quad\quad \hat{\Diamond}(n_l \bullet 1 \downarrow \mathsf{broadcast}_{\mathsf{beb}}(\textsc{Accept}\langle ts, wval(\mathsf{s}'(n_l))(ts)\rangle))$

That is,

(8) $\Gamma \vdash_{\mathsf{ECC}} [(n_l \bullet 0 \downarrow \mathsf{deliver}_{\mathsf{pl}}(n, \textsc{State}\langle ts, vts, val\rangle)) \wedge$
$\quad\quad\quad (|states(\mathsf{s}'(n_l))| \geq |\mathbb{N}|/2)] \wedge$
$\quad\quad\quad [wval(\mathsf{s}(n_l)) = \bot \vee wval(\mathsf{s}(n_l)) \neq \bot] \Rightarrow$
$\quad\quad\quad \Diamond(n_l \bullet 1 \downarrow \mathsf{broadcast}_{\mathsf{beb}}(\textsc{Accept}\langle ts, wval(\mathsf{s}'(n_l))(ts)\rangle)) \vee$
$\quad\quad\quad \hat{\Diamond}(n_l \bullet 1 \downarrow \mathsf{broadcast}_{\mathsf{beb}}(\textsc{Accept}\langle ts, wval(\mathsf{s}'(n_l))(ts)\rangle))$

That is,

(9) $\Gamma \vdash_{\mathsf{ECC}} [(n_l \bullet 0 \downarrow \mathsf{deliver}_{\mathsf{pl}}(n, \textsc{State}\langle ts, vts, val\rangle)) \wedge$
$\quad\quad\quad (|states(\mathsf{s}'(n_l))| \geq |\mathbb{N}|/2) \Rightarrow$
$\quad\quad\quad \Diamond(n_l \bullet 1 \downarrow \mathsf{broadcast}_{\mathsf{beb}}(\textsc{Accept}\langle ts, wval(\mathsf{s}'(n_l))(ts)\rangle)) \vee$
$\quad\quad\quad \hat{\Diamond}(n_l \bullet 1 \downarrow \mathsf{broadcast}_{\mathsf{beb}}(\textsc{Accept}\langle ts, wval(\mathsf{s}'(n_l))(ts)\rangle))$

### 5.3.8 Epoch Change

**Definition 21.**
$\Gamma =$
$\quad \mathsf{PL}'_1, \mathsf{PL}'_2, \mathsf{PL}'_3, \mathsf{BEB}'_1; \mathsf{BEB}'_2; \mathsf{BEB}'_3; \mathsf{ELE}'_1$

$\mathsf{PL}'_1 = \mathsf{lower}(0, \mathsf{PL}_1) =$
$\quad n \in \mathsf{Correct} \wedge n' \in \mathsf{Correct} \to$
$\quad (n \bullet 0 \downarrow \mathsf{send}_{\mathsf{pl}}(n', m) \rightsquigarrow$
$\quad n \bullet 0 \uparrow \mathsf{deliver}_{\mathsf{pl}}(n, m))$

$\mathsf{PL}'_2 = \mathsf{lower}(0, \mathsf{PL}_2) =$
$\quad [n' \bullet 0 \downarrow \mathsf{send}_{\mathsf{pl}}(n, m) \Rightarrow$
$\quad\quad \hat{\boxminus} \neg (n' \bullet 0 \downarrow \mathsf{send}_{\mathsf{pl}}(n, m))] \to$
$\quad [n \bullet 0 \uparrow \mathsf{deliver}_{\mathsf{pl}}(n', m) \Rightarrow$
$\quad\quad \hat{\boxminus} \neg (n \bullet 0 \uparrow \mathsf{deliver}_{\mathsf{pl}}(n', m))]$

$\mathsf{PL}'_3 = \mathsf{lower}(0, \mathsf{PL}_3) =$
$\quad (n \bullet 0 \uparrow \mathsf{deliver}_{\mathsf{pl}}(n', m)) \leftsquigarrow$
$\quad (n' \bullet 0 \downarrow \mathsf{send}_{\mathsf{pl}}(n, m))$

$\mathsf{BEB}'_1 = \mathsf{lower}(1, \mathsf{BEB}_1) =$
$\quad n \in \mathsf{Correct} \wedge n' \in \mathsf{Correct} \to$
$\quad (n' \bullet 1 \downarrow \mathsf{broadcast}_{\mathsf{beb}}(m)) \rightsquigarrow$
$\quad (n \bullet 1 \uparrow \mathsf{deliver}_{\mathsf{beb}}(n', m))$

$\mathsf{BEB}'_2 = \mathsf{lower}(1, \mathsf{BEB}_2) =$
$\quad [n' \bullet 1 \downarrow \mathsf{broadcast}_{\mathsf{beb}}(n, m) \Rightarrow$
$\quad\quad \hat{\boxminus} \neg (n' \bullet 1 \downarrow \mathsf{broadcast}_{\mathsf{beb}}(n, m))] \to$
$\quad [n \bullet 1 \uparrow \mathsf{deliver}_{\mathsf{beb}}(n', m) \Rightarrow$
$\quad\quad \hat{\boxminus} \neg (n \bullet 1 \uparrow \mathsf{deliver}_{\mathsf{beb}}(n', m))]$

$\mathsf{BEB}'_3 = \mathsf{lower}(1, \mathsf{BEB}_3) =$
$\quad (n \bullet 1 \uparrow \mathsf{deliver}_{\mathsf{beb}}(n', m)) \leftsquigarrow$
$\quad (n' \bullet 1 \downarrow \mathsf{broadcast}_{\mathsf{beb}}(m))$

The property of the eventual leader detector subcomponent:
Eventually every correct process trusts the same correct process.
$\mathsf{ELE}'_1 = \mathsf{lower}(2, \mathsf{ELE}_1) =$
$\quad \exists l.\ l \in \mathsf{Correct} \wedge$
$\quad [n \in \mathsf{Correct} \to$
$\quad\quad \Diamond(n \bullet \top \uparrow \mathsf{trust}(l) \wedge \hat{\Box} \neg (n \bullet \top \uparrow \mathsf{trust}(l')))]$

**Theorem 23.** *($ECH_1$: Monotonicity)*
If a correct process starts an epoch $(ts, n_l)$ and later starts an epoch $(ts', n'_l)$, then $ts' > ts$.

$\vdash_{\mathsf{ECH}} n \in \mathsf{Correct} \to$
$\quad (n \bullet \top \uparrow \mathsf{startEpoch}_{\mathsf{ech}}(ts, n_l)) \Rightarrow$
$\quad \hat{\Box}(n \bullet \top \uparrow \mathsf{startEpoch}_{\mathsf{ech}}(ts', n'_l) \to ts' > ts)$

**Proof.**

By OI,
(1) $\Gamma \vdash_{\mathsf{ECH}} (n \bullet \top \uparrow \mathsf{startEpoch}_{\mathsf{ech}}(ts, n_l) \in \mathsf{ois} \wedge \mathsf{self}) \Rightarrow$
$\qquad \hat{\Diamond}(n \bullet \top \uparrow \mathsf{startEpoch}_{\mathsf{ech}}(ts, n_l))$

By OI',
(2) $\Gamma \vdash_{\mathsf{ECH}} n \bullet \top \uparrow \mathsf{startEpoch}_{\mathsf{ech}}(ts, n_l) \Rightarrow$
$\qquad \hat{\Diamond}(n \bullet \top \uparrow \mathsf{startEpoch}_{\mathsf{ech}}(ts, n_l) \in \mathsf{ois} \wedge \mathsf{self})$

By INVL,
(3) $\Gamma \vdash_{\mathsf{ECH}} n \bullet \mathsf{startEpoch}_{\mathsf{ech}}(ts, n_l) \in \mathsf{ois} \wedge \mathsf{self} \Rightarrow$
$\qquad lastts(\mathsf{s}'(n)) = ts \wedge ts > lastts(\mathsf{s}(n))$

That is,
(4) $\Gamma \vdash_{\mathsf{ECH}} n \bullet \mathsf{startEpoch}_{\mathsf{ech}}(ts, n_l) \in \mathsf{ois} \wedge \mathsf{self} \Rightarrow$
$\qquad lastts(\mathsf{s}'(n)) = ts$

By INVL,
(5) $\Gamma \vdash_{\mathsf{ECH}} n \bullet \mathsf{startEpoch}_{\mathsf{ech}}(ts', n'_l) \in \mathsf{ois} \wedge lastts(\mathsf{s}(n)) = ts \wedge \mathsf{self} \Rightarrow$
$\qquad ts' > ts$

That is,
(6) $\Gamma \vdash_{\mathsf{ECH}} lastts(\mathsf{s}(n)) = ts \wedge \mathsf{self} \Rightarrow$
$\qquad (\mathsf{self} \wedge n \bullet \mathsf{startEpoch}_{\mathsf{ech}}(ts', n'_l) \in \mathsf{ois} \Rightarrow ts' > ts)$

That is,
(7) $\Gamma \vdash_{\mathsf{ECH}} lastts(\mathsf{s}(n)) = ts \wedge \mathsf{self} \Rightarrow$
$\qquad (\mathsf{self} \Rightarrow (n \bullet \mathsf{startEpoch}_{\mathsf{ech}}(ts', n'_l) \in \mathsf{ois} \Rightarrow ts' > ts))$

From [4] and [7] and ASA,
(8) $\Gamma \vdash_{\mathsf{ECH}} n \bullet \mathsf{startEpoch}_{\mathsf{ech}}(ts, n_l) \in \mathsf{ois} \wedge \mathsf{self} \Rightarrow$
$\qquad \hat{\Box}(\mathsf{self} \to (n \bullet \mathsf{startEpoch}_{\mathsf{ech}}(ts', n'_l) \in \mathsf{ois} \Rightarrow ts' > ts))$

That is,
(9) $\Gamma \vdash_{\mathsf{ECH}} (n \bullet \top \uparrow \mathsf{startEpoch}_{\mathsf{ech}}(ts, n_l) \in \mathsf{ois} \wedge \mathsf{self}) \Rightarrow$
$\qquad \circ \Box (\mathsf{self} \to (n \bullet \mathsf{startEpoch}_{\mathsf{ech}}(ts', n'_l) \in \mathsf{ois} \Rightarrow ts' > ts))$

That is,
(10) $\Gamma \vdash_{\mathsf{ECH}} (n \bullet \top \uparrow \mathsf{startEpoch}_{\mathsf{ech}}(ts, n_l) \in \mathsf{ois} \wedge \mathsf{self}) \Rightarrow$
$\qquad \circ (\mathsf{self} \wedge n \bullet \mathsf{startEpoch}_{\mathsf{ech}}(ts', n'_l) \in \mathsf{ois} \Rightarrow ts' > ts)$

From [10] and ExeOrderOR,
(11) $\Gamma \vdash_{\mathsf{ECH}} n \bullet \top \uparrow \mathsf{startEpoch}_{\mathsf{ech}}(ts, n_l) \Rightarrow$
$\qquad \Diamond(n \bullet \top \uparrow \mathsf{startEpoch}_{\mathsf{ech}}(ts', n'_l) \Rightarrow ts' > ts)$

That is,
(12) $\Gamma \vdash_{\mathsf{ECH}} n \bullet \top \uparrow \mathsf{startEpoch}_{\mathsf{ech}}(ts, n_l) \Rightarrow$
$\qquad \hat{\Box}(n \bullet \top \uparrow \mathsf{startEpoch}_{\mathsf{ech}}(ts', n'_l) \to ts' > ts)$

**Theorem 24.** *($ECH_2$: Consistency)*
If a correct process starts an epoch $(ts, n_l)$ and another correct process starts an epoch $(ts, n'_l)$, then $n_l = n'_l$.

$\vdash_{\mathsf{ECH}} n \in \mathsf{Correct} \land n' \in \mathsf{Correct} \to$
$\quad (n \bullet \top \uparrow \mathsf{startEpoch}_{\mathsf{ech}}(ts, n_l)) \Rightarrow$
$\quad (n' \bullet \top \uparrow \mathsf{startEpoch}_{\mathsf{ech}}(ts, n'_l) \Rightarrow n_l = n'_l)$

**Proof.**

By OI',
(1) $\Gamma \vdash_{\mathsf{ECH}} (n \bullet \top \uparrow \mathsf{startEpoch}_{\mathsf{ech}}(ts, n_l)) \Rightarrow$
$\quad\quad \hat{\diamond}(n \bullet \mathsf{startEpoch}_{\mathsf{ech}}(ts, n_l) \in \mathsf{ois} \land \mathsf{self})$

By InvL,
(2) $\Gamma \vdash_{\mathsf{ECH}} (n \bullet \mathsf{startEpoch}_{\mathsf{ech}}(ts, n_l) \in \mathsf{ois} \land \mathsf{self}) \Rightarrow$
$\quad\quad (n \bullet 1 \uparrow \mathsf{deliver}_{\mathsf{beb}}(n_l, \langle \mathrm{NEWEPOCH}, ts \rangle))$

By $\mathsf{BEB}_3$,
(3) $\Gamma \vdash_{\mathsf{ECH}} (n \bullet 1 \uparrow \mathsf{deliver}_{\mathsf{beb}}(n_l, \langle \mathrm{NEWEPOCH}, ts \rangle))$
$\quad\quad \diamond (n_l \bullet 1 \downarrow \mathsf{broadcast}_{\mathsf{beb}}(\langle \mathrm{NEWEPOCH}, ts \rangle))$

From [1] to [3],
(4) $\Gamma \vdash_{\mathsf{ECH}} (n \bullet \top \uparrow \mathsf{startEpoch}_{\mathsf{ech}}(ts, n_l)) \Rightarrow$
$\quad\quad \hat{\diamond}\diamond(n_l \bullet 1 \downarrow \mathsf{broadcast}_{\mathsf{beb}}(\langle \mathrm{NEWEPOCH}, ts \rangle))$

That is,
(5) $\Gamma \vdash_{\mathsf{ECH}} (n \bullet \top \uparrow \mathsf{startEpoch}_{\mathsf{ech}}(ts, n_l)) \Rightarrow$
$\quad\quad \hat{\diamond}(n_l \bullet 1 \downarrow \mathsf{broadcast}_{\mathsf{beb}}(\langle \mathrm{NEWEPOCH}, ts \rangle))$

By OR',
(6) $\Gamma \vdash_{\mathsf{ECH}} (n_l \bullet \downarrow \mathsf{broadcast}_{\mathsf{beb}}(\langle \mathrm{NEWEPOCH}, ts \rangle)) \Rightarrow$
$\quad\quad \hat{\diamond}(n_l \bullet (1, \mathsf{broadcast}_{\mathsf{beb}}(\mathrm{NEWEPOCH}, ts)) \in \mathsf{ors} \land ts(\mathsf{s}'(n_l)) = ts \land \mathsf{self})$

From [5] and [6],
(7) $\Gamma \vdash_{\mathsf{ECH}} (n \bullet \top \uparrow \mathsf{startEpoch}_{\mathsf{ech}}(ts, n_l)) \Rightarrow$
$\quad\quad \hat{\diamond}\hat{\diamond}(n_l \bullet \mathsf{broadcast}_{\mathsf{beb}}(\mathrm{NEWEPOCH}, ts) \in \mathsf{ors} \land ts(\mathsf{s}'(n_l)) = ts \land \mathsf{self})$

By InvUSSe' and $\mathcal{S}(\mathsf{s}(n_l)) : ts(\mathsf{s}(n_l)) \mod \mathbb{N} = rank(n_l)$,
(8) $\Gamma \vdash_{\mathsf{ECH}} \square\, [ts(\mathsf{s}(n_l)) \mod \mathbb{N} = rank(n_l)]$

By PostPre on [8],
(9) $\Gamma \vdash_{\mathsf{ECH}} \square\, [ts(\mathsf{s}'(n_l)) \mod \mathbb{N} = rank(n_l)]$

From [8] and [9],
(10) $\Gamma \vdash_{\mathsf{ECH}} \square\, [ts(\mathsf{s}(n_l)) \mod \mathbb{N} = ts(\mathsf{s}'(n_l)) \mod \mathbb{N} = rank(n_l)]$

From [7] and [10]
(11) $\Gamma \vdash_{\mathsf{ECH}} (n \bullet \top \uparrow \mathsf{startEpoch}_{\mathsf{ech}}(ts, n_l)) \Rightarrow$
$\quad\quad \hat{\diamond}\hat{\diamond}\square[ts(\mathsf{s}'(n_l)) = ts \land ts \mod \mathbb{N} = rank(n_l)]$

That is,
(12) $\Gamma \vdash_{\mathsf{ECH}} (n \bullet \top \uparrow \mathsf{startEpoch}_{\mathsf{ech}}(ts, n_l)) \Rightarrow$
$\quad\quad \square[ts(\mathsf{s}'(n_l)) = ts \land ts \mod \mathbb{N} = rank(n_l)]$

Similarly, we can say,
(13) $\Gamma \vdash_{\mathsf{ECH}} (n' \bullet \top \uparrow \mathsf{startEpoch}_{\mathsf{ech}}(ts, n'_l)) \Rightarrow$

$$\Box[ts(\mathsf{s}'(n'_l)) = ts \wedge ts \mod \mathbb{N} = rank(n'_l)]$$

From [12] and [13]

(14) $\Gamma \vdash_{\mathsf{ECH}} (n \bullet \top \uparrow \mathsf{startEpoch}_{\mathsf{ech}}(ts, n_l)) \wedge$
$(n' \bullet \top \uparrow \mathsf{startEpoch}_{\mathsf{ech}}(ts, n'_l)) \Rightarrow$
$\Box[ts \mod \mathbb{N} = rank(n_l) \wedge ts \mod \mathbb{N} = rank(n'_l)]$

From [14] and the uniqueness of rank function we can easily say,

(15) $\Gamma \vdash_{\mathsf{ECH}} (n \bullet \top \uparrow \mathsf{startEpoch}_{\mathsf{ech}}(ts, n_l)) \wedge$
$(n' \bullet \top \uparrow \mathsf{startEpoch}_{\mathsf{ech}}(ts, n'_l)) \Rightarrow \Box(n_l = n'_l)p$

That is,

(16) $\Gamma \vdash_{\mathsf{ECH}} (n \bullet \top \uparrow \mathsf{startEpoch}_{\mathsf{ech}}(ts, n_l)) \Rightarrow$
$((n' \bullet \top \uparrow \mathsf{startEpoch}_{\mathsf{ech}}(ts, n'_l)) \Rightarrow n_l = n'_l)$

**Theorem 25.** *(ECH$_3$: Eventual leadership)*
There is a timestamp $ts$ and a correct process $n_l$ such that eventually every correct process starts an epoch with $ts$ and $n_l$ and does not start another epoch afterwards.

$\vdash_{\mathsf{ECH}} \exists ts, n_l.\ n_l \in \mathsf{Correct} \wedge$
$\qquad [n \in \mathsf{Correct} \to$
$\qquad\quad \Diamond[(n \bullet \top \uparrow \mathsf{startEpoch}_{\mathsf{ech}}(ts, n_l)) \wedge$
$\qquad\qquad \hat{\Box} \neg (n \bullet \top \uparrow \mathsf{startEpoch}_{\mathsf{ech}}(ts', n_l'))]]$

**Proof.**

From $\mathsf{ELE}_1$,
(1) $\Gamma \vdash_{\mathsf{ECH}} \exists n_l.\ n_l \in \mathsf{Correct} \wedge$
$\qquad\qquad [n \in \mathsf{Correct} \to$
$\qquad\qquad\quad \Diamond[(n \bullet 2 \uparrow \mathsf{trust}_{\mathsf{eld}}(n_l)) \wedge \hat{\Box} \neg (n \bullet 2 \uparrow \mathsf{trust}_{\mathsf{eld}}(n_l'))]]$

By II,
(2) $\Gamma \vdash_{\mathsf{ECH}} n \bullet 2 \uparrow \mathsf{trust}_{\mathsf{eld}}(n_l) \Rightarrow$
$\qquad\qquad trusted(\mathsf{s}'(n)) = n_l \wedge \mathsf{self}$

By InvL,
(3) $\Gamma \vdash_{\mathsf{ECH}} trusted(\mathsf{s}(n)) \neq trusted(\mathsf{s}'(n)) \Rightarrow$
$\qquad\qquad n \bullet 2 \uparrow \mathsf{trust}_{\mathsf{eld}}(n_l')$

The contrapositive of [3],
(4) $\Gamma \vdash_{\mathsf{ECH}} \neg (n \bullet 2 \uparrow \mathsf{trust}_{\mathsf{eld}}(n_l')) \Rightarrow$
$\qquad\qquad \neg (trusted(\mathsf{s}(n)) \neq trusted(\mathsf{s}'(n)))$

That is,
(5) $\Gamma \vdash_{\mathsf{ECH}} \neg (n \bullet 2 \uparrow \mathsf{trust}_{\mathsf{eld}}(n_l')) \Rightarrow$
$\qquad\qquad (trusted(\mathsf{s}(n)) = trusted(\mathsf{s}'(n)))$

From [1], [2] and [5],
(6) $\Gamma \vdash_{\mathsf{ECH}} \exists n_l.\ n_l \in \mathsf{Correct} \wedge$
$\qquad\qquad [n \in \mathsf{Correct} \to$
$\qquad\qquad\quad \Diamond[((n \bullet 2 \uparrow \mathsf{trust}_{\mathsf{eld}}(n_l)) \wedge trusted(\mathsf{s}'(n)) = n_l \wedge \mathsf{self}) \wedge$
$\qquad\qquad\qquad \hat{\Box}(trusted(\mathsf{s}(n)) = trusted(\mathsf{s}'(n)))]]$

By PostPre on [6],
(7) $\Gamma \vdash_{\mathsf{ECH}} \exists n_l.\ n_l \in \mathsf{Correct} \wedge$
$\qquad\qquad [n \in \mathsf{Correct} \to$
$\qquad\qquad\quad \Diamond[(n \bullet 2 \uparrow \mathsf{trust}_{\mathsf{eld}}(n_l)) \wedge trusted(\mathsf{s}'(n)) = n_l \wedge$
$\qquad\qquad\qquad \bigcirc (trusted(\mathsf{s}(n)) = n_l \wedge \mathsf{self}) \wedge$
$\qquad\qquad\qquad \hat{\Box}(trusted(\mathsf{s}(n)) = trusted(\mathsf{s}'(n)))]]$

From [7],
(8) $\Gamma \vdash_{\mathsf{ECH}} \exists n_l.\ n_l \in \mathsf{Correct} \wedge$
$\qquad\qquad [n \in \mathsf{Correct} \to$
$\qquad\qquad\quad \Diamond[(n \bullet 2 \uparrow \mathsf{trust}_{\mathsf{eld}}(n_l)) \wedge trusted(\mathsf{s}'(n)) = n_l \wedge$
$\qquad\qquad\qquad \hat{\Box}(trusted(\mathsf{s}(n)) = trusted(\mathsf{s}'(n)) = n_l \wedge \mathsf{self})]]$

By instantiating $n$ to $n_l$,
(9) $\Gamma \vdash_{\mathsf{ECH}} \exists n_l.\ n_l \in \mathsf{Correct} \wedge$
$\qquad\qquad [n \in \mathsf{Correct} \to$
$\qquad\qquad\quad \Diamond[(n_l \bullet 2 \uparrow \mathsf{trust}_{\mathsf{eld}}(n_l)) \wedge trusted(\mathsf{s}'(n_l)) = n_l \wedge$
$\qquad\qquad\qquad \hat{\Box}(trusted(\mathsf{s}(n_l)) = trusted(\mathsf{s}'(n_l)) = n_l \wedge \mathsf{self})]]$

From [8], [9] and Lemma 127,
- (10) $\Gamma \vdash_{\mathsf{ECH}} \exists n_l.\ n_l \in \mathsf{Correct} \wedge$
  $[n \in \mathsf{Correct} \to$
  $\Diamond[\Diamond\hat\Diamond[(n \bullet 2 \uparrow \mathsf{trust}_{\mathsf{eld}}(n_l)) \wedge trusted(\mathsf{s}'(n_l)) = n_l \wedge$
  $\hat\Box(trusted(\mathsf{s}(n_l)) = trusted(\mathsf{s}'(n_l)) = n_l \wedge \mathsf{self})] \wedge$
  $(n \bullet 2 \uparrow \mathsf{trust}_{\mathsf{eld}}(n_l)) \wedge trusted(\mathsf{s}'(n)) = n_l \wedge$
  $\hat\Box(trusted(\mathsf{s}(n)) = trusted(\mathsf{s}'(n)) = n_l \wedge \mathsf{self})]]$

By IR,
- (11) $\Gamma \vdash_{\mathsf{ECH}} (n_l \bullet 2 \uparrow \mathsf{trust}_{\mathsf{eld}}(n_l)) \Rightarrow$
  $n_l \bullet (1, \mathsf{broadcast}_{\mathsf{beb}}(\textsc{NewEpoch}(t))) \in \mathsf{ors} \wedge \mathsf{self}$

By OR and $\mathsf{BEB}_1$,
- (12) $\Gamma \vdash_{\mathsf{ECH}} n \in \mathsf{Correct} \to$
  $n_l \bullet (1, \mathsf{broadcast}_{\mathsf{beb}}(\textsc{NewEpoch}(t))) \in \mathsf{ors} \wedge \mathsf{self} \Rightarrow$
  $\Diamond(n \bullet 1 \uparrow \mathsf{deliver}_{\mathsf{beb}}(n_l, \textsc{NewEpoch}(t)))$

From [11] and [12],
- (13) $\Gamma \vdash_{\mathsf{ECH}} n \in \mathsf{Correct} \to$
  $(n_l \bullet 2 \uparrow \mathsf{trust}_{\mathsf{eld}}(n_l)) \Rightarrow$
  $\Diamond(n \bullet 1 \uparrow \mathsf{deliver}_{\mathsf{beb}}(n_l, \textsc{NewEpoch}(t)))$

From [10] and [13],
- (14) $\Gamma \vdash_{\mathsf{ECH}} \exists n_l.\ n_l \in \mathsf{Correct} \wedge$
  $[n \in \mathsf{Correct} \to$
  $\Diamond[\Diamond\hat\Diamond\Diamond(n \bullet 1 \uparrow \mathsf{deliver}_{\mathsf{beb}}(n_l, \textsc{NewEpoch}(t))) \wedge$
  $(n \bullet 2 \uparrow \mathsf{trust}_{\mathsf{eld}}(n_l)) \wedge trusted(\mathsf{s}'(n)) = n_l \wedge$
  $\hat\Box(trusted(\mathsf{s}(n)) = trusted(\mathsf{s}'(n)) = n_l \wedge \mathsf{self})]]$

That is (node $n$ can deliver $\textsc{NewEpoch}$ message before or after receiving $\mathsf{trust}$ event),
- (15) $\Gamma \vdash_{\mathsf{ECH}} \exists n_l.\ n_l \in \mathsf{Correct} \wedge$
  $[n \in \mathsf{Correct} \to$
  $\Diamond[[\hat\Diamond(n \bullet 1 \uparrow \mathsf{deliver}_{\mathsf{beb}}(n_l, \textsc{NewEpoch}(t))) \vee$
  $\Diamond(n \bullet 1 \uparrow \mathsf{deliver}_{\mathsf{beb}}(n_l, \textsc{NewEpoch}(t)))] \wedge$
  $[(n \bullet 2 \uparrow \mathsf{trust}_{\mathsf{eld}}(n_l)) \wedge trusted(\mathsf{s}'(n)) = n_l \wedge$
  $\hat\Box(trusted(\mathsf{s}(n)) = trusted(\mathsf{s}'(n)) = n_l \wedge \mathsf{self})]]$

That is,
- (16) $\Gamma \vdash_{\mathsf{ECH}} \exists n_l.\ n_l \in \mathsf{Correct} \wedge$
  $[n \in \mathsf{Correct} \to$
  $\Diamond[[\hat\Diamond(n \bullet 1 \uparrow \mathsf{deliver}_{\mathsf{beb}}(n_l, \textsc{NewEpoch}(t))) \wedge$
  $(n \bullet 2 \uparrow \mathsf{trust}_{\mathsf{eld}}(n_l)) \wedge trusted(\mathsf{s}'(n)) = n_l \wedge$
  $\hat\Box(trusted(\mathsf{s}(n)) = trusted(\mathsf{s}'(n)) = n_l \wedge \mathsf{self})] \vee$
  $[\hat\Diamond(n \bullet 1 \uparrow \mathsf{deliver}_{\mathsf{beb}}(n_l, \textsc{NewEpoch}(t))) \wedge$
  $(n \bullet 2 \uparrow \mathsf{trust}_{\mathsf{eld}}(n_l)) \wedge trusted(\mathsf{s}'(n)) = n_l \wedge$
  $\hat\Box(trusted(\mathsf{s}(n)) = trusted(\mathsf{s}'(n)) = n_l \wedge \mathsf{self})]]]$

That is,
- (17) $\Gamma \vdash_{\mathsf{ECH}} \exists n_l.\ n_l \in \mathsf{Correct} \wedge$
  $[n \in \mathsf{Correct} \to$
  $\Diamond[\hat\Diamond[(n \bullet 1 \uparrow \mathsf{deliver}_{\mathsf{beb}}(n_l, \textsc{NewEpoch}(t))) \wedge$
  $\hat\Diamond\Box(trusted(\mathsf{s}(n)) = n_l \wedge \mathsf{self})] \vee$
  $\hat\Diamond[(n \bullet 1 \uparrow \mathsf{deliver}_{\mathsf{beb}}(n_l, \textsc{NewEpoch}(t))) \wedge$
  $\Box(trusted(\mathsf{s}'(n)) = n_l \wedge \mathsf{self})]]]$

By Lemma 70 and Lemma 71 on [17],

(18) $\Gamma \vdash_{\mathsf{ECH}} \exists l.\ l \in \mathsf{Correct} \land$
$[n \in \mathsf{Correct} \to$
$\Diamond[\Diamond[\Diamond(n \bullet \top \uparrow \mathsf{startEpoch}(ts, n_l) \land$
$\hat{\Box}\ nk(\mathsf{s}(n))(n_l) = \mathsf{true} \land \Box\ trusted(\mathsf{s}(n)) = n_l)] \lor$
$\Diamond[\Diamond(n \bullet \top \uparrow \mathsf{startEpoch}(ts, n_l) \land$
$\hat{\Box}\ nk(\mathsf{s}(n))(n_l) = \mathsf{true} \land \Box\ trusted(\mathsf{s}(n)) = n_l)]]]$

That is,

(19) $\Gamma \vdash_{\mathsf{ECH}} \exists l.\ l \in \mathsf{Correct} \land$
$[n \in \mathsf{Correct} \to$
$\Diamond[(n \bullet \top \uparrow \mathsf{startEpoch}(ts, l) \land$
$\hat{\Box}\ nk(\mathsf{s}(n))(n_l) = \mathsf{true} \land \Box\ trusted(\mathsf{s}(n)) = n_l)]]$

By Lemma 72 on [19],

(20) $\Gamma \vdash_{\mathsf{ECH}} \exists l.\ l \in \mathsf{Correct} \land$
$[n \in \mathsf{Correct} \to$
$\Diamond[\Diamond[(n \bullet \top \uparrow \mathsf{startEpoch}(ts, l) \land \Box \neg(n \bullet \top \uparrow \mathsf{startEpoch}(ts', l'))]]]$

That is,

(21) $\Gamma \vdash_{\mathsf{ECH}} \exists l.\ l \in \mathsf{Correct} \land$
$[n \in \mathsf{Correct} \to$
$\Diamond((n \bullet \top \uparrow \mathsf{startEpoch}(ts, l) \land \Box \neg(n \bullet \top \uparrow \mathsf{startEpoch}(ts', l'))))]$

**Lemma 70.**
$\Gamma \vdash_{\mathsf{ECH}}$
$\exists l.\ l \in \mathsf{Correct} \land$
$[n \in \mathsf{Correct} \to$
$(n \bullet 1 \uparrow \mathsf{deliver}_{\mathsf{beb}}(n_l, \mathrm{NEWEPOCH}(t))) \land$
$\Diamond[trusted(\mathsf{s}(n)) \neq n_l \land trusted(\mathsf{s}'(n)) = n_l \land \hat{\Box}\ trusted(\mathsf{s}(n)) = n_l] \Rightarrow$
$\hat{\Diamond}(n \bullet \top \uparrow \mathsf{startEpoch}_{\mathsf{ech}}(ts, n_l) \land \hat{\Box}\ nk(\mathsf{s}(n))(n_l) = \mathsf{true} \land \Box\ trusted(\mathsf{s}(n)) = n_l)]$

**Proof.**

By INVL
(1) $\Gamma \vdash_{\mathsf{ECH}} (n \bullet 1 \uparrow \mathsf{deliver}_{\mathsf{beb}}(n_l, \mathrm{NEWEPOCH}(t))) \land$
$trusted(\mathsf{s}(n)) \neq n_l \Rightarrow$
$\mathsf{self} \land n \bullet (0, \mathsf{send}_{\mathsf{pl}}(n_l, \mathrm{NACK}(lts))) \in \mathsf{ois}$

By OR
(2) $\Gamma \vdash_{\mathsf{ECH}} \mathsf{self} \land n \bullet (0, \mathsf{send}_{\mathsf{pl}}(n_l, \mathrm{NACK}(lts))) \in \mathsf{ois} \Rightarrow$
$\hat{\Diamond}(n \bullet 1 \downarrow \mathsf{send}_{\mathsf{pl}}(n_l, \mathrm{NACK}(lts)))$

From [1] and [2] and $\hat{\Diamond}\Box\ trusted(\mathsf{s}(n)) = n_l$ from the premise,
(3) $\Gamma \vdash_{\mathsf{ECH}} (n \bullet 1 \uparrow \mathsf{deliver}_{\mathsf{beb}}(n_l, \mathrm{NEWEPOCH}(t))) \land$
$trusted(\mathsf{s}(n)) \neq n_l \land \hat{\Diamond}\Box\ trusted(\mathsf{s}(n)) = n_l \Rightarrow$
$\hat{\Diamond}(n \bullet 1 \downarrow \mathsf{send}_{\mathsf{pl}}(n_l, \mathrm{NACK}(lts)))$

By $\mathsf{PL}_1$ and correctness of $n_l$ and $n$,
(4) $\Gamma \vdash_{\mathsf{ECH}} (n \bullet 1 \downarrow \mathsf{send}_{\mathsf{pl}}(n_l, \mathrm{NACK}(lts))) \Rightarrow$
$\Diamond(n_l \bullet 0 \uparrow \mathsf{deliver}_{\mathsf{pl}}(n_l, \mathrm{NACK}(lts)))$

By INVL,
(5) $\Gamma \vdash_{\mathsf{ECH}} (n_l \bullet 0 \uparrow \mathsf{deliver}_{\mathsf{pl}}(n_l, \mathrm{NACK}(lts))) \Rightarrow$
$\mathsf{self} \land (n_l \bullet (1, \mathsf{broadcast}_{\mathsf{beb}}(\mathrm{NEWEPOCH}(lts')) \in \mathsf{ors}))$

By OR,

(6) $\Gamma \vdash_{\mathsf{ECH}} \mathsf{self} \land (n_l \bullet (1, \mathsf{broadcast}_{\mathsf{beb}}(\mathrm{NewEpoch}(lst'))) \in \mathsf{ors}) \Rightarrow$
$\hat{\Diamond}(n_l \bullet 1 \downarrow \mathsf{broadcast}_{\mathsf{beb}}(\mathrm{NewEpoch}(lst')))$

By $\mathsf{BEB}_1$ and correctness of $n_l$ and $n$,

(7) $\Gamma \vdash_{\mathsf{ECH}} (n_l \bullet 1 \downarrow \mathsf{broadcast}_{\mathsf{beb}}(\mathrm{NewEpoch}(lst'))) \Rightarrow$
$\Diamond(n \bullet 1 \uparrow \mathsf{deliver}_{\mathsf{beb}}(n_l, \mathrm{NewEpoch}(lst')))$

From [3] to [7],

(8) $\Gamma \vdash_{\mathsf{ECH}} (n \bullet 1 \uparrow \mathsf{deliver}_{\mathsf{beb}}(n_l, \mathrm{NewEpoch}(t))) \land$
$trusted(\mathsf{s}(n)) \neq n_l \land \hat{\Diamond}\Box\, trusted(\mathsf{s}(n)) = n_l \Rightarrow$
$\hat{\Diamond}(n \bullet 1 \uparrow \mathsf{deliver}_{\mathsf{beb}}(n_l, \mathrm{NewEpoch}(lts'))) \land \hat{\Diamond}\Box\, trusted(\mathsf{s}(n)) = n_l$

By Lemma 123 on [8],

(9) $\Gamma \vdash_{\mathsf{ECH}} (n \bullet 1 \uparrow \mathsf{deliver}_{\mathsf{beb}}(n_l, \mathrm{NewEpoch}(t))) \land$
$trusted(\mathsf{s}(n)) \neq n_l \land \hat{\Diamond}\Box\, trusted(\mathsf{s}(n)) = n_l \Rightarrow$
$\hat{\Diamond}[(n \bullet 1 \uparrow \mathsf{deliver}_{\mathsf{beb}}(n_l, \mathrm{NewEpoch}(lts'))) \land \hat{\Diamond}\Box\, trusted(\mathsf{s}(n)) = n_l] \lor$
$\hat{\Diamond}[(n \bullet 1 \uparrow \mathsf{deliver}_{\mathsf{beb}}(n_l, \mathrm{NewEpoch}(lts'))) \land \Box\, trusted(\mathsf{s}(n)) = n_l]$

By Lemma 124 on [9],

(10) $\Gamma \vdash_{\mathsf{ECH}} (n \bullet 1 \uparrow \mathsf{deliver}_{\mathsf{beb}}(n_l, \mathrm{NewEpoch}(t))) \land$
$trusted(\mathsf{s}(n)) \neq n_l \land \hat{\Diamond}\Box\, trusted(\mathsf{s}(n)) = n_l \Rightarrow$
$\hat{\Diamond}[(n \bullet 1 \uparrow \mathsf{deliver}_{\mathsf{beb}}(n_l, \mathrm{NewEpoch}(lts'))) \land \Box\, trusted(\mathsf{s}(n)) = n_l]$

By Lemma 71 and [9],

(11) $\Gamma \vdash_{\mathsf{ECH}} (n \bullet 1 \uparrow \mathsf{deliver}_{\mathsf{beb}}(n_l, \mathrm{NewEpoch}(t))) \land$
$trusted(\mathsf{s}(n)) \neq n_l \land \hat{\Diamond}\Box\, trusted(\mathsf{s}(n)) = n_l \Rightarrow$
$\Diamond[\Diamond(n \bullet \top \uparrow \mathsf{startEpoch}_{\mathsf{ech}}(ts, n_l)) \land$
$\hat{\Box}\, nk(\mathsf{s}(n))(n_l) = \mathsf{true} \land \Box\, trusted(\mathsf{s}(n)) = n_l)]$

That is,

(12) $\Gamma \vdash_{\mathsf{ECH}} (n \bullet 1 \uparrow \mathsf{deliver}_{\mathsf{beb}}(n_l, \mathrm{NewEpoch}(t))) \land$
$trusted(\mathsf{s}(n)) \neq n_l \land \hat{\Diamond}\Box\, trusted(\mathsf{s}(n)) = n_l \Rightarrow$
$\Diamond[(n \bullet \top \uparrow \mathsf{startEpoch}_{\mathsf{ech}}(ts, n_l)) \land$
$\hat{\Box}\, nk(\mathsf{s}(n_l))(n_l) = \mathsf{true} \land \Box\, trusted(\mathsf{s}(n)) = n_l)]$

**Lemma 71.**
$\Gamma \vdash_{\mathsf{ECH}}$
$\exists n_l.\ n_l \in \mathsf{Correct} \land$
$[n \in \mathsf{Correct} \to$
$(n \bullet 1 \uparrow \mathsf{deliver}_{\mathsf{beb}}(n_l, \mathrm{NewEpoch}(t))) \land \Box\, trusted(\mathsf{s}(n)) = n_l] \Rightarrow$
$\Diamond(n \bullet \top \uparrow \mathsf{startEpoch}_{\mathsf{ech}}(ts, n_l) \land \hat{\Box}\, nk(\mathsf{s}(n))(n_l) = \mathsf{true} \land \Box\, trusted(\mathsf{s}(n)) = n_l)$

**Proof.**
The First case is that $nk(\mathsf{s}(n))(n_l) = \mathsf{false}$
By INVL,

(1) $\Gamma \vdash_{\mathsf{ECH}} (n \bullet 1 \uparrow \mathsf{deliver}_{\mathsf{beb}}(n_l, \mathrm{NewEpoch}(t))) \land$
$\Box\, trusted(\mathsf{s}(n)) = n_l \land nk(\mathsf{s}(n))(n_l) = \mathsf{false} \Rightarrow$
$\mathsf{self} \land n \bullet (0, \mathsf{send}_{\mathsf{pl}}(n_l, \mathrm{State}(lts))) \in \mathsf{ors} \land$
$lastts(\mathsf{s}(n)) = lts \land nk(\mathsf{s}'(n))(n_l) = \mathsf{true}$

By INVS″,

(2) $\Gamma \vdash_{\mathsf{ECH}} \mathsf{self} \land nk[\mathsf{s}'(n_l)] = \mathsf{true} \Rightarrow \hat{\Box}(\mathsf{self} \to nk(\mathsf{s}(n))(n_l) = \mathsf{true})$

By OR,

(3) $\Gamma \vdash_{\mathsf{ECH}} \mathsf{self} \land n \bullet (0, \mathsf{send}_{\mathsf{pl}}(n_l, \mathrm{STATE}(lts))) \in \mathsf{ois} \Rightarrow$
$\hat{\Diamond}(n \bullet 1 \downarrow \mathsf{send}_{\mathsf{pl}}(n_l, \mathrm{STATE}(lts)))$

By $\mathsf{PL}_1$ and correctness of $n_l$ and $n$,

(4) $\Gamma \vdash_{\mathsf{ECH}} (n \bullet 1 \downarrow \mathsf{send}_{\mathsf{pl}}(n_l, \mathrm{STATE}(lts))) \Rightarrow$
$\Diamond(n_l \bullet 0 \uparrow \mathsf{deliver}_{\mathsf{pl}}(n_l, \mathrm{STATE}(lts)))$

From [1] to [4],

(5) $\Gamma \vdash_{\mathsf{ECH}} (n \bullet 1 \uparrow \mathsf{deliver}_{\mathsf{beb}}(n_l, \mathrm{NEWEPOCH}(t))) \land$
$\Box \, trusted(\mathsf{s}(n)) = n_l \land nk(\mathsf{s}(n))(n_l) = \mathsf{false} \Rightarrow$
$\hat{\Diamond}(n_l \bullet 0 \uparrow \mathsf{deliver}_{\mathsf{pl}}(n_l, \mathrm{STATE}(lts))) \land$
$lastts(\mathsf{s}(n)) = lts \land \hat{\Box} nk(\mathsf{s}(n))(n_l) = \mathsf{true}$

By INVL,

(6) $\Gamma \vdash_{\mathsf{ECH}} (n_l \bullet 0 \uparrow \mathsf{deliver}_{\mathsf{pl}}(n_l, \mathrm{STATE}(lts))) \Rightarrow$
$\mathsf{self} \land (n_l \bullet (1, \mathsf{broadcast}_{\mathsf{beb}}(\mathrm{NEWEPOCH}(lts'))) \in \mathsf{ors}) \land lst' > lst$

By OR,

(7) $\Gamma \vdash_{\mathsf{ECH}} \mathsf{self} \land (n_l \bullet (1, \mathsf{broadcast}_{\mathsf{beb}}(\mathrm{NEWEPOCH}(lts'))) \in \mathsf{ors}) \Rightarrow$
$\hat{\Diamond}(n_l \bullet 1 \downarrow \mathsf{broadcast}_{\mathsf{beb}}(\mathrm{NEWEPOCH}(lts')))$

By $\mathsf{BEB}_1$ and correctness of $n_l$ and $n$,

(8) $\Gamma \vdash_{\mathsf{ECH}} (n_l \bullet 1 \downarrow \mathsf{broadcast}_{\mathsf{beb}}(\mathrm{NEWEPOCH}(lts'))) \Rightarrow$
$\Diamond(n \bullet 1 \uparrow \mathsf{deliver}_{\mathsf{beb}}(n_l, \mathrm{NEWEPOCH}(lts')))$

From [6] to [8],

(9) $\Gamma \vdash_{\mathsf{ECH}} (n_l \bullet 0 \uparrow \mathsf{deliver}_{\mathsf{pl}}(n_l, \mathrm{STATE}(lts))) \Rightarrow$
$\hat{\Diamond}(n \bullet 1 \uparrow \mathsf{deliver}_{\mathsf{beb}}(n_l, \mathrm{NEWEPOCH}(lts'))) \land lst' > lst$

From [5] and [9],

(10) $\Gamma \vdash_{\mathsf{ECH}} (n \bullet 1 \uparrow \mathsf{deliver}_{\mathsf{beb}}(n_l, \mathrm{NEWEPOCH}(t))) \land$
$\Box \, trusted(\mathsf{s}(n)) = n_l \land nk(\mathsf{s}(n))(n_l) = \mathsf{false} \Rightarrow$
$\hat{\Diamond}(\Diamond(n \bullet 1 \uparrow \mathsf{deliver}_{\mathsf{beb}}(n_l, \mathrm{NEWEPOCH}(lts'))) \land lst' > lst) \land$
$lastts(\mathsf{s}(n)) = lts \land \hat{\Box} nk(\mathsf{s}(n))(n_l) = \mathsf{true}$

That is,

(11) $\Gamma \vdash_{\mathsf{ECH}} (n \bullet 1 \uparrow \mathsf{deliver}_{\mathsf{beb}}(n_l, \mathrm{NEWEPOCH}(t))) \land$
$\Box \, trusted(\mathsf{s}(n)) = n_l \land nk(\mathsf{s}(n))(n_l) = \mathsf{false} \Rightarrow$
$\hat{\Diamond}[n \bullet 1 \uparrow \mathsf{deliver}_{\mathsf{beb}}(n_l, \mathrm{NEWEPOCH}(lts')) \land$
$lst' > lst \land nk(\mathsf{s}'(n))(n_l) = \mathsf{true} \land trusted(\mathsf{s}(n)) = n_l] \land$
$lastts(\mathsf{s}(n)) = lts \land \hat{\Box} nk(\mathsf{s}(n))(n_l) = \mathsf{true}$

There are two possibilities for the value of $lastts(\mathsf{s}(n))$ when a node receieves $\mathsf{deliver}_{\mathsf{beb}}(n_l, \mathrm{NEWEPOCH}(lts'))$ event,

(12) $\Gamma \vdash_{\mathsf{ECH}} (n \bullet 1 \uparrow \mathsf{deliver}_{\mathsf{beb}}(n_l, \mathrm{NEWEPOCH}(t))) \land$
$\Box \, trusted(\mathsf{s}(n)) = n_l \land nk(\mathsf{s}(n))(n_l) = \mathsf{false} \Rightarrow$
$\hat{\Diamond}[n \bullet 1 \uparrow \mathsf{deliver}_{\mathsf{beb}}(n_l, \mathrm{NEWEPOCH}(lts')) \land$
$lst' > lst \land nk(\mathsf{s}'(n))(n_l) = \mathsf{true} \land trusted(\mathsf{s}(n)) = n_l \land$
$(lastts(\mathsf{s}(n)) = lst \lor lastts(\mathsf{s}(n)) \neq lst)] \land$
$\Box \, trusted(\mathsf{s}(n)) = n_l lastts(\mathsf{s}(n)) = lts \land \hat{\Box} nk(\mathsf{s}(n))(n_l) = \mathsf{true}$

By Lemma 115,

(13) $\Gamma \vdash_{\mathsf{ECH}} (n \bullet 1 \uparrow \mathsf{deliver}_{\mathsf{beb}}(n_l, \mathrm{NEWEPOCH}(t))) \land$

$$\Box\ trusted(\mathsf{s}(n)) = n_l \wedge nk(\mathsf{s}(n))(n_l) = \mathsf{false} \Rightarrow$$
$$[\hat{\Diamond}[(n \bullet 1 \uparrow \mathsf{deliver_{beb}}(n_l, \mathrm{NEWEPOCH}(lts'))) \wedge nk(\mathsf{s}(n))(n_l) = \mathsf{true} \wedge$$
$$lst' > lst \wedge lastts(\mathsf{s}(n)) = lst \wedge\ trusted(\mathsf{s}(n)) = n_l] \vee$$
$$\hat{\Diamond}[(n \bullet 1 \uparrow \mathsf{deliver_{beb}}(n_l, \mathrm{NEWEPOCH}(lts'))) \wedge nk(\mathsf{s}(n))(n_l) = \mathsf{true} \wedge$$
$$lastts(\mathsf{s}(n)) = lts'' \wedge lts'' \neq lst \wedge\ trusted(\mathsf{s}(n)) = n_l]] \wedge$$
$$\Box\ trusted(\mathsf{s}(n)) = n_l \wedge lastts(\mathsf{s}(n)) = lts \wedge \hat{\Box} nk(\mathsf{s}(n))(n_l) = \mathsf{true}$$

By INVL

(14) $\Gamma \vdash_{\mathsf{ECH}} (n \bullet 1 \uparrow \mathsf{deliver_{beb}}(n_l, \mathrm{NEWEPOCH}(lts'))) \wedge nk(\mathsf{s}(n))(n_l) = \mathsf{true} \wedge$
$$lts' > lastts(\mathsf{s}(n)) \wedge\ trusted(\mathsf{s}(n)) = n_l \Rightarrow$$
$$\mathsf{self} \wedge n \bullet \top \uparrow \mathsf{startEpoch_{ech}}(lts', n_l) \in \mathsf{ois}$$

By OI

(15) $\Gamma \vdash_{\mathsf{ECH}} \mathsf{self} \wedge n \bullet \top \uparrow \mathsf{startEpoch_{ech}}(lts', n_l) \in \mathsf{ois} \Rightarrow$
$$\hat{\Diamond}(n \bullet \top \uparrow \mathsf{startEpoch_{ech}}(lts', n_l))$$

From [14] and [15],

(16) $\Gamma \vdash_{\mathsf{ECH}} (n \bullet 1 \uparrow \mathsf{deliver_{beb}}(n_l, \mathrm{NEWEPOCH}(lts'))) \wedge nk(\mathsf{s}(n))(n_l) = \mathsf{true} \wedge$
$$lts' > lastts(\mathsf{s}(n)) \wedge\ trusted(\mathsf{s}(n)) = n_l \Rightarrow$$
$$\hat{\Diamond}(n \bullet \top \uparrow \mathsf{startEpoch_{ech}}(lts', n_l))$$

By [13], [16] and removing unuseful parameters,

(17) $\Gamma \vdash_{\mathsf{ECH}} (n \bullet 1 \uparrow \mathsf{deliver_{beb}}(n_l, \mathrm{NEWEPOCH}(t))) \wedge$
$$\Box\ trusted(\mathsf{s}(n)) = n_l \wedge nk(\mathsf{s}(n))(n_l) = \mathsf{false} \Rightarrow$$
$$[\hat{\Diamond}[\hat{\Diamond}(n \bullet \top \uparrow \mathsf{startEpoch_{ech}}(lts', n_l)] \vee$$
$$\hat{\Diamond}[lastts(\mathsf{s}(n)) = lts'' \wedge lts'' \neq lst \wedge\ \hat{\Box} trusted(\mathsf{s}(n)) = n_l] \wedge$$
$$\Box\ trusted(\mathsf{s}(n)) = n_l \wedge lastts(\mathsf{s}(n)) = lts \wedge \hat{\Box} nk(\mathsf{s}(n))(n_l) = \mathsf{true}]$$

That is,

(18) $\Gamma \vdash_{\mathsf{ECH}} (n \bullet 1 \uparrow \mathsf{deliver_{beb}}(n_l, \mathrm{NEWEPOCH}(t))) \wedge$
$$\Box\ trusted(\mathsf{s}(n)) = n_l \wedge nk(\mathsf{s}(n))(n_l) = \mathsf{false} \Rightarrow$$
$$[\hat{\Diamond}[(n \bullet \top \uparrow \mathsf{startEpoch_{ech}}(lts', n_l)) \wedge$$
$$\Box\ trusted(\mathsf{s}(n)) = n_l \wedge lastts(\mathsf{s}(n)) = lts \wedge \hat{\Box} nk(\mathsf{s}(n))(n_l) = \mathsf{true}] \vee$$
$$[\hat{\Diamond}[lastts(\mathsf{s}(n)) = lts'' \wedge lts'' \neq lst \wedge\ \hat{\Box} trusted(\mathsf{s}(n)) = n_l] \wedge$$
$$\Box\ trusted(\mathsf{s}(n)) = n_l \wedge lastts(\mathsf{s}(n)) = lts \wedge \hat{\Box} nk(\mathsf{s}(n))(n_l) = \mathsf{true}]]$$

By INVSAG,

(19) $\Gamma \vdash_{\mathsf{ECH}} lastts(\mathsf{s}(n)) \neq lts'' \wedge \Diamond(lastts(\mathsf{s}(n)) = lts'') \Rightarrow$
$$\Diamond(n \bullet \mathsf{startEpoch_{ech}}(lts'', n_l') \in \mathsf{ois} \wedge\ \mathsf{self})$$

From [15], [19] and adding $\Box trusted(\mathsf{s}(n)) = n_l$,

(20) $\Gamma \vdash_{\mathsf{ECH}} lastts(\mathsf{s}(n)) \neq lts'' \wedge \Diamond(lastts(\mathsf{s}(n)) = lts'') \wedge \Box trusted(\mathsf{s}(n)) = n_l \Rightarrow$
$$\hat{\Diamond}(n \bullet \top \uparrow \mathsf{startEpoch_{ech}}(lts'', n_l))$$

From [18] and [20],

(21) $\Gamma \vdash_{\mathsf{ECH}} (n \bullet 1 \uparrow \mathsf{deliver_{beb}}(n_l, \mathrm{NEWEPOCH}(t))) \wedge$
$$\Box\ trusted(\mathsf{s}(n)) = n_l \wedge nk(\mathsf{s}(n))(n_l) = \mathsf{false} \Rightarrow$$
$$[\hat{\Diamond}[(n \bullet \top \uparrow \mathsf{startEpoch_{ech}}(lts', n_l)) \wedge$$
$$\Box\ trusted(\mathsf{s}(n)) = n_l \wedge lastts(\mathsf{s}(n)) = lts \wedge \hat{\Box} nk(\mathsf{s}(n))(n_l) = \mathsf{true}] \vee$$
$$[\hat{\Diamond}(n \bullet \top \uparrow \mathsf{startEpoch_{ech}}(lts'', n_l))] \wedge$$
$$\Box\ trusted(\mathsf{s}(n)) = n_l \wedge lastts(\mathsf{s}(n)) = lts \wedge \hat{\Box} nk(\mathsf{s}(n))(n_l) = \mathsf{true}]]$$

That is,

(22) $\Gamma \vdash_{\mathsf{ECH}} (n \bullet 1 \uparrow \mathsf{deliver_{beb}}(n_l, \mathrm{NEWEPOCH}(t))) \wedge$
$$\Box\ trusted(\mathsf{s}(n)) = n_l \wedge nk(\mathsf{s}(n))(n_l) = \mathsf{false} \Rightarrow$$
$$\hat{\Diamond}[(n \bullet \top \uparrow \mathsf{startEpoch_{ech}}(lts', n_l)) \wedge$$

$$\square\ trusted(\mathsf{s}(n)) = n_l \wedge \hat{\square} nk(\mathsf{s}(n))(n_l) = \mathsf{true}]$$

The other case is that $nk(\mathsf{s}(n))(n_l) = \mathsf{true}$,
By INVSA,

(23) $\Gamma \vdash_{\mathsf{ECH}} nk(\mathsf{s}(n))(n_l) = \mathsf{true} \Rightarrow$
$\hat{\diamondsuit}(\mathsf{self} \wedge n \bullet (0, \mathsf{send}_{\mathsf{pl}}(n_l, \mathrm{STATE}(lts))) \in \mathsf{ors} \wedge$
$lastts(\mathsf{s}(n)) = lts \wedge nk(\mathsf{s}'(n))(n_l) = \mathsf{true} \wedge trusted(\mathsf{s}(n)) = n_l)$

From [23] and and adding premise,

(24) $\Gamma \vdash_{\mathsf{ECH}} (n \bullet 1 \uparrow \mathsf{deliver}_{\mathsf{beb}}(n_l, \mathrm{NEWEPOCH}(t))) \wedge$
$\square\ trusted(\mathsf{s}(n)) = n_l \wedge nk(\mathsf{s}(n))(n_l) = \mathsf{true} \Rightarrow$
$\hat{\diamondsuit}(\mathsf{self} \wedge n \bullet (0, \mathsf{send}_{\mathsf{pl}}(n_l, \mathrm{STATE}(lts))) \in \mathsf{ors} \wedge$
$lastts(\mathsf{s}(n)) = lts \wedge nk(\mathsf{s}'(n))(n_l) = \mathsf{true} \wedge trusted(\mathsf{s}(n)) = n_l)$

By doing the same reasoning we can say,

(25) $\Gamma \vdash_{\mathsf{ECH}} (n \bullet 1 \uparrow \mathsf{deliver}_{\mathsf{beb}}(n_l, \mathrm{NEWEPOCH}(t))) \wedge$
$\square\ trusted(\mathsf{s}(n)) = n_l \wedge nk(\mathsf{s}(n))(n_l) = \mathsf{true} \Rightarrow$
$\hat{\diamondsuit}[\hat{\diamondsuit}[(n \bullet \top \uparrow \mathsf{startEpoch}_{\mathsf{ech}}(lts', n_l)) \wedge$
$\square\ trusted(\mathsf{s}(n)) = n_l \wedge \hat{\square} nk(\mathsf{s}(n))(n_l) = \mathsf{true}]$

Therefore, we can say,

(26) $\Gamma \vdash_{\mathsf{ECH}} (n \bullet 1 \uparrow \mathsf{deliver}_{\mathsf{beb}}(n_l, \mathrm{NEWEPOCH}(t))) \wedge$
$\square\ trusted(\mathsf{s}(n)) = n_l] \Rightarrow$
$\hat{\diamondsuit}[\hat{\diamondsuit}[(n \bullet \top \uparrow \mathsf{startEpoch}_{\mathsf{ech}}(lts', n_l)) \wedge$
$\square\ trusted(\mathsf{s}(n)) = n_l \wedge \hat{\square} nk(\mathsf{s}(n))(n_l) = \mathsf{true}]$

**Lemma 72.**
$\Gamma \vdash_{\mathsf{ECH}}$
$\exists n_l.\ n_l \in \mathsf{Correct} \wedge$
$[n \in \mathsf{Correct} \rightarrow$
$(n \bullet \top \uparrow \mathsf{startEpoch}(ts, n_l) \wedge \hat{\square}\ nk(\mathsf{s}(n))(n_l) = \mathsf{true} \wedge \square\ trusted(\mathsf{s}(n)) = n_l) \Rightarrow$
$\diamondsuit[(n \bullet \top \uparrow \mathsf{startEpoch}(ts', l) \wedge \square \neg (n \bullet \top \uparrow \mathsf{startEpoch}(ts'', l')]$

**Proof.**
By Lemma 73,

(1) $\Gamma \vdash_{\mathsf{ECH}} n \in \mathsf{Correct} \rightarrow$
$nk(\mathsf{s}(n))(n_l) = \mathsf{true} \wedge \mathsf{self} \Rightarrow$
$\hat{\diamondsuit}\diamondsuit[(n_l \bullet 0 \uparrow \mathsf{deliver}_{\mathsf{pl}}(n, \mathrm{STATE}(lts))) \wedge$
$\hat{\square} \neg (n_l \bullet 0 \uparrow \mathsf{deliver}_{\mathsf{pl}}(n, \mathrm{STATE}(lts))) \wedge$
$\hat{\boxminus} \neg (n_l \bullet 0 \uparrow \mathsf{deliver}_{\mathsf{pl}}(n, \mathrm{STATE}(lts)))]$

By Lemma 126 and [1],

(2) $\Gamma \vdash_{\mathsf{ECH}} n \in \mathsf{Correct} \rightarrow$
$(nk(\mathsf{s}(n))(n_l) = \mathsf{true} \wedge \mathsf{self}) \Rightarrow$
$\diamondsuit\hat{\square}\neg (n_l \bullet 0 \uparrow \mathsf{deliver}_{\mathsf{pl}}(n, \mathrm{STATE}(lts)))$

By INVL,

(3) $\Gamma \vdash_{\mathsf{ECH}} (n_l \bullet (1, \mathsf{broadcast}_{\mathsf{beb}}(\mathrm{NEWEPOCH}(t'))) \in \mathsf{ors} \wedge \mathsf{self}) \Rightarrow$
$(n_l \bullet 0 \uparrow \mathsf{deliver}_{\mathsf{pl}}(n, \mathrm{STATE}(lts'))) \vee$
$(n_l \bullet 0 \uparrow \mathsf{deliver}_{\mathsf{pl}}(n, \mathrm{NACK}(lts'))) \vee$
$(n_l \bullet 2 \uparrow \mathsf{trust}_{\mathsf{eld}}(n))$

By INVL,

(4) $\Gamma \vdash_{\mathsf{ECH}} (n_l \bullet 2 \uparrow \mathsf{trust}_{\mathsf{eld}}(n)) \Rightarrow$
$\quad trusted(\mathsf{s}(n_l)) \neq trusted(\mathsf{s}'(n_l))$

From [3] and [4],

(5) $\Gamma \vdash_{\mathsf{ECH}} (n_l \bullet (1, \mathsf{broadcast}_{\mathsf{beb}}(\textsc{NewEpoch}(t'))) \in \mathsf{ors} \land \mathsf{self}) \Rightarrow$
$\quad (n_l \bullet 0 \uparrow \mathsf{deliver}_{\mathsf{pl}}(n, \textsc{State}(lts'))) \lor$
$\quad (n_l \bullet 0 \uparrow \mathsf{deliver}_{\mathsf{pl}}(n, \textsc{Nack}(lts'))) \lor$
$\quad trusted(\mathsf{s}(n_l)) \neq trusted(\mathsf{s}'(n_l))$

The contra-positive of [5],

(6) $\Gamma \vdash_{\mathsf{ECH}} \neg(n_l \bullet 0 \uparrow \mathsf{deliver}_{\mathsf{pl}}(n, \textsc{State}(lts'))) \land$
$\quad \neg(n_l \bullet 0 \uparrow \mathsf{deliver}_{\mathsf{pl}}(n, \textsc{Nack}(lts'))) \land$
$\quad trusted(\mathsf{s}(n_l)) = trusted(\mathsf{s}'(n_l)) \Rightarrow$
$\quad \neg(n_l \bullet (1, \mathsf{broadcast}_{\mathsf{beb}}(\textsc{NewEpoch}(t'))) \in \mathsf{ors} \land \mathsf{self})$

That is,

(7) $\Gamma \vdash_{\mathsf{ECH}} \hat{\Box} \neg (n_l \bullet 0 \uparrow \mathsf{deliver}_{\mathsf{pl}}(n, \textsc{State}(lts'))) \land$
$\quad \Box \neg (n_l \bullet 0 \uparrow \mathsf{deliver}_{\mathsf{pl}}(n, \textsc{Nack}(lts'))) \land$
$\quad \Box (trusted(\mathsf{s}(n_l)) = trusted(\mathsf{s}'(n_l))) \Rightarrow$
$\quad \Box \neg (n_l \bullet (1, \mathsf{broadcast}_{\mathsf{beb}}(\textsc{NewEpoch}(t'))) \in \mathsf{ors} \land \mathsf{self})$

By INVL,

(8) $\Gamma \vdash_{\mathsf{ECH}} (n \bullet (0, \mathsf{send}_{\mathsf{pl}}(n_l, \textsc{Nack}(lts'))) \in \mathsf{ors} \land \mathsf{self}) \Rightarrow$
$\quad trusted(\mathsf{s}(n)) \neq n_l$

The contra-positive of [8],

(9) $\Gamma \vdash_{\mathsf{ECH}} trusted(\mathsf{s}(n)) = n_l \Rightarrow$
$\quad \neg(n \bullet (0, \mathsf{send}_{\mathsf{pl}}(n_l, \textsc{Nack}(lts'))) \in \mathsf{ors} \land \mathsf{self})$

That is,

(10) $\Gamma \vdash_{\mathsf{ECH}} \Box trusted(\mathsf{s}(n)) = n_l \Rightarrow$
$\quad \Box \neg (n \bullet (0, \mathsf{send}_{\mathsf{pl}}(n_l, \textsc{Nack}(lts'))) \in \mathsf{ors} \land \mathsf{self})$

By EXEFEOR and $\mathsf{PL}'_1$

(11) $\Gamma \vdash_{\mathsf{ECH}} \Box trusted(\mathsf{s}(n)) = n_l \Rightarrow$
$\quad \Diamond \Box \neg (n_l \bullet 0 \uparrow \mathsf{deliver}_{\mathsf{pl}}(n, \textsc{Nack}(lts')))$

From [2] and adding a premise,

(12) $\Gamma \vdash_{\mathsf{ECH}} n \in \mathsf{Correct} \rightarrow$
$\quad (nk(\mathsf{s}(n))(n_l) = \mathsf{true} \land \mathsf{self}) \land \Box (trusted(\mathsf{s}(n)) = n_l) \Rightarrow$
$\quad \Diamond \hat{\Box} \neg (n_l \bullet 0 \uparrow \mathsf{deliver}_{\mathsf{pl}}(n, \textsc{State}(lts))) \land$
$\quad \Box (trusted(\mathsf{s}(n)) = trusted(\mathsf{s}'(n)) = n_l) \land$
$\quad \Box (trusted(\mathsf{s}(n)) = n_l)$

From [12] and [11],

(13) $\Gamma \vdash_{\mathsf{ECH}} n \in \mathsf{Correct} \rightarrow$
$\quad (nk(\mathsf{s}(n))(n_l) = \mathsf{true} \land \mathsf{self}) \land \Box (trusted(\mathsf{s}(n)) = n_l) \Rightarrow$
$\quad \Diamond \hat{\Box} \neg (n_l \bullet 0 \uparrow \mathsf{deliver}_{\mathsf{pl}}(n, \textsc{State}(lts))) \land$
$\quad \Box (trusted(\mathsf{s}(n)) = trusted(\mathsf{s}'(n)) = n_l) \land$
$\quad \Diamond \Box \neg (n_l \bullet 0 \uparrow \mathsf{deliver}_{\mathsf{pl}}(n, \textsc{Nack}(lts')))$

That is,

(14) $\Gamma \vdash_{\mathsf{ECH}} n \in \mathsf{Correct} \rightarrow$
$\quad (nk(\mathsf{s}(n))(n_l) = \mathsf{true} \land \mathsf{self}) \land \Box (trusted(\mathsf{s}(n)) = n_l) \Rightarrow$
$\quad \Diamond [\Box \neg (n_l \bullet 0 \uparrow \mathsf{deliver}_{\mathsf{pl}}(n, \textsc{State}(lts))) \land$
$\quad \Box (trusted(\mathsf{s}(n)) = trusted(\mathsf{s}'(n)) = n_l) \land$
$\quad \Box \neg (n_l \bullet 0 \uparrow \mathsf{deliver}_{\mathsf{pl}}(n_l, \textsc{Nack}(lts')))]$

[7] and [14],

(15) $\Gamma \vdash_{\mathsf{ECH}} n \in \mathsf{Correct} \rightarrow$
$\qquad (nk(\mathsf{s}(n))(n_l) = \mathsf{true} \wedge \mathsf{self}) \wedge \Box(trusted(\mathsf{s}(n)) = n_l) \Rightarrow$
$\qquad\qquad \Diamond[\Box\neg(n_l \bullet (1, \mathsf{broadcast}_{\mathsf{beb}}(\mathrm{NEWEPOCH}(t'))) \in \mathsf{ors} \wedge \mathsf{self}]$

By ExEFEOR and $\mathrm{BEB}'_1$

(16) $\Gamma \vdash_{\mathsf{ECH}} n \in \mathsf{Correct} \rightarrow$
$\qquad (nk(\mathsf{s}(n))(n_l) = \mathsf{true} \wedge \mathsf{self}) \wedge \Box(trusted(\mathsf{s}(n)) = n_l) \Rightarrow$
$\qquad\qquad \Diamond[\Diamond\Box\neg(n \bullet 1 \uparrow \mathsf{deliver}_{\mathsf{beb}}(n_l, \mathrm{NEWEPOCH}(t'))]$

By INVL,

(17) $\Gamma \vdash_{\mathsf{ECH}} (n \bullet \mathsf{startEpoch}_{\mathsf{ech}}(t, n')\mathsf{ois} \wedge \mathsf{self}) \Rightarrow$
$\qquad (n \bullet 1 \uparrow \mathsf{deliver}_{\mathsf{beb}}(n_l, \mathrm{NEWEPOCH}(t')))$

The contra-positive of [17],

(18) $\Gamma \vdash_{\mathsf{ECH}} \neg(n \bullet 1 \uparrow \mathsf{deliver}_{\mathsf{beb}}(n_l, \mathrm{NEWEPOCH}(t'))) \Rightarrow$
$\qquad \neg(n \bullet \mathsf{startEpoch}_{\mathsf{ech}}(t, n')\mathsf{ois} \wedge \mathsf{self})$

From [16] and [18],

(19) $\Gamma \vdash_{\mathsf{ECH}} (n \bullet \top \uparrow \mathsf{startEpoch}(ts, n_l) \wedge$
$\qquad \hat{\Box}\, nk(\mathsf{s}(n))[n_l] = \mathsf{true} \wedge \Box\, trusted(\mathsf{s}(n)) = n_l) \Rightarrow$
$\qquad (n \bullet \top \uparrow \mathsf{startEpoch}(ts, n_l) \wedge$
$\qquad \Diamond\Box\neg(n \bullet \top \uparrow \mathsf{startEpoch}(ts'', l'))$

By Lemma 122,

(20) $\Gamma \vdash_{\mathsf{ECH}} (n \bullet \top \uparrow \mathsf{startEpoch}(ts, n_l) \wedge$
$\qquad \hat{\Box}\, nk(\mathsf{s}(n))[n_l] = \mathsf{true} \wedge \Box\, trusted(\mathsf{s}(n)) = n_l) \Rightarrow$
$\qquad \Diamond[(n \bullet \top \uparrow \mathsf{startEpoch}(ts, n_l) \wedge$
$\qquad\qquad \Box\neg(n \bullet \top \uparrow \mathsf{startEpoch}(ts'', l'))]$

**Lemma 73.**
$\Gamma \vdash_{\mathsf{ECH}}$
$\qquad n \in \mathsf{Correct} \rightarrow$
$\qquad\qquad nk(\mathsf{s}(n))[n_l] = \mathsf{true} \wedge \mathsf{self} \Rightarrow$
$\qquad\qquad\qquad \hat{\Diamond}\Diamond[(n_l \bullet 0 \uparrow \mathsf{deliver}_{\mathsf{pl}}(n, \mathrm{STATE}(lts))) \wedge$
$\qquad\qquad\qquad \hat{\Box}\neg(n_l \bullet 0 \uparrow \mathsf{deliver}_{\mathsf{pl}}(n, \mathrm{STATE}(lts))) \wedge$
$\qquad\qquad\qquad \hat{\boxminus}\neg(n_l \bullet 0 \uparrow \mathsf{deliver}_{\mathsf{pl}}(n, \mathrm{STATE}(lts)))]$

By INVSASE with $S : nk(\mathsf{s}(n))(n_l) = \mathsf{true}$

(1) $\Gamma \vdash_{\mathsf{ECH}} \mathbb{S}\, nk(\mathsf{s}(n))[n_l] = \mathsf{true} \wedge \mathsf{self} \Rightarrow$
$\qquad \hat{\Diamond}(n \bullet (0, \mathsf{send}_{\mathsf{pl}}(\mathrm{STATE}(lts))) \in \mathsf{ors} \wedge \mathsf{self} \wedge$
$\qquad lastts(\mathsf{s}(n)) = lts \wedge nk(\mathsf{s}(n))(n_l) = \mathsf{false} \wedge$
$\qquad nk(\mathsf{s}'(n))(n_l) = \mathsf{true} \wedge trusted(\mathsf{s}(n)) = n_l)$

By INVSSE''

(2) $\Gamma \vdash_{\mathsf{ECH}} \mathbb{S}\, nk(\mathsf{s}'(n))(n_l) = \mathsf{true} \Rightarrow \hat{\Box}(nk(\mathsf{s}(n))(n_l) = \mathsf{true})$

By INVUSSE

(3) $\Gamma \vdash_{\mathsf{ECH}} \mathbb{S}\, nk(\mathsf{s}(n))(n_l) = \mathsf{true} \Rightarrow \Box(nk(\mathsf{s}(n))(n_l) = \mathsf{true})$

By Lemma 125,

(4) $\Gamma \vdash_{\mathsf{ECH}} \mathbb{S}\, nk(\mathsf{s}(n))(n_l) = \mathsf{false} \Rightarrow \boxminus(nk(\mathsf{s}(n))(n_l) = \mathsf{false})$

From [1], [2] [4] and POSTPRE,

(5) $\Gamma \vdash_{\mathsf{ECH}} \mathbb{S}\, nk(\mathsf{s}(n))[n_l] = \mathsf{true} \wedge \mathsf{self} \Rightarrow$
$\qquad \hat{\Diamond}[n \bullet (0, \mathsf{send}_{\mathsf{pl}}(\mathrm{STATE}(lts))) \in \mathsf{ors} \wedge \mathsf{self} \wedge$

$$\hat{\boxminus}(nk(\mathsf{s}'(n))(n_l) = \mathsf{false}) \land$$
$$\hat{\Box}(nk(\mathsf{s}(n))(n_l) = \mathsf{true})]$$

By INVL,

(6) $\Gamma \vdash_{\mathsf{ECH}} (n \bullet (0, \mathsf{send}_{\mathsf{pl}}(n_l, \mathrm{STATE}(lts))) \in \mathsf{ors} \land \mathsf{self}) \Rightarrow$
$$nk(\mathsf{s}'(n))(n_l) = \mathsf{true}$$

The contra-positive of [6],

(7) $\Gamma \vdash_{\mathsf{ECH}} nk(\mathsf{s}'(n))(n_l) = \mathsf{false} \Rightarrow$
$$\neg(n \bullet (0, \mathsf{send}_{\mathsf{pl}}(n_l, \mathrm{STATE}(lts))) \in \mathsf{ors} \land \mathsf{self})$$

That is,

(8) $\Gamma \vdash_{\mathsf{ECH}} \hat{\boxminus} nk(\mathsf{s}'(n))(n_l) = \mathsf{false} \Rightarrow$
$$\hat{\boxminus}(n \bullet (0, \mathsf{send}_{\mathsf{pl}}(n_l, \mathrm{STATE}(lts))) \notin \mathsf{ors} \land \mathsf{self})$$

By INVL,

(9) $\Gamma \vdash_{\mathsf{ECH}} (n \bullet (0, \mathsf{send}_{\mathsf{pl}}(n_l, \mathrm{STATE}(lts))) \in \mathsf{ors} \land \mathsf{self}) \Rightarrow$
$$nk(\mathsf{s}(n))(n_l) = \mathsf{false}$$

The contra-positive of [9],

(10) $\Gamma \vdash_{\mathsf{ECH}} nk(\mathsf{s}(n))(n_l) = \mathsf{true} \Rightarrow$
$$\neg(n \bullet (0, \mathsf{send}_{\mathsf{pl}}(n_l, \mathrm{STATE}(lts))) \in \mathsf{ors}) \land \mathsf{self}$$

That is,

(11) $\Gamma \vdash_{\mathsf{ECH}} \hat{\Box}(nk(\mathsf{s}(n))(n_l) = \mathsf{true}) \Rightarrow$
$$\hat{\Box}(n \bullet (0, \mathsf{send}_{\mathsf{pl}}(n_l, \mathrm{STATE}(lts))) \notin \mathsf{ors}) \land \mathsf{self}$$

From [5], [8], [11] and SINV,

(12) $\Gamma \vdash_{\mathsf{ECH}} nk(\mathsf{s}(n))(n_l) = \mathsf{true} \land \mathsf{self} \Rightarrow$
$$\hat{\diamondsuit}[(n \bullet (0, \mathsf{send}_{\mathsf{pl}}(n_l, \mathrm{STATE}(lts))) \in \mathsf{ors}) \land \mathsf{self} \land$$
$$\hat{\Box}((n \bullet (0, \mathsf{send}_{\mathsf{pl}}(n_l, \mathrm{STATE}(lts))) \notin \mathsf{ors}) \land \mathsf{self}) \land$$
$$\hat{\boxminus}((n \bullet (0, \mathsf{send}_{\mathsf{pl}}(n_l, \mathrm{STATE}(lts))) \notin \mathsf{ors}) \land \mathsf{self})]$$

By UNIOR and $\mathsf{PL}'_1$,

(13) $\Gamma \vdash_{\mathsf{ECH}} n \in \mathsf{Correct} \rightarrow$
$$nk(\mathsf{s}(n))(n_l) = \mathsf{true} \land \mathsf{self} \Rightarrow$$
$$\Leftrightarrow \diamondsuit[(n \bullet 0 \downarrow \mathsf{send}_{\mathsf{pl}}(n_l, \mathrm{STATE}(lts))) \land$$
$$\hat{\Box} \neg(n \bullet 0 \downarrow \mathsf{send}_{\mathsf{pl}}(n_l, \mathrm{STATE}(lts))) \land$$
$$\hat{\boxminus} \neg(n \bullet 0 \downarrow \mathsf{send}_{\mathsf{pl}}(n_l, \mathrm{STATE}(lts)))]$$

By $\mathsf{PL}'_2$,

(14) $\Gamma \vdash_{\mathsf{ECH}} n \in \mathsf{Correct} \rightarrow$
$$nk(\mathsf{s}(n))(n_l) = \mathsf{true} \land \mathsf{self} \Rightarrow$$
$$\Leftrightarrow \diamondsuit[(n_l \bullet 0 \uparrow \mathsf{deliver}_{\mathsf{pl}}(n, \mathrm{STATE}(lts))) \land$$
$$\hat{\Box} \neg(n_l \bullet 0 \uparrow \mathsf{deliver}_{\mathsf{pl}}(n, \mathrm{STATE}(lts))) \land$$
$$\hat{\boxminus} \neg(n_l \bullet 0 \uparrow \mathsf{deliver}_{\mathsf{pl}}(n, \mathrm{STATE}(lts)))]$$

### 5.3.9 Uniform Consensus

**Definition 22.**
$\Gamma = \text{ECH}'_1, \text{ECH}'_2, \text{ECH}'_3, \text{EC}'_1, \text{EC}'_2, \text{EC}'_3, \text{EC}'_4$

$\text{ECH}'_1 = \text{lower}(0, \text{ECH}_1) =$
$n \in \text{Correct} \rightarrow$
$(n \bullet 0 \uparrow \text{startEpoch}_{\text{ech}}(ts, n_l)) \Rightarrow$
$\hat{\Box}(n \bullet 0 \uparrow \text{startEpoch}_{\text{ech}}(ts', n'_l) \rightarrow ts' > ts)$

$\text{ECH}'_2 = \text{lower}(0, \text{ECH}_2) =$
$n \in \text{Correct} \land n' \in \text{Correct} \rightarrow$
$(n \bullet 0 \uparrow \text{startEpoch}_{\text{ech}}(ts, n_l)) \Rightarrow$
$(n' \bullet 0 \uparrow \text{startEpoch}_{\text{ech}}(ts, n'_l) \Rightarrow n_l = n'_l)$

$\text{ECH}'_3 = \text{lower}(0, \text{ECH}_3) =$
$\exists ts, n_l.\ n_l \in \text{Correct} \land$
$[n \in \text{Correct} \rightarrow$
$\quad \Diamond [(n \bullet 0 \uparrow \text{startEpoch}_{\text{ech}}(ts, n_l)) \land$
$\quad\quad \hat{\Box} \neg (n \bullet 0 \uparrow \text{startEpoch}_{\text{ech}}(ts', n'_l))]]$

$\text{EC}'_1 = \text{lower}(1, \text{EC}_1) =$
$(n \bullet 1 \uparrow \text{decide}_{\text{ec}}(v)) \Rightarrow$
$\exists n'.\ \Diamondtail(n' \bullet 1 \downarrow \text{propose}_{\text{ec}}(v))$

$\text{EC}'_2 = \text{lower}(1, \text{EC}_2) =$
$n \bullet 1 \uparrow \text{decide}_{\text{ec}}(v) \land$
$\Diamond(n' \bullet 1 \uparrow \text{decide}_{\text{ec}}(v')) \Rightarrow$
$v = v'$

$\text{EC}'_3 = \text{lower}(1, \text{EC}_3) =$
$(n \bullet 1 \uparrow \text{decide}_{\text{ec}}(v)) \Rightarrow$
$\hat{\Box} \neg (n \bullet 1 \uparrow \text{decide}_{\text{ec}}(v'))$

$\text{EC}'_4 = \text{lower}(1, \text{EC}_4) =$
$|\text{Correct}| > |\mathbb{N}|/2 \land n \in \text{Correct} \rightarrow$
$(n \bullet 1 \downarrow \text{propose}_{\text{ec}}(v)) \Rightarrow$
$(n \bullet 1 \downarrow \text{epoch}_{\text{ec}}(n, ts)) \land \Box \neg (n \bullet 1 \downarrow \text{epoch}_{\text{ec}}(n', ts')) \Rightarrow$
$\forall n'.\ n' \in \text{Correct} \rightarrow \Diamondtail \Diamond \exists v'.\ (n' \bullet 1 \uparrow \text{decide}_{\text{ec}}(v'))$

**Theorem 26** ($UC_1$: Termination).
Every correct node eventually decides some value.
$\Gamma \vdash_{\mathsf{UC}} |\mathsf{Correct}| > |\mathbb{N}|/2 \wedge n \in \mathsf{Correct} \to$
$[\forall n'. n' \in \mathsf{Correct} \to \exists v.\ v \neq \bot \wedge \Diamond(n' \bullet \top \downarrow \mathsf{propose}_{\mathsf{uc}}(v))] \Rightarrow$
$\exists v.\ \Diamond\Diamond(n \bullet \top \uparrow \mathsf{decide}_{\mathsf{uc}}(v))$
where $\Gamma$ is defined in Definition 22.

**Proof.**
By IRSE,
  (1) $\Gamma' \vdash_{\mathsf{UCC}} \mathsf{S}\ (n_l \bullet \top \downarrow \mathsf{propose}_{\mathsf{uc}}(v) \wedge v \neq \bot) \Rightarrow$
     $prop(\mathsf{s}'(n_l)) \neq \bot$
By INVSSE,
  (2) $\Gamma' \vdash_{\mathsf{UCC}} \mathsf{S}\ prop(\mathsf{s}(n_l)) \neq \bot \Rightarrow$
     $\Box(prop(\mathsf{s}(n_l)) \neq \bot)$
From [1], [2] and POSTPRE,
  (3) $\Gamma' \vdash_{\mathsf{UCC}} \mathsf{S}\ (n_l \bullet \top \downarrow \mathsf{propose}_{\mathsf{uc}}(v) \wedge v \neq \bot) \Rightarrow$
     $\hat{\Box}(prop(\mathsf{s}(n_l)) \neq \bot)$
By using the premise and [3],
  (4) $\Gamma' \vdash_{\mathsf{UCC}} \mathsf{S}\ |\mathsf{Correct}| > |\mathbb{N}|/2 \wedge n \in \mathsf{Correct} \to$
     $[\forall n'.n' \in \mathsf{Correct} \to$
     $\exists v.\ v \neq \bot \wedge \Diamond(n' \bullet \top \downarrow \mathsf{propose}_{\mathsf{uc}}(v))] \Rightarrow$
     $\hat{\Box}(prop(\mathsf{s}(n')) \neq \bot)$
By $\mathsf{ECH}_3'$ and instantiating $n$ to $n_l$,
  (5) $\Gamma' \vdash_{\mathsf{UCC}} \exists ts, n_l.\ n_l \in \mathsf{Correct} \wedge$
     $\Diamond[(n_l \bullet 0 \uparrow \mathsf{startEpoch}_{\mathsf{ech}}(ts, n_l)) \wedge$
     $\hat{\Box} \neg (n_l \bullet 0 \uparrow \mathsf{startEpoch}_{\mathsf{ech}}(ts', n_l'))]$
From [4] and [5],
  (6) $\Gamma' \vdash_{\mathsf{UCC}} \exists ts, n_l.\ n_l \in \mathsf{Correct} \wedge$
     $\Diamond[(n_l \bullet 0 \uparrow \mathsf{startEpoch}_{\mathsf{ech}}(ts, n_l)) \wedge$
     $\hat{\Box} \neg (n_l \bullet 0 \uparrow \mathsf{startEpoch}_{\mathsf{ech}}(ts', n_l'))] \wedge$
     $\Diamond\Diamond\hat{\Box}(prop(\mathsf{s}(n_l)) \neq \bot)$
That is,
  (7) $\Gamma' \vdash_{\mathsf{UCC}} \exists ts, n_l.\ n_l \in \mathsf{Correct} \wedge$
     $\Diamond[(n_l \bullet 0 \uparrow \mathsf{startEpoch}_{\mathsf{ech}}(ts, n_l)) \wedge$
     $[(prop(\mathsf{s}(n_l)) \neq \bot) \vee \Diamond(prop(\mathsf{s}(n_l)) \neq \bot)] \wedge$
     $\hat{\Box} \neg (n_l \bullet 0 \uparrow \mathsf{startEpoch}_{\mathsf{ech}}(ts', n_l'))]$
That is,
  (8) $\Gamma' \vdash_{\mathsf{UCC}} \exists ts, n_l.\ n_l \in \mathsf{Correct} \wedge$
     $\Diamond[[(n_l \bullet 0 \uparrow \mathsf{startEpoch}_{\mathsf{ech}}(ts, n_l)) \wedge (prop(\mathsf{s}(n_l)) \neq \bot) \wedge$
     $\hat{\Box} \neg (n_l \bullet 0 \uparrow \mathsf{startEpoch}_{\mathsf{ech}}(ts', n_l'))] \vee$
     $[(n_l \bullet 0 \uparrow \mathsf{startEpoch}_{\mathsf{ech}}(ts, n_l)) \wedge (prop(\mathsf{s}(n_l)) = \bot) \wedge \hat{\Diamond}(prop(\mathsf{s}(n_l)) \neq \bot) \wedge$
     $\hat{\Box} \neg (n_l \bullet 0 \uparrow \mathsf{startEpoch}_{\mathsf{ech}}(ts', n_l'))]]$
By INVSA,
  (9) $\Gamma' \vdash_{\mathsf{UCC}} prop(\mathsf{s}(n_l)) \neq \bot \wedge \mathsf{self} \Rightarrow$
     $\hat{\Diamond}(n_l \bullet (1, \mathsf{propose}_{\mathsf{ec}}(v)) \in \mathsf{ors} \wedge \mathsf{self})$
By using Lemma 74 and Lemma 74 on [8] and [9],
  (10) $\Gamma' \vdash_{\mathsf{UCC}} \exists ts, n_l.\ n_l \in \mathsf{Correct} \wedge$

$$[n_l \in \mathsf{Correct} \to \Diamond[\hat{\Diamond}(n_l \bullet (1, \mathsf{propose}_{\mathsf{ec}}(v)) \in \mathsf{ors} \land \mathsf{self}) \land$$
$$(n_l \bullet (1, \mathsf{epoch}_{\mathsf{ec}}(n_l, ts)) \in \mathsf{ors} \land \mathsf{self}) \land$$
$$\Box \neg (n_l \bullet (1, \mathsf{epoch}_{\mathsf{ec}}(n_l', ts')) \in \mathsf{ors} \land \mathsf{self})]]$$

That is,

(11) $\Gamma' \vdash_{\mathsf{UCC}} \exists ts, n_l.\ n_l \in \mathsf{Correct} \land$
$$[n_l \in \mathsf{Correct} \to \Diamond \hat{\Diamond}[(n_l \bullet (1, \mathsf{propose}_{\mathsf{ec}}(v)) \in \mathsf{ors} \land \mathsf{self}) \land$$
$$\Diamond[(n_l \bullet (1, \mathsf{epoch}_{\mathsf{ec}}(n_l, ts)) \in \mathsf{ors} \land \mathsf{self}) \land$$
$$\Box \neg (n_l \bullet (1, \mathsf{epoch}_{\mathsf{ec}}(n_l', ts')) \in \mathsf{ors} \land \mathsf{self})]]]$$

By [11] and ExeOrderOR and ExeFEOR,

(12) $\Gamma' \vdash_{\mathsf{UCC}} \exists ts, n_l.\ n_l \in \mathsf{Correct} \land$
$$[n_l \in \mathsf{Correct} \to \Diamond \hat{\Diamond}[(n_l \bullet \downarrow \mathsf{propose}_{\mathsf{ec}}(v)) \land$$
$$\hat{\Diamond}[(n_l \bullet 1 \downarrow \mathsf{epoch}_{\mathsf{ec}}(n_l, ts)) \land \Diamond \Box \neg (n_l \bullet 1 \downarrow \mathsf{epoch}_{\mathsf{ec}}(n_l', ts'))]]]$$

by IR,

(13) $\Gamma' \vdash_{\mathsf{UCC}} (n \bullet \top \downarrow \mathsf{propose}_{\mathsf{uc}}(v)) \Rightarrow$
$$(n \bullet (1, \mathsf{propose}_{\mathsf{ec}}(v)) \in \mathsf{ors} \land \mathsf{self})$$

By OR

(14) $\Gamma' \vdash_{\mathsf{UCC}} (n \bullet (1, \mathsf{propose}_{\mathsf{ec}}(v)) \in \mathsf{ors} \land \mathsf{self}) \Rightarrow$
$$\hat{\Diamond}(n \bullet 1 \downarrow \mathsf{propose}_{\mathsf{ec}}(v))$$

From [13] and [14],

(15) $\Gamma' \vdash_{\mathsf{UCC}} (n \bullet \top \downarrow \mathsf{propose}_{\mathsf{uc}}(v)) \Rightarrow$
$$\hat{\Diamond}(n \bullet 1 \downarrow \mathsf{propose}_{\mathsf{ec}}(v))$$

From the premise and [15] we know that,

(16) $\Gamma' \vdash_{\mathsf{UCC}} \forall n'.n' \in \mathsf{Correct} \to \exists v.\ \Diamond(n \bullet \top \downarrow \mathsf{propose}_{\mathsf{uc}}(v)) \Rightarrow$
$$\Diamond \hat{\Diamond}(n \bullet 1 \downarrow \mathsf{propose}_{\mathsf{ec}}(v))$$

From [12] and [16],

(17) $\Gamma' \vdash_{\mathsf{UCC}} n \in \mathsf{Correct} \to$
$$\forall n'.n' \in \mathsf{Correct} \to \exists v.\ \Diamond(n \bullet \top \downarrow \mathsf{propose}_{\mathsf{uc}}(v)) \Rightarrow$$
$$\Diamond \hat{\Diamond}(n_l \bullet 1 \downarrow \mathsf{propose}_{\mathsf{ec}}(v)) \land$$
$$\Diamond[(n_l \bullet 1 \downarrow \mathsf{epoch}_{\mathsf{ec}}(n_l, ts)) \land \Box \neg (n_l \bullet 1 \downarrow \mathsf{epoch}_{\mathsf{ec}}(n_l', ts'))]$$

From $\mathsf{EC}_4'$,

(18) $\Gamma' \vdash_{\mathsf{UCC}} |\mathsf{Correct}| > |\mathbb{N}|/2 \land n \in \mathsf{Correct} \to$
$$(n \bullet 1 \downarrow \mathsf{propose}_{\mathsf{ec}}(v)) \Rightarrow$$
$$(n \bullet 1 \downarrow \mathsf{epoch}_{\mathsf{ec}}(n, ts)) \land \Box \neg (n \bullet 1 \downarrow \mathsf{epoch}_{\mathsf{ec}}(n', ts')) \Rightarrow$$
$$\forall n'.\ n' \in \mathsf{Correct} \to \Diamond \Diamond \exists v'.\ (n' \bullet 1 \uparrow \mathsf{decide}_{\mathsf{ec}}(v'))$$

From [17] and [18],

(19) $\Gamma' \vdash_{\mathsf{UCC}} n \in \mathsf{Correct} \to$
$$\forall n'.n' \in \mathsf{Correct} \to \exists v.\ \Diamond(n \bullet \top \downarrow \mathsf{propose}_{\mathsf{uc}}(v)) \Rightarrow$$
$$\Diamond \Diamond [\Diamond \Diamond \exists v'.\ (n' \bullet 1 \uparrow \mathsf{decide}_{\mathsf{ec}}(v'))]$$

That is,

(20) $\Gamma' \vdash_{\mathsf{UCC}} n \in \mathsf{Correct} \to$
$$\forall n'.n' \in \mathsf{Correct} \to \exists v.\ \Diamond(n \bullet \top \downarrow \mathsf{propose}_{\mathsf{uc}}(v)) \Rightarrow$$
$$\Diamond \Diamond [\exists v'.\ (n' \bullet 1 \uparrow \mathsf{decide}_{\mathsf{ec}}(v'))]$$

**Lemma 74.**
$\Gamma \vdash_{\mathsf{UC}}$
$$\exists ts, n_l.\ n_l \in \mathsf{Correct} \land$$
$$\Diamond[[(n_l \bullet 0 \uparrow \mathsf{startEpoch}_{\mathsf{ech}}(ts, n_l)) \land (prop(\mathsf{s}(n_l)) \ne \bot) \land$$
$$\hat{\Box}\neg(n_l \bullet 0 \uparrow \mathsf{startEpoch}_{\mathsf{ech}}(ts', n_l'))] \Rightarrow$$
$$\hat{\Diamond}\hat{\Diamond}[[(n_l \bullet (1, \mathsf{epoch}_{\mathsf{ec}}(n_l, ts)) \in \mathsf{ors} \land started(\mathsf{s}'(n_l)) = \mathsf{true} \land \mathsf{self})] \land$$
$$\hat{\Box}\neg(n_l \bullet (1, \mathsf{epoch}_{\mathsf{ec}}(n_l, ts)) \in \mathsf{ors} \land \mathsf{self})]$$

By II,
(1) $\Gamma' \vdash_{\mathsf{UCC}} (n_l \bullet 0 \uparrow \mathsf{startEpoch}_{\mathsf{ech}}(ts, n_l)) \land prop(\mathsf{s}(n_l)) \ne \bot \Rightarrow$
$$[n_l \bullet (1, \mathsf{epoch}_{\mathsf{ec}}(n_l, ts)) \in \mathsf{ors} \land started(\mathsf{s}'(n_l)) = \mathsf{true} \land \mathsf{self}]$$

By usinng premise and [1],
(2) $\Gamma' \vdash_{\mathsf{UCC}} \exists ts, n_l.\ n_l \in \mathsf{Correct} \land$
$$\Diamond[[(n_l \bullet 0 \uparrow \mathsf{startEpoch}_{\mathsf{ech}}(ts, n_l)) \land (prop(\mathsf{s}(n_l)) \ne \bot) \land$$
$$\hat{\Box}\neg(n_l \bullet 0 \uparrow \mathsf{startEpoch}_{\mathsf{ech}}(ts', n_l'))] \Rightarrow$$
$$\Diamond[[n_l \bullet (1, \mathsf{epoch}_{\mathsf{ec}}(n_l, ts)) \in \mathsf{ors} \land started(\mathsf{s}'(n_l)) = \mathsf{true} \land \mathsf{self}] \land$$
$$\hat{\Box}\neg(n_l \bullet 0 \uparrow \mathsf{startEpoch}_{\mathsf{ech}}(ts', n_l'))]$$

By INVL,
(3) $\Gamma' \vdash_{\mathsf{UCC}} started(\mathsf{s}(n_l)) = \mathsf{true} \land started(\mathsf{s}'(n_l)) = \mathsf{false} \Rightarrow$
$$(n_l \bullet 0 \uparrow \mathsf{startEpoch}_{\mathsf{ech}}(ts', n_l'))$$

The contra-positive of [3],
(4) $\Gamma' \vdash_{\mathsf{UCC}} \neg(n_l \bullet 0 \uparrow \mathsf{startEpoch}_{\mathsf{ech}}(ts', n_l')) \Rightarrow$
$$started(\mathsf{s}(n_l)) = \mathsf{false} \lor started(\mathsf{s}'(n_l)) = \mathsf{true}$$

From [2] and [4],
(5) $\Gamma' \vdash_{\mathsf{UCC}} \exists ts, n_l.\ n_l \in \mathsf{Correct} \land$
$$\Diamond[[(n_l \bullet 0 \uparrow \mathsf{startEpoch}_{\mathsf{ech}}(ts, n_l)) \land (prop(\mathsf{s}(n_l)) \ne \bot) \land$$
$$\hat{\Box}\neg(n_l \bullet 0 \uparrow \mathsf{startEpoch}_{\mathsf{ech}}(ts', n_l'))] \Rightarrow$$
$$\Diamond[[n_l \bullet (1, \mathsf{epoch}_{\mathsf{ec}}(n_l, ts)) \in \mathsf{ors} \land started(\mathsf{s}'(n_l)) = \mathsf{true} \land \mathsf{self}] \land$$
$$\hat{\Box}(started(\mathsf{s}(n_l)) = \mathsf{false} \lor started(\mathsf{s}'(n_l)) = \mathsf{true})]$$

By using induction and POSTPRE on [5] (considering the conjunction in $started(\mathsf{s}'(n_l)) = \mathsf{true}$ and $\hat{\Box}(started(\mathsf{s}(n_l)) = \mathsf{false} \lor started(\mathsf{s}'(n_l)) = \mathsf{true})$ for the base case of the induction),
(6) $\Gamma' \vdash_{\mathsf{UCC}} \exists ts, n_l.\ n_l \in \mathsf{Correct} \land$
$$\Diamond[[(n_l \bullet 0 \uparrow \mathsf{startEpoch}_{\mathsf{ech}}(ts, n_l)) \land (prop(\mathsf{s}(n_l)) \ne \bot) \land$$
$$\hat{\Box}\neg(n_l \bullet 0 \uparrow \mathsf{startEpoch}_{\mathsf{ech}}(ts', n_l'))] \Rightarrow$$
$$\Diamond[[n_l \bullet (1, \mathsf{epoch}_{\mathsf{ec}}(n_l, ts)) \in \mathsf{ors} \land started(\mathsf{s}'(n_l)) = \mathsf{true} \land \mathsf{self}] \land$$
$$\hat{\Box}(started(\mathsf{s}'(n_l)) = \mathsf{true})]$$

By INVL,
(7) $\Gamma' \vdash_{\mathsf{UCC}} (n \bullet (1, \mathsf{epoch}_{\mathsf{ec}}(n_l, ts)) \in \mathsf{ors} \land n = n_l \land \mathsf{self}) \Rightarrow$
$$started(\mathsf{s}'(n)) \ne started(\mathsf{s}(n))$$

That is,
(8) $\Gamma' \vdash_{\mathsf{UCC}} (n_l \bullet (1, \mathsf{epoch}_{\mathsf{ec}}(n_l, ts)) \in \mathsf{ors} \land \mathsf{self}) \Rightarrow$
$$started(\mathsf{s}'(n_l)) \ne started(\mathsf{s}(n_l))$$

The contra-pasitive of [8],
(9) $\Gamma' \vdash_{\mathsf{UCC}} started(\mathsf{s}'(n_l)) = started(\mathsf{s}(n_l)) \Rightarrow$
$$\neg(n_l \bullet (1, \mathsf{epoch}_{\mathsf{ec}}(n_l, ts)) \in \mathsf{ors} \land \mathsf{self})$$

From [6] and [9]
  (10) $\Gamma' \vdash_{\mathsf{UCC}} \exists ts, n_l.\ n_l \in \mathsf{Correct} \land$
  $\Diamond[[(n_l \bullet 0 \uparrow \mathsf{startEpoch}_{\mathsf{ech}}(ts, n_l)) \land (prop(\mathsf{s}(n')) \ne \bot) \land$
  $\hat{\Box}\neg(n_l \bullet 0 \uparrow \mathsf{startEpoch}_{\mathsf{ech}}(ts', n'_l))] \Rightarrow$
  $\hat{\Diamond}[[n_l \bullet (1, \mathsf{epoch}_{\mathsf{ec}}(n_l, ts)) \in \mathsf{ors} \land started(\mathsf{s}'(n_l)) = \mathsf{true} \land \mathsf{self}] \land$
  $\hat{\Box}\neg(n_l \bullet (1, \mathsf{epoch}_{\mathsf{ec}}(n_l, ts)) \in \mathsf{ors} \land \mathsf{self})]$

**Lemma 75.**
$\Gamma \vdash_{\mathsf{UC}}$
  $\exists ts, n_l.\ n_l \in \mathsf{Correct} \land$
  $[(n_l \bullet 0 \uparrow \mathsf{startEpoch}_{\mathsf{ech}}(ts, n_l)) \land (prop(\mathsf{s}(n_l)) = \bot) \land \hat{\Diamond}(prop(\mathsf{s}(n_l)) \ne \bot) \land$
  $\hat{\Box}\neg(n_l \bullet 0 \uparrow \mathsf{startEpoch}_{\mathsf{ech}}(ts', n'_l))] \Rightarrow$
  $\hat{\Diamond}\hat{\Diamond}[[(n_l \bullet (1, \mathsf{epoch}_{\mathsf{ec}}(n_l, ts)) \in \mathsf{ors} \land started(\mathsf{s}'(n_l)) = \mathsf{true} \land \mathsf{self}] \land$
  $\Box\neg(n_l \bullet (1, \mathsf{epoch}_{\mathsf{ec}}(n_l, ts)) \in \mathsf{ors} \land \mathsf{self})]$

By INVL,
  (1) $\Gamma' \vdash_{\mathsf{UCC}} (n \bullet 0 \uparrow \mathsf{startEpoch}_{\mathsf{ech}}(ts, n_l)) \land prop(\mathsf{s}(n)) = \bot \land n = n_l \land \mathsf{self} \Rightarrow$
  $started(\mathsf{s}'(n)) = \mathsf{false}$
That is,
  (2) $\Gamma' \vdash_{\mathsf{UCC}} (n_l \bullet 0 \uparrow \mathsf{startEpoch}_{\mathsf{ech}}(ts, n_l)) \land prop(\mathsf{s}(n_l)) = \bot \land \mathsf{self} \Rightarrow$
  $started(\mathsf{s}'(n_l)) = \mathsf{false}$
By using premise and [2],
  (3) $\Gamma' \vdash_{\mathsf{UCC}} \exists ts, n_l.\ n_l \in \mathsf{Correct} \land$
  $[(n_l \bullet 0 \uparrow \mathsf{startEpoch}_{\mathsf{ech}}(ts, n_l)) \land (prop(\mathsf{s}(n_l)) = \bot) \land \hat{\Diamond}(prop(\mathsf{s}(n_l)) \ne \bot) \land$
  $\hat{\Box}\neg(n_l \bullet 0 \uparrow \mathsf{startEpoch}_{\mathsf{ech}}(ts', n'_l))] \Rightarrow$
  $started(\mathsf{s}'(n_l)) = \mathsf{false} \land \hat{\Diamond}(prop(\mathsf{s}(n_l)) \ne \bot) \land$
  $\hat{\Box}\neg(n_l \bullet 0 \uparrow \mathsf{startEpoch}_{\mathsf{ech}}(ts', n'_l))]]$
By INVL,
  (4) $\Gamma' \vdash_{\mathsf{UCC}} started(\mathsf{s}(n_l)) = \mathsf{true} \land started(\mathsf{s}'(n_l)) = \mathsf{false} \Rightarrow$
  $(n_l \bullet 0 \uparrow \mathsf{startEpoch}_{\mathsf{ech}}(ts', n'_l))$
The contra-positive of [4],
  (5) $\Gamma' \vdash_{\mathsf{UCC}} \neg(n_l \bullet 0 \uparrow \mathsf{startEpoch}_{\mathsf{ech}}(ts', n'_l)) \Rightarrow$
  $started(\mathsf{s}(n_l)) = \mathsf{false} \lor started(\mathsf{s}'(n_l)) = \mathsf{true}$
That is,
  (6) $\Gamma' \vdash_{\mathsf{UCC}} \exists ts, n_l.\ n_l \in \mathsf{Correct} \land$
  $[(n_l \bullet 0 \uparrow \mathsf{startEpoch}_{\mathsf{ech}}(ts, n_l)) \land (prop(\mathsf{s}(n_l)) = \bot) \land \hat{\Diamond}(prop(\mathsf{s}(n_l)) \ne \bot) \land$
  $\hat{\Box}\neg(n_l \bullet 0 \uparrow \mathsf{startEpoch}_{\mathsf{ech}}(ts', n'_l))] \Rightarrow$
  $started(\mathsf{s}'(n_l)) = \mathsf{false} \land \hat{\Diamond}(prop(\mathsf{s}(n_l)) \ne \bot) \land$
  $\hat{\Box}(started(\mathsf{s}(n_l)) = \mathsf{false} \lor started(\mathsf{s}'(n_l)) = \mathsf{true})]]$
That is,
  (7) $\Gamma' \vdash_{\mathsf{UCC}} \exists ts, n_l.\ n_l \in \mathsf{Correct} \land$
  $[(n_l \bullet 0 \uparrow \mathsf{startEpoch}_{\mathsf{ech}}(ts, n_l)) \land (prop(\mathsf{s}(n_l)) = \bot) \land \hat{\Diamond}(prop(\mathsf{s}(n_l)) \ne \bot) \land$
  $\hat{\Box}\neg(n_l \bullet 0 \uparrow \mathsf{startEpoch}_{\mathsf{ech}}(ts', n'_l))] \Rightarrow$
  $\hat{\Diamond}[[[started(\mathsf{s}(n_l)) = \mathsf{false} \land (prop(\mathsf{s}(n_l)) \ne \bot)] \lor$
  $[started(\mathsf{s}(n_l)) = \mathsf{true} \land (prop(\mathsf{s}(n_l)) \ne \bot)]] \land$

$$\Box(started(\mathsf{s}(n_l)) = \mathsf{false} \lor started(\mathsf{s'}(n_l)) = \mathsf{true})]$$

By PE,

(8) $\Gamma' \vdash_{\mathsf{UCC}} n_l \in \mathsf{Correct} \land prop(\mathsf{s}(n_l)) \neq \bot \land started(\mathsf{s}(n_l)) = \mathsf{false} \Rightarrow$
$$(n_l \bullet (1, \mathsf{epoch}_{\mathsf{ec}}(n_l, ts)) \in \mathsf{ors} \land started(\mathsf{s'}(n_l)) = \mathsf{true} \land \mathsf{self})$$

By INVSA,

(9) $\Gamma' \vdash_{\mathsf{UCC}} started(\mathsf{s}(n_l)) = \mathsf{true}) \Rightarrow$
$$\hat{\diamond}(n_l \bullet (1, \mathsf{epoch}_{\mathsf{ec}}(n_l, ts)) \in \mathsf{ors} \land started(\mathsf{s'}(n_l)) = \mathsf{true} \land \mathsf{self})$$

From [7], [8] and [9],

(10) $\Gamma' \vdash_{\mathsf{UCC}} \exists ts, n_l.\ n_l \in \mathsf{Correct} \land$
$$[(n_l \bullet 0 \uparrow \mathsf{startEpoch}_{\mathsf{ech}}(ts, n_l)) \land (prop(\mathsf{s}(n_l)) = \bot) \land \hat{\diamond}(prop(\mathsf{s}(n_l)) \neq \bot) \land$$
$$\hat{\Box}\neg(n_l \bullet 0 \uparrow \mathsf{startEpoch}_{\mathsf{ech}}(ts', n_l'))] \Rightarrow$$
$$\hat{\diamond}[[[(n_l \bullet (1, \mathsf{epoch}_{\mathsf{ec}}(n_l, ts)) \in \mathsf{ors} \land started(\mathsf{s'}(n_l)) = \mathsf{true} \land \mathsf{self})] \lor$$
$$\hat{\diamond}[(n_l \bullet (1, \mathsf{epoch}_{\mathsf{ec}}(n_l, ts)) \in \mathsf{ors} \land started(\mathsf{s'}(n_l)) = \mathsf{true} \land \mathsf{self}]] \land$$
$$\Box(started(\mathsf{s}(n_l)) = \mathsf{false} \lor started(\mathsf{s'}(n_l)) = \mathsf{true})]$$

By using induction and POSTPRE on [10] (considering the conjunction for the base case of induction),

(11) $\Gamma' \vdash_{\mathsf{UCC}} \exists ts, n_l.\ n_l \in \mathsf{Correct} \land$
$$[(n_l \bullet 0 \uparrow \mathsf{startEpoch}_{\mathsf{ech}}(ts, n_l)) \land (prop(\mathsf{s}(n_l)) = \bot) \land \hat{\diamond}(prop(\mathsf{s}(n_l)) \neq \bot) \land$$
$$\hat{\Box}\neg(n_l \bullet 0 \uparrow \mathsf{startEpoch}_{\mathsf{ech}}(ts', n_l'))] \Rightarrow$$
$$\hat{\diamond}[[[(n_l \bullet (1, \mathsf{epoch}_{\mathsf{ec}}(n_l, ts)) \in \mathsf{ors} \land started(\mathsf{s'}(n_l)) = \mathsf{true} \land \mathsf{self})] \lor$$
$$\hat{\diamond}[(n_l \bullet (1, \mathsf{epoch}_{\mathsf{ec}}(n_l, ts)) \in \mathsf{ors} \land started(\mathsf{s'}(n_l)) = \mathsf{true} \land \mathsf{self}]] \land$$
$$\Box(started(\mathsf{s'}(n_l)) = \mathsf{true})]$$

By adding the premise again,

(12) $\Gamma' \vdash_{\mathsf{UCC}} \exists ts, n_l.\ n_l \in \mathsf{Correct} \land$
$$[(n_l \bullet 0 \uparrow \mathsf{startEpoch}_{\mathsf{ech}}(ts, n_l)) \land (prop(\mathsf{s}(n_l)) = \bot) \land \hat{\diamond}(prop(\mathsf{s}(n_l)) \neq \bot) \land$$
$$\hat{\Box}\neg(n_l \bullet 0 \uparrow \mathsf{startEpoch}_{\mathsf{ech}}(ts', n_l'))] \Rightarrow$$
$$\hat{\diamond}[[[(n_l \bullet (1, \mathsf{epoch}_{\mathsf{ec}}(n_l, ts)) \in \mathsf{ors} \land started(\mathsf{s'}(n_l)) = \mathsf{true} \land \mathsf{self})] \lor$$
$$\hat{\diamond}[(n_l \bullet (1, \mathsf{epoch}_{\mathsf{ec}}(n_l, ts)) \in \mathsf{ors} \land started(\mathsf{s'}(n_l)) = \mathsf{true} \land \mathsf{self}]] \land$$
$$\Box(started(\mathsf{s'}(n_l)) = \mathsf{true}) \land \Box\neg(n_l \bullet 0 \uparrow \mathsf{startEpoch}_{\mathsf{ech}}(ts, n_l))]$$

By INVL,

(13) $\Gamma' \vdash_{\mathsf{UCC}} (n_l \bullet (1, \mathsf{epoch}_{\mathsf{ec}}(n_l, ts)) \in \mathsf{ors} \land started(\mathsf{s}(n_l)) = \mathsf{true} \land \mathsf{self}) \Rightarrow$
$$(n_l \bullet 0 \uparrow \mathsf{startEpoch}_{\mathsf{ech}}(ts, n_l))$$

The contra-positive of [13],

(14) $\Gamma' \vdash_{\mathsf{UCC}} \neg(n_l \bullet 0 \uparrow \mathsf{startEpoch}_{\mathsf{ech}}(ts, n_l)) \Rightarrow$
$$\neg(n_l \bullet (1, \mathsf{epoch}_{\mathsf{ec}}(n_l, ts)) \in \mathsf{ors} \land \mathsf{self}) \lor (started(\mathsf{s}(n_l)) = \mathsf{false})$$

By using [12] and [14]

(15) $\Gamma' \vdash_{\mathsf{UCC}} \exists ts, n_l.\ n_l \in \mathsf{Correct} \land$
$$[(n_l \bullet 0 \uparrow \mathsf{startEpoch}_{\mathsf{ech}}(ts, n_l)) \land (prop(\mathsf{s}(n_l)) = \bot) \land \hat{\diamond}(prop(\mathsf{s}(n_l)) \neq \bot) \land$$
$$\hat{\Box}\neg(n_l \bullet 0 \uparrow \mathsf{startEpoch}_{\mathsf{ech}}(ts', n_l'))] \Rightarrow$$
$$\hat{\diamond}\hat{\diamond}[[[(n_l \bullet (1, \mathsf{epoch}_{\mathsf{ec}}(n_l, ts)) \in \mathsf{ors} \land started(\mathsf{s'}(n_l)) = \mathsf{true} \land \mathsf{self})] \land$$
$$\Box(started(\mathsf{s'}(n_l)) = \mathsf{true}) \land$$
$$\Box[\neg(n_l \bullet (1, \mathsf{epoch}_{\mathsf{ec}}(n_l, ts)) \in \mathsf{ors} \land \mathsf{self}) \lor (started(\mathsf{s}(n_l)) = \mathsf{false})]]$$

That is,

(16) $\Gamma' \vdash_{\mathsf{UCC}} \exists ts, n_l.\ n_l \in \mathsf{Correct} \land$
$$[(n_l \bullet 0 \uparrow \mathsf{startEpoch}_{\mathsf{ech}}(ts, n_l)) \land (prop(\mathsf{s}(n_l)) = \bot) \land \hat{\diamond}(prop(\mathsf{s}(n_l)) \neq \bot) \land$$
$$\hat{\Box}\neg(n_l \bullet 0 \uparrow \mathsf{startEpoch}_{\mathsf{ech}}(ts', n_l'))] \Rightarrow$$
$$\hat{\diamond}\hat{\diamond}[[[(n_l \bullet (1, \mathsf{epoch}_{\mathsf{ec}}(n_l, ts)) \in \mathsf{ors} \land started(\mathsf{s'}(n_l)) = \mathsf{true} \land \mathsf{self})] \land$$

$$\Box \neg (n_l \bullet (1, \mathsf{epoch}_{\mathsf{ec}}(n_l, ts)) \in \mathsf{ors} \wedge \mathsf{self})]$$

**Theorem 27** ($UC_2$: Validity).
If a node decides $v$, then $v$ was proposed by some node.
$\Gamma \vdash_{\mathsf{UC}} (n \bullet \top \uparrow \mathsf{decide}_{\mathsf{uc}}(v)) \leadsto \exists n'. (n \bullet \top \downarrow \mathsf{propose}_{\mathsf{uc}}(v))$
where $\Gamma$ is defined in Definition 22.

**Proof.**
By OI',
(1) $\Gamma' \vdash_{\mathsf{UCC}} (n \bullet \top \uparrow \mathsf{decide}_{\mathsf{uc}}(v)) \Rightarrow$
    $\hat{\diamondsuit}(n \bullet \mathsf{decide}_{\mathsf{uc}}(v) \in \mathsf{ois} \wedge \mathsf{self})$

By INVL,
(2) $\Gamma' \vdash_{\mathsf{UCC}} (n \bullet \mathsf{decide}_{\mathsf{uc}}(v) \in \mathsf{ois} \wedge \mathsf{self}) \Rightarrow$
    $n \bullet 1 \uparrow \mathsf{decide}_{\mathsf{ec}}(v)$

By $\mathsf{EC}'_1$
(3) $\Gamma' \vdash_{\mathsf{UCC}} (n \bullet \top \uparrow \mathsf{decide}_{\mathsf{ec}}(v)) \Rightarrow$
    $\exists n'. \diamondsuit(n' \bullet \top \downarrow \mathsf{propose}_{\mathsf{ec}}(v))$

By OR',
(4) $\Gamma' \vdash_{\mathsf{UCC}} (n' \bullet \top \downarrow \mathsf{propose}_{\mathsf{ec}}(v)) \Rightarrow$
    $\hat{\diamondsuit}(n' \bullet (1, \mathsf{propose}_{\mathsf{ec}}(v)) \in \mathsf{ors} \wedge \mathsf{self})$

By INVL,
(5) $\Gamma' \vdash_{\mathsf{UCC}} (n' \bullet (1, \mathsf{propose}_{\mathsf{ec}}(v)) \in \mathsf{ors} \wedge \mathsf{self}) \Rightarrow$
    $(n' \bullet 1 \downarrow \mathsf{propose}_{\mathsf{uc}}(v))$

From [1] to [5] and Lemma 86,
(6) $\Gamma' \vdash_{\mathsf{UCC}} (n \bullet \top \uparrow \mathsf{decide}_{\mathsf{uc}}(v)) \Rightarrow$
    $\exists n'. \diamondsuit(n' \bullet \top \downarrow \mathsf{propose}_{\mathsf{uc}}(v))$

That is,
(7) $\Gamma' \vdash_{\mathsf{UCC}} (n \bullet \top \uparrow \mathsf{decide}_{\mathsf{uc}}(v)) \leadsto$
    $\exists n'. (n' \bullet \top \downarrow \mathsf{propose}_{\mathsf{uc}}(v))$

**Theorem 28** ($UC_3$: Integrity).
No node decides twice.
$\Gamma \vdash_{\mathsf{UC}} (n \bullet \top \uparrow \mathsf{decide}_{\mathsf{uc}}(v)) \Rightarrow \hat{\Box} \neg (n \bullet \top \uparrow \mathsf{decide}_{\mathsf{uc}}(v'))$
where $\Gamma$ is defined in Definition 22.

**Proof.**
By OI',
   (1)   $\Gamma' \vdash_{\mathsf{UCC}} (n \bullet \top \uparrow \mathsf{decide}_{\mathsf{uc}}(v)) \Rightarrow$
                $\hat{\diamondsuit}(n \bullet \mathsf{decide}_{\mathsf{uc}}(v) \in \mathsf{ois} \wedge \mathsf{self})$
By InvL,
   (2)   $\Gamma' \vdash_{\mathsf{UCC}} (n \bullet \mathsf{decide}_{\mathsf{uc}}(v) \in \mathsf{ois} \wedge \mathsf{self}) \Rightarrow$
                $(n \bullet 1 \uparrow \mathsf{decide}_{\mathsf{ec}}(v)) \wedge \mathsf{occ}(\mathsf{ois}, \mathsf{decide}_{\mathsf{uc}}(v)) \leq 1$
From [2],
   (3)   $\Gamma' \vdash_{\mathsf{UCC}} \neg (n \bullet 1 \uparrow \mathsf{decide}_{\mathsf{ec}}(v)) \Rightarrow$
                $(\mathsf{n} = n \wedge \mathsf{self} \rightarrow \neg (\mathsf{decide}_{\mathsf{uc}}(v) \in \mathsf{ois}))$
From $\mathsf{EC}'_3$, we have
   (4)   $\Gamma' \vdash_{\mathsf{UCC}} (n \bullet 1 \uparrow \mathsf{decide}_{\mathsf{ec}}(v)) \Rightarrow$
                $\hat{\Box} \neg (n \bullet 1 \uparrow \mathsf{decide}_{\mathsf{ec}}(v'))$
By Lemma 108 on [4], we have
   (5)   $\Gamma' \vdash_{\mathsf{UCC}} (n \bullet 1 \uparrow \mathsf{decide}_{\mathsf{ec}}(v)) \Rightarrow$
                $\hat{\Box} \neg (n \bullet 1 \uparrow \mathsf{decide}_{\mathsf{ec}}(v')) \wedge$
                $\hat{\boxminus} \neg (n \bullet 1 \uparrow \mathsf{decide}_{\mathsf{ec}}(v'))$
From [1]-[5], we have
   (6)   $\Gamma' \vdash_{\mathsf{UCC}} (n \bullet \top \uparrow \mathsf{decide}_{\mathsf{uc}}(v)) \Rightarrow \hat{\diamondsuit}[$
                $\mathsf{occ}(\mathsf{ois}, \mathsf{decide}_{\mathsf{uc}}(v)) \leq 1 \wedge$
                $\hat{\Box}(\mathsf{n} = n \wedge \mathsf{self} \rightarrow \neg (\mathsf{decide}_{\mathsf{uc}}(v) \in \mathsf{ois})) \wedge$
                $\hat{\boxminus}(\mathsf{n} = n \wedge \mathsf{self} \rightarrow \neg (\mathsf{decide}_{\mathsf{uc}}(v) \in \mathsf{ois}))]$
From UniOR on [6]
   (7)   $\Gamma' \vdash_{\mathsf{UCC}} (n \bullet \top \uparrow \mathsf{decide}_{\mathsf{uc}}(v)) \Rightarrow \hat{\diamondsuit}[$
                $(n \bullet \top \uparrow \mathsf{decide}_{\mathsf{uc}}(v)) \Rightarrow$
                    $\hat{\Box} \neg (n \bullet \top \uparrow \mathsf{decide}_{\mathsf{uc}}(v)) \wedge$
                    $\hat{\boxminus} \neg (n \bullet \top \uparrow \mathsf{decide}_{\mathsf{uc}}(v))]$
By Lemma 109 on [7], we have
   (8)   $\Gamma' \vdash_{\mathsf{UCC}} (n \bullet \top \uparrow \mathsf{decide}_{\mathsf{uc}}(v)) \Rightarrow$
                $\hat{\Box} \neg (n \bullet \top \uparrow \mathsf{decide}_{\mathsf{uc}}(v))$

**Theorem 29** ($UC_4$: Uniform Agreement).
No two nodes decide differently.
$\Gamma \vdash_{\text{UC}} (n \bullet \top \uparrow \text{decide}_{\text{uc}}(v)) \wedge \Diamond(n' \bullet \top \uparrow \text{decide}_{\text{uc}}(v')) \Rightarrow$
$\quad (v = v')$
where $\Gamma$ is defined in Definition 22.

**Proof.**

By OI$'$,
(1) $\Gamma' \vdash_{\text{UCC}} (n \bullet \top \uparrow \text{decide}_{\text{uc}}(v)) \Rightarrow$
$\quad\quad \hat{\Diamond}(n \bullet \text{decide}_{\text{uc}}(v) \in \text{ois} \wedge \text{self})$
By INVL,
(2) $\Gamma' \vdash_{\text{UCC}} (n \bullet \text{decide}_{\text{uc}}(v) \in \text{ois} \wedge \text{self}) \Rightarrow$
$\quad\quad n \bullet 1 \uparrow \text{decide}_{\text{ec}}(v)$
From [1] and [2],
(3) $\Gamma' \vdash_{\text{UCC}} (n \bullet \top \uparrow \text{decide}_{\text{uc}}(v)) \Rightarrow$
$\quad\quad \hat{\Diamond}(n \bullet 1 \uparrow \text{decide}_{\text{ec}}(v))$
From [3] and instantiating $v$ to $v'$ and $n$ to $n'$,
(4) $\Gamma' \vdash_{\text{UCC}} (n \bullet \top \uparrow \text{decide}_{\text{uc}}(v)) \wedge \Diamond(n' \bullet \top \uparrow \text{decide}_{\text{uc}}(v')) \Rightarrow$
$\quad\quad \hat{\Diamond}(n \bullet 1 \uparrow \text{decide}_{\text{ec}}(v)) \wedge \Diamond(\hat{\Diamond}(n' \bullet 1 \uparrow \text{decide}_{\text{ec}}(v')))$
By Lemma 112 on [4],
(5) $\Gamma' \vdash_{\text{UCC}} (n \bullet \top \uparrow \text{decide}_{\text{uc}}(v)) \wedge \Diamond(n' \bullet \top \uparrow \text{decide}_{\text{uc}}(v')) \Rightarrow$
$\quad\quad \hat{\Diamond}(n \bullet 1 \uparrow \text{decide}_{\text{ec}}(v)) \wedge$
$\quad\quad\quad [\Diamond(n' \bullet 1 \uparrow \text{decide}_{\text{ec}}(v')) \vee \Diamond(n' \bullet 1 \uparrow \text{decide}_{\text{ec}}(v'))]$
That is,
(6) $\Gamma' \vdash_{\text{UCC}} (n \bullet \top \uparrow \text{decide}_{\text{uc}}(v)) \wedge \Diamond(n' \bullet \top \uparrow \text{decide}_{\text{uc}}(v')) \Rightarrow$
$\quad\quad [\hat{\Diamond}(n \bullet 1 \uparrow \text{decide}_{\text{ec}}(v)) \wedge \Diamond(n' \bullet 1 \uparrow \text{decide}_{\text{ec}}(v'))] \vee$
$\quad\quad [\hat{\Diamond}(n \bullet 1 \uparrow \text{decide}_{\text{ec}}(v)) \wedge \Diamond(n' \bullet 1 \uparrow \text{decide}_{\text{ec}}(v'))]$
By Lemma 105 on [6],
(7) $\Gamma' \vdash_{\text{UCC}} (n \bullet \top \uparrow \text{decide}_{\text{uc}}(v)) \wedge \Diamond(n' \bullet \top \uparrow \text{decide}_{\text{uc}}(v')) \Rightarrow$
$\quad\quad\quad \Diamond[(n \bullet 1 \uparrow \text{decide}_{\text{ec}}(v)) \wedge \Diamond(n' \bullet 1 \uparrow \text{decide}_{\text{ec}}(v'))] \vee$
$\quad\quad\quad \Diamond[\Diamond(n \bullet 1 \uparrow \text{decide}_{\text{ec}}(v)) \wedge (n' \bullet 1 \uparrow \text{decide}_{\text{ec}}(v'))] \vee$
$\quad\quad\quad \Diamond[(n \bullet 1 \uparrow \text{decide}_{\text{ec}}(v)) \wedge \Diamond(n' \bullet 1 \uparrow \text{decide}_{\text{ec}}(v'))]$
By Lemma 119 on [7],
(8) $\Gamma' \vdash_{\text{UCC}} (n \bullet \top \uparrow \text{decide}_{\text{uc}}(v)) \wedge \Diamond(n' \bullet \top \uparrow \text{decide}_{\text{uc}}(v')) \Rightarrow$
$\quad\quad\quad \Diamond[\Diamond(n \bullet 1 \uparrow \text{decide}_{\text{ec}}(v)) \wedge (n' \bullet 1 \uparrow \text{decide}_{\text{ec}}(v'))] \vee$
$\quad\quad\quad \Diamond[(n \bullet 1 \uparrow \text{decide}_{\text{ec}}(v)) \wedge \Diamond(n' \bullet 1 \uparrow \text{decide}_{\text{ec}}(v'))] \vee$
$\quad\quad\quad \Diamond[(n \bullet 1 \uparrow \text{decide}_{\text{ec}}(v)) \wedge \Diamond(n' \bullet 1 \uparrow \text{decide}_{\text{ec}}(v'))]$
By $\text{EC}'_2$ and [8],
(9) $\Gamma' \vdash_{\text{UCC}} (n \bullet \top \uparrow \text{decide}_{\text{uc}}(v)) \wedge \Diamond(n' \bullet \top \uparrow \text{decide}_{\text{uc}}(v')) \Rightarrow$
$\quad\quad \Diamond[v = v']$
That is,
(10) $\Gamma' \vdash_{\text{UCC}} (n \bullet \top \uparrow \text{decide}_{\text{uc}}(v)) \wedge \Diamond(n' \bullet \top \uparrow \text{decide}_{\text{uc}}(v')) \Rightarrow (v = v')$

## 5.4 Temporal Logic

The following subsections show axioms, rules and lemmas from the basic temporal logic of Manna and Pnueli [1] that we use.

### 5.4.1 Axioms and Inference Rules

**Future Axioms.**

**Axiom 1** (FX0).
$\Box p \to p$

**Axiom 2** (FX1).
$\bigcirc \neg p \Leftrightarrow \neg \bigcirc p$

**Axiom 3** (FX2).
$\bigcirc (p \to q) \Leftrightarrow (\bigcirc p \to \bigcirc q)$

**Axiom 4** (FX3).
$\Box (p \to q) \Rightarrow (\Box p \to \Box q)$

**Axiom 5** (FX4).
$\Box p \to \Box \bigcirc p$

**Axiom 6** (FX5).
$(p \Rightarrow \bigcirc p) \to (p \Rightarrow \Box p)$

**Axiom 7** (FX6).
$p \; \omega \; q \Leftrightarrow [q \vee (p \wedge \bigcirc (p \; \omega \; q))]$

**Axiom 8** (FX7).
$\Box p \Rightarrow p \; \omega \; q$

**Past Axioms.**

**Axiom 9** (PX1).
$\ominus p \Rightarrow \tilde{\ominus} p$

**Axiom 10** (PX2).
$\tilde{\ominus}(p \to q) \Leftrightarrow (\tilde{\ominus} p \to \tilde{\ominus} q)$

**Axiom 11** (PX3).
$\boxminus (p \to q) \Rightarrow (\boxminus p \to \boxminus q)$

**Axiom 12** (PX4).
$\boxminus p \to \boxminus \tilde{\ominus} p$

**Axiom 13** (PX5).
$(p \Rightarrow \tilde{\ominus} p) \to (p \Rightarrow \boxminus p)$

**Axiom 14** (PX6).
$p \; \beta \; q \Leftrightarrow (q \vee [p \wedge \tilde{\ominus}(p \; \beta \; q)])$

**Axiom 15** (PX7).
$\tilde{\ominus} F$

**Mixed Axioms.**

**Axiom 16** (FX8).
$p \Rightarrow \bigcirc \ominus p$

**Axiom 17** (PX8).
$p \Rightarrow \tilde{\ominus} \bigcirc p$

**Rule 1** (Generalization).
$$\text{Gen} \quad \frac{p}{\vdash \Box p}$$

**Axiom 18** ($\forall \bigcirc$-Comm (Universal commutation)).

$\forall x. \; \bigcirc p(x) \Leftrightarrow \bigcirc \forall x. \; p(x)$

### 5.4.2 Derived Rules and Lemmas

**Lemma 76** (CM1).
$\forall x : \ominus p(x) \Leftrightarrow \ominus \forall x : p(x)$

**Lemma 77** (CM2).
$\exists x : \ominus p(x) \Leftrightarrow \ominus \exists x : p(x)$

**Lemma 78** (CM3).
$\exists x : \bigcirc p(x) \Leftrightarrow \bigcirc \exists x : p(x)$

**Lemma 79** (Rule E-MP (for n = 1)).
$p \Rightarrow q, \Box p \vdash \Box q$

**Lemma 80** ($\Rightarrow$T).
$p \Rightarrow q, q \Rightarrow r \vdash p \Rightarrow r$

**Lemma 81** ($\bigcirc$ M).
$$p \Rightarrow q \vdash \bigcirc p \Rightarrow \bigcirc q$$

$$p \Leftrightarrow q \vdash \bigcirc p \Leftrightarrow \bigcirc q$$

**Lemma 82** (CI).
$$p \Rightarrow \bigcirc p \vdash p \Rightarrow \Box p$$

**Lemma 83** (T4).
$$\Box p \Rightarrow p$$

**Lemma 84** (T9).
$$\Box(p \wedge q) \Leftrightarrow \Box p \wedge \Box q$$

**Lemma 85** (T25).
$$\boxminus(p \wedge q) \Leftrightarrow (\boxminus p \wedge \boxminus q)$$
$$\hat{\boxminus}(p \wedge q) \Leftrightarrow (\hat{\boxminus} p \wedge \hat{\boxminus} q)$$

**Lemma 86** (T27).
$$\Diamondleft p \Leftrightarrow \Diamondleft \Diamondleft p$$
$$\hat{\Diamondleft} \hat{\Diamondleft} p \Rightarrow \hat{\Diamondleft} p$$
$$\Diamondleft \hat{\Diamondleft} p \Rightarrow \hat{\Diamondleft} p$$
$$\hat{\Diamondleft} \Diamondleft p \Rightarrow \hat{\Diamondleft} p$$

**Lemma 87** (T11).
$$\Diamond p \Leftrightarrow \Diamond \Diamond p$$

**Lemma 88** ($\Diamondleft$ T).
$$p \Rightarrow \Diamondleft q, q \Rightarrow \Diamondleft r \vdash p \Rightarrow \Diamondleft r$$

**Lemma 89** ($\Diamond$ T).
$$p \Rightarrow \Diamond q, q \Rightarrow \Diamond r \vdash p \Rightarrow \Diamond r$$

**Lemma 90** (T26).
$$p \Rightarrow \Diamondleft p$$

**Lemma 91.**
$$\neg \Box p \Leftrightarrow \Diamond \neg p$$
$$\neg \Diamond p \Leftrightarrow \Box \neg p$$

**Lemma 92** (T6).
$$\Box p \to \bigcirc p$$

**Lemma 93** (CN1).
$$\bigcirc (\neg p) \Leftrightarrow \neg \bigcirc p$$

**Lemma 94** (FP8).
$$p \to \tilde{\ominus} \bigcirc p$$

**Lemma 95** ($\Diamond$M).
$$p \Rightarrow q \vdash \Diamond p \Rightarrow \Diamond q$$

**Lemma 96** ($\Diamondleft$M).
$$p \Rightarrow q \vdash \Diamondleft p \Rightarrow \Diamondleft q$$

### 5.4.3 Extra Temporal Logic Lemmas

**Lemma 97.**
$$((p \Rightarrow q) \land p) \rightarrow q$$

**Proof.**
By Axiom 1.

**Lemma 98.**
$$\diamondsuit\!\!\!\!\cdot\, p \rightarrow \bigcirc \diamondsuit\!\!\!\!\cdot\, p$$

**Proof.**
We assume
 (1) $\diamondsuit\!\!\!\!\cdot\, p$
We prove
 $\bigcirc \diamondsuit\!\!\!\!\cdot\, p$
From rule T28 and [1],
 $\bigcirc \diamondsuit\!\!\!\!\cdot\, p$

**Lemma 99.**
$$p \Rightarrow \diamondsuit q \;\land\; q \Rightarrow r \;\rightarrow$$
$$p \Rightarrow \diamondsuit r$$

$$p \Rightarrow \diamondsuit\!\!\!\!\cdot\, q \;\land\; q \Rightarrow r \;\rightarrow$$
$$p \Rightarrow \diamondsuit\!\!\!\!\cdot\, r$$

$$p \Rightarrow \hat{\diamondsuit} q \;\land\; q \Rightarrow r \;\rightarrow$$
$$p \Rightarrow \hat{\diamondsuit} r$$

**Proof.**
Immediate from Rule ◇M ($\diamondsuit\!\!\!\!\cdot\,$M) and Rule ⇒T.

**Lemma 100.**
$$p \Rightarrow \Box q \;\land\; q \Rightarrow r \;\rightarrow$$
$$p \Rightarrow \Box r$$

**Proof.**
Immediate from Rule □M and Rule ⇒T.

**Lemma 101.**
$$\diamondsuit\!\!\!\!\cdot\, \Box p \Rightarrow p$$
$$\hat{\diamondsuit} \hat{\Box} p \Rightarrow p$$
$$\hat{\diamondsuit} \Box p \Rightarrow p$$

**Lemma 102.**
$$\diamondsuit\!\!\!\!\cdot\, \hat{\Box} p \Rightarrow \hat{\Box} p$$
$$\hat{\diamondsuit} \Box p \Rightarrow \Box p$$
$$\diamondsuit\!\!\!\!\cdot\, \Box p \Rightarrow \Box p$$

**Lemma 103.**
$$p \rightarrow (\hat{\boxminus}(p \Rightarrow \bigcirc p) \Rightarrow p)$$

**Lemma 104.**
$$p \Rightarrow q \Rightarrow r$$
$$\leftrightarrow$$
$$(\diamondsuit\!\!\!\!\cdot\, p \land q) \Rightarrow r$$
$$\leftrightarrow$$
$$(p \land \diamondsuit q) \Rightarrow \diamondsuit r$$

**Lemma 105.**
$$(\diamondsuit p \land \diamondsuit q) \Rightarrow$$
$$\quad \diamondsuit(p \land \diamondsuit q) \lor$$
$$\quad \diamondsuit(q \land \diamondsuit p)$$

$$(\diamondsuit\!\!\!\!\cdot\, p \land \diamondsuit\!\!\!\!\cdot\, q) \Rightarrow$$
$$\quad \diamondsuit\!\!\!\!\cdot\,(p \land \diamondsuit\!\!\!\!\cdot\, q) \lor$$
$$\quad \diamondsuit\!\!\!\!\cdot\,(q \land \diamondsuit\!\!\!\!\cdot\, p)$$

$$\diamondsuit\!\!\!\!\cdot\, p \land \diamondsuit q \Rightarrow$$
$$\quad \diamondsuit\!\!\!\!\cdot\,(p \land \diamondsuit q)$$

**Lemma 106.**
$$(\diamondsuit p \land \hat{\Box} \neg p) \Rightarrow p$$
$$(\diamondsuit\!\!\!\!\cdot\, p \land \hat{\boxminus} \neg p) \Rightarrow p$$

**Lemma 107.**
$$\diamondsuit\!\!\!\!\cdot\, p \Rightarrow \Box \diamondsuit\!\!\!\!\cdot\, p$$

**Lemma 108.**
  $(p \Rightarrow \hat{\Box}\neg p) \rightarrow$
  $(p \Rightarrow \hat{\boxminus}\neg p)$
  *and*
  $(p \Rightarrow \hat{\boxminus}\neg p) \rightarrow$
  $(p \Rightarrow \hat{\Box}\neg p)$

**Proof.**
We assume
  (2)  $(p \Rightarrow \hat{\Box}\neg p)$
We prove
  $(p \Rightarrow \hat{\boxminus}\neg p)$
That is,
  $\Diamondminus p \Rightarrow \neg p$

From [2]
  (3)  $(\hat{\Diamondminus} p \Rightarrow \Diamondminus \hat{\Box}\neg p)$
By Lemma 101,
  (4)  $\hat{\Diamondminus}\hat{\Box}\neg p \Rightarrow \neg p$
From [3] and [4],
  (5)  $(\hat{\Diamondminus} p \Rightarrow \neg p)$

**Lemma 109.**
  $p \Rightarrow \Diamondminus(p \Rightarrow q) \rightarrow$
  $(p \Rightarrow q)$

**Lemma 110.**
  $\hat{\boxminus} p \Rightarrow p \rightarrow$
  $\Box p$

**Lemma 111.**
  $\hat{\boxminus} p \wedge \hat{\boxminus} q \Rightarrow p \wedge q \rightarrow$
  $\Box(p \wedge q)$

**Proof.**
Immediate from Lemma 110 and Lemma 93.

**Lemma 112.**
  $\Diamondminus \Diamond p \Leftrightarrow \Diamond p \vee \Diamondminus p$
  $\hat{\Diamondminus} \Diamond p \Rightarrow \Diamond p \vee \Diamondminus p$
  $\Diamond \Diamondminus p \Rightarrow \Diamond p \vee \Diamondminus p$
  $\Diamond \hat{\Diamondminus} p \Rightarrow \Diamond p \vee \Diamondminus p$
  $\hat{\Diamondminus} \hat{\Diamondminus} p \Rightarrow \Diamond p \vee \Diamondminus p$

**Lemma 113.**
  $\Diamondminus p \vee \Diamond p \Leftrightarrow \hat{\Diamondminus} p \vee \Diamond p$

**Lemma 114.**
  $\Diamondminus p \Leftrightarrow \bigcirc \hat{\Diamondminus} p$

**Lemma 115.**
  $\Diamond p \wedge \Box q \Rightarrow \Diamond(p \wedge \Box q)$

**Lemma 116.**
  $\hat{\Diamond} \Diamond p \Rightarrow \Diamond p$

**Lemma 117.**
  $\neg \hat{\Diamondminus} p \Rightarrow \hat{\boxminus}\neg p$

**Lemma 118.**
  $\hat{\boxminus} p \Leftrightarrow \Box \tilde{\ominus} p$

**Lemma 119.**
  $p \wedge \hat{\Diamondminus} q \Rightarrow$
  $\hat{\Diamondminus}(q \wedge \Diamond p)$

**Lemma 120.**
$$\boxminus p \wedge \Box p \Rightarrow$$
$$\boxminus \hat{\boxminus} p \wedge \Box \hat{\boxminus} p$$

**Lemma 121.**
$$\hat{\Diamondminus} \circ p \Leftrightarrow \Diamondminus p$$

**Lemma 122.**
$$p \wedge \Diamond \hat{\Box} \neg p \Rightarrow$$
$$\Diamond (p \wedge \hat{\Box} \neg p)$$

**Lemma 123.**
$$\Diamond p \wedge \hat{\Diamond} \Box q \Rightarrow$$
$$\Diamond (p \wedge \hat{\Diamond} \Box q) \vee$$
$$\Diamond (p \wedge \Box q)$$

**Lemma 124.**
$$\Diamond p \wedge \hat{\Diamond} \Box q \Rightarrow$$
$$\Diamond (\Diamond p \wedge \hat{\Diamond} \Box q) \vee$$
$$\Diamond (p \wedge \Box q)$$
$$\rightarrow$$
$$\Diamond p \wedge \hat{\Diamond} \Box q \Rightarrow \Diamond (p \wedge \Box q)$$

**Lemma 125.**
$$p \Rightarrow \Box p$$
$$\rightarrow$$
$$\neg p \Rightarrow \hat{\boxminus} \neg p$$

**Lemma 126.**
$$\hat{\Diamondminus} \Diamond \hat{\Box} p \rightarrow \Diamond \hat{\Box} p$$

**Lemma 127.**
$$\Diamond p \wedge \Diamond q \Rightarrow \Diamond (\hat{\Diamondminus} \Diamond p \wedge q)$$

# References

[1] Zohar Manna and Amir Pnueli. *The Temporal Logic of Reactive and Concurrent Systems*. Springer-Verlag New York, Inc., New York, NY, USA, 1992.